\documentclass{tADP2e}

\usepackage{subfigure}
\usepackage{color} 
\usepackage{comment}

\newcommand{\be}{\begin{equation}}
\newcommand{\ee}{\end{equation}}
\newcommand{\beq}{\begin{eqnarray}}
\newcommand{\eeq}{\end{eqnarray}}
\newcommand{\bem}{\begin{multline}}
\newcommand{\eem}{\end{multline}}

\theoremstyle{plain}

\theoremstyle{definition}

\begin{document}
\theoremstyle{plain}


\articletype{REVIEW}

\title{Universal High-Frequency Behavior of Periodically Driven Systems: from Dynamical Stabilization to Floquet Engineering}

\author{Marin Bukov$^{\rm a}$$^{\ast}$\thanks{$^\ast$Corresponding author. Email: mbukov@bu.edu
\vspace{6pt}}, Luca D'Alessio $^{\rm a,b}$, and Anatoli Polkovnikov$^{\rm a}$ \\\vspace{6pt}   $^{a}${\em{Department of Physics, Boston University, 590 Commonwealth Ave., Boston, MA 02215, USA}}, $^{b}${\em{Department of Physics, The Pennsylvania State University, University Park, PA 16802, USA}}; 
}

\maketitle

\begin{abstract}
We give a general overview of the high-frequency regime in periodically driven systems and identify three distinct classes of driving protocols in which the infinite-frequency Floquet Hamiltonian is not equal to the time-averaged Hamiltonian. These classes cover systems, such as the Kapitza pendulum, the Harper-Hofstadter model of neutral atoms in a magnetic field, the Haldane Floquet Chern insulator and others. In all setups considered, we discuss both the infinite-frequency limit and the leading finite-frequency corrections to the Floquet Hamiltonian. We provide a short overview of Floquet theory focusing on the gauge structure associated with the choice of stroboscopic frame and the differences between stroboscopic and non-stroboscopic dynamics. In the latter case one has to work with dressed operators representing observables and a dressed density matrix. We also comment on the application of Floquet Theory to systems described by static Hamiltonians with well-separated energy scales and, in particular, discuss parallels between the inverse-frequency expansion and the Schrieffer-Wolff transformation extending the latter to driven systems. 

\begin{classcode}
        05.45.-a Nonlinear dynamics and chaos, 67.85.-d	Ultracold gases, trapped gases, 67.85.Hj	Bose-Einstein condensates in optical potentials, 71.10.-w	Theories and models of many-electron systems
		\end{classcode}

\begin{keywords}
	Floquet theory, effective Hamiltonian, Magnus expansion, high-frequency limit, quantum simulation, dynamical stabilisation and localisation, artificial gauge fields, topological insulators, spin systems.
\end{keywords}

\centerline{\bfseries Index to information contained in this review}\vspace{12pt}

\hbox to \textwidth{\hsize\textwidth\vbox{\hsize18pc
\hspace*{-12pt} {1.}    Introduction.\\
{2.}  Floquet Theory. Stroboscopic and Non-Stroboscopic Time Evolution.\\
\hspace*{10pt}{2.1.}  The Stroboscopic Floquet Hamiltonian and the P-Operator.\\
\hspace*{10pt}{2.2.}  The Non-Stroboscopic Floquet Hamiltonian and the Kick Operators.\\
\hspace*{10pt}{2.3.}  A Two-Level System in a Circularly Driven Magnetic Field \\
\hspace*{10pt}{2.4.}  Stroboscopic versus Non-Stroboscopic Dynamics.\\
\hspace*{10pt}{2.4.1.}  Stroboscopic and Non-Stroboscopic Dynamics for an Adibataic Ramping of the Drive.\\
\hspace*{10pt}{2.4.2.}  Non-Stroboscopic Evolution in the Two-Level-System.\\
{3.}    Inverse Frequency Expansions for the Floquet Hamiltonian. \\
\hspace*{10pt}{3.1.}  The Magnus Expansion for the Stroboscopic Floquet Hamiltonian. \\
\hspace*{10pt}{3.2.}  The High Frequency Expansion for the Effective Floquet Hamiltonian. \\
\hspace*{10pt}{3.3.}  Magnus vs. High-Frequency Expansions: the Two-Level System in a Circularly Driven Magnetic Field Revisited. \\
\hspace*{10pt}{3.4.}  The Inverse Frequency Expansion in the Rotating Frame. \\
\hspace*{10pt}{3.5.}  Convergence of the Magnus Expansion.\\
{4.}    The Rotating Wave Approximation and the Schrieffer-Wolff Transformation.\\
\hspace*{10pt}{4.1.} A Two-level System.\\
\hspace*{10pt}{4.2.} The High-Frequency Expansion vs. the Schrieffer-Wolff Transformation. \\
\hspace*{10pt}{4.3.}  The Rabi Model.\\
{5.}    The Kapitza Class \\ 
\hspace*{10pt}{5.1.}  The Kapitza Pendulum.\\
\hspace*{10pt}{5.2.}   The Kapitza Hamiltonian in the Rotating Frame. \\ }
\hspace{-24pt}\vbox{\noindent\hsize18pc  
\hspace*{10pt}{5.3.}   Finite-Frequency Corrections. \\
\hspace*{10pt}{5.4.}   Dressed Observables and Dressed Density Matrix.\\
\hspace*{10pt}{5.5.}   Multi-Dimensional and Multi-Particle Generalization of the Kapitza Pendulum. \\
{6.}    The Dirac Class. \\
\hspace*{10pt}{6.1.}  Periodically Driven Magnetic Fields.\\ 
\hspace*{10pt}{6.2.}  Periodically Driven External Potentials.\\ 
{7.}    The Dunlap-Kenkre (DK) Class.\\
\hspace*{10pt}{7.1.}   Noninteracting Particles in a Periodically Driven Potential: Floquet Theory and Experimental Realization. \\
\hspace*{10pt}{7.2.}   Cold Atoms Realization of the Harper-Hofstadter Hamiltonian. \\
\hspace*{10pt}{7.3.}   The Periodically Driven Fermi-Hubbard Model. Floquet Topological Insulators. \\
\hspace*{10pt}{7.4.}   Periodically Driven Spin Systems.\\
{8.}    Summary and Outlook.\\
{Appendix:}    \\
{A.}    Outline of the Derivation of the Inverse-Frequency Expansions.\\
{A.1.} The Magnus Expansion.\\
{A.2.} The High-Frequency Expansion.\\
{B.}    Lattice vs.~Continuum Models.\\
{C.}    Corrections to the Stroboscopic Floquet Hamiltonian $H_F[0]$.\\
{C.1.}  First-order Coefficients for the 1D Driven Boson Model.\\
{C.2.}  First-order Coefficients for the Harper-Hofstadter Model.\\
{D.}    Corrections to the Effective Floquet Hamiltonian $H_\text{eff}$.\\
{D.1.}  First-order Coefficients for the 1D Driven Boson Model.\\
{D.2.}  First-order Coefficients for the Harper-Hofstadter Model.\\
      }}
\end{abstract}

\section{\label{sec:intro}Introduction} 

Periodically driven systems have a long history, one paradigmatic example
being the kicked-Rotor model of a particle moving on a ring subject to time-periodic `kicks' 
\cite{casati_79}, which realizes the famous Chirikov standard map~\cite{chirikov_71} and the Kapitza pendulum~\cite{kapitza_51}. The behavior of such systems is very rich - they can display interesting
integrability-to-chaos transitions, as well as counter-intuitive effects, such as dynamical localization~\cite{chirikov_81,rahav_03_pra,rahav_03,reichl_04,grossmann_91,grossmann_92,gavrila_02,grifoni_98,bavli_92,neu_96} and dynamical stabilization~\cite{kapitza_51,broer_04, landau_lifshitz_1}. The latter manifests itself in reduced ionisation rates in atomic systems irradiated by electromagnetic fields in the regime of high frequencies and high intensities~\cite{su_90,bialynicki-birula_94,eberly_93,piraux_98,pont_90,zakrzewski_95}, or as diminished spreading of wave packets in systems subject to periodic driving~\cite{holthaus_95,buchleitner_02}. The consequences of an AC-drive for quantum phase transitions have been investigated in a variety of models~\cite{bastidas_Dicke_12,bastidas_Ising_12, engelhardt_13, bastidas_14}, such as the Dicke model~\cite{bastidas_Dicke_12} and the Ising model~\cite{bastidas_Ising_12}. The modification of transport properties in periodically driven systems has been the subject of multiple studies, too~\cite{denisov_07,auerbach_07,poletti_09,salger_13,denisov_14,grossert_14}. 

In the recent years, it has been shown that periodic perturbations can be used as a flexible experimental knob to realize new phases not accessible in equilibrium systems~\cite{ovadyahu_12,iwai_03,kaiser_14,goldstein_14}, synthetic (engineered) matter~\cite{hafezi_07,zenesini_09, oka_09,kitagawa_11,creffield_11,hauke_12,iadecola_13,lindner_11,wang_13,aidelsburger_13, miyake_13,jotzu_14,aidelsburger_14,baur_14,struck_11,goldman_res_14,eckardt_15}, and quantum motors, which are similar to a quantum ratchet~\cite{ponomarev_09,ponomarev_10}.
This new line of research, which can be termed `Floquet engineering', has motivated a renaissance of interest in periodically driven systems. Floquet systems also naturally appear in digital quantum computation schemes, where one implements a continuous unitary evolution by effectively `trotterizing' it~\cite{blatt_12, zoller_private}.

\begin{figure}
	\centering
	\includegraphics[width = 0.7\columnwidth]{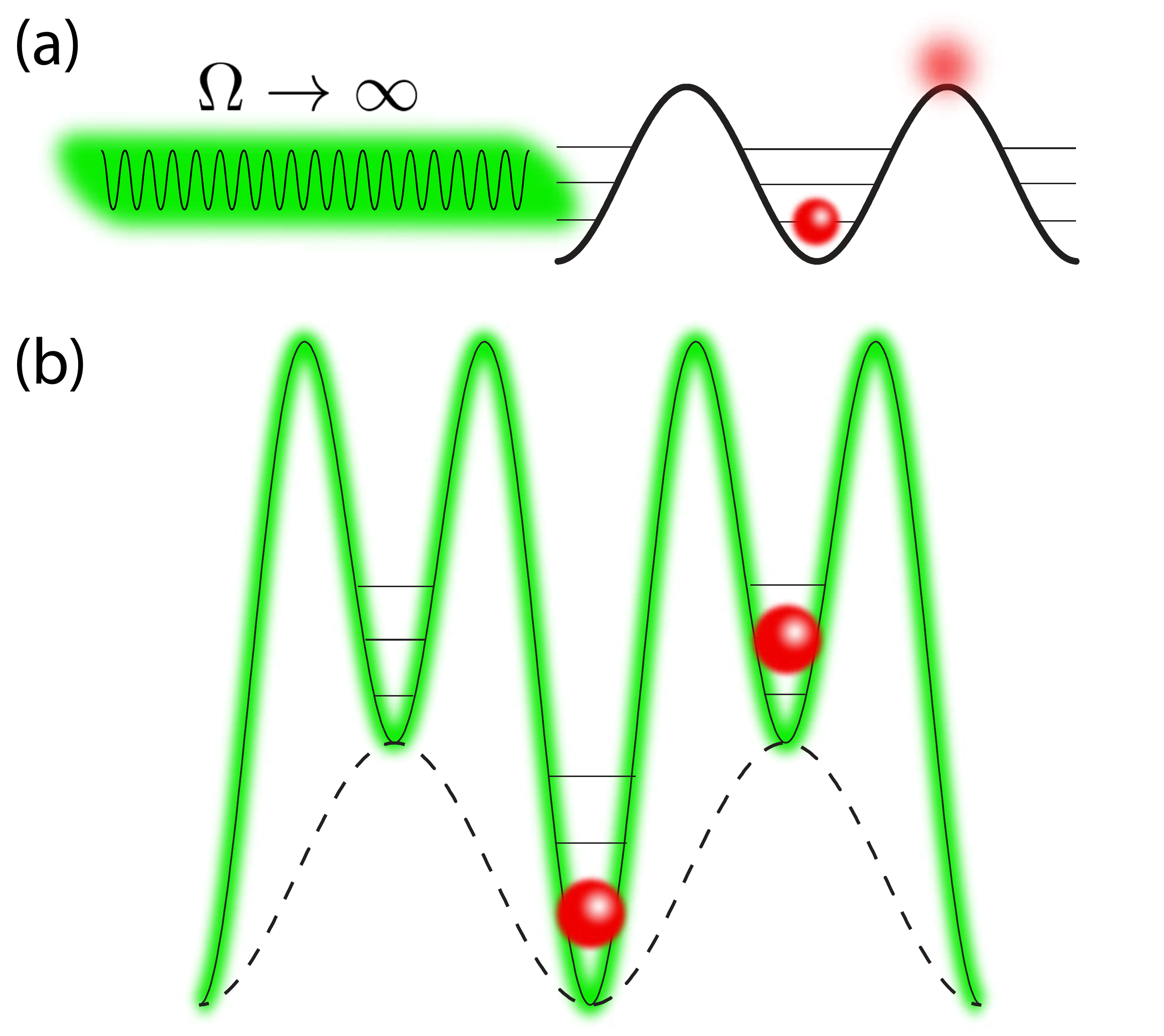}
	\caption{\label{fig:intro_fig}(Color online) The application of a high-frequency periodic perturbation to a static Hamiltonian, (a), may lead to the emergence of an effective high-frequency Hamiltonian with renormalized parameters (b). The purpose of this review is to discuss under which conditions this is possible, and what types of effective Hamiltonians can be engineered in this way.}
\end{figure}

Periodic perturbations arise naturally in many experimental setups. Examples 
include irradiating materials with electromagnetic waves, testing the 
response of a system to periodic currents or to mechanical shaking and deformations. Periodic driving protocols also appear in non-driven systems, after a transformation into a rotating frame, which typically results in the emergence of fast oscillating terms in the Hamiltonian. 

In the simplest case, one considers a single monochromatic driving protocol, characterized by a coupling 
strength (driving amplitude), and a single frequency $\Omega = 2\pi/T$. The dynamics of the periodically driven systems can be highly complex even in few-body systems. Usually, it can be analyzed in the two extreme regimes of slow and fast driving. In the former regime, the system almost adiabatically follows the instantaneous Hamiltonian. In the latter regime, where the driving frequency is fast compared to the natural frequencies of the non-driven model, the system typically feels an effective static potential, which can depend on the driving amplitude, c.f.~Fig.~\ref{fig:intro_fig}. If one deviates from either of these limits, one expects that sufficiently complex systems would heat up, and eventually reach infinite temperature in the absence of a coupling to a heat bath. This has been confirmed numerically and analytically in different setups~\cite{prosen_98a, prosen_99, bukov_12,dalessio_13, dalessio_14, ponte_14,lazarides_14}.

Away from the adiabatic limit, the analysis of periodically driven systems often relies on the Floquet theorem, which is very similar to the Bloch theorem in quantum mechanics. In its most general form, it states that one can write the evolution operator as
\be
U(t_2, t_1)=\mathrm e^{-i K(t_2)} \mathrm e^{-i \hat{H}_F (t_1-t_2)}\mathrm e^{i K(t_1)}, 
\label{floquet2}
\ee
where $K(t) = K(t+T)$ is a periodic hermitian operator, and $\hat{H}_F$ is the time-independent Floquet Hamiltonian. In fact, the choices of the periodic operator $K$ and the Floquet Hamiltonian $\hat{H}_F$ are not unique, and there is some freedom in defining them. As we shall discuss in the next section, different choices correspond to different gauges. There can be several convenient gauge choices, depending on the details of the setup. Despite being equivalent, these gauge choices can lead to different approximation schemes.

\begin{figure}
	\centering
	\includegraphics[width = 0.7\columnwidth]{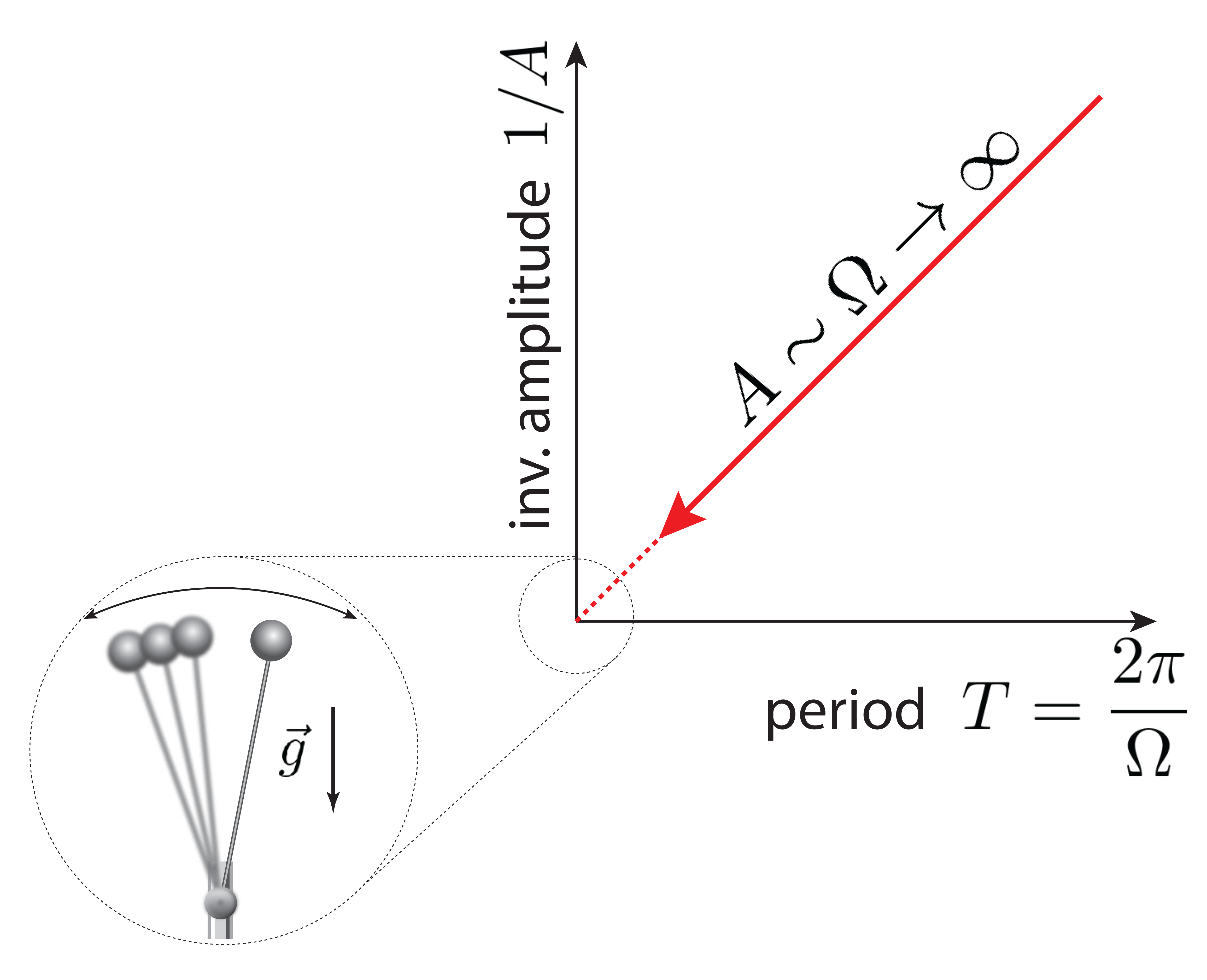}
	\caption{\label{fig:intro_fig2}(Color online). Schematic representation of the parameter space of periodically driven systems. In this work we consider a setup in which the amplitude of the driving scales with the frequency, i.e.~$A\sim\Omega$. In the infinite-frequency limit we obtain a well-defined local Floquet Hamiltonian which is qualitatively different from the time-averaged Hamiltonian. For example, in the case of the Kapitza pendulum, the Floquet Hamiltonian allows for oscillations around the upright position, a phenomenon known as dynamical stabilization (see Sec.~\ref{sec:kapitza}).}
\end{figure}

The formulation of Floquet's theorem simplifies if one observes the system stroboscopically, i.e.~at times $t_2=t_1+nT$, where $nT$ is the stroboscopic time measured in units of the driving period. In this case the operators $K(t_2)$ and $K(t_1)$ are identical, and the full evolution operator is equivalent to the evolution of the system generated by the static Hamiltonian $H_F[t_1]=\exp[-i K(t_1)]\hat{H}_F\exp[i K(t_1)]$. 

In general, it is not possible to evaluate $\hat{H}_F$, and $K(t)$ explicitly, and one has to rely on approximations~\cite{ maricq_82,maricq_88,magnus_54,verdeny_13,breuer_91,rahav_03_pra,goldman_14,goldman_res_14, eckardt_15}. Moreover, in macroscopic systems, there is no guarantee that $\hat{H}_F$ is a local physical Hamiltonian. In fact, in the case of generic interacting systems, a local $\hat{H}_F$ might not exist~\cite{dalessio_14}. In such situations, the dynamics of the system can be completely chaotic and the Floquet theorem is not particularly useful. 

An important limit, where the Floquet Hamiltonian can be defined at least perturbatively, corresponds to the fast driving regime, in which the driving frequency is larger than any natural energy scale in the problem. Then the driving does not couple resonantly to the slow degrees of freedom, but rather results in renormalization and dressing of the low-energy Hamiltonian. In many instances the Floquet Hamiltonian in the high-frequency limit is simply the time-averaged Hamiltonian, $\frac{1}{T}\int_0^T H(t)\mathrm{d} t$. But there are important exceptions, in which the Floquet Hamiltonian is \emph{not} given by $\frac{1}{T}\int_0^T H(t)\mathrm{d} t$, even in the infinite-frequency limit. These situations are of particular interest since the system can display interesting and counterintuitive behavior, such as dynamical stabilization, as it happens in the Kapitza pendulum~\cite{landau_lifshitz_1}. Such situations naturally occur, for instance, when the amplitude of the driving is proportional to a power of the driving frequency, c.f.~Fig.~\ref{fig:intro_fig2}. This was the case in recent experimental realisation of the Harper-Hofstadter Hamiltonian~\cite{jaksch_03,aidelsburger_13,miyake_13,aidelsburger_14}, and the Haldane Chern insulator~\cite{kitagawa_11}\footnote{We note that the key equilibrium property of topological states, namely robustness against various small perturbations, is not guaranteed to hold due to generic heating in ergodic driven systems~\cite{dalessio_13, dalessio_14, ponte_14}.} 
using cold atoms. A general understanding of such nontrivial limits is the main purpose of the present work.

Of course, in real systems the infinite-frequency limit is a mathematical abstraction. Typically, as one increases the driving frequency, new degrees of freedom can enter the game. Examples include internal molecular or atomic resonances in solid state systems or intra-band transitions in cold atom systems confined in optical lattices. Thus, one always deals with finite driving frequencies, which could still be larger than any natural frequency of the non-driven system. In such situations, the infinite-frequency limit of the Floquet Hamiltonian can be a good reference point, but finite-frequency corrections can still be significant. For this reason, in this work we discuss both the infinite-frequency limit of various model Hamiltonians, and the leading $\Omega^{-1}$-corrections~\cite{blanes_09}. 

The main purpose of this review is to discuss different generic scenarios, where one can engineer non-trivial Floquet Hamiltonians in the high-frequency limit. While these scenarios are not exhaustive, they cover a large class of driving protocols, and identify possible routes for finding new interesting Floquet systems. We shall refer to the different classes of driving protocols corresponding to these scenarios as (i) \emph{Kapitza class}: the Hamiltonian is quadratic in momentum, and the driving potential couples only to the coordinates of the particles (either as an external potential or through the interaction term). (ii) \emph{Dirac class}: same as the Kapitza class but for the system with relativistic linear dispersion such as graphene. (iii) \emph{Dunlap-Kenkre (DK) class}: the periodic drive couples to a single particle potential such as a periodically driven external electric or magnetic field. In the DK class the dispersion relation between particles is not restricted. These classes are not mutually exclusive, e.g.~there is a clear overlap between the \emph{Kapitza} class and the \emph{DK} class if one drives a system of non-relativistic particles by an external field, and a similar overlap exists between the \emph{Dirac} class and the \emph{DK} class for particles with a relativistic dispersion. 

We shall argue that, in models belonging to these three classes, the Floquet Hamiltonian has a nontrivial high-frequency limit, which is different from the time-averaged Hamiltonian allowing the systems to display new, qualitatively different features. These non-trivial limits can be used as a tool to realize synthetic matter, i.e.~matter with specific engineered properties. On the theoretical side, we justify the existence of stable high-frequency fixed points in $\Omega$-space, whose physics is governed by a well-defined effective (local) Hamiltonian. Although such fixed-point Hamiltonians may never be accessible experimentally, they provide a good reference point in many realistic situations. Moreover, the corrections to the effective Hamiltonian, which we also discuss in detail, allow one to estimate the finite-frequency effects for particular setups, and find the regimes where these corrections are negligible. We stress that these non-trivial limits exist even for driven ergodic interacting many-particle systems, though interactions often lead to additional finite-frequency corrections to the effective Hamiltonian, which may ultimately result in faster heating rates.\\

\emph{This review is organized as follows.}

\begin{itemize}
	\item In Section~\ref{sec:floquet} we review some general properties of Floquet's theory. We define the stroboscopic Floquet Hamiltonian and the associated concept of the Floquet gauge. Then we introduce the more general non-stroboscopic Floquet Hamiltonian and the notion of the kick operator. We illustrate these concepts using an exactly solvable model of a two-level system in a circularly polarized periodic drive. Finally we introduce the concept of the Floquet non-stroboscopic (FNS) and Floquet stroboscopic (FS) dynamics, and compare them. In particular, we explain how Floquet theory extends to systems where the initial phase of the drive and/or the measurement time fluctuate within the driving period.
	
	\item In Section~\ref{sec:magnus_rotframe} we briefly review the inverse-frequency Magnus expansion for the stroboscopic Floquet Hamiltonian and a related but not equivalent expansion for the effective Floquet Hamiltonian. We present the discussion both in the laboratory (lab) and in the rotating frames. At the end of this section we briefly comment on the convergence properties of the inverse-frequency expansion. 
	
	\item Section~\ref{sec:ME_RWA} discusses applications of Floquet theory to static and driven Hamiltonians with large separation between energy levels. We show how one can derive the rotating wave approximation (RWA) as the leading term in the inverse-frequency expansion and how one can find systematic corrections to RWA. After discussing the toy model of a static two-level system, we show that one can apply this expansion to derive the Kondo model from the Anderson impurity model, and discuss its relation to the well-established Schrieffer-Wolff transformation extending it to driven systems. We conclude this section with the discussion of the RWA and leading finite-frequency corrections applied to the Rabi model, going beyond the Jaynes-Cummings Hamiltonian.
	
	\item In Section~\ref{sec:kapitza} we define the Kapitza driving class. We thoroughly analyze the prototypical example of dynamical stabilization - the Kapitza pendulum. We derive the leading corrections to the infinite-frequency Hamiltonian as well as the dressed observables and the dressed density matrix appearing in FNS dynamics. At the end of this section we discuss higher-dimensional and many-body generalizations of the Kapitza model.
	
	\item In Section~\ref{sec:diraclimit} we define and study the Dirac class, which describes relativistic systems with a linear dispersion. We derive the infinite-frequency Hamiltonian and describe some interesting effects, such as a dynamically generated spin-orbit coupling. We show that in this class the leading $1/\Omega$ corrections to the infinite-frequency Hamiltonian vanish and the first non-zero corrections are of order $1/\Omega^2$, suggesting that the systems in this class are more robust against heating. 
	
	\item In Section~\ref{sec:driven_external_fields}, we define the Dunlap-Kenkre (DK) driving class, which includes periodically driven tight-binding models. We begin by studying the shaken bosonic chain, and demonstrate the consequences of FS and FNS dynamics for different observables. Afterwards, we derive the leading corrections to the Harper-Hofstadter Hamiltonian, and the driven Fermi-Hubbard model relevant for Floquet topological insulators. Finally, we briefly discuss some spin Hamiltonians which can be implemented in existing nuclear magnetic resonance setups.
	
	\item In Section~\ref{sec:conclusions} we give the summary of this review and an outlook to some open problems.
	
\end{itemize}


\section{\label{sec:floquet} Floquet Theory. Stroboscopic and Non-Stroboscopic Time Evolution.}

In this section we review Floquet's Theorem. We shall use the language of quantum mechanics but, as it becomes apparent later on in the next section, all results have a well-defined classical limit. Unless otherwise stated, we shall work in units of $\hbar=1$.

\subsection{\label{subsec:HF&P} The Stroboscopic Floquet Hamiltonian and the $P$-Operator.}

\begin{figure}
	\centering
	\includegraphics[width = 0.7\columnwidth]{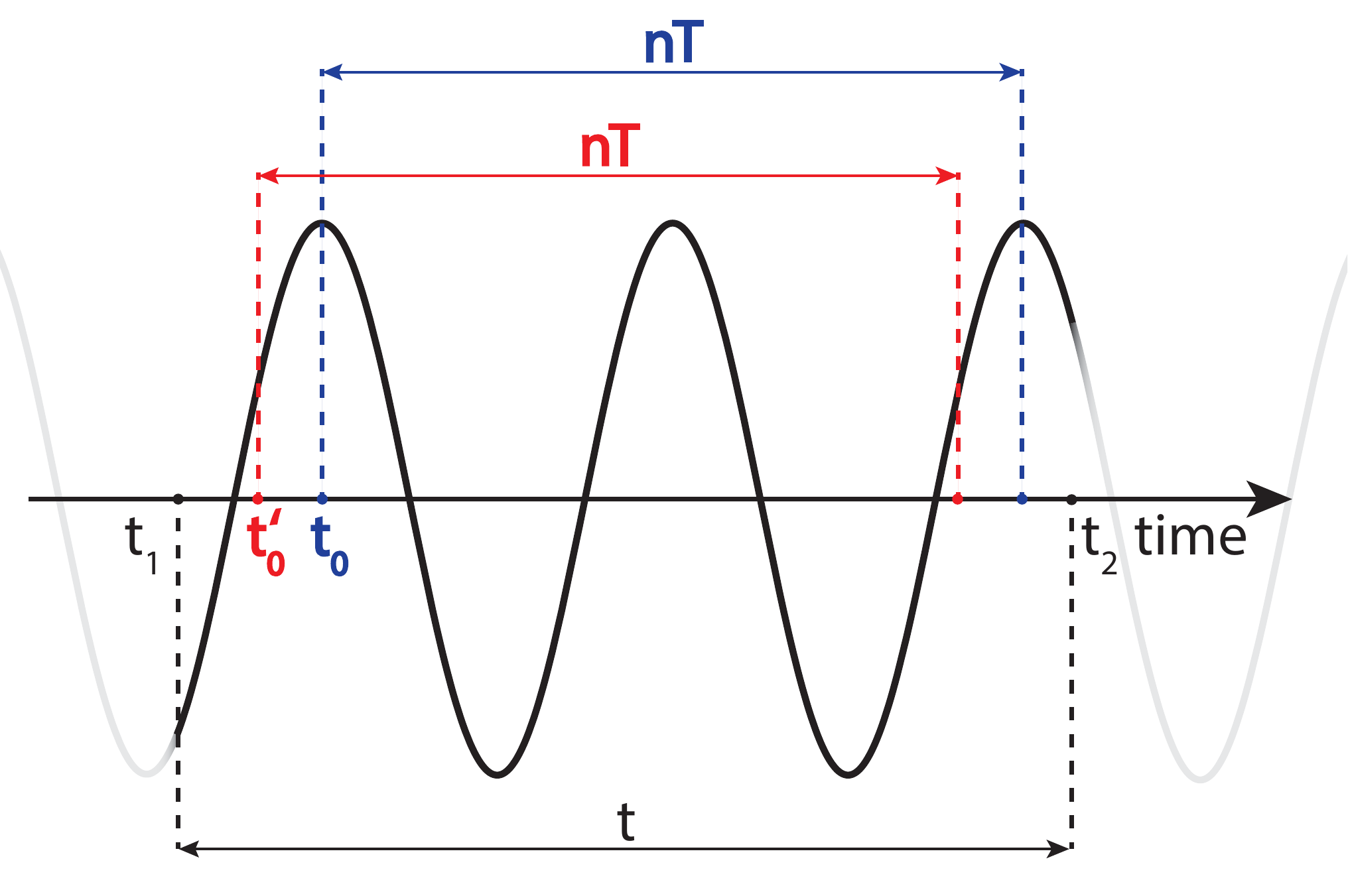}
	\caption{\label{fig:Floquet_gauge}(Color online) Floquet gauge: the system evolves from time $t_1$ to time $t_2$. The stroboscopic evolution starts at time $t_0$ which can be chosen to be anywhere within the first period $[t_1,t_1+T)$. The choice of the Floquet gauge, i.e.~the choice of $t_0$, in general affects the form of the stroboscopic Floquet Hamiltonian $H_F[t_0]$ (see text).}
\end{figure}

Let us consider a dynamical process in which the Hamiltonian depends periodically on time, e.g.~through a periodically modulated coupling constant. This means that the evolution operator defined as

\be
U(t_2,t_1)=\mathcal{T}_t \exp\left[-{i} \int_{t_1}^{t_2} H(\tilde t) \mathrm{d}\tilde t\right] = \prod_{j:\;t_1\leq \tilde t_j\leq t_2} \mathrm e^{-i H(\tilde t_j)(\tilde t_{j+1} - \tilde t_j)}
\ee
is invariant under discrete translations in time $(t_1,\,t_2)\to (t_1+n T,\,t_2+nT)$, where $n$ is an integer. The factorization (group) property of the evolution operator, $U(t_2,t_1)=U(t_2, t') U(t',t_1)$ for arbitrary $t'$, implies that $U(t_0+2T,t_0)=U(t_0+2T, t_0+T) U(t_0+T,t_0)=U(t_0+T,t_0)^2$, which generalizes to
\be
U(t_0+nT,t_0)=U(t_0+T,t_0)^n.
\ee
It is convenient to formally define the evolution within one period as an evolution with the time-independent stroboscopic Floquet Hamiltonian $H_F[t_0]$:
\be
U(t_0+T,t_0)=\exp\left[-{i} H_F[t_0] T\right].
\label{floquetansatz}
\ee
This representation is always possible because $U(t_0+T,t_0)$ is a unitary operator. The stroboscopic Floquet Hamiltonian $H_F[t_0]$ defined in this way depends on the choice of the time $t_0$ which defines the beginning of the stroboscopic driving period. This is a gauge choice, and it is completely arbitrary. To avoid confusion with general gauge transformations, related to the arbitrary choice of basis, we shall term the gauge associated with the choice of $t_0$ the \emph{Floquet gauge}. Very often, one chooses a particular Floquet gauge, in which the Floquet Hamiltonian assumes its simplest form. This often happens when $t_0$ is a symmetric point of the driving protocol. For example, if the driving field is $\cos\Omega t$, it is often convenient to choose $t_0=0$. 

It becomes clear from Fig.~\ref{fig:Floquet_gauge} that for arbitrary times $t_1$ and $t_2$ the evolution operator can always be written as
\be
U(t_2,t_1)=U( t_2,t_0+nT ) \exp\left[-i H_F[t_0] n T \right] U(t_0,t_1).
\label{evolution1}
\ee
The initial and final evolution operators $U(t_0,t_1)$ and $U( t_2,t_0+nT )$ effectively occur during small intervals of time $\delta t_1=(t_1-t_0)$ and $\delta t_2=(t_2-nT-t_0)$ which can always be chosen such that $\delta t_1\in[-T,0]$ and $\delta t_2\in[0,T]$. The operators $U(t_0,t_1)$ and $U( t_2,t_0+nT )$ are necessary to bring the time from the initial point of the evolution $t_1$ to $t_0$, and from the last full period $t_0+nT$ to the final point of evolution $t_2$. By construction $U(t_0,t_0)={\bm 1}$ and $U(t_0+T,t_0)=\exp[-i H_F[t_0] T]$. Now we can easily rewrite Eq.~(\ref{evolution1}) in the form of Floquet's theorem by doing a simple trick
\beq
\begin{split}
U(t_2,t_1)=& U(t_2,t_0+nT) e^{i H_F[t_0] (t_2-t_0 - nT) } e^{-i H_F[t_0] (t_2-t_0 - nT) } e^{-i H_F[t_0] n T }\times  \\
           & \times e^{-iH_F[t_0] (t_0-t_1) } e^{i H_F[t_0] (t_0-t_1) } U(t_0,t_1) = \\
	  =& P(t_2,t_0) e^{-i H_F[t_0] (t_2-t_1)} P^\dagger(t_1,t_0) = e^{-i K_F[t_0](t_2)} e^{-i H_F[t_0] (t_2-t_1)} e^{i K_F[t_0](t_1)}
\label{evolution2}
\end{split}
\eeq
where we have defined the fast motion unitary operator $P$:
\begin{equation}
P(t_2,t_0)\equiv U(t_2,t_0) e^{i H_F[t_0] (t_2-t_0) } \equiv  e^{-i K_F[t_0](t_2)},
\label{eq:kick_p}
\end{equation}
and the last equality defines the \emph{stroboscopic kick operator} $K_F[t_0](t)$ which depends explicitly on the Floquet gauge $t_0$. Note that, with the above definition, the operator $P$ is periodic $P(t_2+nT,t_0)=P(t_2,t_0+nT)=P(t_2, t_0)$ and, by construction, it also satisfies the property $P(t_0+nT,t_0)={\bf 1}$, for an arbitrary integer $n$. This means that the stroboscopic kick operator reduces to zero at stroboscopic times, i.e.~$K_F[t_0](t_0+nT)=\bf{0}$. From Eq.~\eqref{eq:kick_p} it immediately follows that $K_F[t_0](t)=i \log[P(t,t_0)]$. 

Floquet's theorem can be simplified by choosing $t_0$ to coincide with either $t_1$ or $t_2$, thus eliminating one of the two $P$-operators. However, this simplification can be somewhat deceptive, because in these cases, the Floquet Hamiltonian is tied to the initial (final) times of the evolution, and its definition continuously changes with one of those times. Since in experiments, especially in the high-frequency limit, both the initial time and the final (measurement) time often fluctuate within a period, it is more convenient to tie the Floquet Hamiltonian to some fixed Floquet gauge $t_0$, independent of both $t_1$ and $t_2$.

From Eqs.~\eqref{floquetansatz} and (\ref{evolution1}) together with the  factorization property of the evolution operator it becomes clear that the choice of $t_0$, defining the Floquet Hamiltonian, is indeed a gauge choice. To see this, we write the evolution operator $U(t_0+nT,t_0)$ in two different (but equivalent) ways:
\begin{equation}
U(t_0+ n T, t_0)=\mathrm \mathrm e^{-i H_F[t_0] n T } = U^\dagger(t_0+\delta t_0,t_0) \mathrm \mathrm e^{-i H_F[t_0+\delta t_0] n T} U(t_0+\delta t_0,t_0).
\end{equation}
This is equivalent to the gauge transformation of the Floquet Hamiltonian
\begin{eqnarray}
H_F[t_0+\delta t_0] &=& U(t_0+\delta t_0,t_0) H_F[t_0] U^\dagger(t_0+\delta t_0,t_0),\nonumber\\
&=&  P(t_0+\delta t_0,t_0)H_F[t_0] P^\dagger( t_0+\delta t_0,t_0). 
\label{eq:gauge_floquet}
\end{eqnarray}
As expected, this Floquet gauge is periodic and continuous, such that $H_F[t_0+T]=H_F[t_0]$. 

Let us also point out that one can rewrite Floquet's theorem in a differential form~\cite{rahav_03,grozdanov_88,verdeny_13}. Indeed, on the one hand, for any Hamiltonian evolution one can write
\be
i \partial_{t_2} U(t_2,t_1)=H(t_2) U(t_2,t_1).
\ee
On the other hand, using Eq.~(\ref{evolution2}) we arrive at
\begin{multline}
i \partial_{t_2} U(t_2,t_1)=\left(i  \partial_{t_2} P(t_2,t_0)\right) \mathrm \mathrm e^{-i H_F[t_0] (t_2-t_1)} P^\dagger( t_1, t_0)\nonumber\\
 + P(t_2,t_0) H_F[t_0] \mathrm \mathrm e^{-i H_F[t_0] (t_2-t_1)} P^\dagger(t_1,t_0).
\end{multline}
Equating these two expressions, we find
\be
H_F[t_0]=P^\dagger (t_2,t_0) H(t_2) P(t_2,t_0)-iP^\dagger(t_2,t_0) \partial_{t_2} P( t_2,t_0) 
\label{F2}
\ee	
or, equivalently, 
\be
H(t_2)=P(t_2,t_0) H_F[t_0] P^\dagger(t_2,t_0)+ i\left( \partial_{t_2} P(t_2,t_0)\right)\, P^\dagger(t_2,t_0) . 
\label{F1}
\ee 

Equation~(\ref{F2}) can be viewed as a statement of the existence of a periodic operator $P$ such that the RHS of this equation is time-independent. Very often, in the literature this equation is used as a starting point to find the Floquet Hamiltonian iteratively~\cite{maricq_82,rahav_03,grozdanov_88,verdeny_13}. Due to the gauge freedom associated with the choice of $t_0$, the solution of this equation is not unique, but all solutions are gauge-equivalent. Another possible application of Eq.~\eqref{F1} is that it allows one to do `reverse-engineering'. Once the Floquet Hamiltonian $H_F[t_0]$ and the periodic operator $P(t,t_0)$ with interesting properties are chosen, one can use Eq.~\eqref{F1} to determine which time-dependent driving protocol $H(t)$ needs to be experimentally implemented to realize those properties.

\subsection{\label{subseq:Heff&kick} The Non-Stroboscopic Floquet Hamiltonian and the Kick Operators.}

In the previous section, we showed that one can choose a family of stroboscopic Floquet Hamiltonians $H_F[t_0]$, each one of which exactly describes the evolution operator at stroboscopic times $U(t_0+n T, t_0)=\exp[-i H_F[t_0] nT]$. As we discussed in Sec.~\ref{subsec:HF&P}, the choice of $t_0$ is the Floquet gauge choice, and different stroboscopic Floquet Hamiltonians are gauge equivalent. In other words, by choosing one member of this family and applying to it the gauge transformation using the $P$-operator, one can obtain all other Floquet Hamiltonians from this family, c.f.~Eq.~(\ref{eq:gauge_floquet}). 

On the other hand, because $t_0$ is a gauge choice, all these Hamiltonians are also gauge-equivalent to some fixed Floquet Hamiltonian $\hat H_F$, which is $t_0$ independent. Therefore, there exists a family of Hermitian operators $K(t_0)$ which, following Ref.~\cite{goldman_14}, we call \emph{kick operators}, such that
\be
\hat H_F= \mathrm{e}^{iK(t_0)}H_F[t_0]\ \mathrm{e}^{-iK(t_0)},
\label{eq:HF-Heff0}
\ee
or equivalently
\begin{eqnarray}
H_F[t_0] = \mathrm{e}^{-iK(t_0)}\hat H_F\ \mathrm{e}^{iK(t_0)}.
\label{eq:HF-Heff}
\end{eqnarray} 
From now on we shall always reserve the hat to indicate some fixed Floquet Hamiltonian $\hat H_F$ and the notation $H_F[t_0]$ to indicate the stroboscopic Floquet Hamiltonian introduced in the previous section. By construction, the kick operator carries all the Floquet-gauge dependence, and it is periodic in time: $K(t_0+nT)=K(t_0)$. Moreover the Hamiltonian $\hat H_F$ is explicitly Floquet-gauge independent. If $K(t_0)$ is the zero operator ${\bm 0}$ at any time $t_0\in[0,T)$ then, following Eq.~(\ref{eq:HF-Heff0}), $\hat H_{F}$ coincides with the stroboscopic Floquet Hamiltonian $H_F[t_0]$. However, if this is not the case then $\hat H_F$ does not describe the stroboscopic evolution for any choice of $t_0$. It is clear from Eqs.~\eqref{eq:HF-Heff0} and \eqref{eq:HF-Heff} that the kick operator and the fixed Floquet Hamiltonian are not completely independent. Usually, one uses the freedom in the definition of the kick operator to obtain $\hat H_F$ in its simplest form. Developing separate inverse-frequency expansions for $\hat H_{F}$ and the kick operator $K(t_0)$ allows one to separate the Floquet-gauge independent terms, which determine the fixed Floquet Hamiltonian, from the Floquet-gauge dependent terms, which are all part of the kick operator. The latter are also responsible for the effect of the Floquet gauge on the initial state and the observables under consideration. There is a particularly convenient choice of $\hat H_{F}$, which is typically termed the \emph{effective Hamiltonian}~\cite{rahav_03_pra,rahav_03,goldman_14,goldman_res_14,eckardt_15} and it is indicated by $H_{\rm eff}$, which we discuss below in Sec.~\ref{sec:magnus_rotframe}. 

It is straightforward to find the relation between the kick operator $K$ and the fast-motion operator $P$. Namely, substituting Eq.~(\ref{eq:HF-Heff}) into Eq.~(\ref{eq:gauge_floquet}) we find
\begin{eqnarray*}
H_F[t_0+\delta t_0] &=& P(t_0+\delta t_0, t_0) H_F[t_0] P^\dagger(t_0+\delta t_0,t_0)\\
&=& P(t_0+\delta t_0,t_0) \mathrm{e}^{-iK(t_0)}\hat H_F\ \mathrm{e}^{iK(t_0)} P^\dagger(t_0+\delta t_0, t_0).
\end{eqnarray*}
On the other hand, by construction
\[
H_F[t_0+\delta t_0] = \mathrm{e}^{-iK(t_0+\delta t_0)}\hat H_F\ \mathrm{e}^{iK(t_0+\delta t_0)}.
\]
Since $\delta t_0$ is arbitrary, we see from these two equations that the kick operator and the fast-motion operator are not independent:
\be
P(t,t_0)= \mathrm{e}^{-iK(t)}\mathrm{e}^{iK(t_0)} = \mathrm{e}^{-i K_F[t_0](t)}.
\label{rel:P_K}
\ee
Hence, similarly to the fast-motion operator $P$, the kick operator $K$ describes the dynamics of the system within the driving period~\cite{goldman_14}. We stress that the stroboscopic kick operator $K_F[t_0](t)$ is different from the kick operator $K(t)$. From Eq.~\eqref{rel:P_K} it follows that if the kick operator $K(t_0)$ vanishes for some particular value of $t_0$ (and as a consequence $\hat H_F$ describes stroboscopic dynamics) then the kick operator becomes identical to the stroboscopic kick operator, i.e.~$K(t)= K_F[t_0](t)$.
Using Eq.~\eqref{rel:P_K} and Eq.~\eqref{eq:HF-Heff0} we can rewrite the evolution operator in Eq.~(\ref{evolution2}) as
\beq
\begin{split}
U(t_2,t_1)=& \mathrm{e}^{-i K(t_2)} \mathrm{e}^{iK(t_0)} \mathrm{e}^{-i H_F[t_0] (t_2-t_1)} \mathrm{e}^{-iK(t_0)} \mathrm{e}^{iK(t_1)} \\ 
          =& \mathrm{e}^{-i K(t_2)} \mathrm{e}^{-i \hat{H}_F (t_2-t_1)} \mathrm{e}^{iK(t_1)},
\end{split}
\label{eq:floquet_HF}
\eeq
which is precisely the form of Floquet's theorem introduced in Eq.~(\ref{floquet2}).

In the next section, we discuss an example and calculate explicitly both the stroboscopic Floquet Hamiltonian $H_F[t_0]$ and the effective Floquet Hamiltonian $H_\text{eff}$.

\subsection{\label{subsec:foquet_circ_pol_drive} A Two-Level System in a Circularly Driven Magnetic Field.}

Let us illustrate the construction above using the simple example of a two-level system in a rotating magnetic field:
\begin{equation}
H(t)=B_z \sigma_z + B_\parallel \left( \sigma_x\cos\Omega t +  \sigma_y\sin\Omega t \right).
\label{eq:circ_B_field_Hlab}
\end{equation}
Not surprisingly this problem becomes time-independent after a transformation to a rotating frame.
The evolution operator in the original (lab) frame can be evaluated by first going into the rotating reference frame, where the Hamiltonian is time-independent (and therefore the evolution is simple), and then transforming back to the lab reference frame:
\be
U(t_2,t_1)=V(t_2,t_0)\,\,\mathrm e^{-{i} H^\text{rot}\,[t_0](t_2-t_1)}\,\,V^\dagger(t_1,t_0).
\label{eq:evol_rot}
\ee
where $V(t,t_0)=\exp\left[ - i \frac{\sigma_z}{2} \Omega (t-t_0) \right]$ is the operator which transforms from the rotating frame into the lab frame and the Hamiltonian in the rotating frame is:
\begin{eqnarray}
H^\text{rot}[t_0]&=&V^\dagger(t,t_0)H(t)V(t,t_0) - i  V^\dagger(t,t_0)\dot{V}(t,t_0) \nonumber \\
&=& B_z \sigma_z + B_\parallel \left( \sigma_x\cos\Omega t_0 + \sigma_y\sin\Omega t_0 \right) - \frac{\Omega }{2} \sigma_z.  
\label{transf}
\end{eqnarray}
Equation~\eqref{eq:evol_rot} for the evolution operator resembles the Floquet ansatz (\ref{evolution2}) with the only caveat that the function $V(t,t_0)$ is periodic with twice the period of the driving. This is however not a problem since the correct periodicity can be easily restored by redefining the operator $V$: $V(t,t_0)\to \tilde V(t,t_0)= V(t,t_0)\exp[-i\Omega t/2]$. More importantly, the eigenvalues of $H^\text{rot}[t_0]$, $\pm \epsilon_\text{rot}$, where
\be
\label{energy_2LS}
\epsilon_\text{rot} = \sqrt{ (B_z-\Omega/2)^2 + B_\parallel^2},
\ee
diverge in the high frequency limit, while naively one would expect that for $\Omega\to\infty$ the Floquet Hamiltonian reduces to the time-averaged Hamiltonian $\frac{1}{T}\int_0^T H(t)\mathrm{d} t=B_z\sigma_z$ whose energies do not diverge as $\Omega\to\infty$. 

Before showing how to fix this issue, we mention that the discussion here is not limited to two-level systems and the transformation to the rotating frame can be performed for any spin
using the operator $V(t)=\exp\left[-i\, {\bm L}\cdot {\bm\Omega}\,t \right]$, where ${\bm L}$ is the total angular momentum.
Obviously, doing a transformation to the rotating frame helps only, if the stationary part of the Hamiltonian is rotationally invariant. Otherwise, Floquet's theory tells us that Eq.~(\ref{evolution2}) still applies but the stroboscopic Floquet Hamiltonian $H_F[t_0]$ is not directly related to the Hamiltonian in the rotating reference frame.

In certain situations one can completely eliminate the time dependence of the lab-frame Hamiltonian, and find $H_F[t_0]$ by performing two consecutive transformations in two rotating frames~\cite{xu_14}. In general, however, $H_F[t_0]$ can only be written through an infinite series of transformations. In Refs.~\cite{rahav_03,grozdanov_88,maricq_82} it was realized that the operator $P$ can be interpreted as a quantum analogue of the generating function of a canonical transformation, and $H_F[t_0]$ - as the Hamiltonian in the new reference frame (see Eq.~\eqref{F2}). Therefore, Floquet's theorem could be stated as follows. For \textit{any} time-periodic Hamiltonian, there exist infinitely many reference frames in which the time evolution is generated by a time-independent Hamiltonian. Unfortunately, in general, it is not possible to find the transformation from the lab to these new reference frames explicitly.  

We now show how to obtain the exact Floquet Hamiltonian and fast-motion operator $P$ for this problem. We start by noting the identity 
\begin{equation}
\exp\left(-iH_F[t_0] 2T\right) = U(t_0+2T,t_0) = \exp\left(-iH^\text{rot}[t_0] 2T\right),
\label{eq:HF_no_div}
\end{equation}
from which it is clear that $H_F[t_0]$ and $H^{\rm rot}[t_0]$ share the same eigenstates while their eigenvalues can only differ by a shift $\pm \Omega/2$. 
We fix the Floquet energies $\pm\epsilon_F$ by requiring that they do not diverge when $\Omega\to\infty$, i.e.:
\[
\epsilon_F=\left(\epsilon_\text{rot} - \frac{\Omega}{2} \right)=\epsilon_\text{rot} \left(1 - \frac{\Omega}{2\epsilon_\text{rot}} \right).
\]
If $S$ is a unitary matrix diagonalizing $H^{\rm rot}[t_0]$, such that $SH^{\rm rot}[t_0]S^\dagger=\epsilon_\text{rot} \sigma_z$ then it is clear that the stroboscopic Floquet Hamiltonian, which does not suffer from the infinite-frequency divergence is
\begin{equation}
H_F[t_0]=S^\dagger\,\epsilon_{F}\,\sigma_z\,S=\left(1-\frac{\Omega}{2\epsilon_\text{rot}}\right)\,H^{\rm rot}[t_0].
\label{eq:2LS_HF}
\end{equation}
From Eq.~\eqref{eq:evol_rot}, the Floquet ansatz in Eq.~\eqref{evolution2}, and the relation between $H_F[t_0]$ and $H^{\rm rot}[t_0]$ in Eq.~\eqref{eq:HF_no_div} it immediately follows that the fast-motion operator $P$ is a composition of two rotations in spin space:
\be
P(t,t_0) = \exp\left[ - i \frac{\sigma_z}{2} \Omega (t-t_0) \right] \exp\left[-i\frac{H^\text{rot}[t_0]}{2\epsilon_\text{rot}}\Omega (t-t_0)\right].
\label{P:circular}
\ee
Equations~\eqref{eq:2LS_HF} and \eqref{P:circular}, together with Eqs.~\eqref{transf} and \eqref{energy_2LS} provide the explicit solution for the Floquet Hamiltonian and the fast-motion operator.
Note that, as required, $P$ is periodic with period $T=2\pi/\Omega$, and reduces to the identity at stroboscopic times, i.e.~$P(t_0+nT,t_0)=\bf{1}$.
The stroboscopic kick operator is $K_F[t_0](t)=i\log(P(t,t_0))$.

Finally, as we explain in Sec.~\ref{subseq:Heff&kick}, there exists yet another natural choice for $\hat H_F$. It is equivalent to choosing the effective Hamiltonian: $H_{\rm eff}$, where
\begin{eqnarray}
H_\text{eff} &=& \left(\frac{\Omega}{2} - \epsilon_\text{rot}\right)\sigma_z,\nonumber\\
K_\text{eff}(t) &=& \frac{\alpha-\pi}{2}\left( -\sigma_x\sin\Omega t + \sigma_y\cos\Omega t \right),\ \ \cos\alpha = \frac{B_z-\Omega/2}{\epsilon_\text{rot}},\,\, 
\sin\alpha=\frac{B_\parallel}{\epsilon_\text{rot}}.
\label{eq:2LS_Heff}
\end{eqnarray}
Note that in the high frequency limit, $\Omega>2 B_z$, with our convention $\alpha\to\pi$ as $B_\parallel\to 0$. One can check that the operator $K_{\rm eff}(t_0)$ is the generator of the gauge transformation between $H_F[t_0]$ and $H_{\rm eff}$, (c.f. Eq.~\eqref{eq:HF-Heff}), i.e.~$H_{\rm eff}=\mathrm e^{i K_{\rm eff}(t_0)} H_F[t_0] \mathrm e^{-i K_{\rm eff}(t_0)}$. According to Eq.~(\ref{floquet2}), in this representation (which we call the effective picture) the evolution operator reads
\begin{eqnarray}
U(t_2,t_1) = \mathrm{e}^{-iK_\text{eff}(t_2)}\mathrm{e}^{-i H_\text{eff}(t_2-t_1)}\mathrm{e}^{iK_\text{eff}(t_1)}.
\end{eqnarray}
The effective Hamiltonian $H_\text{eff}$ is clearly non-stroboscopic because $K(t)$ does not vanish at any moment of time. This is a general feature of the effective kick operator for circularly-polarised drives.

The analysis above can be extended to more complex rotating setups. However, finding the properly folded Floquet Hamiltonian can, in general, be a formidable task, since it requires the knowledge of the spectrum of $H^{\rm rot}[t_0]$, which may be quite complicated if the system is interacting. In the next section, we shall discuss how one can perturbatively construct Floquet Hamiltonians, which have well-behaved infinite-frequency limits.

\subsection{\label{subsec:Floquet_experiment} Stroboscopic versus Non-stroboscopic Dynamics.}

Following the discussion in Secs.~\ref{subsec:HF&P} and \ref{subseq:Heff&kick}, the evolution operator can be written as the exponential of the Floquet Hamiltonian, sandwiched between two periodic unitary operators in two equivalent ways, c.f.~Eq.~\eqref{evolution2} and Eq.~\eqref{eq:floquet_HF}:
\[
U(t_2,t_1)=e^{-i K(t_2)} e^{-i \hat{H}_F (t_2-t_1)} e^{i K(t_1) } =e^{-i K_F[t_0](t_2)} e^{-i H_F[t_0] (t_2-t_1)} e^{i K_F[t_0](t_1)}. 
\]
We now use this observation to find the expectation values of observables. To simplify the discussion, we shall focus only on equal-time expectation values. The generalization to nonequal-time correlation functions is straightforward\footnote{We focus on the representation of the evolution operator through the kick operators $K(t)$ and the Floquet Hamiltonian $\hat{H}_F$. The equivalent expressions in terms of $K_F[t_0](t)$ and $H_F[t_0]$ can be obtained by the simple replacement $K(t)\rightarrow K_F[t_0](t)$ and $\hat{H}_F\rightarrow H_F[t_0]$. Moreover, using Eq.~\eqref{rel:P_K}, it is immediate to transform all the expressions in the language of the fast motion operator $P(t,t_0)$.}.

\begin{figure}
	\centering
	\subfigure{
		\resizebox*{10cm}{!}{\includegraphics[width = 0.6\columnwidth]{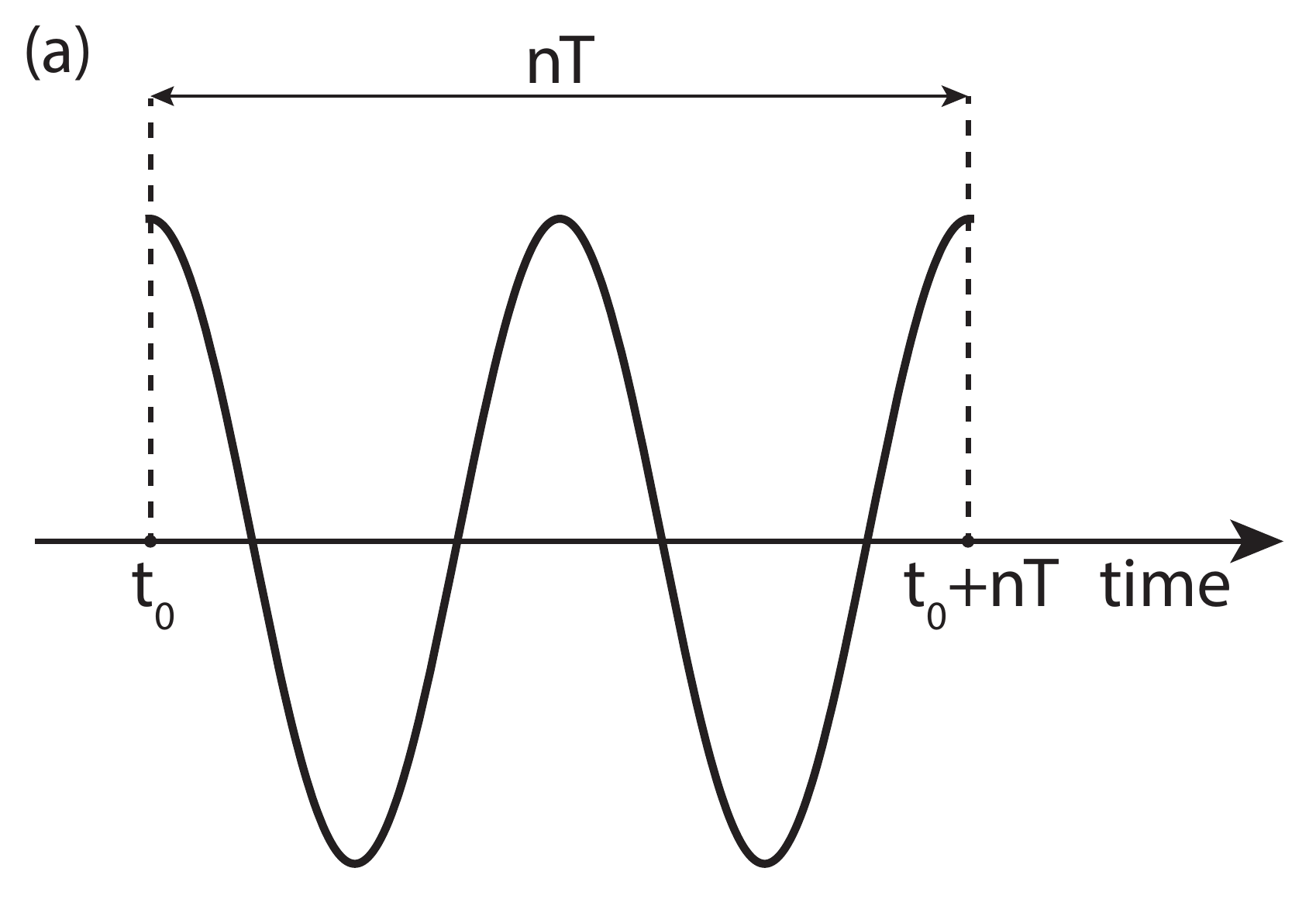}}}
	\hspace{6pt}
	\subfigure{
		\resizebox*{10cm}{!}{\includegraphics[width = 0.6\columnwidth]{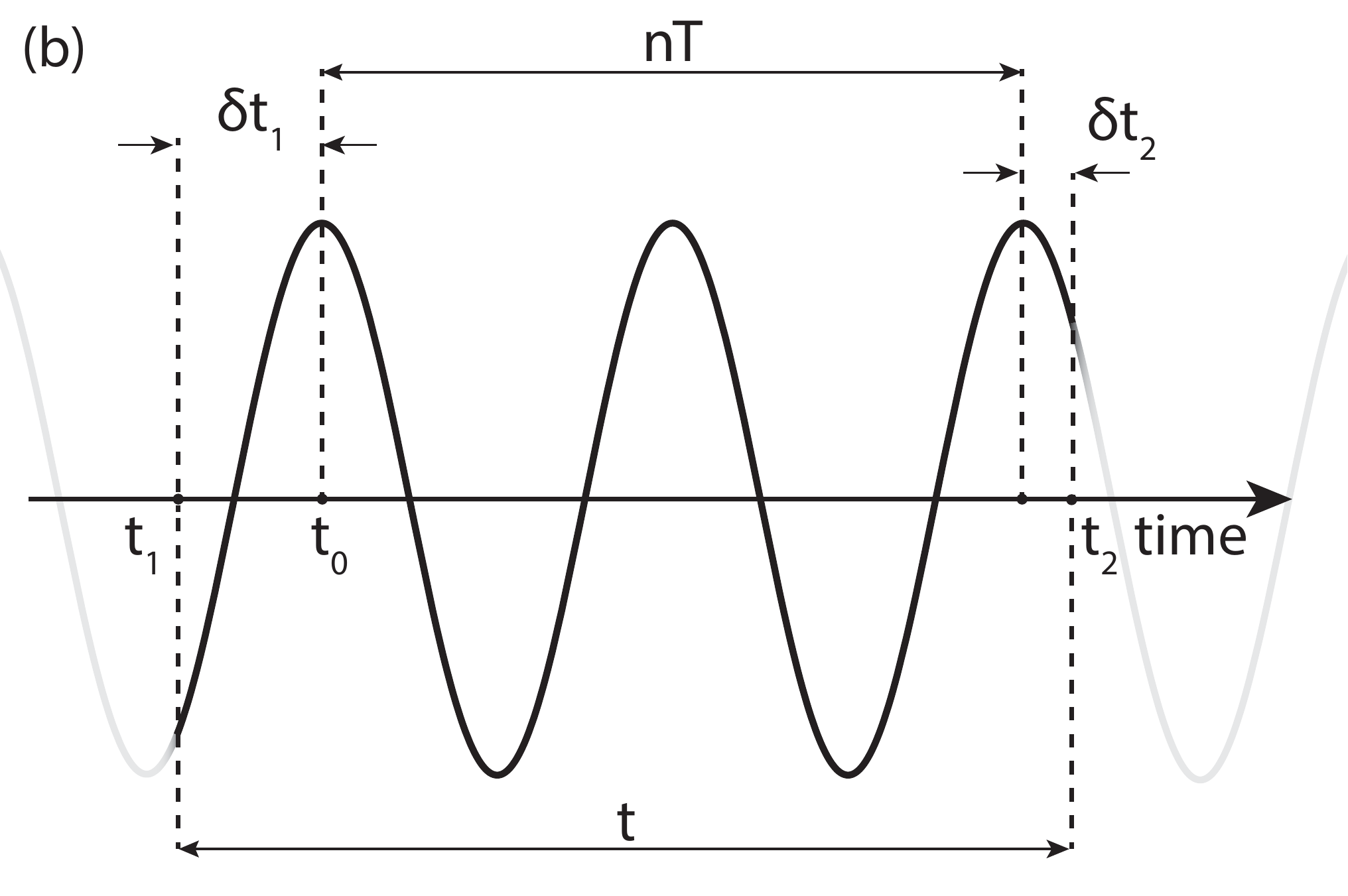}}}
	\caption{\label{fig:FNS_vs_FS}Floquet stroboscopic (FS) vs.~Floquet non-stroboscopic (FNS) evolution. (a) In the FS scheme the driving is initialized at time $t_0=0$ when the stroboscopic frame starts and the measurement is performed after exactly $n$ periods at time $n T$. (b) In the FNS scheme the driving is initiated at time $t_1$, the stroboscopic evolution begins at $t_0$ and the measurement is carried out at time $t_2$ within the $(n+1)$-st driving period. In the FNS scheme, the initial density matrix and the expectation values of the observables are obtained by averaging over $\delta t_1$ and $\delta t_2$.}
\end{figure}

Consider an observable $\mathcal O(t_2)$ in the Heisenberg picture, where it explicitly depends on time. Also, let us assume that initially (at time $t_1$) the system is prepared in some state described by the density matrix $\rho$, which in the Heisenberg picture remains time-independent. Then
\begin{eqnarray}
	\langle \mathcal O(t_2)\rangle &=& {\rm Tr} \left(\rho\, U^\dagger(t_2, t_1) \mathcal O U(t_2,t_1) \right) \nonumber\\
	&=&{\rm Tr} \left(\rho \mathrm e^{-i K(t_1)} \mathrm \mathrm e^{i \hat H_F (t_2-t_1)} \mathrm e^{i K(t_2)}  \mathcal O \mathrm e^{-i K(t_2)} \mathrm \mathrm e^{-i\hat H_F (t_2-t_1)}  \mathrm e^{i K(t_1)}\right)\nonumber\\
&=& {\rm Tr}\left( \mathrm{e}^{iK(t_1)} \rho \mathrm{e}^{-iK(t_1)}  \mathrm \mathrm e^{i \hat H_F(t_2-t_1)}  \mathrm{e}^{iK(t_2)}  \mathcal O \mathrm{e}^{-iK(t_2)} \mathrm \mathrm e^{-i \hat H_F(t_2-t_1)} \right).\nonumber\\
	\label{floquet:evolution}
\end{eqnarray}
We see that the dynamics of the system is solely generated by the Floquet Hamiltonian if we properly identify a new density matrix and a new observable as
\be
\rho \to \mathrm{e}^{iK(t_1)} \rho\ \mathrm{e}^{-iK(t_1)} ,\; \mathcal O\to \mathrm{e}^{iK(t_2)}  \mathcal O\ \mathrm{e}^{-iK(t_2)}.
\label{dressed}
\ee
The operators $K(t_1)$ and $K(t_2)$ can be viewed as time-dependent gauge transformations applied to the initial density matrix (wave function) and the observable. The main difference with the conventional gauge transformation is that the density matrix and the observable are transformed differently unless $t_2-t_1=nT$. In Ref.~\cite{goldman_14} the generators of these transformations were termed the initial and the final `kicks'. As we showed, these operators are periodic.

The simplest case, which is often discussed in the literature, is when the initial time $t_1$ coincides with $t_0$, and the final time is $t_2=t_0 + nT$ (see Fig.~\ref{fig:FNS_vs_FS}, Panel (a)). This condition defines the so-called \emph{Floquet stroboscopic (FS)} dynamics. In this setup, it is convenient to use the stroboscopic kick operators $K_F[t_0](t_1)$ and $K_F[t_0](t_2)$ which vanish identically at times $t_1$ and $t_2$, as defined above. As a consequence, the operators $\rho$ and $\mathcal O$ are not modified and the evolution is generated by the stroboscopic Hamiltonian $H_F[t_0]$. Sometimes, analyzing such FS dynamics is sufficient for describing the whole time evolution. This happens, for example, when the observable and the initial density matrix do not change significantly within a period. The FS dynamics can also be described in terms of the non-stroboscopic kick operator $K(t)$ and non stroboscopic Floquet Hamiltonian $\hat{H}_F$. Then one has to include the effect of the kick operator $K(t_0)$ on the observables and the density matrix [see Eq.~\eqref{dressed}]. In this case, the gauge transformations for the density matrix and the observable are identically given by the same operator $K(t_0)$, which is equivalent to the standard static global gauge transformation.

Another common setup, which naturally occurs in experiments, emerges if the initial time of the driving $t_1$ (which is equivalent to the driving phase) and the measurement time $t_2$ are random variables which fluctuate independently. We name this scenario \emph{Floquet non-stroboscopic (FNS)} dynamics.
In this case, one has to average the expectation value of $\mathcal O(t)$ over the intervals $\delta t_1$ and $\delta t_2$ (c.f.~Fig.~\ref{fig:FNS_vs_FS}, Panel (b)). From Eq.~(\ref{floquet:evolution}) it becomes clear that this averaging procedure affects both the density matrix and the observable. One can also consider other schemes, where e.g.~the initial phase is deterministic but the measurement time is random, or conversely the initial phase is random but the measurement is locked to the phase of the drive. We shall not consider such situations but from our discussion it will become clear how one can find the appropriate density matrix and the observable.

In order to obtain an accurate description of the FNS evolution, one needs to average the density matrix and the operator in Eq.~(\ref{floquet:evolution}) with respect to the uncertainty in $t_1$ and $t_2$:
\be
\overline{\rho}=\overline{\mathrm{e}^{iK(t_1)} \rho\ \mathrm{e}^{-iK( t_1)}},\; \overline{\mathcal O}=\overline{\mathrm{e}^{iK(t_2)} \mathcal O\ \mathrm{e}^{-iK(t_2)}},
\label{bar_O}
\ee
where the bar implies averaging over some, say Gaussian, distribution for $t_{1,2}$. Further, because $K(t)$ is a periodic operator, the averaging over $t_1$ and $t_2$ becomes equivalent to averaging over one period if the width of the distribution becomes larger than the driving period. In the following, we shall focus on this situation. Then the whole time evolution is effectively described by the quench to the Floquet Hamiltonian starting from the dressed density matrix  $\overline{\rho}$ instead of $\rho$ and measuring the dressed operator $\overline{\mathcal O}$ instead of $\mathcal O$. There is a certain care needed in precisely understanding this statement. We assumed that $t_1$, $t_2$ and $t=t_2-t_1$ are statistically independent variables, which is clearly not the case. However, they become effectively independent when the total time $t$ is much larger than the uncertainty in both $t_1$ and $t_2$. Intuitively, one can understand this averaging procedure using a time-scale separation argument. In the high-frequency limit, the periodic kick operator $K(t)$ is responsible for the fast dynamics, while the Floquet Hamiltonian $\hat H_F$ governs the slow dynamics. Therefore when averaging over $t_{1,2}$, provided that  $t=t_2-t_1$ is much larger than the uncertainty in both $t_1$ and $t_2$,  one can assume that the operator $\exp[-i \hat H_F (t_2-t_1)]$ in \eqref{floquet:evolution} is practically unchanged and the averaging procedure only affects the observables and the density matrix (see \eqref{bar_O}). Finally, notice that, by construction, $\overline{ \rho}$ does not depend on the initial phase of the drive since it is averaged over a full cycle. Similarly $\overline{\mathcal O}$ does not depend on the final measurement time.

Note that, even if one starts from a pure state described by a wave function, in the FNS scheme, averaging over $t_1$ typically generates a mixed state. In this sense, the initial uncertainty in the initial time $t_1$ plays a similar role to temperature since both broaden the initial density matrix. Intuitively, the difference between $\rho$ and $\overline{ \rho}$ is determined by how much the density matrix changes within one period. Similarly, the difference between $\mathcal O$ and $\overline{\mathcal O}$ can be large or small, depending on how much the observable changes within one period. 


The dressed operators have some unusual properties. In particular, from the definition it becomes clear that $\overline{\mathcal{O}^2}\neq \left({\overline{\mathcal O}}\right)^2$, e.g.~in the rotating spin example $\overline{\sigma_x^2}=\bm{1}\neq (\overline\sigma_x)^2$, c.f.~Sec.~\ref{subsubsec:FNS_2LS}. Another example illustrating this property of dressed operators is discussed in detail in Sec.~\ref{kapitza:floquet_meas}, Eq.~\eqref{eq:kapitza_dressed_p^2_vs_p}. We also observe that, in the high-frequency limit, the  dressed operators satisfy the Heisenberg equations of motion with the Floquet Hamiltonian. Indeed, let us consider the Heisenberg equation of motion for some operator $\mathcal O(t)$. Using the Floquet ansatz~\eqref{eq:floquet_HF} and ignoring the kick operator at $t_1$ because it only dresses the density matrix, we obtain: 
\begin{eqnarray}
&&i \partial_t \mathcal O(t)=i \partial_t \left(\mathrm{e}^{i \hat H_F (t-t_1)} \mathrm e^{i K(t)} \mathcal O \mathrm e^{-i K(t)} \mathrm{e}^{-i \hat H_F (t-t_1)}\right)\nonumber\\
&&=\mathrm{e}^{i \hat H_F (t-t_1)}   [ \mathrm e^{i K(t)} \mathcal O  \mathrm e^{-i K(t)}, \hat H_F] \mathrm e^{-i \hat H_F (t-t_1)} + i \mathrm e^{i \hat H_F (t-t_1)}  \partial_t \left(\mathrm e^{i K(t)} \mathcal O \mathrm e^{-i K(t)} \right)\mathrm e^{- i \hat H_F(t-t_1)}.\nonumber\\
\label{eq:derivative_heisenberg}
\end{eqnarray}
We can average both sides of this equation over a period w.r.t.~the time $t$ assuming, as before, that 
it is independent of the total time interval $t-t_1$. The last term in Eq.~\eqref{eq:derivative_heisenberg} vanishes, since the average of a derivative of a periodic function is zero. As a result we find 
\be
i \partial_t  \overline{\mathcal O} = i \overline{ \partial_t  \mathcal O(t)}= [\overline {\mathcal O}(t),\hat H_F].
\label{eq:floquet_heisenberg}
\ee
where we have defined the Heisenberg picture of the dressed operator 
\[
\overline {\mathcal O}(t)=\mathrm{e}^{i \hat H_F (t-t_1)} \overline {\mathcal O} \mathrm{e}^{-i \hat H_F (t-t_1)}
\]
This equation is the Heisenberg equation of motion for the dressed operator. The left equality in Eq.~\eqref{eq:floquet_heisenberg} is similar to the Hellmann-Feynman theorem, in which the average over the quantum state plays a role analogous to the average over the period. 

If $\mathcal O$ represents a conserved quantity, then we can define an associated current $ \bm{J}_{\mathcal O}$ through
\be
\partial_t{\mathcal O(t)}+ \bm{\nabla}\cdot \bm{ J}_{\mathcal O} (t)=0.
\ee
Averaging both sides of this equation over time and using Eq.~(\ref{eq:floquet_heisenberg}), we see that the time average of the current operator must represent the dressed current $\overline{ \bm{J}}_{\mathcal O}$ governing the slow evolution of $\overline{\mathcal O}$:
\be
\partial_t\overline {\mathcal O}+ \bm{\nabla}\cdot \overline{\bm{ J}}_{\mathcal O} (t)=0,
\ee
where
\be
\overline{\bm{ J}}_{\mathcal O}=\overline {\mathrm{e}^{iK (t)}  \bm{J}_{\mathcal O}\ \mathrm{e}^{-iK (t)}},
\label{bar_current}
\ee
Thus, both in numerical simulations and in experiments, in order to measure the current associated with the Floquet Hamiltonian one has to appropriately dress the current operator using the FNS averaging. In contrast, the current evaluated at some fixed stroboscopic time $t=nT$ will be a different object, involving both information about the Floquet evolution governed by $\hat H_F$, and an additional contribution related to the derivatives of the kick operator (the last term in Eq.~(\ref{eq:derivative_heisenberg})). We shall return to this issue as well as to general differences between FS and FNS dynamics later on, when we discuss specific examples. We shall show that, using stroboscopic measurements, one {\textit{cannot}} obtain the current corresponding to the Floquet Hamiltonian at any driving frequency whenever the latter contains a gauge field. On the other hand, implementing the FNS scheme and averaging the expectation values over the driving period, the Floquet current can be obtained in the high-frequency limit, c.f.~Sec.~\ref{subsec:Floquet_measurement_bosons}. Recently, it was proposed to detect the topological character of the ground state in fermionic systems by measuring the magnetisation of a finite-size sample due to the chiral currents flowing at the edges~\cite{dahlhaus_14}. This proposal explicitly made use of the FNS measurement.

\subsubsection{\label{subsubsec:adiabaticity} Stroboscopic and Non-Stroboscopic Dynamics for an Adiabatic Ramping of the Drive.}

Although in this review we focus on ``sudden Floquet quenches'', where the system is prepared at some definite state at the initial time $t_1$, let us briefly discuss what happens in the ``Floquet adiabatic ramp'', where e.g.~the amplitude of the drive is slowly increased from zero to the finite value. While in the thermodynamic limit, adiabaticity in interacting periodically-driven systems is conjectured to be absent due to the appearance of densely distributed avoided crossings in the Floquet spectrum~\cite{hone_97}, the general understanding of adiabaticity in the experimentally-relevant finite-size systems is still a subject of an active research, because the local Floquet Hamiltonian in systems with unbounded energy spectrum might not exist (see Sec.~\ref{subsec:Magnus_convergence}). However, in simple setups of finite-size systems with few degrees of freedom, or noninteracting systems it is possible to show that the adiabatic limit is well defined~\cite{weinberg_15}. In particular, in Sec.~\ref{subsec:ME_vs_FHE_2LS} we show that in the case of a driven two-level system, a slow ramping-up of the driving amplitude, starting from the ground state\footnote{By Floquet ground state we mean the adiabatically-connected Floquet state.} of the non-driven Hamiltonian, results in the system following the ground state of the instantaneous stroboscopic Floquet Hamiltonian $H_F[t]$ with a very good accuracy:
\be
|\psi(t)\rangle=|\psi_{GS}(H_F[t])\rangle.
\label{eq:inst_gs}
\ee
Thus, all observables evaluated stroboscopically at times $t_0+nT$ in this case are given by the ground state expectation values of $H_F[t_0]$:
\be
\langle \psi(t_0+nT)| \mathcal O|\psi(t_0+nT)\rangle=\langle \psi_{GS}(H_F[t_0])| \mathcal O| \psi_{GS}(H_F[t_0])\rangle.
\ee
This statement has immediate consequences for the FNS dynamics, where the measurement times are fluctuating within a period. We assume that, either the measurement is done after the ramp is over, or that the dynamical phase accumulated due to the Floquet quasi-energies is small. From the gauge equivalence of the Floquet Hamiltonians in Eq.~\eqref{eq:HF-Heff0} we see that, up to an unimportant phase factor, Eq.~(\ref{eq:inst_gs}) implies that
\[
|\psi(t)\rangle=\mathrm e^{-i K(t)} |\psi_{GS}(\hat H_F)\rangle,
\]
where $\hat H_F$ is an arbitrary fixed gauge Floquet Hamiltonian and $K(t)$ is the corresponding kick operator. Using this result it is straightforward to calculate the average over one period of the expectation value of an observable $\mathcal O$:
\begin{eqnarray}
{1\over T} \int_{t_0}^{t_0+T}\mathrm dt \langle \psi(t) |\mathcal O|\psi(t) \rangle&=&
{1\over T}\int_{t_0}^{t_0+T} \mathrm dt \langle \psi_{GS}(\hat H_F)| e^{i K(t)}\mathcal O  e^{-i K(t)} | \psi_{GS}(\hat H_F) \rangle\nonumber\\
&=&\langle \psi_{GS}(\hat H_F)| \overline {\mathcal O}|  \psi_{GS}(\hat H_F) \rangle.
\label{eq:adiabatic_FNS}
\end{eqnarray}
Since we average over the period, the result does not depend on $t_0$. Note that instead of the ground state in Eq.~(\ref{eq:adiabatic_FNS}) one can use any other Floquet eigenstate. In other words, if the system is in an eigenstate of the stroboscopic Floquet Hamiltonian, the FNS expectation value of any observable can be found by evaluating the expectation value of the dressed observable in the eigenstate of the fixed-gauge Floquet Hamiltonian. 

We note that this statement should be understood with certain care. For example, as we show in several examples, the dressed spin operators in the case of a driven two-level system (Eq.~\eqref{dressed_spin}), and the dressed coordinate and momentum operators in the example of the Kapitza pendulum (Eq.~\eqref{eq:bar_p2}), do not necessarily satisfy the canonical commutation relations, nor the uncertainty principle. In Sec.~\ref{subsec:ME_vs_FHE_2LS}, we illustrate the FS and FNS evolution after an adiabatic ramping of the driving amplitude using the circularly driven two-level system. 

\subsubsection{\label{subsubsec:FNS_2LS} Non-Stroboscopic Evolution in the Two-Level-System.} 

Let us briefly illustrate the implications of FNS evolution for the driven spin example of Sec.~\ref{subsec:foquet_circ_pol_drive}. Consider first the stroboscopic Floquet Hamiltonian
\begin{eqnarray}
H_F[0]&&=\left(1-{\Omega\over 2\epsilon_{\rm rot}}\right)H_{\rm rot},\quad H_{\rm rot}=\left[(B_z-\Omega/2)\sigma_z+B_{\parallel} \sigma_x\right], \nonumber \\
\mathrm e^{-iK_F[0](t)}&&=\exp\left(-i\frac{\sigma_z}{2}\Omega t \right)\exp\left(-i\frac{H^\text{rot}}{2\epsilon_\text{rot}}\Omega t\right).
\end{eqnarray}
We discuss two representative initial states $|\psi_1\rangle$ and $|\psi_2\rangle$ defined by
\[
|\psi_1\rangle=|\downarrow\rangle\;\rightarrow\; \rho_1={1\over 2}(1-\sigma_z),\quad |\psi_2\rangle={1\over \sqrt{2}} (|\uparrow \rangle+|\downarrow\rangle)\;\rightarrow\; \rho_2={1\over 2}(1+\sigma_x).
\]
Then the corresponding dressed density matrices and dressed operators (the dressed Pauli matrices) are found according to Eq.~\eqref{bar_O} to be:
\begin{eqnarray}
\overline{\sigma}_{x,F} &=&\overline{\mathrm e^{i K_F[0](t)}\sigma_x \mathrm e^{-i K_F[0](t)}}= -\cos\alpha\sin^2\frac{\alpha}{2}\ \sigma_x + \sin\alpha \sin^2{\alpha\over 2}\ \sigma_z,\nonumber\\
\overline{\sigma}_{y,F} &=&\overline{\mathrm e^{i K_F[0](t)}\sigma_y \mathrm e^{-i K_F[0](t)}}= \sin^2\frac{\alpha}{2}\ \sigma_y,\nonumber\\
\overline{\sigma}_{z,F} &=& \overline{\mathrm e^{i K_F[0](t)}\sigma_z \mathrm e^{-i K_F[0](t)}}= \cos^2\alpha\ \sigma_z + \sin\alpha\cos\alpha\, \sigma_x,
\end{eqnarray} 
where
\[
\cos\alpha = \frac{B_z-\Omega/2}{\epsilon_\text{rot}},\,\,\sin\alpha=\frac{B_\parallel}{\epsilon_\text{rot}},\,\,\epsilon_\text{rot}=\sqrt{(B_z-\Omega/2)^2+B_\parallel^2}
\]
and
\[
\bar \rho_1=\frac{1}{2}\left( {\bm 1} - \overline{\sigma}_z\right),\quad \bar \rho_2=\frac{1}{2}\left( {\bm 1} + \overline{\sigma}_x\right).
\]
In the high frequency limit, $\Omega\gg B_z, B_\parallel$, we have $\alpha\approx \pi$ and the dressed operators 
are approximately equal to the original operators $\bar \sigma_j\approx \sigma_j$. This is expected since the rapidly rotating magnetic field averages to zero without having any significant effect on the spin operators. One can obtain non-trivial dressed operators if $B_\parallel/\Omega$ is kept constant as $\Omega$ gets large and hence $\tan\alpha\approx -2B_\parallel/\Omega$ is fixed. As we discuss in subsequent sections, this is precisely the key idea behind obtaining non-trivial Floquet Hamiltonians, namely to scale the amplitude of the drive with the driving frequency. Let us also point that, in the low-frequency regime $\Omega<2 B_z$, in the limit $B_\parallel\to 0$ we have $\bar \sigma_z\to \sigma_z$ and $\bar \sigma_x, \bar \sigma_y\to 0$. This result might look a bit counter-intuitive (a zero dressed operator $\bar \sigma_x$ means that the outcome of any FNS measurement with any initial conditions of $\sigma_x$ will be zero), but one has to keep in mind that in the low-frequency regime there is no time scale separation. For example if $\Omega\ll B_z$, then averaging over one period necessarily implies averaging over many precession periods in a static magnetic field. Usually dressed operators are useful in the high-frequency regime, if there is a clear time scale separation between the fast dynamics governed by the kick operators and slow effective dynamic governed by the Floquet Hamiltonian.

Similarly one can find the dressed spin operators and the density matrices in the effective Hamiltonian picture (see Eq.~\eqref{eq:2LS_Heff}. In this case we have 
\be
\hat {H}_F=H_{\rm eff}=-\left(1-{\Omega\over 2\epsilon_{\rm rot}}\right)\epsilon_{\rm rot}\sigma_z,\quad K_{\rm eff}(t)= \frac{\alpha-\pi}{2}\left( -\sigma_x\sin\Omega t + \sigma_y\cos\Omega t \right),
\ee
and the averaging over the period gives
\be
\overline{\sigma}_{x,\rm eff} = \sin^2\frac{\alpha}{2}\sigma_x,\quad
\overline{\sigma}_{y,\rm eff} = \sin^2\frac{\alpha}{2}\sigma_y,\quad
\overline{\sigma}_{z,\rm eff} = -\cos\alpha\ \sigma_z.
\label{dressed_spin}
\ee
As expected, the effective Hamiltonian picture gives qualitatively similar asymptotic expressions for the dressed operators as the stroboscopic Floquet picture in the high frequency limit ($\alpha\to \pi$), where the dressed operators approach the bare operators. The main difference is that the effective Hamiltonian picture, unlike the stroboscopic picture, preserves the rotational symmetry around $z$-axes, while the stroboscopic Floquet picture breaks this symmetry. Thus the difference between stroboscopic and effective Floquet descriptions is similar to the difference between  Landau and symmetric gauges for a particle in a magnetic field. Both gauges are completely equivalent. One breaks the rotational symmetry, while the other preserves it. It might seem that the symmetric (effective) gauge is more convenient, but the Floquet gauge also has its own advantages giving a more intuitive picture of the spin dynamics in the lab frame. One can check that the kick operator $K_{\rm eff}(0)$ defines the gauge transformation between the two representations as (c.f. Eq.~(\ref{eq:HF-Heff0})):
\[
H_F[0]=\mathrm e^{-i K_{\rm eff}(0)} H_{\rm eff}\ \mathrm e^{i K_{\rm eff}(0)},\quad \overline{ \sigma_i}_F[0]=\mathrm e^{-i K_{\rm eff}(0)}\overline{ \sigma_i}_{\rm eff}\ \mathrm e^{i K_{\rm eff}(0)}
\]

It is interesting to note that the dressed Pauli matrices no longer obey the commutation relations $[\overline{\sigma}_i,\overline{\sigma}_j] \neq 2i\epsilon_{ijk}\overline{\sigma}_k$. Also it is straightforward to check that the dressed density matrices represent mixed states: $\bar \rho^2\neq \bar \rho$, unless $\alpha=\pi$.

\section{\label{sec:magnus_rotframe} Inverse Frequency Expansions for the Floquet Hamiltonian.}

With very few exceptions, like uniform rotations or driven harmonic systems where the evolution operator can be found exactly, it is impossible to obtain the Floquet Hamiltonian in a closed form. Moreover, in situations where the periodic driving leads to chaotic dynamics at a single particle level~\cite{casati_79, chirikov_71} or to heating to infinite temperatures for many-particle systems~\cite{dalessio_13, dalessio_14, ponte_14, lazarides_14} local Floquet Hamiltonians do not exist. An important limit where one can define the Floquet Hamiltonian at least perturbatively corresponds to the situations of fast driving, where the driving frequency is much faster than all natural frequencies of the system. For example, for a pendulum the driving should be fast compared to the oscillation period, for particles in a periodic potential the driving should be faster than the band width or a typical interaction scale. In such situations, the system has a hard time absorbing energy from the drive, which results in virtual processes dressing the low-energy Hamiltonian.

\subsection{\label{subsec:Magnus_expansion} The Magnus Expansion for the Stroboscopic Floquet Hamiltonian.}

A very efficient tool to compute the Floquet Hamiltonian in the high-frequency limit is the Magnus expansion (ME), which is a perturbative scheme in the driving period $T$ to compute $H_F[t_0]$. We refer to Ref.~\cite{bandyopadhyay_08} for a summary of other perturbative methods to find Floquet Hamiltonians in the high-frequency limit. In general, it is not known whether the Magnus expansion is asymptotic or has a finite radius of convergence, especially in the thermodynamic limit. The issue of the convergence of the Magnus expansion is important for understanding the behavior of the system in the limit $t\to\infty$. However, if one is interested in describing a finite-time evolution, then the short period expansion is well behaved and the Magnus expansion can be safely used. The evolution operator over a full driving cycle is, in general, given by the time-ordered exponential of $H(t)$:
\[
U(T+t_0,t_0)=\mathcal{T}_t\exp\left(-\frac{i}{\hbar}\int_{t_0}^{T+t_0}\mathrm{d}t H(t)\right)=\exp\left(-\frac{i}{\hbar}H_F[t_0] T\right), 
\]
where we have used Floquet's theorem~\eqref{floquetansatz}. In this section we explicitly insert the factors of $\hbar$ to highlight that the limit $\hbar\to 0$ is well-defined, and the expansion applies both to quantum and classical systems. Taking the logarithm of both sides of the equation above and expanding the exponents in a Taylor series (c.f.~App.~\ref{app:ME_HEF_derivation}), which is justified if the period is sufficiently short, one can represent $H_F[t_0]$ as~\cite{blanes_09}:
\begin{equation}
H_F[t_0] = \sum_{n=0}^\infty H_F^{(n)}[t_0],\ \ \ K_F[t_0](t) = \sum_{n=0}^\infty K_F^{(n)}[t_0](t).
\end{equation}
The superindex $^{(n)}$ means that $H_F^{(n)}[t_0]$ is of order $\Omega^{-n}$, and similarly for the stroboscopic kick operator $K_F^{(n)}[t_0](t)$. 
The first few terms are given by  
\be
\begin{split}
H_F^{(0)} &= \frac{1}{T}\int_{t_0}^{T+t_0}\mathrm{d}t\,H(t) = H_0,\\
H_F^{(1)}[t_0] &=  \frac{1}{2!Ti\hbar}\int_{t_0}^{T+t_0}\mathrm{d}t_1\int_{t_0}^{t_1}\mathrm{d}t_2\,[H(t_1),H(t_2)],\\ 
&= \frac{1}{\hbar\Omega}\sum_{l=1}^\infty \frac{1}{l}\left( [H_l,H_{-l}] - \mathrm e^{i l \Omega t_0}[H_l,H_0]  + \mathrm e^{-i l \Omega t_0}[H_{-l},H_0] \right) \\
H_F^{(2)}[t_0] &= \frac{1}{3!T(i\hbar)^2}\int_{t_0}^{T+t_0}\mathrm{d}t_1\int_{t_0}^{t_1}\mathrm{d}t_2\int_{t_0}^{t_2}\mathrm{d}t_3\bigg([H(t_1),[H(t_2),H(t_3)]] + (1\leftrightarrow 3)\bigg),
\end{split}
\label{eq:magnus_series}
\ee
where we have expanded the time-periodic Hamiltonian in its Fourier harmonics as:
\be
H(t) = \sum_{l\in\mathbb{Z}}H_l \mathrm e^{ i l \Omega t }. 
\label{eq:H(t)_harmonics} 
\ee
Similarly, the leading terms in the series for the stroboscopic kick operator are given by
\be
\begin{split}
K_F^{(0)}[t_0](t) &= {\bm 0},\\
K_F^{(1)}[t_0](t) &=  \frac{1}{\hbar}\int_{t_0}^t dt'\left( H(t')-H_F^{(1)}[t_0] \right)\\
&=-\frac{1}{2\hbar}\left[\int_{t}^{T+t}\mathrm{d}t'H(t')\left(1 + 2\frac{t-t'}{T}\right) -\int_{t_0}^{T+t_0}\mathrm{d}t'H(t')\left(1 + 2\frac{t_0-t'}{T}\right) \right] \\
&=\frac{1}{i\hbar\Omega}\sum_{l\neq 0}\,\,H_l\,\,\frac{\mathrm{e}^{i l \Omega t} - \mathrm{e}^{i l \Omega t_0}}{l}.
\end{split}
\label{eq:kick_operator_ME}
\ee
Higher-order terms can be obtained directly, e.g.~following Appendix~\ref{app:ME_derivation}. The zeroth-order term in the Floquet Hamiltonian is simply the time-averaged Hamiltonian while the zeroth-order stroboscopic kick operator is identically zero. Obviously both terms are Floquet-gauge invariant, i.e.~independent of $t_0$. On the contrary, the corrections to the stroboscopic Hamiltonian $H_F[t_0]$ and kick operator $K_F[t_0](t)$ depend on the Floquet gauge $t_0$. This gauge dependence is not always convenient especially for FNS dynamics. As we discussed in the previous section using the circularly driven two-level system, fixing $t_0$ in the stroboscopic Floquet Hamiltonian is similar to using the Landau gauge for a particle in a constant magnetic field, which explicitly breaks the U(1) symmetry of the Hamiltonian (rotations around the magnetic field). In the Floquet Hamiltonian this U(1) symmetry corresponds to the symmetry with respect to the phase shift of the drive and is equivalent to the translations of $t_0$. In many situations, it might be preferable to work with a Floquet Hamiltonian which does not break this U(1) symmetry. This can be achieved by doing a different expansion for the effective Hamiltonian~\cite{rahav_03,rahav_03_pra,goldman_14,goldman_res_14,eckardt_15} which we discuss in Sec~\ref{subsec:eff_expansion}.

For classical systems, the equivalent ME expansion can be obtained by substituting the commutators between the operators with the Poisson brackets of the corresponding classical functions: $[\cdot,\cdot]/i\hbar \rightarrow \{\cdot,\cdot \}$. It is interesting to note that there is no formal Floquet theorem for classical non-linear systems. Nevertheless, there is a well-defined classical limit for the high frequency expansion of the Floquet Hamiltonian. So if this expansion has a finite radius of convergence, effectively the results of Floquet theory applies to classical systems as well. For a numerical algorithm to implement the Magnus series, see Ref.~\cite{alvermann_11}.

\subsection{\label{subsec:eff_expansion} The High Frequency Expansion for the Effective Floquet Hamiltonian.}

As mentioned in Sec.~\ref{subseq:Heff&kick}, it is possible to change basis and work with the manifestly Floquet-gauge invariant effective Hamiltonian $H_\text{eff}$ and the kick operator $K_\text{eff}(t_0)$. The latter carries all the dependence on the Floquet gauge $t_0$, and describes the micromotion. This approach offers the advantage that the dependence on the Floquet gauge will not enter the inverse-frequency expansion of $H_\text{eff}$, and is enabled by the fact that the unitary change-of-basis transformation generated by $K_\text{eff}(t_0)$ effectively re-organises the terms in the perturbative series expansions. Such an expansion is provided by the High-Frequency Expansion (HFE) for the effective Hamiltonian~\cite{rahav_03,rahav_03_pra,goldman_14,goldman_res_14,eckardt_15}.

In a similar fashion to the Magnus expansion, we can decompose the effective Hamiltonian and the kick operator as
\begin{eqnarray}
H_\text{eff} = \sum_{n=0}^\infty H_\text{eff}^{(n)},\ \ \ K_\text{eff}(t) = \sum_{n=0}^\infty K_\text{eff}^{(n)}(t),
\end{eqnarray}
where $H_\text{eff}^{(n)}\sim\Omega^{-n}$ and $K_\text{eff}^{(n)}(t)\sim\Omega^{-n}$. Then, using the Fourier decomposition of the time-dependent Hamiltonian $H(t)$ in Eq.~\eqref{eq:H(t)_harmonics}, one has~\cite{rahav_03_pra,goldman_14,eckardt_15} (see also Appendix~\ref{app:ME_HEF_derivation}):
\begin{eqnarray}
H_\text{eff}^{(0)} &=& H_0 = \frac{1}{T}\int_0^T\mathrm{d}t\,H(t),\nonumber\\
H_\text{eff}^{(1)} &=& \frac{1}{\hbar\Omega}\sum_{l=1}^\infty \frac{1}{l} [H_l,H_{-l}] = \frac{1}{2!Ti\hbar}\int_{0}^{T}\mathrm{d}t_1\int_{0}^{t_1}\mathrm{d}t_2\, \left(1 - 2\frac{t_1-t_2}{T} \right) [H(t_1),H(t_2)],\nonumber\\
H_\text{eff}^{(2)} &=& \frac{1}{\hbar^2\Omega^2}\sum_{l\neq 0} \left( \frac{[H_{-l},[H_0,H_{l}]]}{2l^2} + \sum_{l'\neq 0,l} \frac{[H_{-l'},[H_{l'-l},H_{l}]]}{3ll'} \right).
\label{eq:HFE}
\end{eqnarray} 
The expansion for the kick operator is given by~\cite{rahav_03_pra,goldman_14,eckardt_15}
\begin{eqnarray}
K_\text{eff}^{(0)}(t) &=& \bm{0},\nonumber\\
K_\text{eff}^{(1)}(t) &=& \frac{1}{i\hbar\Omega}\sum_{l\neq 0}\frac{\mathrm{e}^{il\Omega t}}{l} H_l = -\frac{1}{2\hbar}\int_{t}^{T+t}\mathrm{d}t'H(t')\left(1 + 2\frac{t-t'}{T}\right).
\label{eq:kick_operator_HFE}
\end{eqnarray} 
The relation between the stroboscopic and the effective (non-stroboscopic) Floquet Hamiltonian and kick operator is given by
\begin{eqnarray}
H_F[t_0] &=& H_\text{eff}^{(0)}+ H_\text{eff}^{(1)} -i\left( \left[ K^{(1)}_\text{eff}(t_0),H_\text{eff}^{(0)} \right] + \left[ K^{(0)}_\text{eff}(t_0),H_\text{eff}^{(1)} \right] \right) + \mathcal{O}(\Omega^{-2})\nonumber\\
H_F^{(1)}[t_0]&=& H_\text{eff}^{(1)} -i \left[ K^{(1)}_\text{eff}(t_0),H_\text{eff}^{(0)} \right] ,\nonumber\\
K^{(1)}_F[t_0](t) &=& K^{(1)}_\text{eff}(t) - K^{(1)}_\text{eff}(t_0). 
\label{eq:F_vs_eff_1st_order}
\end{eqnarray}

Whenever the parameters in the Hamiltonian do not scale with the driving frequency, the $n$-th order term in both expansions is proportional to $T^n$. Thus, the higher-order terms get more and more suppressed as the period $T\to 0$. It then follows that in the infinite-frequency limit both $H_F[t_0]$ and $H_\text{eff}$ reduce to the time-averaged Hamiltonian, as one would intuitively expect. As we shall discuss later in greater detail, very interesting non-trivial limits can occur when some couplings in the Hamiltonian scale with frequency. In this case, terms in different orders in the above expansions can scale with the same power of the period $T$. Then in the infinite-frequency limit, one can obtain nontrivial Floquet Hamiltonians, very different from the time-averaged Hamiltonian, as it is the case for the Kapitza pendulum. The ME and HFE help one to identify both the leading and subleading terms in the driving period $T$ for different models. They also allow one to understand the required scaling behaviour of the driving amplitude with the frequency to obtain interesting infinite-frequency limits. And finally, they can be used to design protocols suitable for engineering synthetic Floquet Hamiltonians with prescribed properties. In the next section, we discuss the differences and similarities between the two expansions using an exactly-solvable model.

\subsection{\label{subsec:ME_vs_FHE_2LS} Magnus vs.~High-Frequency Expansions: the Two-Level System in a Circularly Driven Magnetic Field Revisited. }
	
Although the  Magnus expansion (ME) and the High Frequency expansion (HFE) share many common properties, there are also some very distinctive features between the two. In order to illustrate them intuitively, we shall briefly revisit the exactly solvable model of a two-level system in a circularly-polarised magnetic field.  

We want to compare the exact expression for $H_\text{F}[t_{0}]$ and $H_\text{eff}$
to the approximated Hamiltonians $\tilde{H}_\text{F}[t_0]$ and $\tilde{H}_\text{eff}$ obtained from the ME and HFE, respectively, up to order $\Omega^{-2}$. Using Eq.~\eqref{eq:magnus_series} we find:
\begin{eqnarray}
H_\text{F}^{(0)} &=& B_z \sigma_z \nonumber \\
H_\text{F}^{(1)}[t_{0}] &=& -\frac{1}{\Omega} \left[ B_{\parallel}^{2} \sigma_{z} + 2B_{\parallel}B_{z} \left(\sigma_{x}\cos(\Omega t_{0})+\sigma_{y}\sin(\Omega t_{0})\right)\right] \nonumber \\
H_\text{F}^{(2)}[t_{0}] &=& \frac{1}{\Omega^2} \bigg[ \left(2B_{\parallel}^{3} -4B_{\parallel}B_{z}^{2}\right) \left(\sigma_{x}\cos(\Omega t_{0})+\sigma_{y}\sin(\Omega t_{0})\right) - 4B_{\parallel}^{2}B_{z} \sigma_{z} \bigg]. \nonumber
\end{eqnarray}
Similarly, from Eq.~\eqref{eq:HFE} we derive:
\be
H_\text{eff}^{(0)} = B_z \sigma_z,\quad H_\text{eff}^{(1)} = -\frac{B_{\parallel}^{2}}{\Omega}\sigma_{z},\quad
H_\text{eff}^{(2)} = -\frac{2B_{\parallel}^{2}B_{z}}{\Omega^{2}}\sigma_{z}.
\ee
One can check that these expansions are consistent with expansions of Eqs.~\eqref{eq:2LS_HF} and~\eqref{eq:2LS_Heff}, respectively. 

We now compute the approximated spectra $\tilde{\epsilon}_F$ and $\tilde{\epsilon}_\text{eff}$ obtained by summing the ME and HFE up to order $\Omega^{-2}$.
The inverse-frequency expansions of $\tilde{\epsilon}_F$, $\tilde{\epsilon}_\text{eff}$ as well as the exact Floquet spectrum $\epsilon_F$ read: 
\begin{eqnarray}
\tilde{\epsilon}_\text{F} &=&\pm\left(-B_{z}+\frac{B_{\parallel}^{2}}{\Omega}+2\frac{B_{\parallel}^{2}B_{z}}{\Omega^{2}} +  \frac{2\left(B_\parallel^4 - 4B_\parallel^2B_z^2\right)}{\Omega^3} \right)  +\mathcal{O}(\Omega^{-4}) ,\nonumber\\
\tilde{\epsilon}_\text{eff} &=& \pm\left(-B_{z}+\frac{B_{\parallel}^{2}}{\Omega}+2\frac{B_{\parallel}^{2}B_{z}}{\Omega^{2}} +  0\times\frac{1}{\Omega^3} \right) ,\nonumber\\
\epsilon_\text{F}&=&\pm\left(-B_{z}+\frac{B_{\parallel}^{2}}{\Omega}+2\frac{B_{\parallel}^{2}B_{z}}{\Omega^{2}} +  \frac{4B_\parallel^2B_z^2 - B_\parallel^4}{\Omega^3} \right) +\mathcal{O}(\Omega^{-4}).
\label{eq:all_spectra}
\end{eqnarray} 
Clearly, the spectra of the approximated Hamiltonians agree to each other, and to the exact spectrum up to order $\Omega^{-2}$, i.e.~within the validity of the approximation. They differ starting from order $\Omega^{-3}$. This is not surprising since we have computed $\tilde{H}_\text{F}[t_0]$ and $\tilde{H}_\text{eff}$ to order $\Omega^{-2}$ and, therefore, all terms in the spectrum to higher order should be considered spurious. If we include the $\Omega^{-3}$-correction in both expansions, the spectra will agree to order $\Omega^{-3}$ and disagree starting from order $\Omega^{-4}$. Other quantities, which are invariant under a change of basis, are expected to display similar behaviour.

For general models (but not this one), it is possible that the spectrum of the approximated Floquet Hamiltonian contains phase-dependent corrections. However these corrections always appear beyond the order of the validity of the approximation and should not be taken into consideration~\cite{eckardt_15}.

\begin{figure}
	\includegraphics[width=0.8\columnwidth]{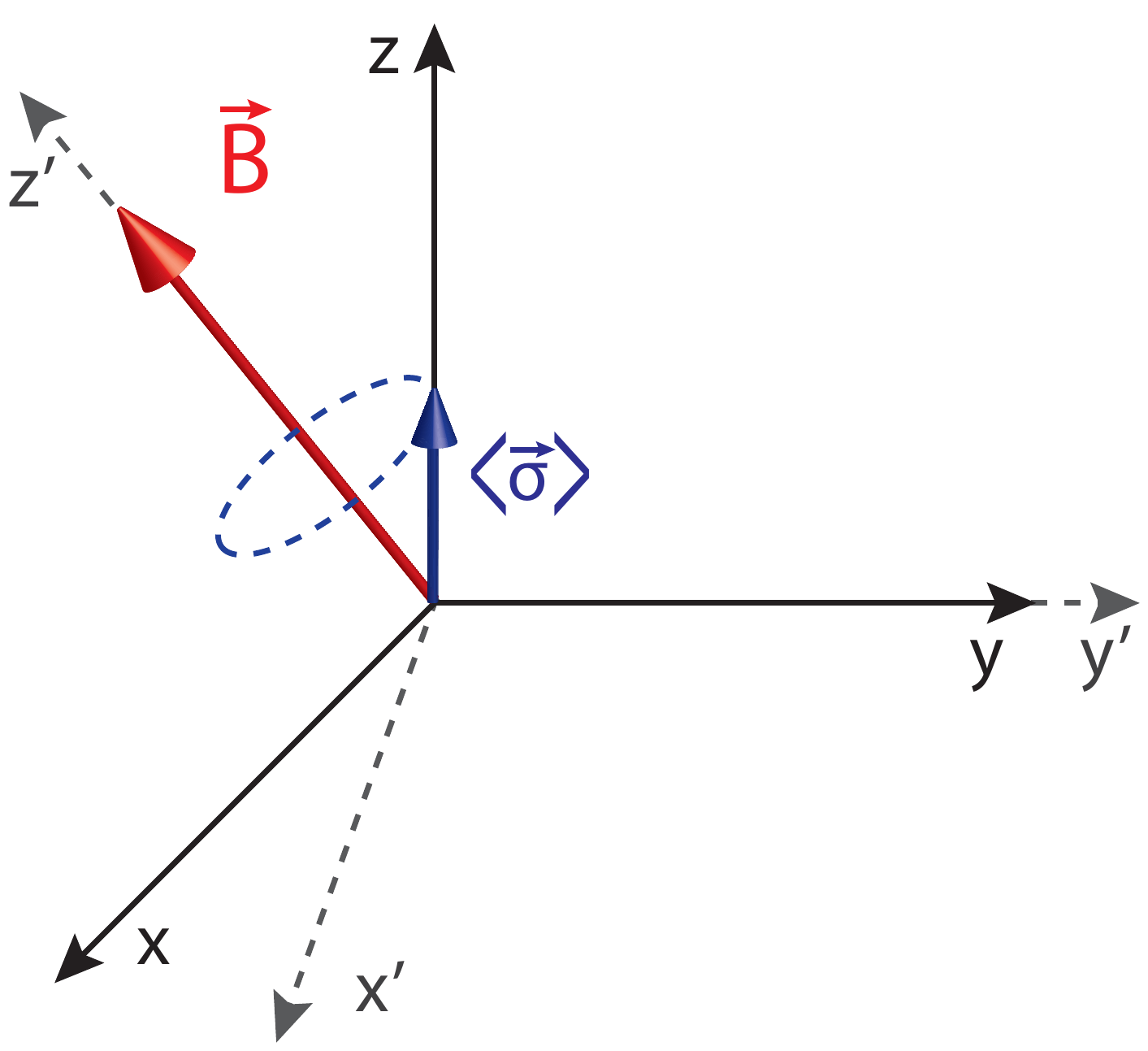}
	\caption{\label{fig:Bfield}(Color online). Precession of the initial state (blue) around the effective magnetic field (red) ${\bf B}$ described by the Hamiltonian  $H_{F}^{(0)}[0] + H_{F}^{(1)}[0] = (B_z - B_\parallel^2/\Omega)\sigma_z -(2 B_\parallel B_z/\Omega)\sigma_x$. The non-primed coordinate frame is the lab-frame, where the dynamics is described by $K_F[0](t)$ and $H_{F}[0]$. The primed coordinates correspond to the frame where the dynamics is governed by $K_\text{eff}(t)$ and $H_\text{eff}$. The transformation between the two is given by the initial kick $K_\text{eff}(0)$. This image assumes that the Floquet gauge is $t_0 = 0$, so that $K_\text{eff}(0)$ is the generator of rotations along the $y$-axis, c.f.~Eq.~\eqref{eq:2LS_Heff}. }
\end{figure} 

\emph{Equivalence of the two descriptions.} Within this example it is easy to understand the difference between the stroboscopic $H_F[t_0]$ and effective $H_\text{eff}$ Hamiltonian.  For simplicity, let us approximate both Hamiltonians to order $\Omega^{-1}$. In the stroboscopic Hamiltonian the Floquet-gauge dependent term represents a small magnetic field of magnitude $-2B_{\parallel}B_z/\Omega$, confined to the $xy$-plane. Its direction with respect to the $x$-axis is determined by the angle $\phi=\Omega t_0$, i.e.~it explicitly depends on the Floquet gauge. In particular, for $t_0=0$, it points along the $x$-axis (see Fig.~\ref{fig:Bfield}). On the contrary, such a term is not present in $H_\text{eff}$, which is explicitly $t_0$-independent. Instead, the gauge dependence is encoded in the kick operator $K_\text{eff}(t)$, which defines the direction of the instantaneous magnetic field, c.f.~Fig.~\ref{fig:Bfield}. 

For instance, if we are interested in stroboscopic dynamics with $t_1 = t_0 =0$ and $t_2=t_1+n T$ (i.e.~we initialize the system at $t_1=0$ and measure observables at the final time $t_2=t_1+nT$), we can either use the stroboscopic Floquet Hamiltonian, which contains a small $x$-magnetic field, or the effective Hamiltonian, whose magnetic field is purely along the $z$-direction. However, in the latter case one has to apply the kick operator $K_\text{eff}(t_0)$ to both the initial state and the measured observables. This kick operator transforms the initial state and the observables into the new coordinate system (see Fig.~\ref{fig:Bfield}). Similar considerations apply to FNS evolution where the stroboscopic and effective descriptions are completely equivalent.

Whenever one is interested in stroboscopic dynamics only, the ME can be preferable to the HFE, as one needs to evaluate only the stroboscopic Floquet Hamiltonian $H_F[t_0]$. Conversely, in the effective description, one has to compute both the effective Hamiltonian $H_\text{eff}$ and the effective kick operator $K_\text{eff}(t)$. If one is interested in FNS dynamics, or in the spectral properties of the Floquet Hamiltonian, then the effective description offers an advantage, since it gives a Hamiltonian which does not contain terms that depend on the phase of the drive. One has to keep in mind, though, that it is crucial to use the proper dressed operators and the dressed initial density matrix in both the stroboscopic and effective description for FNS evolution.

\emph{Adiabatic turning on of the drive.} Finally, we would like to briefly comment on what happens in both descriptions if one turns on the driving adiabatically. Although, this topic is still a subject of active investigation and is beyond the scope of this review, we would like to show numerical results for this simple example, and highlight how one should choose the correct gauge transformation. Let us consider the system to be initially prepared in the ground state $|\downarrow\ \rangle$ of the non-driven Hamiltonian $H_0$ (see Eq.~\eqref{eq:circ_B_field_Hlab} with $B_\parallel=0$). We then slowly turn on the the driving amplitude $B_{\parallel}$, using the ramp:
\[
B_\parallel(t) = B_\parallel^{\text{max}} \times \left\{ \begin{array}{ccc}
0 & \text{for} & t\le0\\
\cos^2\left(\frac{\pi}{2}\frac{t - t_R}{t_R}\right) & \text{for} & 0<t<t_R\\
1 & \text{for} & t\ge t_R
\end{array}\right.
\]
This ramp protocol is chosen because it starts and ends smoothly, i.e.~$\dot{B_\parallel}(t=0)=\dot{B_\parallel}(t=t_R)=0$ and, therefore, it is expected to minimize non-adiabatic effects related to the discontinuities in the velocity during the ramp (see e.g.~Ref.~\cite{degrandi_13}). Here $t_R$ is the ramp time which is taken to be an integer multiple of the driving period, i.e.~$t_R=n T$. It is also convenient to define the ramp rate $v$ as the rate of change of the magnetic field in the middle point of the ramp, i.e.~$v = \dot B_\parallel(t_R/2) = \pi B_\parallel^\text{max}/(2 t_R)$. 

\begin{figure}[ht]
	\centering
	\subfigure{
		\resizebox*{12cm}{!}{\includegraphics[width = 0.9\columnwidth]{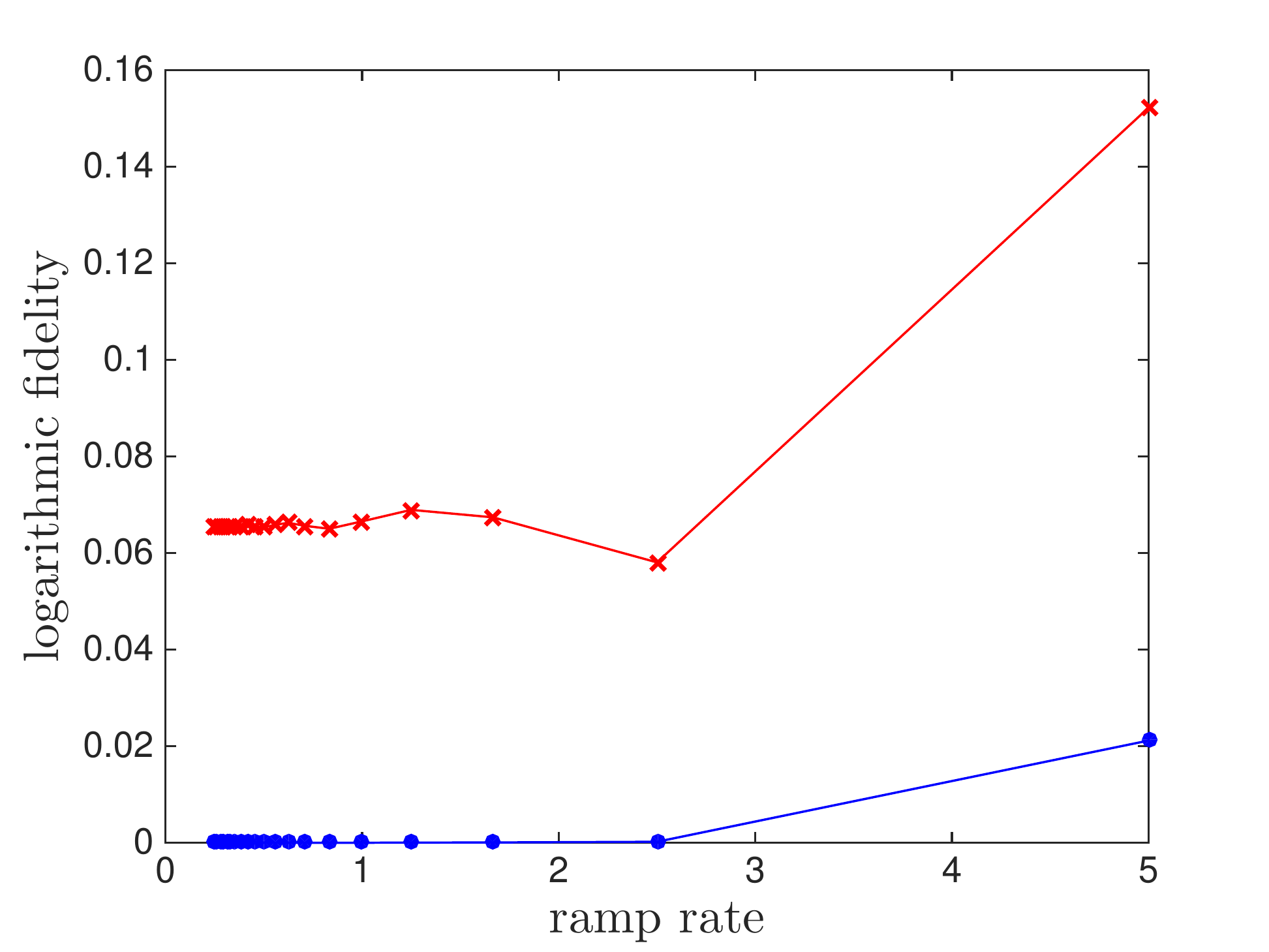}}}
	\caption{\label{fig:2LS_adiabatic}(Color online) Logarithmic fidelity as a function of the ramp rate to find the system in the GS of  the Floquet Hamiltonian $H_{F}[t_R]$, $-\log |\langle\psi(t_R)|\psi_{GS}(H_{F})\rangle|^2$ (blue dots) and the GS of the effective Hamiltonian $H_\text{eff}$, $-\log |\langle\psi(t_R)|\psi_{GS}(H_\text{eff})\rangle|^2$ (red crosses). The ramp rate is $v = \pi B_\parallel^\text{max}/(2 t_R)$. In order to reproduce the blue line in the effective picture, one needs to apply the kick operator at time $t_R$ to rotate the state: $|\psi_{GS}(H_\text{eff})\rangle=e^{i K_\text{eff}(t_R)}\,|\psi_{GS}(H_F[t_R])\rangle$. }
\end{figure}

The question we are interested in is whether the system, initially prepared in the ground state of $H_0$, evolves into the GS of $H_{F}$ or the GS of $H_\text{eff}$. In other words, we would like to know which ground state is adiabatically connected to the ground state of the non-driven Hamiltonian. In Fig.~\ref{fig:2LS_adiabatic}(a) we show the value, at the end of the ramp, of the logarithmic fidelity to find the system in each ground state: $-\log |\langle\psi(t_R)|\psi_{GS}(H_{\text{eff},F})\rangle|^2$. We find that, in the adiabatic limit (for large $t_R$ or equivalently small ramp rate $v$), the system is in the ground state of the stroboscopic Floquet Hamiltonian $H_F[t_R]$ with unity probability, i.e.~the GS of $H_0$ is adiabatically connected to the GS of $H_F[t_R]$. From this fact it immediately follows that the system cannot be in the GS of $H_\text{eff}$ since the two ground states are different:
\[
|\psi_{GS}(H_\text{eff})\rangle=e^{i K_\text{eff}(t_R)}\,|\psi_{GS}(H_F[t_R])\rangle
\]
Therefore, in the effective picture, the effect of the final kick at time $t_R$, which is responsible for a change of basis from the lab frame to the basis of $H_\text{eff}$, cannot be eliminated by the adiabatic ramp. 

This finding also has very simple consequences for FNS dynamics, e.g.~if we are interested in the average over one period of any observable $\mathcal O$ after the ramp. Using the relation between the ground states above and Eq.~\eqref{eq:adiabatic_FNS} we find 
\begin{eqnarray}
{1\over T} \int_{t_R}^{t_R+T}\mathrm dt \langle \mathcal O(t) \rangle&=&{1\over T}\int_{t_R}^{t_R+T}\mathrm dt \langle \psi_{GS}(H_F[t]) |\mathcal O | \psi_{GS}(H_F[t]) \rangle\nonumber\\
&=&\langle \psi_{GS}(H_F[t_R])|\ \overline {\mathcal O}_F[t_R]\ | \psi_{GS}(H_F[t_R]) \rangle\nonumber\\
&=&\langle \psi_{GS}( H_\text{eff})|\ \overline {\mathcal O}_\text{eff}\ |\psi_{GS}( H_\text{eff}) \rangle.
\label{eq:adiabatic_FNS_2LS}
\end{eqnarray}
As discussed in Sec.~\ref{subsec:Floquet_experiment}, if the system is in an eigenstate of the Floquet Hamiltonian (here the ground state), the FNS expectation value of any observable is given by the expectation value of the corresponding dressed observable, computed in the eigenstate of the Floquet Hamiltonian. One can freely choose whether one works in the stroboscopic or the effective pictures.

\subsection{\label{subsec:magnus_resumsubseries_pf} The Inverse Frequency Expansion in the Rotating Frame.}

In some cases, for example when the driving amplitude scales with frequency, one needs to re-sum an infinite sub-series in the ME (HFE) to obtain the proper infinite-frequency stroboscopic (effective) Hamiltonian. For example, let us imagine the simplest protocol 
\be
H(t)=H_0+\Omega\lambda(t) H_1,
\label{h_drive}
\ee
where $\lambda(t)$ is a periodic function with zero mean, whose period is $T$. To emphasize that the amplitude of the driving protocol is proportional to the driving frequency $\Omega$ we made this explicit in Eq.~\eqref{h_drive}. Using Eq.~\eqref{eq:magnus_series} we infer that the $n$-th order term in the inverse-frequency expansion involves a nested commutator containing $n+1$ terms (e.g.~$H_F^{(1)}[t_0]$ has a commutator containing the Hamiltonian $H(t)$ twice, $H^{(2)}_F[t_0]$ has a nested commutator containing $H(t)$ three times, etc.). It then becomes clear that the terms in the $n$-th order of the expansion containing once $H_0$ and $n$-times $H_1$ are all independent of the driving frequency, since the power-law divergence with $\Omega$ is precisely cancelled by the $\Omega$-dependent factor coming from the measure when the time-ordered integral is made dimensionless. On the other hand, the other terms, which contain $H_0$ more than once are subleading and vanish in the limit $T\to 0$ ($\Omega\to\infty$). This may look like a special case, but it is precisely the setup necessary to obtain interesting and counter-intuitive behavior in the high-frequency limit.

One can significantly simplify the analysis if the Hamiltonian can be written in the form~\eqref{h_drive} or more generally as
\[
H(t)=H_0+\Omega\sum_{j=1}^n \lambda_j(t) H_j,
\]
where $\lambda_j(t)$ are periodic functions with the same common period and $H_j$, $j=1,\dots n$, are mutually commuting terms (but not commuting with $H_0$). Notice that since the driving amplitude scales with the driving frequency, it is not immediately clear what the infinite-frequency limit is. In such situations, it is convenient to first make a transformation into a rotating frame (rot frame). Focusing on the Hamiltonian~\eqref{h_drive} we define the rotation operator as
\be
V(t)=\exp\left[-i\Omega \int_{t_0}^t \lambda(t') dt'\, H_1\right]=\exp\left[-i\,F(t)\,H_1\right],
\label{eq:V(t)_general}
\ee
where $F(t)=\int_{t_0}^t \Omega\lambda(t') dt'$ only depends on the driving frequency via its the time-periodic argument. As before, the choice of $t_0$ is the Floquet gauge choice. The transformation to the rotating frame $V(t)$ explicitly depends on the choice $t_0$ and hence the $t_0$-dependent part of it represents a Floquet gauge transformation. We adopt the convention that $V(t)$ transforms from the rotating frame into the lab frame. Then the wave function, the density matrix and the operators transform as
\begin{eqnarray}
|\psi^{\rm rot}(t_1)\rangle &=& V^\dagger(t_1) |\psi(t_1)\rangle,\;\rho^{\rm rot}(t_1)=V^\dagger(t_1) \rho(t_1) V(t_1),\nonumber\\
\mathcal O^{\rm rot}(t_2) &=& V^\dagger(t_2) \mathcal O(t_2) V(t_2).
\label{O_rot}
\end{eqnarray}
The Hamiltonian in the rotating frame acquires an extra Galilean term due to the fact that the transformation is time-dependent:
\be
\begin{split}
	H^{\rm rot}(t)&=V^\dagger (t) \left[ H_0+\Omega\lambda(t) H_1 \right] V(t)-i V^\dagger (t) \partial_t V(t) \\
	&=V^\dagger (t) H_0 V(t).
\end{split}
\ee
Thus, the transformation to the rotating frame removes the oscillating term with a divergent amplitude $H_1$, effectively replacing it by a Hamiltonian with a fast oscillating phase. Note that $H^{\rm rot}(t)$ is a periodic function of time if $\lambda(t)$ has a zero mean (if the mean is nonzero the rot frame Hamiltonian can be still periodic in special cases, e.g.~when the spectrum of $H_1$ is quantized in integers, as will be the case for the realisation of the Harper-Hofstadter Hamiltonian discussed in Sec.~\ref{subsec:harper}, or when we discuss the static and the dynamic Schrieffer-Wolff transformation in Sec.~\ref{subsec:Schrieffer_Wolff}). If $V^\dagger (t) H_0 V(t)$ is a local operator, one can find the evolution in the rotating frame by applying the Magnus (High-Frequency) expansion. But unlike in the original lab frame, there are no more divergent terms in the transformed Hamiltonian. Hence, the infinite-frequency limit is simply determined by the time average of $H^{\rm rot}(t)$, and the $n$-th order corrections in the inverse frequency are precisely given by the $n$-th order of Magnus (High-Frequency) expansion in the rotating frame. Evaluating $H^{\rm rot}(t)$ explicitly is only possible when $V(t)$ is simple. This is the case, for example, when $H_1$ is a single-particle operator. These are precisely the situations, in which one can do a partial resummation of the ME (HFE) in the lab frame. We note in passing that for $F(-t) = F(t)$ the driving protocol in the rot frame is an even function of time, and hence all odd-order corrections in the Magnus expansion in the symmetric gauge vanish identically~\cite{blanes_09}.

It is straightforward to find the relation between the Floquet Hamiltonians as well as the kick operators in the original lab frame and the rot frame. Recall from the general Floquet theory (c.f.~Eq.~\eqref{evolution2}) that in the lab and the rot frames the evolution operator reads as:
\beq
&& U(t_2, t_1)= \mathrm e^{-i K(t_2)} \mathrm e^{-i \hat H_F (t_2-t_1)} \mathrm e^{i K(t_1) } \nonumber \\
&& U^{\rm rot}(t_2, t_1)=\mathrm e^{-i K^\text{rot}(t_2)} \mathrm e^{-i \hat H_F^{\rm rot} (t_2-t_1)} \mathrm e^{i K^\text{rot}(t_1)}.
\eeq
On the other hand, the evolution operators in the two frames are related by
\be
U(t_2, t_1)=V(t_2)U^{\rm rot}(t_2,t_1) V^\dagger (t_1).
\ee
Comparing the three expressions above and noting that $V(t)$ is periodic with period $T$ by construction, we see that:
\be
\mathrm e^{-i K(t)}=V(t) \mathrm e^{-iK_{\rm rot}(t)}=\mathrm e^{-i F(t) H_1} \mathrm e^{-iK_{\rm rot}(t)},\quad\hat H_F=\hat H^{\rm rot}_{F},
\label{P_rot}
\ee
where we used Eq.~\eqref{eq:V(t)_general}. The expression above allows one to transform the kick operator from the lab to the rotating frame using the operator $V(t)$. We can calculate the expansions for the kick operator and the Floquet Hamiltonian directly in the rotating frame 
by replacing $H(t)\rightarrow H^{\rm rot}(t)=V^\dagger (t) H_0 V(t)$ in Eqs.~\eqref{eq:magnus_series}, \eqref{eq:kick_operator_ME}, \eqref{eq:HFE} and \eqref{eq:kick_operator_HFE}.
Using specific examples, we shall illustrate that a successful strategy for finding the Floquet Hamiltonian and the dressed operators is: i) first to perform the transformation to the rotating frame w.r.t.~the driving Hamiltonian in order to remove the terms which diverge with the driving frequency, and ii) then use the ME (HFE) to find the stroboscopic (effective) Floquet Hamiltonian as well as the dressed operators. Finally, iii), (if needed) we return back to the lab frame. 
Going to the rotating frame can offer the same benefits for calculating dressed operators (including the density matrix) as for calculating Floquet Hamiltonians. Namely, if the amplitude of the driving diverges with the frequency, going to the rot frame and evaluating a simple time-average of the corresponding operator (or the density matrix) is equivalent to a re-summation of an infinite sub-series for $\overline{\mathcal{O}}$ in the lab frame. So both for the Hamiltonian and for the dressed observables the ME and HFE are the proper $1/\Omega$ expansions even if the driving amplitude scales with the driving frequency.

Let us also emphasize that the exact dressed operators and the exact dressed density matrix are the same in the lab and in the rotating frames both in the stroboscopic and the effective pictures:
\begin{equation}
\overline{ \rho^{\rm rot}}=\overline{ \rho},\; \overline{ {\mathcal O}^{\rm rot}}=\overline{ \mathcal O}.
\label{eq:dressed_ops_rot_vs_lab}
\end{equation}
Obviously, this is not ture in general for the bare operators and the bare density matrix
\[
\rho^{\rm rot}(t) \ne \rho,\; \mathcal{O}^{\rm rot}(t) \ne \mathcal{O}.
\]
Equation~\eqref{eq:dressed_ops_rot_vs_lab} follows from an observation that $V(t)$ entering the new kick operator (Eq.~\eqref{P_rot}) exactly cancels the corresponding transformation of the operator $\mathcal O$ into the rotating frame (Eq.~\eqref{O_rot}). 

As anticipated above, it is often convenient to compute the the dressed operators and dressed density matrix in the rotating frame where the driving amplitude does not scale with the driving frequency. The leading terms in $\Omega^{-1}$ are given by
\begin{eqnarray}
\overline{\mathcal O}^{\rm rot}&=&{1\over T}\int_0^T dt\, \left( \mathcal O^{\rm rot}(t)-i \left[K^\text{rot,(1)}(t), \mathcal O^{\rm rot}(t)\right] \right)+\mathcal{O}(\Omega^{-2}) \nonumber\\
\overline{ \rho}^{\rm rot}&=&{1\over T}\int_0^T dt \left( \rho^{\rm rot}(t)- i \left[K^\text{rot,(1)}(t), \rho^{\rm rot}(t)\right] \right)+\mathcal{O}(\Omega^{-2})
\label{bar_O_magnus}
\end{eqnarray}
where we recall that, by construction, $K^{\rm rot,(1)}\sim\Omega^{-1}$. If an observable commutes with the operator $H_1$ to which the driving couples, then it is left unchanged by the transformation to the rotating frame, i.e.~the observable $\mathcal O^{\rm rot}(t)$ is time-\emph{in}dependent and equal to the observable in the lab frame $\mathcal O^{\rm rot}(t)=\mathcal O^{\rm lab}$. As a consequence, all time dependence in the integral comes from the kick operator $K^{(1)}(t)$. If the kick operator has a zero average (as it is the case in the effective picture, $K^{(1)}(t) = K^{(1)}_\text{eff}(t)$, c.f.~Eq.~\eqref{eq:kick_operator_HFE}), then the dressed observable does not have a $\Omega^{-1}$-correction. A similar reasoning applies to the density matrix, $\overline{\rho}^{\rm rot}(t)$. Notice, however, that the $\Omega^{-1}$-corrections are in general present if: i) the observables and/or the density matrix do not commute with $H_1$, or ii) if one chooses the stroboscopic picture since, in this case, $K^{(1)}(t) = K^{(1)}_F[t_0](t)$ does not have a zero average, c.f.~Eq~\eqref{eq:kick_operator_ME}. We demonstrate this explicitly in Sec.~\ref{subsec:corr_kapitza} using the example of the Kapitza pendulum.

Finally, the dressed observables in the stroboscopic and effective pictures are related by the transformation
\begin{equation}
\overline{ \mathcal{O}}_F[0]=\mathrm e^{-i K_{\rm eff}(0)}\overline{ \mathcal{O} }_{\rm eff}\ \mathrm e^{i K_{\rm eff}(0)}, \ \ \ \ 
\overline{ \rho}_F[0]=\mathrm e^{-i K_{\rm eff}(0)}\overline{ \rho }_{\rm eff}\ \mathrm e^{i K_{\rm eff}(0)}.
\end{equation}
Expanding these equations to leading order in $\Omega^{-1}$ and using $K_\text{eff}^{(0)} = 0$, we find
\begin{equation}
\overline{\mathcal{O}}_F^{(1)}[t_0] = -i \left[ K_\text{eff}^{(1)}(t_0), \overline{\mathcal{O}}_\text{eff}^{(0)}  \right], \ \ \ \ 
\overline{\rho}_F^{(1)}[t_0] = -i \left[ K_\text{eff}^{(1)}(t_0), \overline{\rho}_\text{eff}^{(0)}  \right].
\label{eq:F_vs_eff_dressed_1st_oder}
\end{equation}

\subsection{\label{subsec:Magnus_convergence} Convergence of the Magnus Expansion.}

In this subsection, we summarise a collection of results about the convergence of the Magnus expansion (ME). To present date, little is known about the convergence of the High-Frequency Expansion (HFE), which is related to the ME by a $\Omega$-dependent unitary transformation. Therefore, in this subsection, we focus on the convergence of the ME. There are arguments in literature that in general the HFE has better convergence properties than ME since it is manifestly $t_0$-invariant~\cite{eckardt_15}, but, to the best of our knowledge, there are no rigorous statements available so far.

As we discussed, the ME is a very powerful tool to compute the Floquet Hamiltonian in the high-frequency limit. However, as it often happens in physics, perturbative expansions can be asymptotic, i.e.~can have a zero radius of convergence. This does not mean that these expansions are useless because they still can give very accurate predictions for the behavior of the system e.g.~for finite evolution times, but eventually such asymptotic expansions inevitably break down. 

In the context of the Magnus expansion, the question of true vs.~asymptotic convergence is ultimately related to the question of heating in the driven system. A convergent ME implies that the Floquet Hamiltonian is a local operator and, thus, the evolution of the system (up to the kick operators) is governed by a local static Hamiltonian so the total energy of the system is conserved~\cite{dalessio_13}. This leads to a dynamical localization transition where the system does not absorb energy from the external drive  even in the infinite-time limit. On the other hand, a divergent Magnus series indicates that there exists no local Floquet Hamiltonian, and the system heats up indefinitely. 

From a mathematical point of view, this issue has been extensively investigated in the literature, and a few different theorems are known (see Ref.~\cite{blanes_09} and references therein). In particular, the Magnus expansion is guaranteed to converge to the Floquet Hamiltonian if:
\be
\int_0^T dt\, |\epsilon_\text{max}(t)-\epsilon_\text{min}(t)| <  \,\xi
\label{eq:convergence_general}
\ee
where $\epsilon_\text{max}(t)$, $\epsilon_\text{min}(t)$ are the largest and smallest eigenvalues of the Hamiltonian $H(t)$, and $\xi$ is a number of order one. 

While this result is exact, it is not particularly useful for many-particle systems. It only guarantees the convergence if the driving frequency scales with the system size, while the relevant time scales, separating the fast and slow driving regimes, are never extensive. This condition for the convergence of the Magnus expansion is only sufficient. It does not give much insight about what happens at longer periods. To the best of our knowledge, there are no rigorous results about the convergence of the ME for interacting systems in the thermodynamic limit, and it is unknown whether in this case the radius of convergence of the ME is zero or finite. The Magnus expansion can be definitely convergent even in the thermodynamic limit, if the time-dependent Hamiltonian can be mapped to a static one, by going to some rotating frame. In Sec.~\ref{sec:ME_RWA} we discuss such situations in the context of the Schrieffer-Wolff transformation and show that in this case the ME (or more accurately the HFE) reproduces the conventional static perturbation theory, which is known to converge.

In more generic situations, where a local transformation to the constant Hamiltonian does not exist, the situation is not well understood. In Refs.~\cite{prosen_98a, prosen_99, dalessio_13, citro_15},
a numerical evidence indicated that for particular driving protocols in one-dimensional fermionic or spin chains, the radius of convergence of the Magnus expansion is finite even in the thermodynamic limit. In other words, there exists a critical period $T_c$ separating regimes of finite and infinite heating. At the critical period there is a dynamical transition between these two regimes, which can be interpreted as a many-body localization transition~\cite{altshuler_97,gornyi_05,basko_06,oganesyan_07,pal_10,imbrie_14} in energy space. This finding is also consistent with previous numerical results obtained for periodically kicked spinless fermions in one dimension~\cite{prosen_98a,prosen_98b,prosen_99} equivalent to a periodically kicked XXZ spin chain. In these works two qualitatively different regimes were found. In the first one the evolution is well described by random matrices from the circular ensemble (see also Ref.~\cite{dalessio_14}) strongly suggesting that the ME is divergent, while in the other regime the system displays features consistent with the ME being convergent to a local Hamiltonian.

At the same time a numerical study of a different driving protocol in a spin chain indicated a zero radius of convergence~\cite{dalessio_14}, i.e.~$T_c=0$ in the thermodynamic limit. In Ref.~\cite{ponte_14}, using the Eigenstate Thermalization Hypothesis, it was argued that an ergodic system with a local driving term always heats up to infinite temperature in the thermodynamic limit, while the energy can stay localized (and thus the Magnus expansion converges) if the system is in the many-body localized phase, i.e.~non-ergodic. In Ref.~\cite{heyl_10} it was shown that the Magnus expansion has zero radius of convergence for a Kondo model if the driving frequency is smaller than the bandwidth of the conduction electrons, though for faster driving the numerical results seem to indicate convergence of the Magnus expansion~\cite{heyl_private}. There is no contradiction with Ref.~\cite{ponte_14} because in the Kondo model the conduction band electrons were considered non-interacting (i.e.~non-ergodic). 

The Magnus expansion can be rigorously shown to have a finite radius of convergence for integrable systems, which can be factorized into uncoupled sectors, e.g.~in momentum space. Then the extensivity of the system is not important and the criterion~\eqref{eq:convergence_general} can be applied to each sector independently. Such systems do not heat up indefinitely and, in the long-time limit, effectively reach a steady state with respect to the Floquet Hamiltonian~\cite{russomanno_12, lazarides_PRL_14}.

From a physical point of view, the divergence of the ME, and the corresponding heating of the system can be traced back to the existence of resonances~\cite{eckardt_08}. When the frequency of the drive $\Omega$ matches a single particle energy scale $J$, i.e.~$\Omega \approx J$, the system can efficiently absorb energy from the periodic drive leading to fast heating. Here $J$ can represent, for example, the energy associated with a single spin flip in a spin system, and then the process described above corresponds to the absorption of a photon of the driving field and a subsequent spin flip in the system. When the driving frequency is increased, $\Omega\gg J$, the photon energy can be absorbed, only if many spins are flipped simultaneously. These \textit{many-body} processes are described by higher-order perturbation theory and, therefore, occur with small probabilities. Hence, they can become important only at very long times. The same large-energy absorption processes determine whether in the \textit{off-resonance} regime, i.e.~$\Omega\gg J$, the heating is slow and finite, or completely absent.  It is unclear at the moment whether the heating can be understood through including more and more terms in the Magnus expansion, or if it is a non-perturbative phenomenon in the driving frequency.

Even in the situations, where the ME formally diverges, the heating remains slow at fast driving frequencies~\cite{maricq_82} . Then the Magnus expansion truncated to the first few orders can accurately describe the transient dynamics of the system for many periods of oscillations. In particular, in Ref.~\cite{maricq_82} it was shown that, for a dipolar-coupled periodically driven spin systems, the magnetization quickly approaches a quasi-stationary value predicted by the Magnus expansion truncated at second order. Then, at much longer times, the magnetization decays to zero due to slow heating processes which are not captured by the ME. Therefore, in this context, an important question is not whether the Magnus expansion has a finite radius of convergence or only asymptotic, but whether there is a time-scale separation between interesting transient dynamics described by the local Floquet Hamiltonian truncated to some low order, and heating phenomena at longer times. While this issue is also not very well understood in general, there is sufficient evidence that such a time scale separation always exists at high driving frequencies~\cite{mori_14}. For this reason, in Secs.~\ref{subsec:harper}, \ref{subsec:Floquet_top_ins} and \ref{spins}, the approximate Floquet Hamiltonians obtained by truncating the Magnus expansion to order $\Omega^{-1}$ are, at the very least, expected to describe the transient dynamics and the relaxation to a quasi-steady state.

A natural way to prevent infinite heating is to couple the driven system to a thermal bath. In this case it is expected that the system will eventually approach a non-equilibrium steady state in which the energy absorbed from the driving is balanced by the energy dissipated into the environment~\cite{breuer_00,ketzmerick_10,vorberg_13,iadecola_14}. The value of measurable quantities (such as transport coefficients and correlation functions) will depend crucially on the nature of the (putative) non-equilibrium steady state~\cite{oka_09,kitagawa_11,fregoso_14} which, for this reason, has been the focus of intense research~\cite{langemeyer_14,dehghani_14}. Despite this intense effort, a general understanding of the non-equilibrium steady state is still missing but it seems clear that the steady state will, in general, be non-thermal~\cite{iadecola_14,iadecola_15,seetharam_15}. Therefore the thermodynamic behavior of periodically driven systems is expected to be qualitatively different from those of non-driven systems~\cite{kohn_01,hone_09}.


\section{\label{sec:ME_RWA}The Rotating Wave Approximation and the Schrieffer-Wolff Transformation.}

In this section we show that the Magnus (High-Frequency) expansion can be used even in static, i.e.~non-driven, systems by first going into the interaction picture (see for example Refs.~\cite{parra-murillo_PRA_13} and \cite{parra-murillo_13}). In particular, it can be used to eliminate highly excited states, which are never populated but nevertheless lead to renormalization and modification (dressing) of the low-energy Hamiltonian. These ideas are also behind the widely used Schrieffer-Wolff transformation~\cite{schrieffer_66} which, as we shall show, is closely related to the High-Frequency (Magnus) expansion\footnote{We are grateful to A.~Rosch for pointing this out.}. The formalism introduced in Sec.~\ref{sec:magnus_rotframe} can be applied to find the leading behavior and the first subleading correction to the Floquet Hamiltonian, the dressed operators and the density matrix. As we illustrate below, this framework has an additional advantage allowing one to extend the Schrieffer-Wolff transformation to periodically driven systems.

\subsection{\label{subsec:dble_well} A Two-level System.}

\begin{figure}
	\centering
	\includegraphics[width = 0.55\columnwidth]{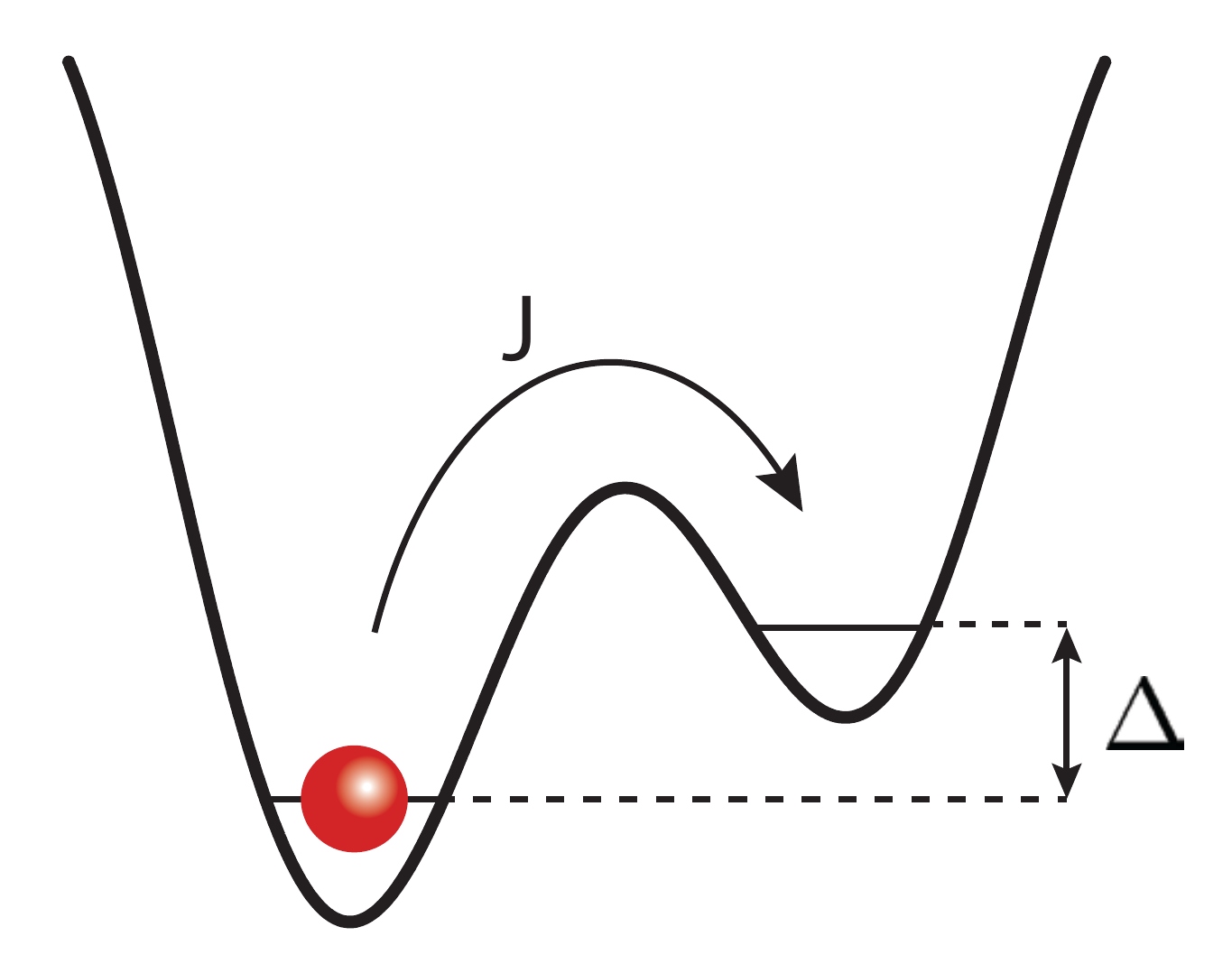}
	\caption{\label{fig:dblewell}(Color online) A single particle in a tilted double well: the two wells have an energy difference $\Delta$ and are connected by the matrix element $J$.}
\end{figure} 

\emph{Non-driven two-level system.} To understand the relation between the Magnus (High-Frequency) expansion and time-independent perturbation theory, consider first a time-independent problem of a single particle hopping in a tilted double well potential, c.f.~Fig.~\ref{fig:dblewell}. This model is exactly solvable and, with the correct identification of the lab and rotating frames, it is equivalent to the two-level system in a circularly driven magnetic field described in Sec.~\ref{subsec:foquet_circ_pol_drive}. Here we revisit this model paying a specific attention to the convergence properties of the Magnus (High-Frequency) expansion. The system is described by the Hamiltonian
\begin{equation}
H = -J\left(d^\dagger_2 d_1 + d^\dagger_1 d_2 \right) + \Delta n_2,
\label{eq:ham_tilt}
\end{equation}
where the operator $d^\dagger_m$ creates a particle on site $m$ and $n_m = 
d^\dagger_md_m$ is the particle number operator. The tilt is given 
by the parameter $\Delta$, while the hopping matrix element is denoted by $J$. We are interested in the limit $\Delta\gg J$. The exact single-particle eigenenergies of this Hamiltonian are $E_{\pm} = 1/2\left(\Delta \pm\sqrt{\Delta^2 + 4J^2}\right)$.

The Hamiltonian in Eq.~\eqref{eq:ham_tilt} does not have any explicit time 
dependence, let alone a periodic one. In order to make use of the Magnus (High-Frequency) expansion, we 
apply a unitary transformation into the interaction picture w.r.t.~the diagonal 
part $H_1 = \Delta n_2$, given by $V(t) = 
\text{diag}\left(1,\exp(-i\Delta t)\right)$ in the Fock basis. We thus obtain a time-dependent Hamiltonian in the rotating frame (interaction picture): 
\begin{equation}
H^\text{rot}(t) = -J\left(\mathrm e^{i\Delta t}d^\dagger_2 d_1 + \mathrm e^{-i\Delta 
	t}d^\dagger_1 d_2 \right),
\label{eq:H_two_level_rot}
\end{equation} 
which is similar to Eq.~\eqref{eq:circ_B_field_Hlab} with the identification $B_z\rightarrow 0$, $B_\parallel\rightarrow -J$, and $\Delta\to\Omega$, whenever the system is populated by a single particle. Notice that $H^\text{rot}(t)$ contains only the harmonics $H_{l=\pm1}$, see Eq.~\eqref{eq:H(t)_harmonics}. Therefore, for this example the Magnus expansion coincides with the High-Frequency expansion to order $\Omega^{-1}$, c.f. Eqs.~\eqref{eq:magnus_series} and \eqref{eq:HFE}. In the following, we choose the Floquet gauge $t_0=0$.

Observe that by doing the transformation to the interaction picture we eliminate the high-energy level from the problem at the expense of introducing an explicit time dependence. This transformation is identical to the gauge transformation in electromagnetism, where a static scalar potential can be traded for a linear in time vector potential. Now we can apply Floquet theory to the Hamiltonian in Eq.~\eqref{eq:H_two_level_rot}. Using Eqs.~\eqref{transf} and \eqref{eq:2LS_HF} we find that the full time-independent Floquet Hamiltonian coincides with the original Hamiltonian, i.e.~$H_F[0]=H$, as expected. Moreover, from Eqs.~\eqref{eq:kick_p} and \eqref{P_rot} we also see that the fast-motion operator is $P(t,0)=e^{-i K_F[0](t)}=V^\dagger(t)$. This implies that, if we are interested in the time evolution at scales slower than $1/\Delta$, we can compute the dressed density matrix for the initial state and the dressed operator for the observable of interest and evolve them in time with the Floquet Hamiltonian. If we are interested in the high-frequency (i.e.~large $\Delta$) structure of the dynamics, we can fully recover it from the operator $P(t,0)$.

For the Hamiltonian in Eq.~(\ref{eq:H_two_level_rot}) the leading few orders in the Magnus (High-Frequency) expansion result in
\begin{eqnarray}
H_F^{(0)} &=&\ 0, \ \ \ \ \ \ \ \ \ \ \ \ \ \ \ \ \ \ \ \ \ \ \ \ \ \  \ \ \ H_\text{eff}^{(0)} = 0,        \nonumber\\
H_F^{(1)}[0] &=&\ \frac{J^2}{\Delta}(n_2 - n_1), \ \ \ \ \ \ \ \ \ \ \ \ \ \  H_\text{eff}^{(1)} = \frac{J^2}{\Delta}(n_2 - n_1),               \nonumber\\
H_F^{(2)}[0] &=&\ 2\frac{J^3}{\Delta^2}\left(d^\dagger_2 d_1 + d^\dagger_1 d_2 \right)  , \ \ \ \  H_\text{eff}^{(2)} = 0,
\label{eq:tilt_corr}
\end{eqnarray}
with the following kick operators in the rotating frame:
\begin{eqnarray}
K_F^{\text{rot},(1)}[0](t) &=& -\frac{J}{i\Delta}\left( (\mathrm e^{i\Delta t} - 1) d^\dagger_2 d_1 - (\mathrm e^{-i\Delta t} - 1) d^\dagger_1 d_2 \right),\nonumber\\
K_\text{eff}^{\text{rot},(1)}(t) &=& -\frac{J}{i\Delta}\left( \mathrm e^{i\Delta t} d^\dagger_2 d_1 - \mathrm e^{-i\Delta t} d^\dagger_1 d_2 \right).
\end{eqnarray}
It is easy to see that the stroboscopic Floquet Hamiltonian up to order $\Omega^{-2}$, i.e.~$H_F^{(0)}+H_F^{(1)}[0]+H_F^{(2)}[0]$, is equivalent to the original static Hamiltonian when one rescales the original couplings $\Delta$ and $J$ by a factor of $ 2 J^2/\Delta^2$. In fact, for this simple problem one can re-sum the entire series to obtain the Hamiltonian~\eqref{eq:ham_tilt}, i.e.~$H=\sum_{n=0}^\infty H_F^{(n)}[0]$. However, for more complicated Hamiltonians, the Magnus series is not guaranteed to converge (c.f.~Sec.~\ref{subsec:Magnus_convergence}). As was the case for a two-level system in the circularly-polarized field (see Sec.~\ref{subsec:ME_vs_FHE_2LS}), the effective and stroboscopic Floquet Hamiltonian correctly reproduces the exact spectrum to the order of $1/\Delta^2$ (recall that $\Delta=\Omega$). The effective and stroboscopic Floquet Hamiltonians differ at order $\Delta^{-2}$, i.e.~$H_F^{(2)}[0]\ne H_\text{eff}^{(2)}$ but this difference becomes manifest in the spectrum to order $\Delta^{-3}$, i.e.~beyond the validity of the approximation. The hopping (mixing) between the two levels is encoded into $H_F^{(2)}[0]$ but it is absent in $H_\text{eff}^{(2)}$. However, the kick operator $K_\text{eff}^{\text{rot},(1)}(t)$ precisely compensates for this and introduces hopping (mixing) between the two levels. Hence, in order to describe the dynamics (FS or FNS) of the system using $H_{\rm eff}$, one has to take into account the transformation of orbitals encoded in the kick operator. Conversely, the stroboscopic Hamiltonian correctly describes the full evolution of the system if we are interested in the stroboscopic times $nT$ since the stroboscopic kicks vanish at those times, i.e.~$K_F[0](nT)=0$.

Let us briefly comment on the physical meaning of the different terms in the Hamiltonian. In the leading approximation, the Floquet Hamiltonian is zero, which indicates that, in the infinite-frequency limit i.e.~when the energy offset of the two wells is larger than the hopping ($\Delta\gg J$), the system remains frozen since the two levels are effectively uncoupled. The first correction is responsible for the (opposite) energy shifts of the ground and the excited states and introduces a level repulsion. The second correction, in turn, leads to renormalisation of the eigenstates, since it represents a hopping (mixing) between the two levels.

\begin{figure}
	\centering
	\includegraphics[width = 1\columnwidth]{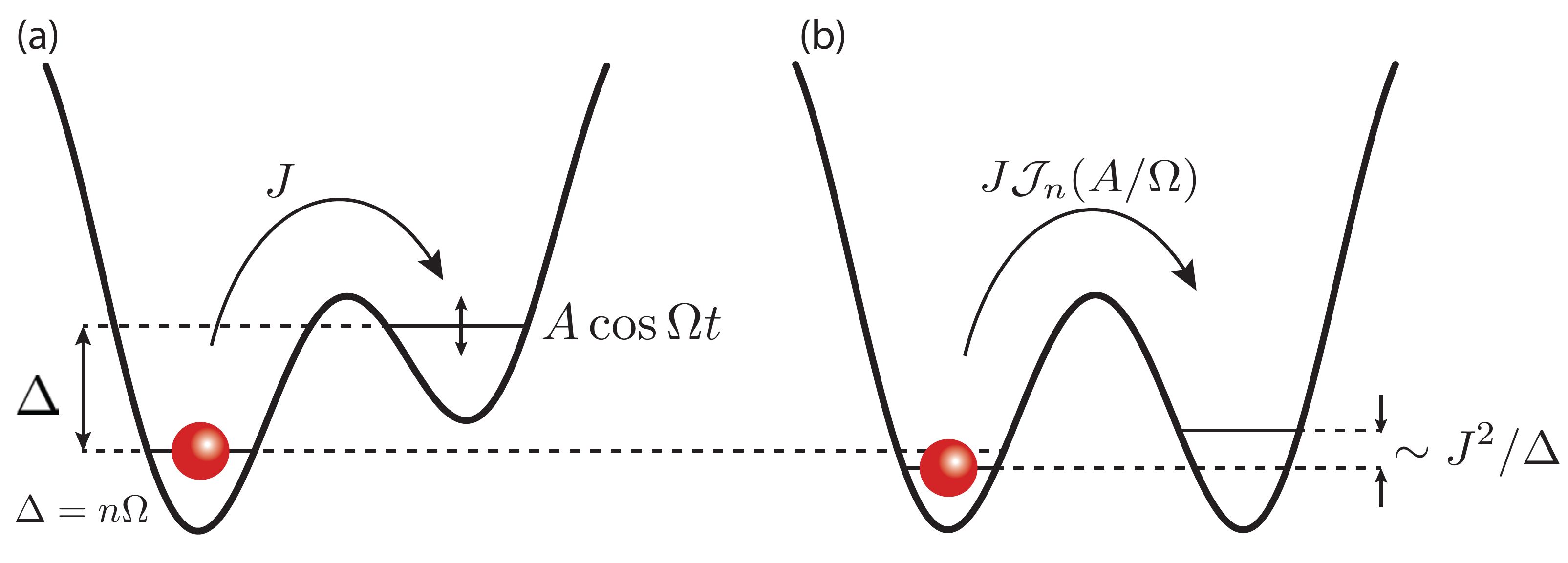}
	\caption{\label{fig:drivendblewell}(Color online) The periodically driven two-level system. (a) The higher energy level is modulated periodically w.r.t.~the lower level in the lab frame. (b) As a result, the Floquet Hamiltonian governing the slow dynamics features a mixing term between the two levels to leading order (absent in the non-driven system), while the first $\Omega^{-1}$-correction has the physical meaning of a small level repulsion.  }
\end{figure}

\emph{Periodically Driven Two-Level System.} The application of the Magnus (High-Frequency) expansion in the previous example might not be the easiest way to study this simple system, but it paves the way towards studying the behavior of a 
more complicated system containing both a high-energy level and a periodic driving. Suppose that we now shake the higher-energy level with an amplitude $A$ and a frequency $\Omega$, c.f.~Fig.~\ref{fig:drivendblewell}. The Hamiltonian becomes

\begin{equation}
H = -J\left(d^\dagger_2 d_1 + d^\dagger_1 d_2 \right) + (\Delta - A\cos\Omega t\; ) n_2.
\label{eq:2LS_driven}
\end{equation}
We are interested in the limit $\Delta,\Omega\gg J$. Our first goal is to understand how the drive changes the physics compared to the non-driven case. Extending the procedure from the static example to the driven case above, we eliminate both the higher-energy level and the driving term altogether by a transformation to a rotating frame:
\begin{eqnarray}
V(t) &=& \exp\left[-i\left(\Delta t - \frac{A}{\Omega}\sin\Omega t\;  \right) n_2 \right],\nonumber\\
d_2 &\to& \mathrm e^{i\Delta t - i \zeta\sin\Omega t\; }d_2 ,\ \ \ d_1\to d_1,
\label{eq:V(t)_2LS_driven}
\end{eqnarray}
where we defined $\zeta =  A/\Omega$. The Hamiltonian in the rotating frame is given by
\begin{equation}
H^\text{rot}(t) = -J\left(\mathrm e^{i\Delta t - i\zeta\sin\Omega t\;  }d^\dagger_2 d_1 + \mathrm e^{-i\Delta 
	t + i \zeta\sin\Omega t\; }d^\dagger_1 d_2 \right).
\label{eq:H_2LS_driven_rot}
\end{equation}
Note that, as before, in the rotating frame there is no energy offset between the two levels, but the hopping term has a more complex time dependence encoding both static and dynamic information.

In general, the new Hamiltonian~(\ref{eq:H_2LS_driven_rot}) is not strictly periodic in time, since $\Delta$ and $\Omega$ are arbitrary real numbers. One can deal with this in several ways. One possibility is to find co-prime integers $n$ and $m$ and a frequency $\Omega_0$ such that $n\Omega_0\approx \Delta$ and $m\Omega_0\approx\Omega$. As long as $\Omega_0\gg J$ the Floquet analysis should hold. If the frequencies are not exactly commensurate then one can define a commensurate $\tilde\Delta=n\Omega_0$ and make the transformation to the rotating frame using $\tilde \Delta$ instead of $\Delta$ in Eq.~(\ref{eq:V(t)_2LS_driven}). It is easy to see that this will result in a small extra static term $(\Delta-\tilde\Delta)d_2^\dagger d_2$ in the rot-frame Hamiltonian~(\ref{eq:H_2LS_driven_rot}). And finally, one can take the commensurate limit, e.g.~$\Delta=n\Omega$ and make an analytic continuation in the final result to non-integer $n$. This should work if the result, e.g.~the  Floquet Hamiltonian is a simple analytic function of $n$. We shall show below that this trick works indeed in the case of the Schrieffer-Wolff transformation. It is intuitively clear that especially in many-particle systems with continuous spectra the exact commensurability of the driving should not play a crucial role.

We now leave all these subtleties aside and assume that $\Delta$ and $\Omega$ are commensurate such that $\Delta=n\Omega_0$ and $\Omega=m\Omega_0$, where $n$ and $m$ are positive co-prime integers. We shall also assume that $\Omega_0\gg J$. First, let us understand the leading time averaged Hamiltonian $H_F^{(0)}$, which was strictly zero in the non-driven case. We note the mathematical identity:
\begin{equation}
{1\over T_0} \int_0^{T_0} dt \mathrm \mathrm e^{i\Delta t - i\zeta\sin\Omega t\;  } = \sum_{l\in\mathbb{Z}}\mathcal{J}_l(\zeta)\frac{1}{T_0}\int_0^{T_0} dt \mathrm e^{i(-l m +n)\Omega_0 t},
\label{eq:driven_2LS_averaged_drive}
\end{equation}
where $\mathcal{J}_l(\zeta)$ is the $l$-th Bessel function of the first kind and $T_0=2\pi/\Omega_0$ is the common period. The integral above is nonzero if and only if there is a solution to the equation $lm=n$, or in other words $n$ is a multiple integer of $m$. Because by assumption $m$ and $n$ are co-prime this equality can only be satisfied when $m=1$ (and hence $\Omega_0=\Omega$ and $\Delta=n\Omega$). This means that the leading Floquet Hamiltonian $H_F^{(0)}$ simply corresponds to the $n$-photon resonance. Let us focus on this resonant scenario. Clearly in this case
\begin{equation}
H_F^{(0)} = H_\text{eff}^{(0)} = -J\mathcal{J}_n(\zeta)\left(d^\dagger_2 d_1 + d^\dagger_1 d_2 \right).
\end{equation}  
In the infinite-frequency limit (at fixed $n$) $H_F^{(0)}$ determines the  Floquet Hamiltonian. It splits the two levels into the symmetric and antisymmetric combinations. This is very different from the non-driven or non-resonantly driven case, where $H_F^{(0)}=0$ and the leading order contribution in the Magnus (High-Frequency) expansion, i.e.~$H_F^{(1)}[0]$, gives the energy splitting between the levels (c.f.~$H_F^{(1)}$ in Eq.~\eqref{eq:tilt_corr}) and hence keeps the eigenstates essentially unmixed (up to a small correction of the order $J/\Delta$). This observation already hints toward possible heating mechanisms in the Floquet system. For example, if one prepares the two-level system in the lower-energy state then in the resonant case this state is equally projected on the symmetric and antisymmetric Floquet eigenstates, resulting in an equal population of the two levels. This is equivalent to heating to an infinite-temperature state. In the non-resonant case, conversely, the Floquet eigenstates are still predominantly the eigenstates of the non-driven Hamiltonian, and thus the initial state is only weakly perturbed, while the excited state is only weakly populated. Admittedly, this example is too simple to understand real heating mechanisms in more complex interacting systems, but it shows how resonant periodic driving can fundamentally change the nature of the Floquet eigenstates~\cite{mori_14}.
  
If the amplitude of the driving is small, then $\zeta\ll 1$ and we find $\mathcal{J}_n(\zeta)\sim\zeta^n$ such that the effective hopping is proportional to the $n$-th power of the driving amplitude. This is not surprising, since it means that the $n$-photon absorption processes are exponentially suppressed. This result can also be obtained using time-dependent perturbation theory. However, for $\zeta\sim 1$, i.e.~in the strong-coupling regime, this term 
becomes non-perturbative and the multi-photon absorption processes are not suppressed. 

For completeness, we give the leading correction term in the Magnus and the High-Frequency expansions for this resonant case:
\begin{eqnarray}
H_F^{(1)}[0] &=&  \frac{J^2}{\Delta} g_n(\zeta)(n_2-n_1),\nonumber\\
g_n(\zeta) &=& -n\,\text{Im}\left\{\int_0^{2\pi}\frac{\mathrm{d}\tau_1}{2\pi}\int_0^{\tau_1}\mathrm{d}\tau_2 \text{e}^{ -in(\tau_1-\tau_2) + i\zeta\left(\sin\tau_1 -\sin\tau_2\right) } \right\},\nonumber\\
H_\text{eff}^{(1)} &=&  \frac{J^2}{\Delta} \tilde g_n(\zeta)(n_2-n_1),\nonumber\\
\tilde g_n(\zeta) &=& -n\,\text{Im}\left\{\int_0^{2\pi}\frac{\mathrm{d}\tau_1}{2\pi}\int_0^{\tau_1}\mathrm{d}\tau_2 \left( 1 - \frac{\tau_1-\tau_2}{\pi} \right)  \text{e}^{ -in(\tau_1-\tau_2) + i\zeta\left(\sin\tau_1 -\sin\tau_2\right) } \right\}.
\label{eq:ME_RWA_SW_gtilde_m}
\end{eqnarray}
Similarly to the non-driven case this correction gives the level repulsion term, but with a renormalized coefficient. If the driving is not very strong, $\zeta\lesssim 1$, then $g_n(\zeta),\tilde g_n(\zeta)\approx 1$ for all $n\ne1$, i.e.~the presence of the driving results in a small modification of the non-driven Floquet Hamiltonian. For $n=1$ the functions $g_1(\zeta),\tilde g_1(\zeta)$ are small, oscillate and even become zero at special values of $\zeta$. This indicates that the driving can have a strong effect if $n=1$. We want to emphasize again that the $\Omega^{-1}$-term is now only a sub-dominant correction, provided that $\mathcal{J}_n(\zeta)\gg J/\Delta=J/(n\Omega)$. For completeness we also show leading order approximation for the kick operators:
\begin{eqnarray}
H_l &=& -J\left(\mathcal{J}_{n-l}(\zeta)d^\dagger_2d_1 + \mathcal{J}_{n+l}(\zeta)d^\dagger_1d_2\right),\nonumber\\
K_F^{\text{rot},(1)}[0](t) &=& \frac{1}{i\Omega}\sum_{l=1}^\infty \frac{1}{l}\left( (\mathrm e^{il\Omega t}-1) H_l - (\mathrm e^{-il\Omega t}-1) H_{-l}\right),\nonumber\\
K_\text{eff}^{\text{rot},(1)}(t) &=& \frac{1}{i\Omega}\sum_{l=1}^\infty \frac{1}{l}\left( \mathrm e^{il\Omega t} H_l - \mathrm e^{-il\Omega t} H_{-l}\right).
\end{eqnarray}

\subsection{\label{subsec:Schrieffer_Wolff} The High-Frequency Expansion vs.~the Schrieffer-Wolff Transformation.}

The Schrieffer-Wolff (SW) transformation is a standard way to eliminate high energy states in static systems. The SW transformation is a perturbative unitary rotation of the Hamiltonian, which eliminates the coupling to the high-energy states at the expense of creating effective low-energy terms. A famous example of the SW transformation, where it was first applied, is the reduction of the Anderson impurity model to the Kondo model~\cite{schrieffer_66}. In this section we show that the SW transformation is essentially equivalent to the High-Frequency (Magnus) expansion. Namely, the SW transformation coincides with the HFE at least up to the second order, and agrees with the ME up to a static gauge transformation. An approach similar to the one described here has been used to study topological Floquet bands in Ref.~\cite{nakagawa_14}. Following the simple case of a two-level system, we shall show how one can extend the SW transformation to periodically-driven systems. To make the discussion more transparent we shall analyze three specific examples of increasing complexity. First, we describe the non-driven non-interacting system, then a driven non-interacting one, and finally an interacting, non-driven one. The driven interacting system is very similar to the driven non-interacting one, and we only comment on the result.

\begin{figure}
	\centering
	\includegraphics[width = 0.7\columnwidth]{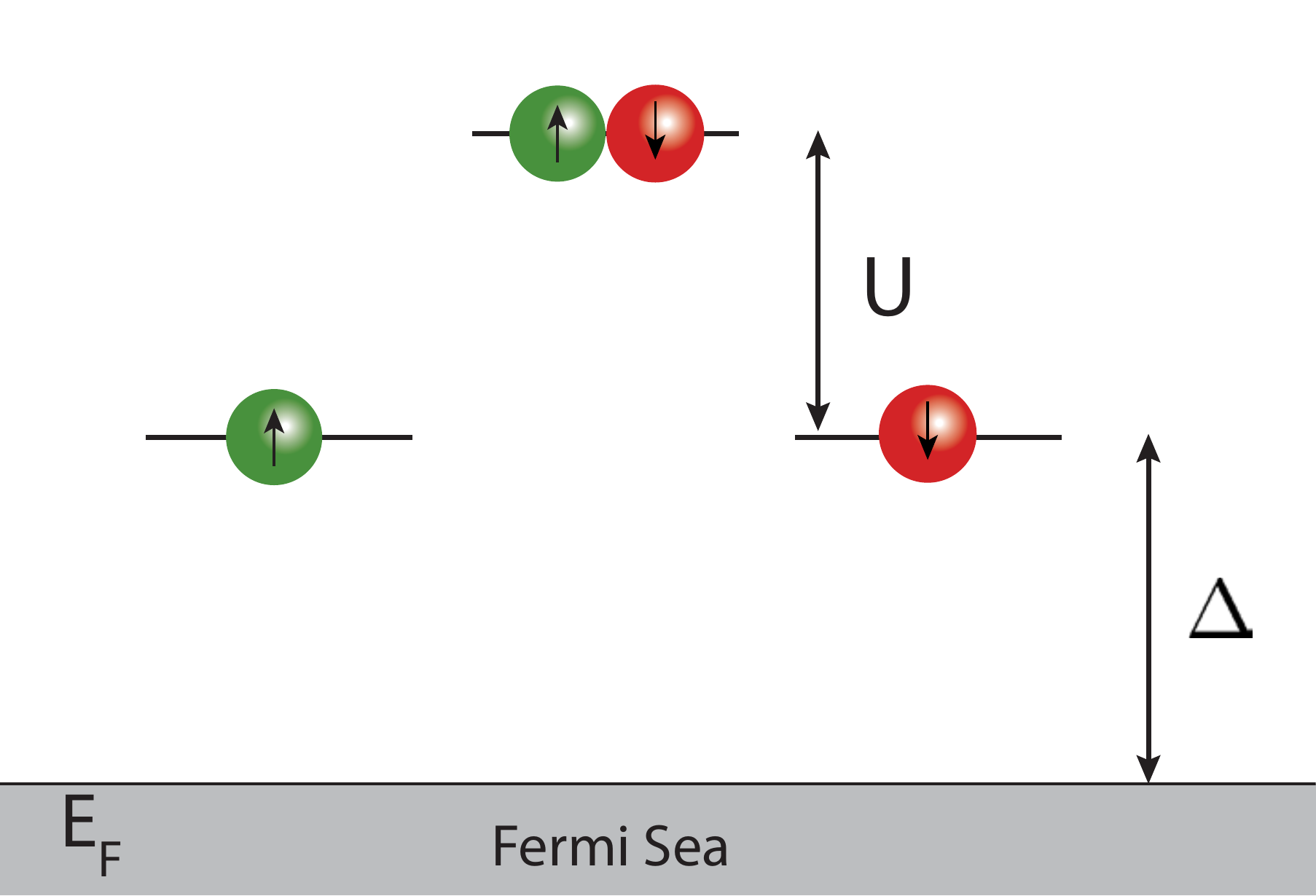}
	\caption{\label{fig:anderson}(Color online) The Anderson Model: spinful fermions can occupy an impurity level, separated from the Fermi sea by an energy $\Delta$. The coupling between the impurity atom and the conducting electrons is $V_d$ (not shown). Moreover, in the presence of interactions, an additional energy cost $U$ has to be paid for the double occupancy of the impurity atom.}
\end{figure}

\emph{A non-interacting impurity coupled to a conducting band.} We begin by generalizing the simple two-level setup. Consider the non-interacting Anderson model (also called the Resonant Level model or the Friedrichs model~\cite{friedrichs_48} describing a single impurity coupled to free electrons (cf.~Fig~\ref{fig:anderson}): 
\begin{eqnarray}
H &=&\ H_0 + H_1,\nonumber\\
H_0 &=&\ \sum_{k,s}\epsilon_k n_{ks} + \Delta\sum_s  n_{ds}, \nonumber\\
H_1 &=& \frac{1}{\sqrt{L^D} }\sum_{ks}\left(V_{d}c^\dagger_{ks}d_{s} + \text{h.c.}\right).
\label{eq:Anderson_model_H}
\end{eqnarray}
Here $d$ refers to the impurity atom with an energy $\Delta$, $s=\uparrow,\ \downarrow$ is the spin index, $\epsilon_k\geq 0$ is the dispersion of the band electrons, $V_d$ is the hybridization strength, $L$ is the linear system size, and $D$ is the dimensionality of the system. The prefactor $1/\sqrt{L^D}$ ensures that in real space the coupling to the impurity $V_d$ is independent of the system size. The fermionic creation and annihilation operators obey the canonical commutation relations $\{c_{ks},c^\dagger_{k's'}\}_+ = \delta_{kk'}\delta_{ss'}$, and $\{ d_s,d^\dagger_{s'} \}_+ = \delta_{ss'}$. As usual, $n_{ks} = c^\dagger_{ks}c_{ks}$ and $n_{ds} = d^\dagger_s d_s$ are the number operators. We are interested in the situation where $\Delta$ is the largest energy scale in the system: $\Delta\gg V_d, \epsilon_k$, and the coupling $V_{d}$ between the conducting band and the impurity is small (compared to the Fermi energy). In this limit, the impurity can only be occupied by virtual processes, which effectively dress the low-energy conduction band electrons.

As Schrieffer and Wolff pointed out~\cite{schrieffer_66}, standard perturbation theory fails to provide an accurate description of the weak-coupling limit, $V_d\to 0$, since higher order terms in $V_{d}$ appear together with energy denominators $\epsilon_k - \epsilon_{k'}$. Near the Fermi surface, the latter can be arbitrarily small, and hence render perturbation theory divergent. To solve this problem, they suggested to perform a unitary transformation, which later became known as the SW transformation~\cite{schrieffer_66}. This transformation eliminates the dependence of the Hamiltonian on $V_{d}$ to linear order. As a result, the limiting procedure $V_d\to 0$ becomes well-defined. 

Here, we show that we can achieve a similar goal by first doing a transformation into a rotating frame with respect to the operator $\Delta\sum_s n_{ds}$, and subsequently applying the High Frequency expansion (HFE) to this new periodic Hamiltonian. This is a direct extension of the procedure we used for the two-site single-particle problem discussed above. By doing this transformation, we are eliminating the energy scale $\Delta$ from the effective description at the expense of introducing a fast periodic time dependence in the hybridization term:
\begin{eqnarray}
&& H^\text{rot}(t) =  H_\text{band} + \mathrm e^{-i\Delta t}H^{-} + \mathrm e^{i\Delta t}H^{+},\nonumber\\
&& H_\text{band} = \sum_{k,s}\epsilon_k n_{ks}, \; H^{-} = \frac{1}{\sqrt{L^D} }\sum_{ks}V_{d}c^\dagger_{ks}d_{s},\ \ H^{+} = \left(H^{-}\right)^\dagger.\nonumber\\
\label{anderson}
\end{eqnarray}   
We can now apply the HFE, since we have a periodic Hamiltonian. Clearly, the time-averaged Hamiltonian, $H_\text{eff}^{(0)} = H_\text{band}$, so the linear terms in $V_d$ average to zero. Notice how the absence of linear terms, which can be considered as the main requirement for the choice of the generator of the SW transformations, arises naturally in this setup. The $\Delta^{-1}$-correction includes the following commutator $[H^+,H^-]$ here we do not consider the commutators $[H^\pm,H_0]$ since we are discussing the HFE, c.f.~Eq.~\eqref{eq:HFE}, 
not the ME, c.f.~Eq.\eqref{eq:magnus_series}). This commutator leads to scattering between band electrons to order $1/\Delta$, and thus has to be taken into account for finite $\Delta$. If we restrict the discussion to order $1/\Delta$, we find
\be
H_\text{eff} =  H_{\rm band}-\frac{|V_d|^2}{\Delta L^D}\sum_{kk'}\Psi^\dagger_k\Psi_{k'} +\frac{|V_d|^2}{\Delta} \Psi^\dagger_d\Psi_d + \mathcal{O}\left(\Delta^{-2}\right),
\ee
where we introduced the compact spinor notation:
\begin{eqnarray}
\Psi_k = \left( \begin{array}{c}
c_{k\uparrow}\\
c_{k\downarrow}
\end{array} \right), \ \ \ 
\Psi_d = \left( \begin{array}{c}
c_{d\uparrow}\\
c_{d\downarrow}
\end{array} \right).                
\end{eqnarray}
and the sum over spin indices is assumed. For example $\Psi^\dagger_d\Psi_d=c^\dagger_{d\uparrow}c_{d\uparrow}+c^\dagger_{d\downarrow}c_{d\downarrow}=n_{d\uparrow}+n_{d\downarrow}$. The second term in this effective Hamiltonian represents the static scattering from the impurity atom, while the third term is the new impurity potential. As in the two-level system from Sec.~\ref{subsec:dble_well}, the kick operator $K_{\rm eff}(t)$ governing the micromotion can be calculated explicitly using Eqs.~\eqref{eq:kick_operator_HFE}:

\be
K_{\rm eff}(t)= {1\over i \Omega}\left(\mathrm e^{i\Delta t} H^{+}-\mathrm e^{-i\Delta t} H^{-}\right)
\label{eq:kick_pm}
\ee
In particular, if we evaluate it at stroboscopic times $nT$ we find
\be
K_{\rm eff}(nT)={1\over i\Omega}{V_d\over \sqrt{L^D}}\sum_{ks} [d_s^\dagger c_{ks} -c_{ks}^\dagger d_s].
\label{eq:kick_pm0}
\ee
In the language of the SW transformation the effective Hamiltonian $H_{\rm eff}$ keeps track of the spectrum of the system and the kick operator $K_{\rm eff}(t)$ realises the rotation of the basis. As we discussed many times already, the dynamics of the system can be studied using either the effective Hamiltonian and the effective kick, or the stroboscopic Hamiltonian and the stroboscopic kick. As usual, the stroboscopic Hamiltonian $H_F[t_0]$ is less symmetric than the effective Hamiltonian $H_\text{eff}$ and the stroboscopic kick operator is identically zero at times $t=t_0+nT$, i.e.~$K_F[t_0](t_0+nT)=0$ signifying that in the stroboscopic picture there is no need to rotate the basis states.  

Usually, in the context of the SW transformation, the subtleties associated with the kick operators are not discussed. Moreover, the SW transformation can become quite cumbersome if one needs to go to higher order. On the other hand, the HFE naturally allows us to deal with the kick operators and go to higher order in $\Delta^{-1}$ if necessary.

\emph{Schrieffer-Wolff transformation for periodically driven systems.}
	We now extend the SW transformation to periodically driven systems by adding an extra term
\[
H_1(t)=-A\cos\Omega t\;  \sum_s n_{d,s}
\]
to the Hamiltonian~(\ref{eq:Anderson_model_H}). This system was studied from the point of view of Floquet theory in Refs.~\cite{yamada_12,noba_14,mori_14}. As in the example of the two-level system, we assume a commensurate driving frequency and impurity energy: $\Delta = n\Omega$, where $n\in\mathbb{N}$ (see the discussion in the two level case about the motivation for this assumption and how to relax it). Furthermore, as before we assume that $\Omega$ and hence $\Delta$ are the largest energy scales in the problem. 
	
We eliminate the impurity level and the driving altogether, by going to the rotating frame defined by $V(t) = \exp\left[i(\zeta\sin\Omega t\;  - n\Omega) \sum_s n_{d,s}\right]$, $\zeta = A/\Omega$. This leads to
\begin{eqnarray}
	H^\text{rot}(t) &=& \sum_{k}\epsilon_k \Psi^\dagger_{k}\Psi_{k} + \frac{1}{\sqrt{L^D} }\sum_{k} V_d \mathrm \mathrm e^{i\zeta \sin\Omega t\;  -in\Omega t}\Psi^\dagger_{k} \Psi_d + \text{h.c.}
\end{eqnarray}	
We can now apply the High-Frequency expansion. The derivation of the effective Hamiltonian follows the same guidelines as that of the driven two-level system. The resulting time-averaged Hamiltonian and the leading correction are given by
\begin{eqnarray}
	H^{(0)}_\text{eff} &=& \sum_{k} \epsilon_k\Psi^\dagger_{k} \Psi_{k} + \frac{1}{\sqrt{L^D} }\sum_{k} V_d\mathcal{J}_n\left(\zeta\right)\Psi^\dagger_{k} \Psi_d+\text{h.c.},\nonumber\\
	H^{(1)}_\text{eff} &=& -\frac{|V_d|^2}{\Delta L^D}\tilde g_n(\zeta)\sum_{k,k'}\Psi^\dagger_k \Psi_{k'} + \frac{|V_d|^2}{\Delta}\tilde g_n(\zeta) \Psi_d^\dagger \Psi_d,
    \label{eq:nonint_impurity_driven}
\end{eqnarray}
where $\mathcal{J}_n$ is the $n$-th Bessel Function of first kind, and the function $\tilde g_n(\zeta)$ is defined in Eq.~\eqref{eq:ME_RWA_SW_gtilde_m}.
	
Contrary to the situation in the non-driven case, here in the infinite-frequency limit, the hybridization terms which mix the band and the impurity levels do not vanish. This is very similar to the effect we already observed for the driven two-level system. So unlike the static case this linear coupling has direct physical implications, because the impurity level in the rotating frame is resonant with the bottom of the band. It then follows that the population of the impurity will be significant at any finite driving frequency as long as $\mathcal{J}_n(\zeta)$ is not too small. 
Physically, this corresponds to multi-photon absorption processes.  

Let us point out that one can similarly analyze the limits where $\Delta=n\Omega+\delta\Delta$, where the off-set $|\delta\Delta|<\Omega/2$. As we discussed earlier, in the rotating frame, this off-set leads to an extra (small) static impurity potential $\delta \Delta n_d$. It is intuitively clear that the occupation of the impurity in the steady state will be sensitive to the position of this potential with respect to the Fermi level. A large impurity occupation is possible for $0\leq \delta \Delta\leq E_F$ (where $E_F$ is the Fermi energy). This mechanism of populating the higher level is expected to open up the way towards studying heating in the high-frequency regime if one replaces the impurity atom by an entire excited band. The issue of heating requires a separate careful analysis, which is beyond this review. We also refer the reader to recent works, where this issue was partially addressed for the Kondo model~\cite{heyl_10}.

\emph{The Anderson Model.} Let us now go back to the static model and add an interaction term to the lab-frame Hamiltonian.  The Hamiltonian describing repulsion between the electrons on the impurity is given by:
\begin{equation}
H_\text{int} = Un_{d\uparrow}n_{d\downarrow}.
\end{equation}
For large interactions this term effectively penalizes the double occupancy of the impurity site. As is well known, this leads to the effective low-energy Kondo Hamiltonian~\cite{schrieffer_66}. 

To show the relation between the High-Frequency (Magnus) expansion and the SW transformation, we assume that $\Delta$ and $U$ are the largest energy scales and once again we eliminate both of them together by going to the rotating frame:
\begin{equation}
V(t) = \exp\left( -i\Delta t\sum_s n_{ds}\right)\exp\left( -iU t\; n_{d\uparrow}n_{d\downarrow}\right).
\end{equation}
Note that this transformation consists of the product of two commuting operators and it is a direct generalization of the transformation used in the non-interacting non-driven model above. The Hamiltonian in the rotating frame gets modified according to:
\begin{eqnarray}
H^\text{rot}(t)&\longrightarrow& H^\text{rot}(t) + \mathrm e^{-i\Delta t}(\mathrm e^{-iU t} - 1)W^- +\mathrm e^{i\Delta t}(\mathrm e^{iU t} - 1)W^+,\nonumber\\
W^- &=& \frac{1}{\sqrt{L^D} }\sum_{ks}V_d c^\dagger_{ks}d_sn_{d,\bar s},
\end{eqnarray}
where $H^\text{rot}(t)$ on the RHS above is the Hamiltonian~\eqref{anderson}, and $\bar s$ denotes the opposite spin species to $s$. The new terms $W^\pm$ represent an interaction-dependent hopping from the conducting band to the impurity. In general, the interaction $U$ and the impurity energy $\Delta$ need not be commensurate, and thus the transformation to the rotating frame is not periodic. In the spirit of our previous discussion we assume commensurability, $U=m\Delta$, and moreover choose $m=1$. One can check that the resulting Kondo Hamiltonian is correctly reproduced for any $m$ and by taking analytic continuation to non-integer $m$'s one obtains the correct result for any values of $U$ and $\Delta$. In addition, one can easily convince themselves that by choosing a common frequency $\Omega$ from the two energies $U$ and $\Delta$: $U=n\Omega$, $\Delta=m\Omega$, one also reproduces exactly the result of the standard SW transformation.

The interaction-dependent hopping $W^\pm$ does not contribute to the time-averaged Hamiltonian $H_\text{eff}^{(0)}$. However, it gives an important contribution to the first-order correction to the effective Hamiltonian coming from the commutators $[W^+,W^-]$, $[W^+, H^-]$, and $[W^-, H^+]$. Evaluating these explicitly, we find
\begin{eqnarray}
H_\text{eff} &=& H_{\rm band}-\frac{|V_d|^2}{4\Delta L^D}\sum_{kk'} \left(\Psi^\dagger_k \bm{\sigma} \Psi_{k'}\right)\cdot\left( \Psi^\dagger_d \bm{\sigma} \Psi_d \right)\nonumber\\
&&\ + \frac{|V_d|^2}{\Delta L^D} \sum_{kk'}\left[ -1 +\frac{1}{4} \Psi^\dagger_d\Psi_{d}\right]  \Psi^\dagger_k\Psi_{k'} + \frac{|V_d|^2}{\Delta}\Psi^\dagger_d\Psi_d + \mathcal{O}\left(\Omega^{-2}\right). \nonumber\\
\end{eqnarray}
For our choice of parameters we have $U=\Delta$. Here $\bm{\sigma}$ is the vector of Pauli matrices and the summation over the spin indices is taken care of using the spinor notation. It is immediate to recognize that we have reproduced precisely the Kondo coupling, which one otherwise derives from the original SW transformation~\cite{schrieffer_66}. 

As before, we note that the High-Frequency (Magnus) expansion, allows one to explicitly take into account both the slow dynamics of the system through the Floquet Hamiltonian, and the fast dynamics through the operator $K(t)$. Finally, we note in passing that including the driving in the interacting model is straightforward. The new term appearing in $H_\text{eff}^{(0)}$ will be identical to the one in Eq.~(\ref{eq:nonint_impurity_driven}) for the non-interacting model, while the other terms will be modified by functions similar to $\tilde g_n(\zeta)$. Last, we also point out that from Eq.~\eqref{eq:kick_operator_HFE} it becomes clear that, to leading order in $\Omega^{-1}$, the kick operator $K_{\rm eff}^{(1)}(t)$ is modified by the interactions, accordingly, both in the driven and the non-driven Anderson model.

\subsection{\label{subsec:dicke}The Rabi Model.}

\begin{figure}
	\centering
	\includegraphics[width = 0.7\columnwidth]{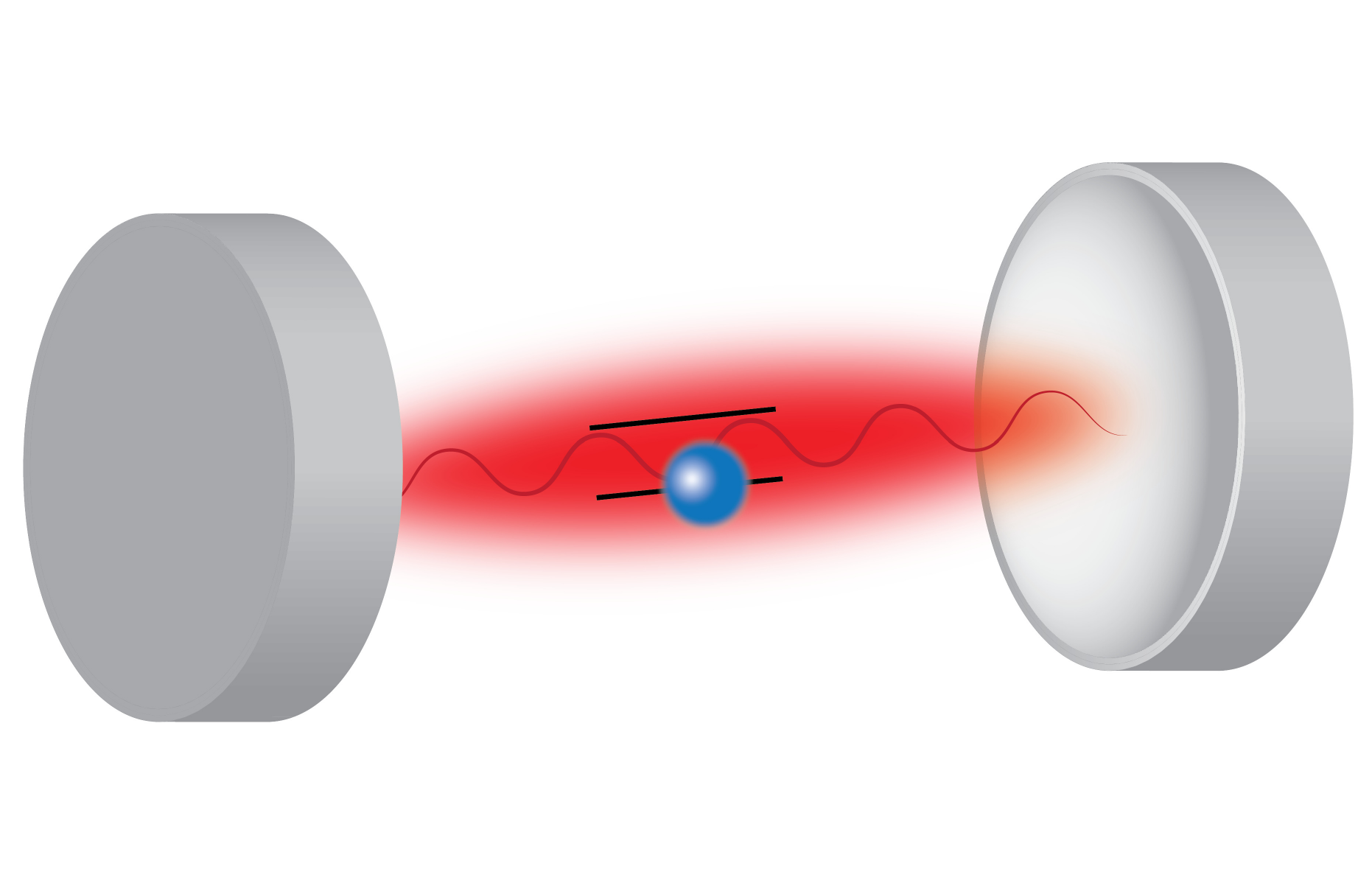}
	\caption{\label{fig:dicke}(Color online) The Rabi Model: a two-level system interacts with a quantized electromagnetic field.}
\end{figure}

Another famous example in which a transformation to the rotating frame eliminates a highly excited state, is the Rabi model, which describes a two-level atom coupled to a quantized electromagnetic field. In the rotating frame the Rabi Hamiltonian is
\be
H=H_0+H^+(t)+H^-(t), 
\ee
where 
\beq
&& H_0 =g\left(a^\dagger\sigma^- + a\sigma^+\right),\nonumber\\
&& H^-(t) = g' \mathrm e^{-i\Omega t} a \sigma^-,\, H^+(t)=(H^-(t))^\dagger,
\eeq
and we have chosen the laser frequency $\omega_L$ to match resonantly the difference of the energies of the two levels of the atom, in which case $\Omega = 2\omega_L$ (see Fig.~\ref{fig:dicke}). We mention in passing that the solution of the Rabi model for $g=g'$ can be expressed in terms of a functional differential equation~\cite{braak_11}. Moreover by adding a magnetic field along the $z$-direction one can obtain an entire line of integrable points, where the generalized Rabi model is supersymmetric~\cite{tomka_13}.

Applying the Magnus (High-Frequency) expansion to order $\Omega^{-1}$ to the Rabi Hamiltonian gives:
\begin{eqnarray}
H_F[0] &=& g\left(a^\dagger\sigma^- + a\sigma^+\right) + \frac{g'^2}{\Omega}\left(a^\dagger a\sigma^z - \sigma^+\sigma^-\right)
+\frac{gg'}{\Omega}\left(a^2 + 
\left(a^\dagger\right)^2\right)\sigma^z + \mathcal{O}\left(\Omega^{-2}\right),\nonumber\\
H_\text{eff} &=& g\left(a^\dagger\sigma^- + a\sigma^+\right) + \frac{g'^2}{\Omega}\left(n\sigma^z - \sigma^+\sigma^-\right) + \mathcal{O}\left(\Omega^{-2}\right),\nonumber\\
K_F^{(1)}[0](t) &=& \frac{g'}{i\Omega}\left( (\mathrm e^{i\Omega t}-1) a^\dagger \sigma^+ -(\mathrm e^{-i\Omega t}-1) a\sigma^- \right),\nonumber\\
K_\text{eff}^{(1)}(t) &=& \frac{g'}{i\Omega}\left(\mathrm e^{i\Omega t} a^\dagger \sigma^+ -\mathrm e^{-i\Omega t} a\sigma^- \right).
\label{eq:JC_H}
\end{eqnarray}
It is also straightforward to obtain the second-order correction to the Floquet Hamiltonian:
\begin{eqnarray}
H_F^{(2)}[t_0] &=&  -\frac{2g^2 g'}{\Omega^2}(a^2a^\dagger\sigma^- - 2a\sigma^- - a^3\sigma^+ + \text{h.c.})\nonumber\\
&-& 2\frac{g'^3}{\Omega^2}(na\sigma^- - 2a\sigma^- + \text{h.c.})\nonumber\\
&+& 2\frac{g g'^2}{\Omega^2}(2a^\dagger\sigma^- + a^\dagger n\sigma^- - \frac{1}{2}a^3\sigma^- + \text{h.c.}),\nonumber\\
H_\text{eff}^{(2)} &=&  \frac{g g'^2}{\Omega^2}(a^\dagger\sigma^-\sigma^z  + \text{h.c.}).
\end{eqnarray}
It follows that, up to order $\Omega^{-2}$, the effective Hamiltonian conserves the sum of the total number of photons and the $z$-component of the spin and, therefore, it only couples pairs of states such as $|1\rangle=|n,+\rangle$ and $|2\rangle=|n+1,-\rangle$.\footnote{One has to keep in mind though that the photons and the spins are dressed by the kick operator and this conservation law breaks down for the bare (undressed) operators.} Here $n$ is the number of photons and $\pm$ indicates the values of the spin-projection along the $z$-axis. Thus, to obtain the spectrum to order $\Omega^{-2}$, we simply need to diagonalize a collection of independent two-by-two Hamiltonians. The matrix elements of the effective Hamiltonian in the sector with $n$ and $n+1$ photons coupled read:
\begin{equation*}
\begin{split}
\langle 1 | H_\text{eff}^{(0)}+H_\text{eff}^{(1)}+H_\text{eff}^{(2)} | 1\rangle &= -\frac{g'^2}{\Omega} (2-n_p),\quad \langle 2 | H_\text{eff}^{(0)}+H_\text{eff}^{(1)}+H_\text{eff}^{(2)} | 2\rangle = -\frac{g'^2}{\Omega} n_p \\
\langle 1 | H_\text{eff}^{(0)}+H_\text{eff}^{(1)}+H_\text{eff}^{(2)} | 2\rangle &= \sqrt{n_p} \left(g + \frac{g g'^2}{\Omega^2} \right),
\end{split}
\end{equation*}
where $n_p=n+1$. From this it is immediate to compute the spectrum:
\begin{eqnarray}
E_{n_p} = -\frac{g'^2}{\Omega} \pm g \sqrt{ n_p } \left(1+ \left(\frac{g'}{g}\right)^2\frac{(1-n_p)^2}{2n_p} +\frac{g'^2}{\Omega^2} \right) + \mathcal{O}\left(\Omega^{-3}\right)
\end{eqnarray}
In the case of $g'=0$, the spectrum reduces to the Jaynes-Cummings one, as it should, and exhibits the hallmark feature of a quantized Rabi frequency $\Omega_{\text{R}} = g\sqrt{n_p}=g\sqrt{n+1}$. The Magnus (High-Frequency) expansion captures both the Bloch-Siegert shift $g'^2/\Omega$, which has been known to be the leading correction to the spectrum for a long time, as well as the subleading correction of order $\Omega^{-2}$. Many terms in the stroboscopic Hamiltonian break the conservation of the total number of photons and spins, and significantly complicate the analysis of the Floquet Hamiltonian. However, these terms do not appear in the effective Hamiltonian, and hence their effect must be captured by the kick operator $K_\text{eff}(t)$ (see Eqs.~\eqref{eq:kick_operator_HFE} and~\eqref{eq:kick_pm}). Recently, it was argued that this type of terms can important for the stabilization of finite-density quantum phases~\cite{schiro_12}. 

The Rabi model can be realised experimentally using highly controllable optical cavities, whose size determines the mode frequency $\Omega$ through the quantisation/boundary conditions. In the same spirit as in the two-level system, and the Anderson model, one can imagine shaking the cavity boundaries out of phase periodically, which would induce a periodic modulation of the frequency $\Omega$. To study the physics of this model, one could go to a rotating frame and apply the Magnus (High-Frequency) expansion. In this case, the counter-rotating (particle non-conserving) terms, $g'a\sigma^- + \text{h.c.}$, will not vanish in the zeroth order, in analogy with the emergent hybridisation terms at the level of the time-averaged Hamiltonian in the models discussed in Secs.~\ref{subsec:dble_well} and~\ref{subsec:Schrieffer_Wolff} potentially leading to new qualitative phenomena. 

Let us conclude this section by pointing out that, through the leading terms in the Magnus (High-Frequency) expansion, one can formally understand the generation of stationary optical lattice potentials used to trap neutral atoms~\cite{bloch_05}. It is then not difficult to find subleading terms including those responsible to various heating processes~\cite{pichler_10}.


\section{\label{sec:kapitza}The Kapitza Class.}

We shall now move on to apply the formalism developed in Secs.~\ref{sec:floquet} and~\ref{sec:magnus_rotframe} to specific examples. In the remainder of this paper, we review various models, in which the  Floquet Hamiltonian exhibits a non-trivial high-frequency limit. By `non-trivial' we mean not equal to the time-averaged lab-frame Hamiltonian. We shall also discuss leading corrections in the inverse driving frequency to the infinite-frequency limit, which are important for experimental realizations. As we show below, different setups leading to non-trivial infinite-frequency Hamiltonians can be classified according to three generic classes of driving protocols. While this classification might not be exhaustive, it covers most of the examples known in the literature, and suggests possible routes for engineering new  Floquet Hamiltonians in various types of systems. 

Let us open the discussion analyzing the Kapitza class which is defined as a non-relativistic system with a quadratic in momentum kinetic energy and arbitrary (momentum-independent) interactions. The driving protocol couples only to operators which depend on the coordinates. In other words, the Hamiltonian should be of the form:
\be
H(p_j,x_j)=H_\text{kin}(\{p_j\})+H_{\rm int}(\{x_j\})+\Omega f(t)  H_1 (\{x_j\}),
\label{h_kapitza}
\ee
where 
\[
H_\text{kin}(\{p_j\})=\sum_{j=1}^N {p_j^2\over 2m_j},
\]
and $f(t)$ is some periodic function of time with period $T$ and zero mean. Note that the driving term $H_1$ can include both a single particle external potential and interactions. Hence, the Kapitza class comprises the periodically driven nonlinear Schr\"odinger equation. For instance, in has been shown that when the interaction strength in the Gross-Pitaevskii equation is shaken strongly and at a high frequency, it is possible to stabilise the solitonic solution against critical collapse~\cite{towers_02,abdullaev_03,saito_03,matuszewski_05,malomed_06}. When we say that the Hamiltonian should be of the form~(\ref{h_kapitza}), we imply that it should be gauge-equivalent to this form. For instance, any time-dependent scalar potential can be absorbed into a vector potential by choosing a different electromagnetic gauge, as it is well-known from classical electromagnetism. While we do not explicitly consider here systems in the presence of an orbital magnetic field, the Kapitza class can be extended to such situations as well. Such extension will simply result in few additional terms in the infinite-frequency Hamiltonian and the leading inverse frequency corrections. We made the prefactor $\Omega=2\pi/T$ explicit in Eq.~\eqref{h_kapitza} to emphasize that, in order to get a non-trivial high-frequency limit, one needs to scale the driving amplitude linearly with the frequency. This scaling guarantees that, when the driving becomes infinitely fast, the system is strongly perturbed, and its evolution cannot be described by the time-averaged Hamiltonian at any frequency.  

To derive the infinite-frequency Floquet Hamiltonian, we employ the inverse frequency expansion in the lab frame up to second order. We show explicit expressions only for the stroboscopic Hamiltonian $H_F[t_0]$ obtained using the Magnus Expansion, Eq.~\eqref{eq:magnus_series}. Similar arguments apply for the effective Hamiltonian using the High Frequency  Expansion, Eq.~\eqref{eq:HFE}. Since, it is easier to work in the rotating frame, as we show in the next section, we shall discuss in detail the comparison between the effective and stroboscopic pictures in there.
	\begin{eqnarray}
	&&H^{(0)} = \frac{1}{2\pi}\int\limits_{0}^{2\pi}\mathrm{d}\tau H(\tau)=H_\text{kin}+H_{\rm int},\label{eq:magnus_kapitza0}\\
	&& H^{(1)}[0] = 
	\frac{[H_\text{kin}, H_1]}{4\pi i}\int\limits_{0}^{2\pi}\mathrm{d}\tau_1\int\limits_{0}^{\tau_1} d\tau_2\; f(\tau_1)-f(\tau_2)
	=
	\frac{[H_\text{kin}, H_1]}{ 2\pi i}\int\limits_{0}^{2\pi}\mathrm{d}\tau_1 (\tau_1-\pi) f(\tau_1),\label{eq:magnus_kapitza1} \\
	&&H^{(2)}[0] = -\frac{[[H_{kin},H_1],H_\text{kin}]}{12\pi \Omega} \int\limits_{0}^{2\pi}\mathrm{d}\tau_1\int\limits_{0}^{\tau_1}\mathrm{d}\tau_2\int\limits_{0}^{\tau_2}\mathrm{d}\tau_3\; 2f(\tau_2)-f(\tau_1)-f(\tau_3)\label{eq:magnus_kapitza2} \\
	&&~~~~~~~~~-\frac{[[H_\text{kin},H_1],H_1]}{12\pi} \int\limits_{0}^{2\pi}\mathrm{d}\tau_1\int\limits_{0}^{\tau_1}\mathrm{d}\tau_2\int\limits_{0}^{\tau_2
	}\mathrm{d}\tau_3\; f(\tau_2) f(\tau_3)+f(\tau_2) f(\tau_1)-2 f(\tau_1) f(\tau_3),\nonumber
	\end{eqnarray} 
where $\tau_i = \Omega t_i$. In order to keep the notation consistent, we drop the subindex $_F$ in the Magnus Expansion of the Floquet Hamiltonian in the lab frame: $H_F[t_0] = \sum_n H^{(n)}[t_0]$, to contrast  with the proper inverse-frequency Magnus expansion $H_F[t_0] = \sum_n H_F^{(n)}[t_0]$, defined in Sec.~\ref{sec:magnus_rotframe}. The difference between the two expansions is due to the non-trivial scaling of the driving amplitude with frequency. For instance, $H^{(2)}[0]$ contains both the term scaling as the first power of the inverse frequency, and the term which survives the infinite-frequency limit. We reserve, the subindex $_F$ in  $H_F^{(n)}[t_0]$ for terms which scale strictly as $\Omega^{-n}$. The term $H_F^{(n)}[t_0]$ can be viewed as a result of either finite or infinite resummation of a lab-frame subseries. 

It becomes clear that, for $\Omega\to\infty$, the first term in $H^{(2)}[0]$ vanishes (it represents one of the subleading $1/\Omega$ corrections) while the other term in $H^{(2)}[0]$ together with $H^{(0)}$ and $H^{(1)}[0]$ give the correct Floquet Hamiltonian in the infinite-frequency limit. The term $H^{(1)}[0]$ in the Magnus expansion can always be set to zero by choosing an appropriate Floquet gauge $t_0$, such that the time-integral appearing in Eq.~(\ref{eq:magnus_kapitza1}) vanishes. For example, if the protocol is symmetric around the middle of the period: $f(t)=f(T-t)$, e.g.~$f(t)=\cos\Omega t\; $, then this integral is identically zero. One has to be cautious, though, that this may not be the case in other gauges. For instance, if $f(t)=\sin\Omega t\; $ then the integral in Eq.~(\ref{eq:magnus_kapitza1}) is non-zero, and one either has to shift the stroboscopic point $t_0$ to $T/2$, or deal with this term. Choosing the symmetric Floquet gauge, the time-ordered integral in the last term in Eq.~(\ref{eq:magnus_kapitza2}) is finite, and has a well-defined non-zero infinite-frequency limit. Note that because the kinetic energy is quadratic in momentum this term depends only on the coordinates $\{x_j\}$, and hence represents an additional external potential or an interaction. Indeed,
\[
[[H_\text{kin},H_1],H_1]=-\sum_{j=1}^N {1\over m_j} \left({\partial H_1\over \partial x_j}\right)^2,
\]
and thus, for symmetric driving protocols, the infinite-frequency limit of the Floquet Hamiltonian reads as:
\be
H_F^{(0)}=H_\text{kin}+H_{\rm int}+A \sum_j {1\over m_j} \left({\partial H_1\over \partial x_j}\right)^2,
\label{h_f_kapitza}
\ee
where
\begin{equation}
A={1\over 12\pi}\iiint \limits_{0<\tau_3<\tau_2<\tau_1<2\pi}\mathrm{d}\tau_1\mathrm{d}\tau_2\mathrm{d}\tau_3 (f(\tau_2) f(\tau_3)+f(\tau_2) f(\tau_1)-2 f(\tau_1) f(\tau_3)).
\end{equation}
The time integral here depends on the details of the periodic function $f(\tau)$. For instance, if $f(\tau)=\lambda \cos(\tau)$ then $A=\lambda^2/ 4$. If the time average of $f(\tau)$ is zero then one can show that
\be
A={1\over 4\pi}\int_0^{2\pi} \Delta^2(\tau) d\tau,\; {\rm where}\; \Delta(\tau)=\int_0^\tau f(\tau') d\tau'. 
\label{eq:A_square}
\ee

Let us argue that the asymptotic form of the Floquet Hamiltonian in the infinite-frequency limit given by Eq.~(\ref{h_f_kapitza}) for the Kapitza class is exact. In other words, there are no other terms in the Magnus expansion which survive as $\Omega\to\infty$. From the structure of the expansion, it is clear that the only non-vanishing terms in the $n$-th order contribution are those which contain $n$-times the driving term $H_1$, and once the kinetic energy. Since the driving amplitude scales with frequency, each extra time integral (giving an extra factor $1/\Omega$) will be precisely compensated for by the extra factor coming from the driving amplitude. So the only terms which survive have the structure of $[[\dots [H_\text{kin}, H_1], H_1],\dots H_1]$ multiplied by some dimensionless number. However, because the kinetic energy is quadratic in momentum, all such terms containing more than two commutators vanish identically. Hence, the only surviving terms beyond the second order must contain the kinetic energy at least twice, so they are at least of order $\Omega^{-1}$. Note that, in principle, one can evaluate the $\Omega^{-1}$-corrections to $H_F^{(0)}$ in a similar way. But the general expressions become very involved so we shall rather show these corrections for a specific case of the Kapitza pendulum. As we shall show, it is much easier to derive these corrections going first to the rotating frame, where there is a systematic and convenient way to count the powers of frequency.

\subsection{\label{subsec:standard_kapitza} The Kapitza Pendulum.}

We now illustrate how the infinite-frequency limit, the leading corrections, the Floquet-gauge freedom, and the dressing of the observables and the density matrix emerge for a specific setup of a single Kapitza pendulum~\cite{kapitza_51,dalessio_13}. At the end of the section, we shall briefly discuss many-particle generalizations of the Kapitza pendulum.

\begin{figure}
	\centering
	\subfigure{
		\resizebox*{7cm}{!}{\includegraphics[width = 0.5\columnwidth]{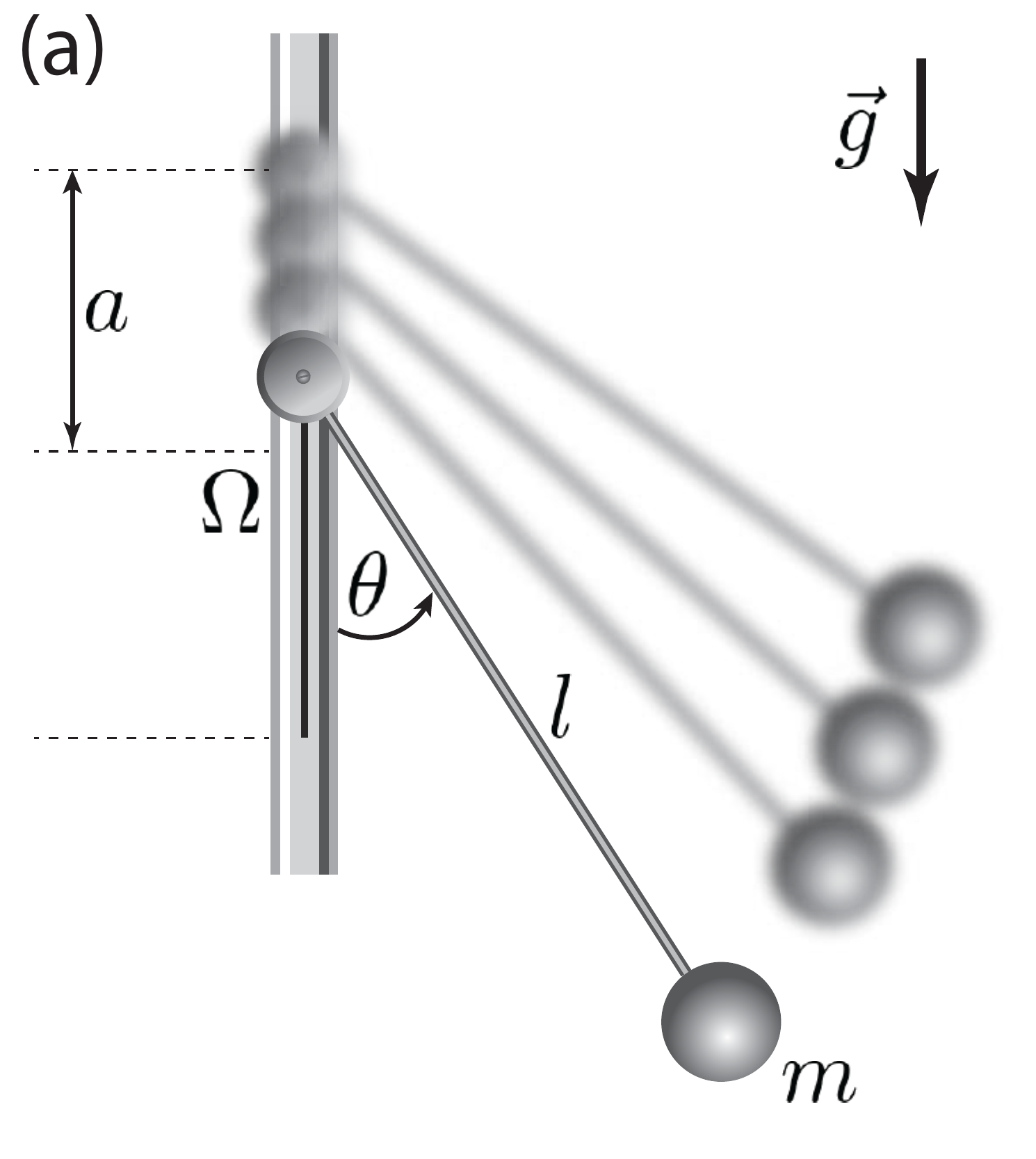}}}
	\hspace{6pt}
	\subfigure{
		\resizebox*{7cm}{!}{\includegraphics[width = 0.5\columnwidth]{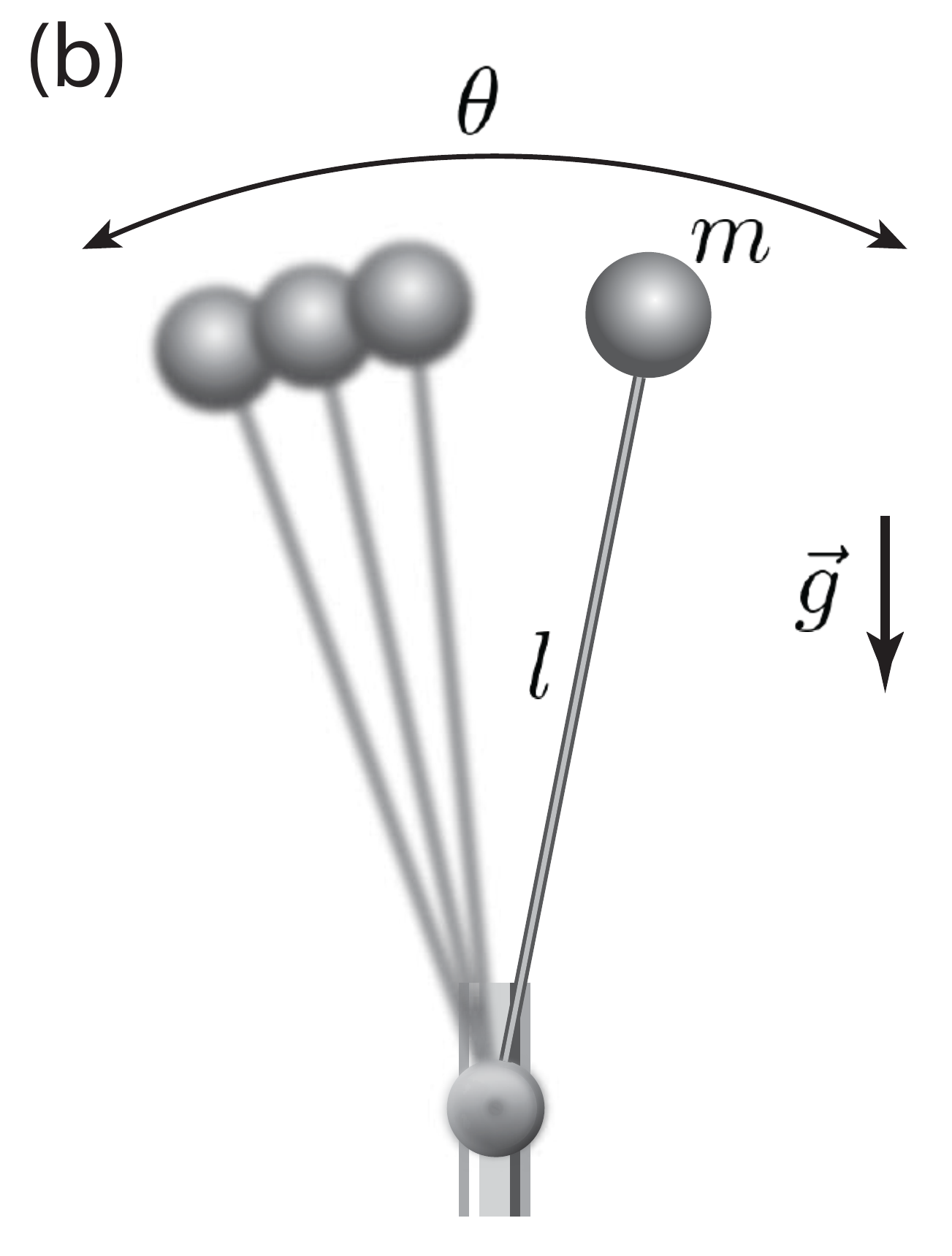}}}
    \caption{\label{fig:kapitza}(Color online) The Kapitza pendulum. (a) The point of suspension of a rigid pendulum of mass $m$ and natural frequency $\omega_0 = \sqrt{g/l}$ is periodically modulated in the vertical direction with an amplitude $a$ and frequency $\Omega$. (b) In the high-frequency limit. the upper equilibrium at $\theta = \pi$ becomes dynamically stable and the system can oscillate around the inverted position.}
\end{figure}

The Kapitza pendulum is a rigid pendulum of length $l$ in which the point of suspension is being displaced periodically along the vertical direction according to the time-dependent protocol $y_0= a\cos\Omega t$. We parametrize the problem in polar coordinates:
\[
x=l\sin\theta,\quad  y=(y-y_0)+y_0=l\cos\theta + a\cos\Omega t
\]
where $\theta$ is the angle measured from the downward direction, c.f.~Fig.~\ref{fig:kapitza}. The Lagrangian is 
\begin{equation}
\mathcal L=\frac{m}{2} \left(\dot{x}^2+\dot{y}^2\right) + m g y = \frac{ m l^2}{2} \left(\dot\theta^2+ \frac{2 a \Omega}{l} \sin\Omega t\,\dot\theta\,\sin\theta\right) + m l^2 \omega_0^2 \cos\theta  
\label{lagrangian-kapitza}
\end{equation}
with $\omega_0=\sqrt{g/l}$. In the last equality we have dropped all terms which are independent of $\theta$ and $\dot\theta$, since they have no physical meaning. Using the standard definitions for the canonical momentum $p_\theta=\partial \mathcal L/\partial \dot\theta$ and the Hamiltonian $H=p_\theta\dot\theta-\mathcal L$ we arrive at~\cite{kapitza_51,dalessio_13}
\be
H={1\over 2 ml^2} \left(p_\theta-ml a\Omega \sin\theta\sin\Omega t\; \right)^2-ml^2 \omega_0^2\cos\theta.
\label{hamiltonian-kapitza}
\ee
The shift in momentum can be removed by a standard gauge transformation in the Hamiltonian, resulting in the scalar potential, which effectively modulates the internal frequency $\omega_0$, so that the Hamiltonian becomes equivalent to
\be
H={p_\theta^2\over 2ml^2}-m l^2 \cos\theta\left(\omega_0^2+{a\Omega\over l}\Omega \cos\Omega t\; \right).
\label{shift-momentum}
\ee
To simplify the notations we re-define $m l^2\to m$, $a\Omega/l\to \lambda$ resulting in the celebrated Kapitza Hamiltonian
\be
H={p_\theta^2\over 2m}-m \cos\theta(\omega_0^2+\lambda \Omega \cos\Omega t\; ).
\label{eq:standard-kapitza}
\ee
In this form the Kapitza Hamiltonian obviously belongs to the Kapitza class (hence its name). As we discussed above, it has a well-defined infinite-frequency limit if we keep $\lambda$ fixed, i.e.~scale the driving amplitude linearly with frequency\footnote{It should be noted that it is the driving amplitude in the Hamiltonian~\eqref{eq:standard-kapitza} which scales linearly in frequency. The shaking amplitude $a$ scales inversely proportional to the frequency.}. Formally one can obtain the Kapitza Hamiltonian by directly modulating the coupling constant in the cosine potential (gravitational constant $g$ in this case). However, notice that the large frequency limit effectively corresponds to changing the sign of this coupling, which is not always easy to achieve experimentally.

The  Floquet Hamiltonian in the infinite-frequency limit, Eq.~(\ref{h_f_kapitza}), is
\be
H_F^{(0)} = H_\text{eff}^{(0)}= \frac{p_\theta^2}{2m}- m\omega_0^2\cos\theta+m{\lambda^2\over 4} \sin^2\theta.
\label{eq:kapitza_Heff}
\ee
When $\lambda>\sqrt{2}\omega_0$ the effective potential in Eq.~\eqref{eq:kapitza_Heff}
supports a stable local minimum at the inverted position $\theta = \pi$. In the absence of the driving, 
the equilibrium position $\theta=\pi$ is clearly unstable. Therefore, by driving the pendulum, it is possible to change the stability of the upper equilibrium. This phenomenon is known as dynamical stabilization and it is widely used in many areas of physics~\cite{casati_94,wiedemann_94,gavrila_92,reinhold_97}.

\subsection{\label{subsec:Kapitza_gauge} The Kapitza Hamiltonian in the Rotating Frame. }

In this section we demonstrate a simpler derivation of the infinite-frequency Floquet Hamiltonian by going to a rotating frame. 
First we transform the Kapitza Hamiltonian~\eqref{eq:standard-kapitza} to the rotating frame by
\begin{eqnarray}
V(t) =&&\ \exp\left(-i\Delta(t)\cos\theta\right)\nonumber\\
\Delta(t) =&&-m\,\lambda\,\Omega \int_{0}^t\,dt\, \cos(\Omega t)= \  -m\,\lambda\,\sin(\Omega t).
\label{eq:kapitza_rot_frame_transf}
\end{eqnarray}
As everywhere else in this paper, $V(t)$ is the transformation which goes from the rotating to the lab frame.

It is often (but not always) convenient to define the rotating frame such that $V(0)={\bm 1}$, so that the initial states in the lab and the rotating frame are the same at $t=0$.  By construction, this transformation eliminates the divergence of the driving protocol with $\Omega$ in the infinite-frequency limit. Hence, as $\Omega\to\infty$, the Floquet Hamiltonian becomes effectively equivalent to the time-averaged Hamiltonian in the rotating frame, as discussed in  Sec.~\ref{subsec:magnus_resumsubseries_pf}. In the rotating frame, the transformed Hamiltonian is given by
\begin{eqnarray}
H^{\text{rot}}(t) &=& V^\dagger(t)\left[\frac{p_\theta^2}{2m}  - m\omega_0^2\cos\theta\right]V(t)\nonumber\\
&=& \frac{p_\theta^2}{2m}  - m\omega_0^2\cos\theta + \frac{\Delta^2(t)}{2m}\sin^2\theta + \frac{\Delta(t)}{2m}\{\sin\theta,p_\theta\}_+,
\label{eq:kapitza_H(t)_rot_frame}
\end{eqnarray}
where $\{\cdot,\cdot\}_+$ denotes the anti-commutator. Noticing that $\frac{1}{T}\int_0^T\,\mathrm{d}t\,\Delta(t) = 0$ and $\frac{1}{T}\int_0^T\mathrm{d}t\Delta^2(t)=m^2\lambda^2/2$, we find for the infinite-frequency Floquet-Hamiltonian:
\be
H_F^{(0)} =H_\text{eff}^{(0)} = \frac{p_\theta^2}{2m}  - m\omega_0^2\cos\theta + m{\lambda^2\over 4}\sin^2\theta.
\label{eq:kapitza_Heff(gamma)}
\ee
This is exactly the Hamiltonian from Eq.~\eqref{eq:kapitza_Heff}, showing explicitly the equivalence of the Floquet Hamiltonians in the lab and rotating frames.

The Floquet Hamiltonian~\eqref{eq:kapitza_Heff(gamma)} is consistent with the predictions based on classical mechanics (see for example Landau and Lifshitz~\cite{landau_lifshitz_1}).
Usually the effective potential (i.e.~the $\theta$-dependent terms in Eq.~\eqref{eq:kapitza_Heff(gamma)}) is obtained by splitting the degrees of freedom into fast and slow modes. One eliminates the fast modes, and derives the effective potential for the slow modes. It then follows that the effective potential is proportional to the time integral of the squared driving protocol~\cite{landau_lifshitz_1}, i.e.~$\propto \int_0^T \,dt\,f(t)^2$. However, according to Eq.~(\ref{eq:A_square}), the effective potential is proportional to the average of its time integral squared, i.e.~$\propto\int_0^T\mathrm{d}t\,\Delta(t)^2$, where $\Delta(t)=\int^t_0\,\mathrm dt\,f(t')$. This makes no difference for a simple $\cos\Omega t $ driving protocol, but will be important for more complex periodic protocols, e.g.~$f(t)=\cos\Omega t\; +\;\cos 2\Omega t$. We showed in Sec.~\ref{sec:magnus_rotframe} that, in the infinite-frequency limit, the stroboscopic and effective Hamiltonians coincide, i.e.~$H_F^{(0)}=H_{\rm eff}^{(0)}$, and the kick operators are equal to zero. One has to keep in mind, though, that the bare observables in the rotating frame are in general different from the bare observables in the lab frame, except for times at which $V(t)$ reduces to the identity operator. For this reason, the dressed observables in the lab frame are generally modified even in the infinite-frequency limit (see Sec.~\ref{kapitza:floquet_meas}).

\subsection{\label{subsec:corr_kapitza}Finite-Frequency Corrections.}

The ME (HFE) expansion allows one to identify leading finite-frequency corrections to the stroboscopic (effective) Hamiltonian $H_F^{(0)}$ ($H_{\rm eff}^{(0)}$). This can be done both in the lab frame and in the rotating frame. However, going to the rotating frame makes the calculations much simpler because, as we discussed, there the Magnus (High-Frequency) expansion coincides with the $\Omega^{-1}$-expansion. On the other hand, in the lab frame, terms from different order in the Magnus (High-Frequency) expansion can have the same scaling with $\Omega$.

To see this explicitly, let us first identify all terms of order $\Omega^{-1}$ appearing in the lab frame. To avoid lengthy expressions, we only state the relevant commutators, which have to be multiplied by the corresponding time integrals. Clearly, two-fold nested commutators appear in the second-order Magnus expansion $H^{(2)}[t_0]$, three-fold nested commutators appear in the third order, and so on. However, each additional commutator comes with an extra $\Omega^{-1}$ suppression coming from the time integral. It is straightforward to see that all the terms which scale as $\Omega^{-1}$ are those containing twice the static Hamiltonian $H_0$
\be
H_0 = {p_\theta^2\over 2m}-m\omega_0^2\cos\theta,\nonumber
\ee
and arbitrary many times the driving term 
\[
H_1(t)= -m\,\lambda\,\Omega\,\cos(\Omega t)\,\cos\theta. 
\]
The relevant corrections are given by
\begin{eqnarray}
&&  [H_0,[H_0,H_1]],\nonumber\\
&&  [H_0,[H_1,[H_0,H_1]]],\;  [H_1,[H_0,[H_0,H_1]]],\nonumber\\
&&  [H_1,[H_0,[H_1,[H_0,H_1]]]],\;[H_1,[H_1,[H_0,[H_0,H_1]]]].\phantom{XXX}
\label{eq:kapitza_corr_terms}
\end{eqnarray} 
These commutators are non-zero because $H_0$ depends on the momentum $p$ while $H_1$ depends on the coordinates $\theta$. Every time the commutator with $H_1$ is applied, the power of the momentum operator is lowered by one. For example, for $H_0\sim p^2$, we have $[H_0,H_1]\sim p$, and $[[H_0,H_1],H_1]$ does not depend on $p$, i.e.~it is a function of the coordinates alone and therefore commutes with $H_1$.
It then immediately follows that all higher-order nested commutators, containing two $H_0$ and four or more times $H_1$, vanish identically. If we work in the symmetric Floquet-gauge $t_0=0$, the driving protocol becomes symmetric w.r.t.~the origin of the time axis, i.e.~$\cos\left(\Omega t\right)=\cos(-\Omega t)$. One can then show that all odd-order terms in the Magnus expansion vanish identically~\cite{blanes_09} and, thus, only the second-order (first line in Eq.~\eqref{eq:kapitza_corr_terms}) and the fourth-order (third line in Eq.~\eqref{eq:kapitza_corr_terms}) terms contribute.

While the evaluation of all these terms and the corresponding time integrals is in principle possible, it is quite cumbersome and computationally heavy. Instead, it is much easier to get the same $1/\Omega$ correction in the rotating frame by simply evaluating the first order Magnus expansion:
\begin{equation}
H_F^{(1)}[ 0] = \frac{1}{4\pi i\Omega}\int_0^{2\pi}\mathrm{d}\tau_1\int_0^{\tau_1}\mathrm{d}\tau_2 \left[H^{\text{rot}}(\tau_1),H^{\text{rot}}(\tau_2) \right],\nonumber\\
\end{equation}
where $H^{\text{rot}}(t)$ is the Hamiltonian of Eq.~\eqref{eq:kapitza_H(t)_rot_frame} and, as before, $\tau = \Omega t$. Likewise one can use Eq.~\eqref{eq:kick_operator_ME} for finding the stroboscopic kick operator. Then the calculation of $H_F^{(1)}[0]$ and $K_F^{{\rm rot}, (1)}[0](t)$ becomes very simple and we find
\begin{eqnarray}
H_F^{(1)}[0] &=& \frac{1}{\Omega}\bigg[ \frac{\lambda}{4m}\left( p_\theta^2\cos\theta + 2p_\theta\cos\theta\, p_\theta + \cos\theta\, p_\theta^2\right) - m\omega_0^2\lambda\sin^2\theta - m\frac{\lambda^3}{2} \cos\theta\sin^2\theta\bigg],\nonumber\\
K_F^{\text{rot},(1)}[0](t) &=& \int_{0}^t \mathrm{d} t' H^\text{rot}(t') - H_F^{(0)}[0]\nonumber\\
 &=&  \frac{1}{\Omega}\left( \frac{\lambda}{2}(\cos\Omega t -1) \{ \sin\theta,p_\theta \}_+ - \frac{m\lambda^2}{8}\sin 2\Omega t \sin^2\theta \right).
\label{eq:kapitza_corr}
\end{eqnarray}  

In parallel, we also give the first correction to the effective Hamiltonian in the rotating frame. 
According to the Eq.~\eqref{eq:HFE} this correction is given by $\sum_l [H_l, H_{-l}]/l$ where $H_l$ are the Fourier harmonics of $H^\text{rot}(t)$ (see Eq.~\eqref{eq:kapitza_H(t)_rot_frame}).
One can convince themselves that $H_l=H_{-l}$ and, therefore, the first order correction to the effective Hamiltonian vanishes:
\be
H_\text{eff}^{(1)} = {\bf 0}.
\ee
The difference between the $\Omega^{-1}$-correction terms $H_F^{(1)}[0]$ and $H_\text{eff}^{(1)}$ means that the $\Omega^{-1}$ terms in the Floquet Hamiltonian only contribute to the Floquet spectrum to order $\Omega^{-2}$. Using Eq.~\eqref{eq:kick_operator_HFE} we find the that the effective kick operator is given by:
\be
K_{\rm eff}^{\text{rot},(1)}(t) = \frac{1}{\Omega}\left( \frac{\lambda}{2}\cos\left(\Omega t\right) \{ \sin\theta,p_\theta \}_+ - \frac{m\lambda^2}{8}\sin \left(2\Omega t\right) \sin^2\theta \right).
\label{eq:kapitza_Heff_K_order_1}
\ee
Using Eq.~\eqref{eq:HFE} and \eqref{eq:kick_operator_HFE} it is also straightforward to calculate higher-order corrections in the rotating frame.

\subsection{\label{kapitza:floquet_meas} Dressed Observables and Dressed Density Matrix.}

Let us now derive the dressed operators and the dressed density matrix which are important to analyze correctly the FNS dynamics of the system, c.f.~Sec.~\ref{subsec:Floquet_experiment}. All calculations can be done again both in the lab and the rotating frames, but we choose the latter for simplicity.

As before, we show the dressed density matrix and observables both in the stroboscopic and the effective pictures. We consider the following natural observables: $\,\sin\theta$, $\sin^2\theta$, $p_\theta$, and $p_\theta^2$, and explicitly consider the initial state characterized by the Gaussian wave-function
\be
\langle\theta|\psi_0\rangle={1\over (2\pi)^{1/4}\sqrt{\sigma}}\mathrm \mathrm e^{-{\sin^2\theta\over 4\sigma^2}}
\ee
with the corresponding density matrix
\be
\rho_0(\theta_1,\theta_2)={1\over \sqrt{2\pi}\sigma}\mathrm \exp\left(-{\sin^2\theta_1+\sin^2\theta_2\over 4\sigma^2}\right).
\label{rho_theta1_theta2}
\ee
We assume that the Gaussian state is well-localised around $\theta = n\pi$, i.e.~$\sigma\ll 1$. In the rotating frame, the operators $\sin\theta$ and $\sin^2\theta$ remain the same as in the lab frame, while the operators $p_\theta$ and $p_\theta^2$, as well as the off-diagonal elements of the density matrix acquire a time dependence:
\begin{eqnarray}
\sin\theta^\text{rot}(t)&=&\sin\theta,\quad\quad \sin^2\theta^{\text{rot}}(t)=\sin^2\theta \nonumber\\
p_\theta^\text{rot}(t) &=& V^\dagger(t)p_\theta V(t) = p_\theta + \Delta(t)\sin\theta,\nonumber\\
p_\theta^{2,\text{rot}}(t) &=& p_\theta^{2} + \Delta(t)^2\sin^2\theta + \Delta(t)\{\sin\theta, p_\theta\}_+,\nonumber\\
\rho^\text{rot}(\theta_1,\theta_2;t) &=& \mathrm e^{i\Delta(t)(\cos\theta_1 - \cos\theta_2)} \rho_0(\theta_1,\theta_2).
\label{eq:op-rotating}
\end{eqnarray} 
Here the operator $V(t)$ and the function $\Delta(t)$ are defined in Eq.~(\ref{eq:kapitza_rot_frame_transf}). 

The dressed operators and density matrix are defined by Eqs.~(\ref{bar_O}). To compute the leading and the first subleading terms in $\Omega^{-1}$ we use Eq.~\eqref{bar_O_magnus}. In the infinite-frequency limit, the dressed operators and density matrix are obtained from the corresponding time-averaged quantities in the rotating frame (this is true both in the stroboscopic and the effective picture). This implies that all operators, which are functions of $\theta$ are unaffected while the operators, which depend on momentum beyond linear order get dressed:
\beq
\begin{split}
\overline{\sin\theta}_F^{(0)} &= \overline{\sin\theta}_\text{eff}^{(0)}= \sin\theta,\quad\quad \overline{\sin^2\theta}_F^{(0)}= \overline{\sin^2\theta}_\text{eff}^{(0)}= \sin^2\theta \\
\overline{ p_\theta}_F^{(0)} &= \overline{ p_\theta}_\text{eff}^{(0)}= p_\theta,\quad \overline{p_\theta^2}_F^{(0)} = \overline{p_\theta^2}_\text{eff}^{(0)}= p_\theta^2+m^2 {\lambda^2\over 2}\sin^2\theta.
\label{eq:bar_p2}
\end{split}
\eeq
The density matrix, being a function of both coordinates and momenta, also gets dressed. In particular
\begin{eqnarray}
\overline{\rho}_F^{(0)}(\theta_1,\theta_2) = \overline{\rho}_\text{eff}^{(0)}(\theta_1,\theta_2) &=& {1\over T} \int_0^T \mathrm{e}^{i\Delta(t) (\cos\theta_1-\cos\theta_2)}\rho_0(\theta_1,\theta_2)\nonumber\\
&=& \mathcal{J}_0(m\lambda(\cos\theta_1-\cos\theta_2))\rho_0(\theta_1,\theta_2),
\label{eq:kapitza_dressed_p^2_vs_p}
\end{eqnarray}
where, as usual, $\mathcal{J}_0$ is the zero-th Bessel function of first kind.
Note that the diagonal elements of $\rho_0$, defining the probabilities of a particular value of $\theta$, are not dressed in the infinite-frequency limit (recall that $\mathcal{J}_0(0)=1$), while the off-diagonal elements, which determine the momentum distribution, get affected. To gain more intuition about this density matrix one can take a partial Fourier transform defining the Wigner function (dropping the subindices $_F$ and $_\text{eff}$ for simplicity):
\be
\overline{W}^{(0)}(\theta,p_\theta)={1\over 2\pi}\int_{-\infty}^{\infty}\, {d\xi}\, \bar{\rho}^{(0)}(\theta +\xi/2,\theta-\xi/2) \mathrm e^{i p_\theta\xi}.
\ee
If the width of the Wigner function, $\sigma$, in Eq.~\eqref{rho_theta1_theta2} is small, the weight of the density matrix is largest for $\theta_1,\theta_2\ll1$ and we can approximate $\cos\theta\approx 1-\theta^2/2$ in the expression above. This immediately leads to:
\begin{eqnarray}
\overline{W}(\theta,p_\theta)&\approx& { \mathrm e^{-\theta^2/( 2\sigma^2)}\over( 2\pi)^{3/2}\sigma}\int_{-\infty}^{\infty} {d\xi} \mathrm e^{-\xi^2/(8\sigma^2)} \mathcal{J}_0(m\lambda\theta\xi) \mathrm{e}^{i p_\theta \xi}\nonumber\\
&\approx& 
{ \mathrm e^{-\theta^2/( 2\sigma^2)}\over( 2\pi)^{3/2}\sigma}\int_{-\infty}^{\infty} {d\xi} \mathrm e^{-\xi^2/(8\sigma^2)} \mathrm e^{-m^2\lambda^2 \theta^2\xi^2/4}\mathrm e^{i p_\theta \xi}\nonumber\\
&\approx& {1\over \pi \sqrt{1+2m^2\lambda^2\sigma^2\theta^2}}\exp\left[-{\theta^2\over 2\sigma^2}-{2p_\theta^2\sigma^2\over 1+2m^2\lambda^2\sigma^2\theta^2}\right]\nonumber\\
&\approx& {1\over \pi \sqrt{1+2m^2\lambda^2\sigma^2\sin^2\theta}}\exp\left[-{\sin^2\theta\over 2\sigma^2}-{2p_\theta^2\sigma^2\over 1+2m^2\lambda^2\sigma^2\sin^2\theta}\right],
\end{eqnarray}
where we used $\sigma\ll1 $, $\theta\ll 1$, and we have approximated the Bessel function with a Gaussian (recall the symmetry of the Bessel function, $\mathcal{J}_0(x)=\mathcal{J}_0(-x)$). In the last line, we made use of the identity $\theta\approx\sin\theta$ for $\theta\ll 1$. As expected, the dressed density matrix features a broadening of the momentum distribution. The new uncertainty in momentum is
\be
\langle p_\theta^2\rangle\approx {1\over 2\sigma^2}+m^2{\lambda^2\over 2}\sin^2\theta,
\ee
which is consistent with Eq.~(\ref{eq:bar_p2}) given that we relied on $|\theta|\ll 1$. Not surprisingly, the momentum uncertainty given by the dressed density matrix is precisely the uncertainty of the dressed $\overline{p_\theta^2}$ operator calculated with the original density matrix (see Eq.~\eqref{eq:bar_p2}).

Using Eq.~\eqref{bar_O_magnus} together with Eqs.~\eqref{eq:kapitza_Heff_K_order_1} and \eqref{eq:op-rotating} it is immediate to compute the $\Omega^{-1}$ corrections to the dressed operators and the density matrix. We find that, in the effective picture, all these corrections are zero:
\be
\overline{\sin\theta}_\text{eff}^{(1)}=0, \quad \overline{\sin^2\theta}_\text{eff}^{(1)}=0,\quad \overline{ p_\theta}_\text{eff}^{(1)}=0, \quad \overline{p_\theta^2}_\text{eff}^{(1)}=0,\quad \overline{\rho}_\text{eff}^{(1)}=0
\label{kapitza_dressed_1}
\ee
This follows from the fact that the time integrals and/or the commutators in Eq.~\eqref{bar_O_magnus} vanish.
The corresponding corrections in the stroboscopic picture can either be computed ab initio using Eq.~\eqref{bar_O_magnus} together with Eqs.~\eqref{eq:kapitza_corr} and \eqref{eq:op-rotating}, or by transforming the zeroth-order (in $\Omega$) dressed observables/density matrix calculated above, from the effective to the stroboscopic picture using Eq.~\eqref{eq:F_vs_eff_dressed_1st_oder}. The two approaches are equivalent and lead to:
	\begin{eqnarray}
	\bar\theta_F^{\text(1)}[0] &=&  - {\lambda\over 2\Omega} \sin2\theta ,\nonumber\\
	\overline{\theta^2}^{(1)}_F[0] &=&  - {2\lambda\over \Omega} \sin^2\theta\cos\theta, \nonumber\\
	\overline{p_\theta}^{(1)}_F[0] &=&  \frac{\lambda}{2\Omega}\{ p_\theta,\cos\theta \}_+,\nonumber\\
	\overline{p_\theta^2}^{(1)}_F[0] &=& \frac{\lambda}{\Omega}\bigg({\{\cos\theta, p_\theta^{2} \}_+\over 2} + p_\theta\cos\theta p_\theta -m\lambda^2\cos\theta\sin^2\theta\bigg) \nonumber\\
	\overline{\rho}^{(1)}_F[0](\theta_1,\theta_2) &=& - \frac{1}{\Omega}\bigg\{ \lambda\mathcal{J}_0(m\lambda(\cos\theta_1-\cos\theta_2))
	\bigg( \frac{1}{2}(\cos\theta_1 + \cos\theta_2) + \left( \frac{\cos\theta_1}{2\sigma^2} \sin^2\theta_1 +  \frac{\cos\theta_2}{2\sigma^2} \sin^2\theta_2 \right)   \bigg)\nonumber\\
	&&\ + m\lambda^2(\sin^2\theta_1 + \sin^2\theta_2) \mathcal{J}_1(m\lambda(\cos\theta_1-\cos\theta_2))\bigg\} \rho_0(\theta_1,\theta_2),\nonumber\\
	&=& -\frac{\lambda}{\Omega}\bigg\{ \left(\sin\theta_1\partial_{\theta_1} + \frac{1}{2}\cos\theta_1\right)\overline{\rho}_F^{(0)}(\theta_1,\theta_2) + (1\leftrightarrow 2) \bigg\}.
\label{kapitza_dressed_2}
\end{eqnarray}

\subsection{\label{subsec:singlebody_kapitza} Multi-Dimensional and Multi-Particle Generalization of the Kapitza Pendulum.}

\begin{figure}
	\centering
	\includegraphics[width=0.8\columnwidth]{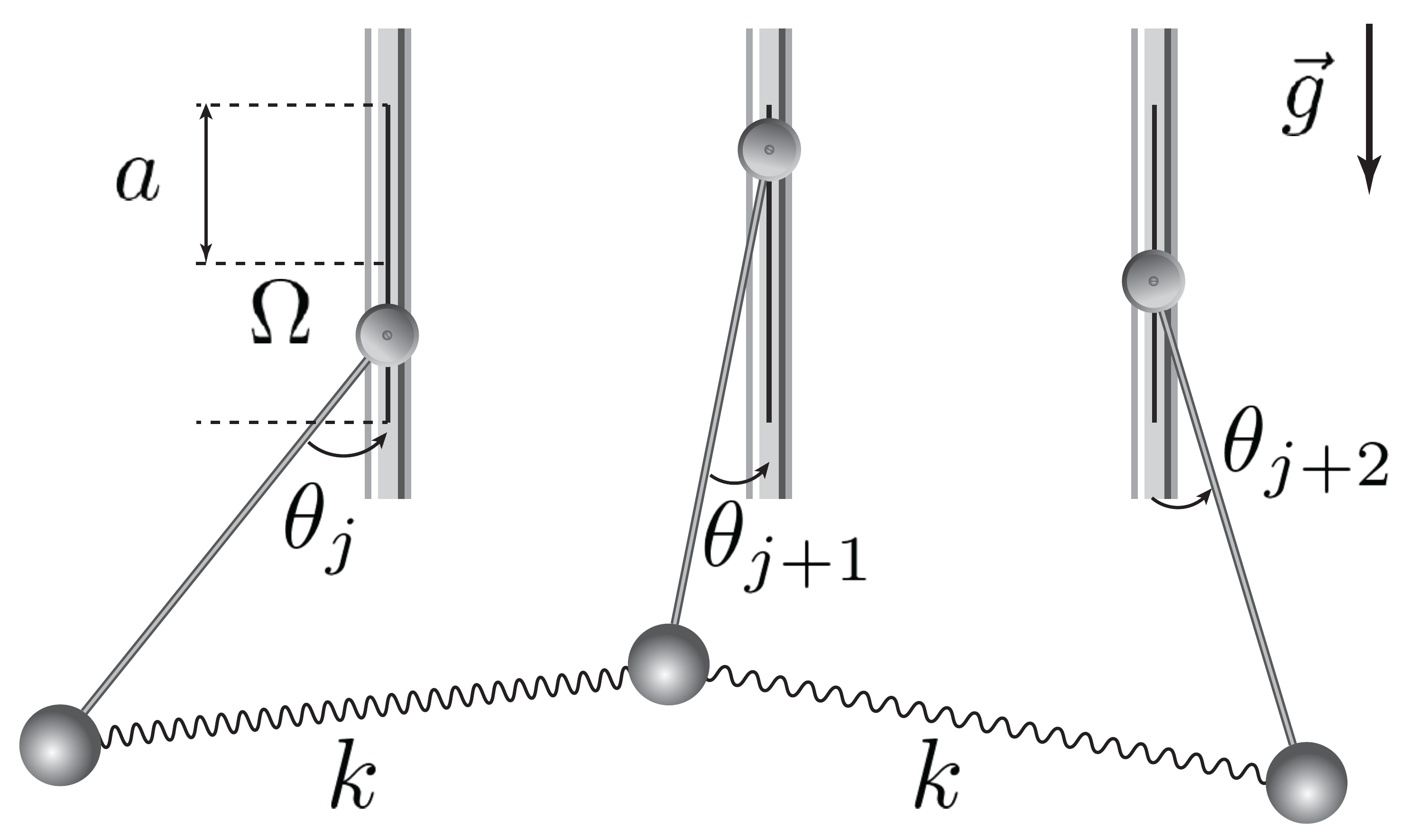}
	\caption{\label{fig:kapitza_coupled} A system of coupled Kapitza pendula: a many-body representative of the Kapitza class.}
\end{figure}

Last, let us discuss two generalizations of the Kapitza pendulum. First, we consider a single-particle multi-dimensional generalization. Namely, we analyze a particle of unit mass whose motion is constrained to a $d$-dimensional hyper-surface embedded in a $D$-dimensional coordinate space. For example, this can be a particle confined to a 2-dimensional sphere or other, more complicated surface. Let this surface be parameterized by the coordinates $\bm r = 
(r_1(\theta_1,\dots,\theta_d),\dots,r_D(\theta_1,\dots,\theta_d))$ with $\theta_1,\dots \theta_d$ being local coordinates. 
Now, suppose we choose a  preferred direction $\bm{e}_i$ in $\mathbb{R}^D$, to shake the entire 
hyper-surface periodically: 
\begin{eqnarray}
\bm{r}(\theta_1,\dots,\theta_d)\to\ \bm{r}(\theta_1,\dots,\theta_d) + a\cos\left(\Omega t\right)\; \bm{e}_i = 
\bm{r}'(t).
\end{eqnarray}    
We follow steps similar to those in Eqs.~\eqref{lagrangian-kapitza} -- \eqref{eq:standard-kapitza} and compute the Lagrangian: 
\[
\mathcal{L}=\frac{m}{2} |\dot{\bm{r}'}|^2 - U_\text{pot} = \frac{m}{2} \left( |\dot{\bm{r}}|^2 + 2\Omega a\sin\left(\Omega t\right) \bm{e}_i\cdot \dot{\bm{r}}\right) - U_\text{pot},
\]
where $U_\text{pot}$ is an unspecified external potential which only depends on the coordinate $\bm{r}$ and, as in Eq.~\eqref{lagrangian-kapitza}, we have dropped the terms independent of $\bm{r}$ and $\dot{\bm{r}}$. Using the standard definitions for the canonical momentum $\bm{p}=\partial \mathcal L/\partial \dot{\bm{r}}$ 
and the Hamiltonian $H=\bm{p}\cdot\dot{\bm{r}}-\mathcal{L}$ we arrive at 
\begin{eqnarray*}
H&=&\frac{1}{2m}\left( \bm{p} - m a \Omega \sin(\Omega t)\bm{e}_i \right)^2 + U_\text{pot} \\
& \rightarrow & \frac{1}{2m} |\bm{p}|^2 + U_\text{pot}  - m a \Omega^2 \cos(\Omega t) \bm{e}_i\cdot \bm{r} \\
& = & \frac{1}{2m} |\bm{p}|^2 + U_\text{pot}  - m \lambda \Omega \cos(\Omega t) \bm{e}_i\cdot \bm{r}
\end{eqnarray*}
where in the second step we cast the vector potential as a scalar potential and we identified $\lambda = a\Omega=\text{const}$.
We then arrive at the conclusion that by shaking the surface at high frequency and small amplitude (i.e.~$\Omega\rightarrow\infty$ and $\lambda=\text{const.}$) we effectively create a large time-dependent ``gravitational-like" potential along the shaking direction. 
This large effect has been achieved by shaking the entire hyper-surface, on which the particle is constrained to move, and could not have been achieved by periodically driving any intrinsic model parameter (such as the gravity $g$), unless one finds a way to scale the driving amplitude with $\Omega$. The Floquet Hamiltonian can be found from Eqs.~\eqref{h_f_kapitza} -- \eqref{eq:A_square}. All finite-frequency corrections as well as the dressed operators can be found by a simple extension of the corresponding results for the Kapitza pendulum.

As a last example we generalize the Kapitza pendulum to a chain of coupled pendula~(see Fig.~\ref{fig:kapitza_coupled}). Consider $N$ coupled pendula, shaken along the $y$-direction using some specific driving protocol. In a way this example can be thought of as a single particle confined to an $N$-dimensional hyper-surface embedded in a $2N$-dimensional space, where $N$ is the number of pendula. One can repeat the derivation of Sec.~\ref{subsec:standard_kapitza} to find that the Hamiltonian of this system reads
\begin{eqnarray}
H&=&\sum_{j=1}^N {p_j^2\over 2m}-J\cos(\theta_j-\theta_{j+1}) -m \omega_0^2\cos\theta_j-m\lambda\Omega\cos\Omega t\;  \cos\theta_j,
\end{eqnarray}
where $J=k l^2$ is the coupling proportional to the spring constant $k$ and, as usual, $\lambda$ is proportional to the product of the driving frequency and the driving amplitude. In the limit of large frequency and $\lambda=\text{const.}$, this Hamiltonian leads to a discretized version of the Sine-Gordon model, which is also very close to the famous Frenkel-Kontorova model~\cite{braun_98}:
\begin{eqnarray}
H_F^{(0)}&=& H_\text{eff}^{(0)} = \sum_{j=1}^N {p_j^2\over 2m}-J\cos(\theta_j-\theta_{j+1}) -m \omega_0^2\cos\theta_j+m{\lambda^2\over 4}\sin^2\theta_j.
\end{eqnarray}
This model can undergo a quantum phase transition, between the gapless and gapped phases, depending on the value of $\lambda$, and the magnitude of the other couplings. It supports various interesting excitations, such as solitons and breathers, and their nature can change with varying $\lambda$~\cite{braun_98}. This model is integrable in the limits $\lambda\ll \omega_0$ and $\lambda\gg \omega_0$ but non-integrable when these couplings are comparable to one another. This opens the possibility of studying interesting thermalization-type dynamics. Additionally, it becomes possible to create interesting infinite-frequency limits by driving different pendulums with different amplitudes and phases. This can be used to generate artificial position-dependent gravitational fields, making the emergent physics even more interesting.


\section{\label{sec:diraclimit} The Dirac Class.}

In this section, we consider periodically driven systems with a kinetic energy linear in momentum. 
According to relativistic quantum mechanics, this requires an additional spin structure in the Hamiltonian~\cite{sakurai_1}. Such systems describe the low-energy physics of graphene~\cite{castro_neto_09}, Weyl semi-metals~\cite{burkov_11}, and other related materials~\cite{uelinger_13,tarruell_13,jotzu_14}.

\subsection{\label{subsec:dirac:driven_B(t)} Periodically Driven Magnetic Fields.} 

The \emph{Dirac} class is defined by the following Hamiltonian
\begin{eqnarray}
H(t) &=& H_0 -\lambda\Omega \sin(\Omega t) H_1,\quad H_1={\bm B}({\bf r})\cdot{\bm \sigma} \nonumber\\
H_0 &=& H_\text{kin} + H_{\rm int}= v_F\,\bm{p}\cdot\bm{\sigma}+ H_{\rm int},\quad ,
\label{eq:dirac_H_densitydriven}
\end{eqnarray}
where $v_F$ is the Fermi velocity, ${\bm B}({\bf r})$ is an external magnetic field and $\bm{\sigma}$ is the vector of $2\times 2$ Pauli matrices (we could similarly analyze a coupling to the $4\times 4$ Dirac $\gamma$-matrices without any need to define a new class). Here $H_{\rm int}$ contains arbitrary spin-independent external potentials and (for many-particle systems) any spin-independent many-body interactions. Taking additional spin-dependent static external potentials into account is straightforward but will unnecessarily aggravate the discussion. Furthermore, to avoid technical complications, our analysis is restricted to situations where the magnetic field does not change its direction in time. To simplify the notations we shall keep the discussion at the single-particle level. 

\begin{figure}
	\centering
	\subfigure{
		\resizebox*{10cm}{!}{\includegraphics[width = 0.6\columnwidth]{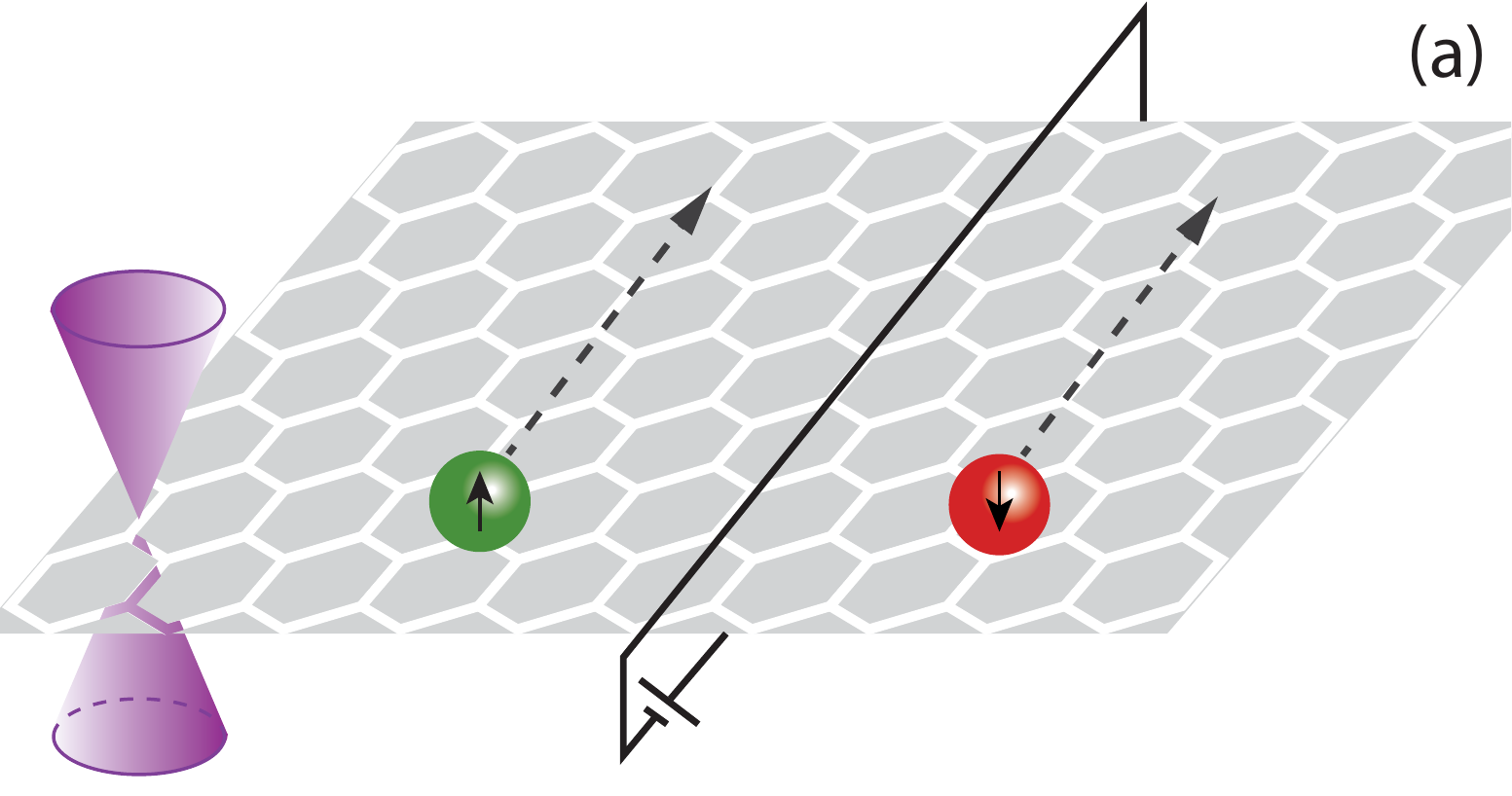}}}
	\hspace{6pt}
	\subfigure{
		\resizebox*{10cm}{!}{\includegraphics[width = 0.6\columnwidth]{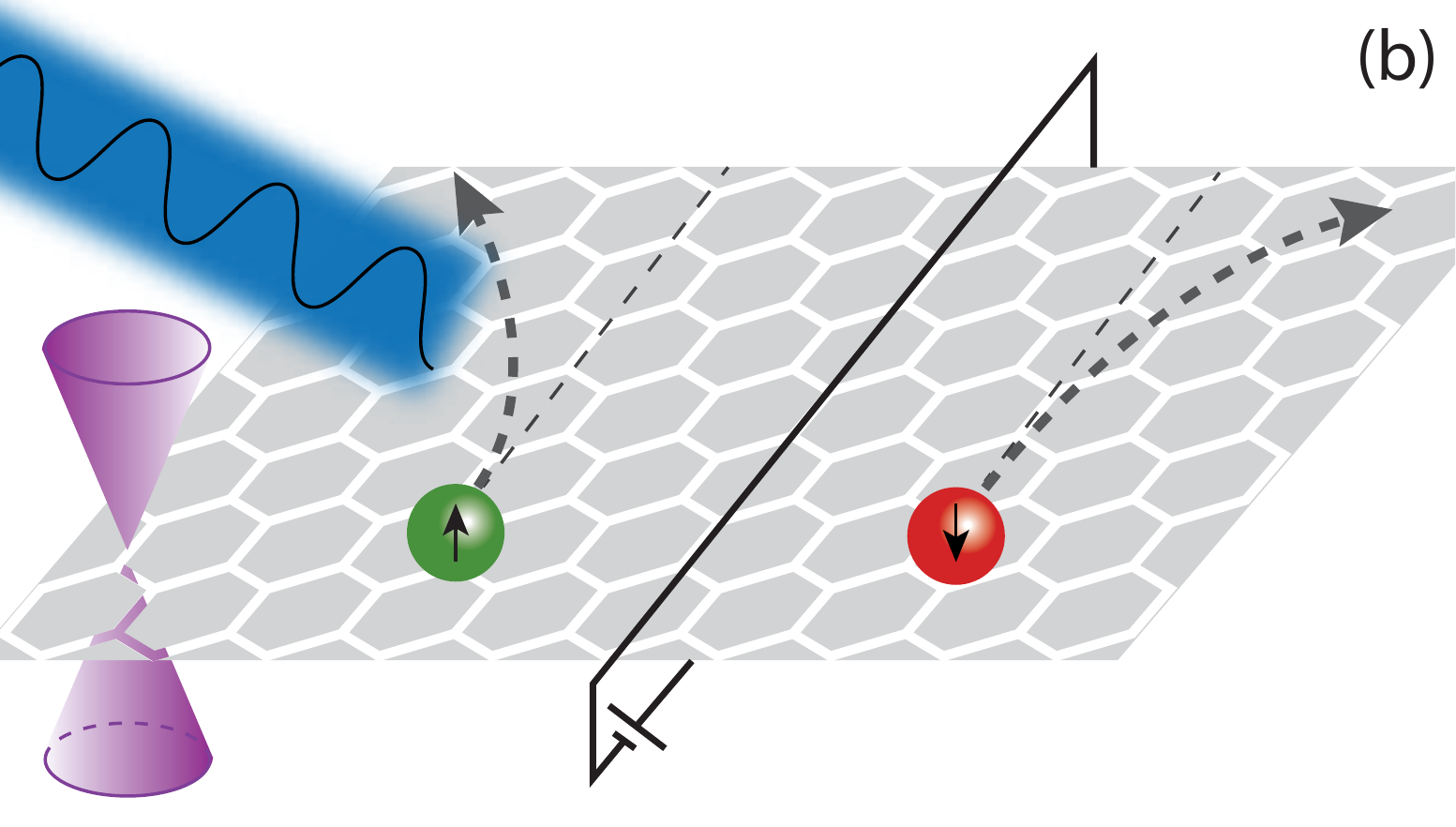}}}
	\caption{\label{fig:dirac_SOC}(Color online) Light induced spin-orbit coupling. Shining light on fermions with a relativistic dispersion, such as graphene close to the neutrality point, leads to spin-orbit coupling whose strength is controlled by the driving amplitude. (a) In absence of the driving, a potential bias generates a longitudinal current. (b) In presence of the driving, a potential bias generates a longitudinal and a {\textit{transverse}} current whose direction depends on the value of the spin.}
\end{figure}

Similarly to the Kapitza class, the analysis of the ME and HFE for the Dirac class can be significantly simplified by performing a transformation to a rotating frame given by
\begin{eqnarray}
V(t) &=& \exp\left(-i\Delta(t){\bm B}({\bf r})\cdot\bm{\sigma} \right),\nonumber\\
\Delta(t) &=& \lambda\cos\Omega t\; .
\label{eq:dirac_VII}
\end{eqnarray}
Clearly, $V(t)$ is a periodic function of time, but with our choice of $\Delta(t)$, it does not satisfy the condition $V(0)={\bm 1}$. Hence, the initial state in the rotating frame is related to the initial state in the lab frame via a unitary rotation by $V(0)$, i.e.~$|\psi^\text{lab}\rangle = V(0)|\psi^\text{rot}\rangle$,~c.f.~Sec.~\ref{subsec:magnus_resumsubseries_pf}. One can of course change $V(t)$ by redefining $\Delta(t)\to \Delta(t)-\Delta(0)$, but this leads to additional gauge-dependent terms in the Floquet Hamiltonian. This is a manifestation of the Floquet-gauge, discussed in Sec.~\ref{sec:floquet}. 

After the transformation to the rotating frame the kinetic energy becomes
\begin{eqnarray}
\frac{H_\text{kin}^{\text{rot}}(t)}{v_F} &=& V^\dagger(t)\,\left(\bm{p}\cdot\bm{\sigma}\right) V(t)\nonumber\\
&=&\frac{1}{2}\{\cos\left(2\Delta(t) B(\bm{r})\right),\bm{p}\cdot\bm{\sigma} - \left(\bm{n}\cdot\bm{p}\right)\left(\bm{n}\cdot\bm{\sigma}\right) \}_+ 
-\frac{1}{2}\{\sin(2\Delta(t)B(\bm{r})),(\bm{n}\times\bm{p})\cdot\bm{\sigma}  \}_+ \nonumber\\
&&-\Delta(t)\bm{n}\cdot\bm{\nabla} B({\bf r}) + \left(\bm{n}\cdot\bm{p}\right)\left(\bm{n}\cdot\bm{\sigma}\right), \nonumber\\
H_\text{int}^{\text{rot}}(t) &=& V^\dagger(t)H_\text{int} V(t)=H_\text{int}\nonumber\\
\label{eq:dirac_Hrot_mostgeneral}
\end{eqnarray}
where $B({\bm r})$ and ${\bm n}$ are the magnitude and the direction of the magnetic field $\bm B({\bm r})$, i.e.~$\bm B({\bm r})=B({\bm r})\,{\bm n}$ with $|{\bm n}|=1$. $H_{\rm int}$ is not affected by the transformation to the rotating frame since it is, by assumption, spin independent. (If the original Hamiltonian contains additional spin-dependent external fields or interactions, then the transformation to the rotating frame will dress $H_{\rm int}$, too.) We can now readily obtain the effective high-frequency Floquet Hamiltonian by taking the time-average of Eq.~\eqref{eq:dirac_Hrot_mostgeneral}: 
\begin{equation}
H_{F}^{(0)} = H_\text{eff}^{(0)} = v_F\,\left(\bm{n}\cdot\bm{p}\right)\left(\bm{n}\cdot\bm{\sigma}\right)
+\frac{v_F}{2}\{\mathcal{J}_0\left(2\lambda B(\bm{r})\right),\bm{p}\cdot\bm{\sigma}
- \left(\bm{n}\cdot\bm{p}\right)\left(\bm{n}\cdot\bm{\sigma}\right) \}_+ + H_\text{int},
\label{eq:dirac_Heff_mostgeneral} 
\end{equation}
where $\mathcal{J}_0$ is the $0$-th order Bessel function of the first kind. One can show that there are no $\Omega^{-1}$-corrections to the Floquet Hamiltonian for the chosen Floquet gauge. This follows from the fact that for a symmetric driving protocol ($\Delta(t)=\Delta(-t)$) all odd-order terms in the ME vanish identically~\cite{blanes_09}, see Eq.~\eqref{eq:magnus_series}. Moreover, there are no $\Omega^{-1}$-corrections to the effective Hamiltonian. This follows from the fact that, for this model, $H_l$ and $H_{-l}$ commute with each other and therefore the first order correction vanishes, c.f.~Eq.~\eqref{eq:HFE}. Hence, the leading non-vanishing correction to this Floquet Hamiltonian is of order $\Omega^{-2}$ suggesting that the infinite-frequency limit in the Dirac class is robust to finite-frequency effects.

\emph{Dresselhaus Spin-Orbit Coupling.} We now consider an example in which we drive a linearly polarized constant magnetic field along a fixed direction in the $xy$-plane. We find an effective Dresselhaus spin-orbit coupling (SOC) in the high-frequency limit. Proposals for Floquet realizations of SOC (see Fig.~\ref{fig:dirac_SOC} for a schematic representation) have already been made for bosons using constant pulse sequences~\cite{anderson_13,xu_13,struck_14,zhang_14}. For fermions, the periodically driven spin-orbit coupling has been studied in graphene~\cite{lopez_12}. 

We consider the Hamiltonian~\eqref{eq:dirac_H_densitydriven} with:
\begin{eqnarray}
B({\bm r})=1,\quad {\bm n}= \frac{1}{\sqrt{2}}\left(1,1, 0\right).
\end{eqnarray}
Specializing Eq.~\eqref{eq:dirac_Heff_mostgeneral} to $B({\bm r})=1$ we arrive at:
\begin{eqnarray}
H_F^{(0)} &=&\ v_F \left(1-\mathcal{J}_0(2\lambda)\right) \left(\bm{n}\cdot\bm{p}\right)\left(\bm{n}\cdot\bm{\sigma}\right) + v_F\,\mathcal{J}_0(2\lambda)\,\left(\bm{p}\cdot\bm{\sigma}\right)+ H_\text{int} = \nonumber\\
&=&\ v'_F \left(\bm{p}\cdot\bm{\sigma}\right) + \frac{v_F}{2} \left(1-\mathcal{J}_0(2\lambda)\right)\left( p_x\sigma_y + p_y\sigma_x  \right)+ H_\text{int},
\end{eqnarray}
where, to obtain the last equality, we have used the explicit form of ${\bf n}$ and we have defined the modified Fermi velocity $v'_F=(v_F/2) \left(1+\mathcal{J}_0(2\lambda)\right)\le v_F$. 
Hence, besides the expected renormalization of the Fermi velocity, one finds an effective Dresselhaus spin-orbit coupling term without affecting the interactions.

\subsection{\label{subsec:dirac:driven_H1(t)} Periodically Driven External Potentials.} 

When one takes into consideration driving systems with a linear dispersion, there exists yet a second possibility in which the driving protocol couples to a scalar external potential. The general form of the lab-frame Hamiltonian is 
\be
H(t) =H_0 +\frac{\lambda}{v_F} \Omega^2\cos\left(\Omega t\right) H_1, 
\label{eq:dirac_H_driven_Omega^2}
\ee
where $H_0$ is defined in Eq.~\eqref{eq:dirac_H_densitydriven} and $H_1$ is an arbitrary spin-independent scalar potential. As we show below, the Hamiltonian above is intimately related to the Hamiltonian~\eqref{eq:dirac_H_densitydriven} defining the Dirac class. For this reason, there is no need to define a new ``class" to accommodate the Hamiltonian~\eqref{eq:dirac_H_driven_Omega^2}. Notice that in the above Hamiltonian the driving amplitude scales with $\Omega^2$ while in Eq.~\eqref{eq:dirac_H_densitydriven} the driving was scaling with $\Omega$. The scaling of the drive with $\Omega^2$ is intimately related to the existence of the additional spin structure in the Hamiltonian and it will be explained from two different perspectives: the lab-frame Magnus expansion, and a transformation to a rotating frame. 

First, we apply the Magnus expansion \emph{in the lab frame}. The zeroth order term gives the time-averaged Hamiltonian $H_\text{kin}+H_{\rm int}$. The $\Omega^{-1}$-corrections vanish identically due to the symmetry of the drive, i.e.~$\cos(\Omega t)=\cos(-\Omega t)$. Therefore the leading contributions the Floquet Hamiltonian are given by the terms:
\[
[H_\text{kin},[H_\text{kin},H_1]],\; [H_1,[H_\text{kin},H_1]].
\]
where each $H_1$ term brings an extra factor of $\Omega^2$ due to the scaling of the driving amplitude and both terms are multiplied by a factor $T^2$ which comes from the double time integral (c.f.~definition of the Magnus expansion in Sec.~\ref{sec:magnus_rotframe}).
For systems with a Dirac dispersion it is easy to verify that the second term vanishes identically. In fact  $\left[H_\text{kin},H_1\right]\propto \left({\bf \nabla} H_1\right) \cdot {\bf\sigma}$ and, therefore, it commutes with $H_1$ which is diagonal in spin space and depends on the position ${\bf r}$ exclusively.
As a result, only the term $[H_\text{kin},[H_\text{kin},H_1]]$ contributes to the Floquet Hamiltonian.
Therefore, to keep this term finite in the infinite-frequency limit, we need to scale the driving amplitude as $\Omega^2$.
Recall that in the non-relativistic Kapitza class the term $[H_1,[H_\text{kin},H_1]]$ was non-zero and dominant. Therefore, to keep the dominant contribution to the Floquet Hamiltonian finite in the infinite-frequency limit, in the Kapitza class the driving amplitude scales only linearly with $\Omega$. 

There are other important differences between the ME (HFE) in the Dirac and Kapitza classes. For example, in the Kapitza class the ME in the limit $\Omega\to\infty$ truncates in the lab frame after the second order, while this is not the case in the Dirac class due to the additional spin structure in the kinetic energy term. For instance, consider the fourth-order commutator $[H_\text{kin},[H_1,[H_\text{kin},[H_1,H_\text{kin}]]]]$, which scales as $\Omega^4$. Taking into account the factor $T^4$ from the time-ordered integrals, we find that this term remains finite as $\Omega\to\infty$. Although the kinetic energy is linear in ${\bm p}$, this 4-nested commutator does not vanish due to the spin structure $[\sigma_\alpha,\sigma_\beta] = 2i\varepsilon_{\alpha\beta\gamma}\sigma_\gamma$. Similar expressions appear in any even higher-order terms in the Magnus expansion (all odd terms being zero due to the symmetry of the drive).

Next, we explain the scaling $\Omega^2$ in Eq.~\eqref{eq:dirac_H_driven_Omega^2} from the point of view of a transformation $\tilde V(t)$ to a preliminary rotating frame:
\begin{eqnarray}
\tilde V(t) = \exp\left(-i \frac{\lambda}{v_F} \Omega\sin\left(\Omega t\right)\; H_1(\bm{r}) \right).
\end{eqnarray} 
In this preliminary rotating frame the Hamiltonian is:
\begin{eqnarray}
\tilde H(t) &=& \tilde H_\text{kin}(\bm{r},t) + H_\text{int}, \nonumber\\
\tilde H_\text{kin}(\bm{r},t) &=& \tilde V^\dagger(t) \left( v_F\,\bm{p}\cdot\bm{\sigma}\right) \tilde V(t)  = v_F\,\bm{p}\cdot\bm{\sigma} - \lambda\Omega\sin\left(\Omega t\right)\; \bm{B}({\bm r})\cdot\bm{\sigma},
\label{eq:dirac_H_Bfielddriven}
\end{eqnarray}
where ${\bm B}({\bm r})=\bm{\nabla} H_1({\bm r})$ is the ``magnetic field'', generated by the spatial gradient of the time-dependent scalar potential. This is only an analogy with real magnetic fields, which are always divergence-free, while an effective magnetic field need not be. For example choosing a parabolic driving potential $H_1({\bm r})= r^2$ induces an effective radial ``magnetic field'' ${\bm B}({\bm r})=2{\bm r}$ in the first rotating frame. The amplitude of this oscillatory ``magnetic field'' scales only linearly with the driving frequency, reflecting the re-summation of an infinite lab-frame Magnus (High-Frequency) subseries. The interaction term $H_{\rm int}$ is not affected by this transformation.
Notice that in this first rotating frame, the Hamiltonian in Eq.~\eqref{eq:dirac_H_Bfielddriven} becomes identical to Eq.~\eqref{eq:dirac_H_densitydriven}. Therefore, we can adopt the entire discussion of Sec.~\ref{subsec:dirac:driven_B(t)} to further analyse this type of models.

This procedure highlights the fact that the Hamiltonians~\eqref{eq:dirac_H_densitydriven} and \eqref{eq:dirac_H_Bfielddriven} are intimately related as we anticipated above. We conclude that, within the Dirac class of systems with a linear dispersion, one can either drive the system via a spatially-dependent scalar potential with an amplitude scaling as $\Omega^2$, or with a spatially-dependent ``magnetic field" with an amplitude scaling linearly with $\Omega$. Using the scalar potential allows one to generate synthetic ``magnetic fields", which may not satisfy the ordinary Maxwell equations (in particular, this might allow one to introduce effective magnetic monopoles into the system). We stress that the Hamiltonian~\eqref{eq:dirac_H_Bfielddriven} can be used as a starting point instead of the Hamiltonian~\eqref{eq:dirac_H_densitydriven}.

\emph{Periodically driven linear potential.} As an illustration let us consider a graphene-type setup in which the momentum of the particle is confined to the $x,y$-plane. The external potential depends linearly on the out-of-plane coordinate $z$ via $H_1(z) = z$. The Hamiltonian is 

\begin{eqnarray}
H(t) &=& H_\text{kin} + H_{\rm int} +\frac{\lambda}{v_F}\Omega^2\cos\left(\Omega t\right)\; z.\nonumber\\
\end{eqnarray} 
Going to the first rotating frame we find a constant in space, time-dependent ``magnetic field'' along the $z$-axis: ${\bm B} = \hat{{\bm z}}$ so that
\begin{equation}
\tilde H(t) =  v_F\,\left(\bm{p}\cdot\bm{\sigma}\right)  +\lambda \Omega\sin\left(\Omega t\right)\; {\bm B}\cdot{\bm\sigma} + H_{\rm int}.\nonumber
\end{equation}
We now do a transformation to a second rotating frame, as discussed in Sec.~\ref{subsec:dirac:driven_B(t)}. For $B({\bf r})=1$ and ${\bf n}=(0,0,1)$ the general  Floquet Hamiltonian in Eq.~\eqref{eq:dirac_Heff_mostgeneral} reduces to
\be
H_F^{(0)} = H_\text{eff}^{(0)} = 
v_F\,\mathcal{J}_0\left(2\lambda\right)\left(\bm{p}\cdot\bm{\sigma}\right) + H_\text{int}=v_F\,\mathcal{J}_0\left(2\lambda\right) (p_x\sigma_x+p_y\sigma_y) + H_{\rm int}.
\label{eq:dirac_Heff_mostgeneral1}
\ee
Note that there are no terms proportional to $p_z$, since the motion of the particles is confined to the two-dimensional $xy$-plane. This driving protocol essentially leads to a renormalised Fermi velocity, which can be tuned to zero by choosing $2\lambda$ to coincide with the zero of the Bessel function $\mathcal J_0$. This dynamical localisation effect can be used for enhancing interaction effects in weakly-interacting many-body systems. 

If in the same setup the effective potential depends linearly on $x$, then the resulting Floquet Hamiltonian becomes anisotropic
\be
H_{F}^{(0)} = H_\text{eff}^{(0)} = v_F\,p_x\sigma_x+
v_F\,\mathcal{J}_0(2\lambda)p_y\sigma_y+ H_{\rm int},
\label{eq:dirac_Heff_mostgeneral2}
\ee
and tuning $\mathcal{J}_0(2\lambda)=0$ makes the kinetic term one-dimensional.


\section{\label{sec:driven_external_fields} The Dunlap-Kenkre (DK) Class.}

As a third class of Hamiltonians, where one can engineer interesting infinite-frequency limits, we consider a setup where the driving couples to a non-interacting term in an arbitrary interacting system. Examples include interacting particles with arbitrary dispersion relation in an external time-dependent electric field, or interacting spin systems in a time-periodic magnetic field, just to name a few. As we shall see in this section, this class of Hamiltonians is paradigmatic for `Floquet engineering'. In this way one can generate Wannier-Stark ladders~\cite{ploetz_11,ploetz_11_interferometry,diener_00,goldman_res_14}, non-trivial tight-binding models with engineered dispersion relations~\cite{gemelke_05,eckardt_05,ciampini_11,goldman_res_14}, including the Harper-Hofstadter Hamiltonian~\cite{jaksch_03, aidelsburger_13,miyake_13,atala_14} and other models exhibiting artificial gauge fields~\cite{creffield_11,struck_12,creffield_14,goldman_gaugefields_14,goldman_res_14}, effective spin Hamiltonians~\cite{struck_13,parker_13, eckardt_10}, quantum Hall states~\cite{hafezi_07}, topologically non-trivial Floquet Hamiltonians~\cite{oka_09,kitagawa_11,iadecola_13, bastidas_13, lindner_11,rechtsman_13,jotzu_14,aidelsburger_14,verdeny_15}, spin-dependent bands~\cite{jotzu_15}, and others. 

To the best of our knowledge, the first theoretical proposal for the realization of a non-trivial high-frequency limit in a tight-binding model with an external periodic electric field was discussed by Dunlap and Kenkre in Refs.~\cite{dunlap_86, dunlap_88}. They discussed the phenomenon of dynamical localization, where the hopping between sites can be completely suppressed in the high frequency limit by choosing an appropriate fixed ratio between the driving amplitude and the driving frequency. Motivated by their idea, we consider the following general class of Hamiltonians
\begin{equation}
H(t) = H_0+ \Omega H_1(t),
\label{eq:limitresumm}
\end{equation}
where $H_0$ represents some (interacting) lattice Hamiltonian, and
\be
H_1=\sum_m f_m(t) n_m
\ee
with $n_m$ being the density operator on the $m$-th lattice site, and $f_m(t)$ is an arbitrary site-dependent periodic function of time with period $T$. Notice that in Eq.~\eqref{eq:limitresumm}, we have explicitly put the $\Omega$-dependence of the driving term $H_1$ to highlight the non-trivial scaling of the driving amplitude with frequency.  

Instead of the lattice system, we could consider a continuum model using $\sum_m f_m(t)n_m\to \int d^dx f(t,{\bf x}) n({\bf x})$ with $f(t+T,{\bf x})=f(t,{\bf x})$. Obviously, in the continuum limit there is an overlap between the DK class and the Kapitza class, if the kinetic energy in $H_0$ is quadratic in momentum, and with the Dirac class if it is linear. The relation between continuum and lattice models is discussed in Appendix~\ref{app:lattice_vs_cont}. In the DK class, we allow for arbitrary dispersion relations at the expense of restricting the driving to couple to single-particle terms. Later on, in Sec.~\ref{spins}, we shall show that the DK class extends to driven spin systems, where $H_0$ describes some arbitrary interacting spin Hamiltonian, while the driving term couples to a spatially dependent, periodic in time magnetic field.

After giving an overview of the general theory of the DK class, we will discuss the recent dynamical realization of the Harper Hamiltonian~\cite{aidelsburger_13,miyake_13}, as a special case of the periodically driven Bose-Hubbard model. We will derive both the infinite-frequency limit, and the leading $\Omega^{-1}$-corrections to the Floquet Hamiltonian. We will also give examples for the dressed operators and density matrix. After that, we will continue with the fermionic case illustrated on the driven Fermi-Hubbard model, and discuss the infinite-frequency limit and the $\Omega^{-1}$-corrections, which are expected to be important for interacting Floquet topological insulators, as realized in Ref.~\cite{jotzu_14}. Finally, we will discuss interacting driven spin chains. 

To be specific, we assume that $H_0$ is a sum of the kinetic energy term represented by the nearest neighbor hopping and additional density-density interactions which can also include a static external potential linearly coupled to the density:
\be
H_0=H_\text{kin}+H_{\rm int},
\ee
where
\[
H_\text{kin}=-J_0\sum_{\langle m,n\rangle}a^\dagger_m a_n + \text{h.c.},\;{\rm and}\; [n_m, H_{\rm int}]=0.
\]
The angular brackets here represent nearest neighbors.

The terms in the inverse-frequency expansion, which do not vanish in the infinite-frequency limit are of the type
\be
H_0,\; [H_1, H_0],\; [H_1,[H_1,H_0]],\; [H_1,[H_1, [H_1, H_0]]]],\dots
\ee
Since each commutator brings an extra factor of $1/\Omega$ from the time integral (see the discussion in Sec.~\ref{sec:magnus_rotframe}), and each $H_1$ term brings an extra factor of $\Omega$ due to the scaling of the driving amplitude, it is easy to see that all these terms are of the same order in $\Omega$. Furthermore, these are the only terms that survive in the infinite-frequency limit. However, unlike in the Kapitza class, this series does not terminate at any finite order and, thus, one has to re-sum an infinite lab-frame subseries to obtain the correct infinite-frequency limit. This is intimately related to the fact that the dispersion relation in $H_0$ is arbitrary and not quadratic in momentum like in the Kapitza class.

From this structure of the inverse-frequency expansion, it is also clear why the Hamiltonian $H_1$ should couple linearly to the density. Only then do these nested commutators not grow both in space (meaning that the resulting effective operators remain local) and in the number of creation and annihilation operators (i.e.~we avoid the generation of three- and higher-body interactions). One can also consider other situations where the commutators do not grow, for example, when the driving couples to the local in space density-density interaction between fermions~\cite{di_liberto_14} or even bosons~\cite{greschner_13,gong_09,wang_14} (though the bosonic case is more subtle), or when the protocol couples to local in space spin interactions for spin models with spin larger than one half. 

While infinite re-summation of a (sub)series is possible and it yields the proper infinite-frequency limit, calculating subleading corrections directly becomes very involved. These complications can be overcome, as before, by going to the rotating frame, which is defined via the transformation
\be
V(t) =\exp\left[ -i\sum_m \Delta_m(t) n_m \right],\quad \Delta_m(t) = \Omega\int_{t_0}^t\mathrm{d}t' f_m(t').
\label{eq:rot_frame_DK}
\ee
The lower limit of the integral defining $\Delta_m(t)$ is a gauge choice, related to the Floquet-gauge, e.g.~one can choose $t_0=0$. Applying this transformation eliminates the term linear in the density operator, which in the lab frame diverges linearly with the frequency. At the same time, in the rotating frame, a periodic drive is imprinted to the kinetic energy:
\begin{eqnarray}
H^{\text{rot}}(t) =&& W(t) + W^\dagger(t) + H_\text{int}\nonumber\\
W(t) =&& -J_0\sum_{\langle mn\rangle}\mathrm e^{ i\left[\Delta_m(t) - \Delta_n(t) \right]} a^\dagger_ma_n.
\label{eq:resummHrot}
\end{eqnarray}
Notice that this transformation leaves the interaction term $H_\text{int}$ invariant. As in the previous classes we discussed, going to the rotating frame generates an effective complex driving protocol, which is well-behaved in the infinite-frequency limit. The infinite-frequency limit of the Floquet Hamiltonian is then simply given by the time average of $H^{\rm rot}(t)$. In the rotating frame, averaging over time is equivalent to a re-summation of an infinite lab-frame inverse-frequency sub-series, in agreement with the general discussion in Sec.~\ref{sec:magnus_rotframe}. Similarly to the Kapitza class, the Magnus expansion (ME) and the High-Frequency expansion (HFE) in the rotating frame can be used to compute the subleading correction in $\Omega^{-1}$ to the Floquet Hamiltonian. Rather than discussing these corrections in the most general form, we will show and analyze them for specific examples.

\subsection{\label{subsec:Floquet_measurement_bosons} Noninteracting Particles in a Periodically Driven Potential: Floquet Theory and Experimental Realization.}

\begin{figure}
	\centering
	\includegraphics[width=.9\columnwidth]{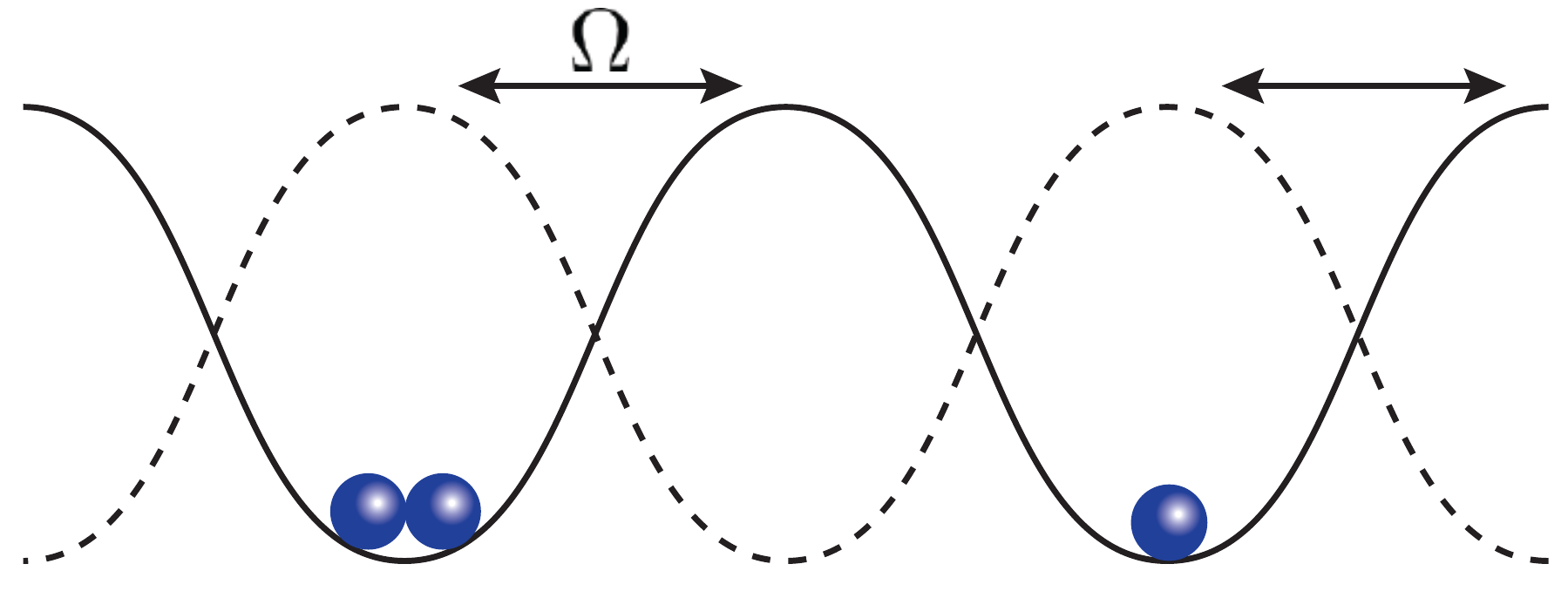}
	\caption{\label{fig:1D_lattice}(Color online). A periodically driven (shaken) optical lattice in which the lattice is shifted periodically at frequency $\Omega$: the prototypical example of the DK class.}
	\label{fig:DK}
\end{figure}

As a first representative of the DK class, we consider a chain of noninteracting, periodically driven spinless particles, which can be either bosons or fermions (see Fig.~\ref{fig:DK}) with the following  Hamiltonian
\begin{eqnarray}
H(t)&=& - J_0\sum_m\left( a^\dagger_{m+1}a_m + \text{h.c.}\right) +  \Omega\sum_m \frac{\zeta}{2}\sin(\Omega t - \Phi m + \Phi/2)n_m,
\end{eqnarray}
where $J_0$ is the hopping, and $a^\dagger_m$ is the operator which creates a particle at site $m$. As anticipated, the driving protocol couples to the density and breaks translational invariance through the site-dependent phase (which may have a more-complicated spatial dependence). The driving amplitude $V_0 = \Omega\zeta$ is constant in space and, in agreement with the general discussion, is proportional to the frequency. The above choice of the Floquet gauge (or the phase lag) ensures a simple form of the infinite-frequency Floquet Hamiltonian. In the next sections we shall generalize our analysis by adding interactions, a second spatial dimension, and finally by adding a spin degree of freedom. 

The transformation to the rotating frame is done using Eq.~(\ref{eq:rot_frame_DK}) with $\Delta_m(t)=-\zeta \cos(\Omega t -\Phi m + \Phi/2)$. We pause to note that, for this particular Floquet gauge choice, $V(0)\neq{\bm 1}$. As a consequence, one needs to transform the initial state to the rotating frame as well. Combining this with Eq.~(\ref{eq:resummHrot}) we find:
\begin{eqnarray}
H^\text{rot}(t) &=& -J_0\sum_m g^{m,m+1}(t) a^\dagger_{m+1} a_m + {\rm h.c.}=\sum_{l\in\mathbb{Z}}H_l\mathrm{e}^{il\Omega t},
\end{eqnarray}
where 
\begin{eqnarray}
g^{m,m+1}(t) &=& \mathrm e^{-i\zeta\sin(\Omega t-\Phi m)},\; \zeta=\lambda\sin(\Phi/2),\nonumber\\
H_l &=& \sum_m \left( \mathrm{e}^{-il\Phi m}\left[\mathcal{J}_{-l}(\zeta)a^\dagger_{m+1}a_m + \mathcal{J}_l(\zeta)a^\dagger_m a_{m+1}\right] \right).
\label{zeta}
\end{eqnarray}
We label the function $g^{m,m+1}(t)$ by two site indices to highlight that it is a link variable, i.e.~defined on the bonds of the lattice. The infinite-frequency  Floquet Hamiltonian and the leading correction are then found from the Magnus (High-Frequency) expansion:
\begin{eqnarray}
H_F^{(0)} &=& H_\text{eff}^{\text{rot},(0)} = -J_\text{eff}(\zeta)\sum_m\left( a^\dagger_{m+1}a_m + \text{h.c.}\right),\nonumber\\ 
H_F^{(1)}[0] &=& -\frac{J_0^2}{\Omega}\sum_m\left(\mathcal{C}_{m,m+2}(\zeta) a^\dagger_{m+2}a_m + \text{h.c.}\right) + \frac{J_0^2}{\Omega}\sum_m \mathcal{G}_{m,m+1}(\zeta)(n_{m} - n_{m+1}),\nonumber\\
H_\text{eff}^{(1)} &=& -\frac{J_0^2}{\Omega}\sum_m\left(\mathcal{\tilde C}_{m,m+2}(\zeta) a^\dagger_{m+2}a_m + \text{h.c.}\right) + \frac{J_0^2}{\Omega}\sum_m \mathcal{\tilde G}_{m,m+1}(\zeta)(n_{m} - n_{m+1}),\nonumber\\
\label{eq:H_F^1_bosons_1D}
\end{eqnarray}
where $J_\text{eff}(\zeta) =J_0\mathcal{J}_0(\zeta)$ is the renormalized hopping parameter and $\mathcal{J}_0$ is the $0$-th order Bessel function of the first kind. 
When the parameter $\zeta$ is tuned to a zero of the Bessel function $\mathcal{J}_0$, the hopping is suppressed showing the phenomenon of dynamical localization as first discussed in Ref.~\cite{dunlap_86}, and experimentally verified in Ref.~\cite{lignier_07}. The thermodynamics of such driven chain has been studied in Ref.~\cite{arimondo_12}.
The leading $\Omega^{-1}$-corrections represent an additional second-nearest-neighbor hopping term, and an extra static potential, which is periodic for any rational $\Phi/\pi$. To order $\Omega^{-1}$ the stroboscopic and effective Hamiltonians are qualitatively the same but the values of the renormalised parameters are different.
The coupling constants for Magnus correction $\mathcal{C}_{m,m+2}(\zeta)$ and $\mathcal{G}_{m,m+1}(\zeta)$ are given in Appendix~\ref{app:DK_corr_coeffs_1D}, while the ones for the High-Frequency expansion are denoted by tilde and are given in Appendix~\ref{app:DK_corr_coeffs_1D_Heff}. 
Higher-order corrections in the inverse frequency appear as longer-range hopping terms, and modifications to the static potential.
The leading correction to the kick operator $K^\text{rot}_\text{eff}(t)$ can be obtained using Eq.~\eqref{eq:kick_operator_HFE}.

Next we discuss the dressed operators emerging in the FNS evolution, i.e.~the operators describing observables averaged over fast oscillations, c.f.~Sec.~\ref{subsec:Floquet_experiment}. Two natural observables are the local density operator on site $m$ and the local current operator flowing from site $m$ to site $m+1$:
\begin{eqnarray}
n_m &=& a^\dagger_m a_m,\nonumber\\
j_{m,m+1} &=& - i J_0(a^\dagger_{m+1}a_m - a^\dagger_m a_{m+1}).
\end{eqnarray}
The transformation to the rotating frame leaves the density operator (commuting with the driving term) invariant, while the current operator (which does not commute with the driving $H_1$) changes in the same way as the hopping term in the Hamiltonian. As we mentioned in Sec.~\ref{subsec:magnus_resumsubseries_pf}, it is convenient to study the finite-frequency corrections to the dressed observables in the rotating frame. One then finds
\begin{eqnarray}
n^\text{rot}_m(t) &=&\ a^\dagger_ma_m = n^\text{lab}_m,\nonumber\\
j^\text{rot}_{m,m+1}(t) &=&\ J_0\left(-ig^{m,m+1}(t) a^\dagger_{m+1}a_m + \text{h.c.}\right),\nonumber\\
\end{eqnarray}
Applying Eq.~\eqref{bar_O_magnus} leads to the following infinite-frequency behavior of the dressed operators in the stroboscopic and effective picture:
\begin{eqnarray}
\overline{n}^{(0)}_{\text{eff},m}=\overline{n}^{(0)}_{F,m} &=& n_{m},\quad \overline{j}^{(0)}_{\text{eff},m,m+1}=\overline{j}^{(0)}_{F,m,m+1}= J_\text{eff}(\zeta)(-i a^\dagger_{m+1}a_m + \text{h.c.}) 
\end{eqnarray}
As in the Kapitza case the difference between the stroboscopic and effective pictures appears in the structure of the subleading $1/\Omega$ corrections to these observables. On one hand, the observables $\mathcal{A}$ which remain invariant under the transformation to the rotating frame $V(t)$ do not get dressed in the infinite-frequency limit, i.e.~$\overline{\mathcal{A}}^{(0)} = \mathcal{A}$. Furthermore, these observables do not possess $\Omega^{-1}$-corrections in the effective picture (but they do have $\Omega^{-1}$-corrections in the stroboscopic picture). On the other hand, all observables which are not invariant with respect to the transformation to the rotating frame get dressed even in the infinite-frequency limit. In agreement with our general results (see Eq.~\eqref{bar_current}) the dressed current operator is precisely the current operator associated with the  Floquet Hamiltonian. In other words, the dressed current describes the slow charge dynamics with respect to $H_F$. Such a dressed chiral current was successfully measured in a recent cold-atom experiment realizing the Harper-Hofstadter model in a ladder geometry~\cite{atala_14}. The $\Omega^{-1}$-corrections to the dressed operators can be calculated with the help of the general expression, Eq.~\eqref{bar_O_magnus}. Since they are quite lengthy, we shall not show them explicitly. Physically the corrections are responsible for delocalization of the corresponding dressed operators, meaning that the operator support on the lattice grows as $\Omega$ deviates from infinity. For example, the corrections to the dressed density involve terms which involve hopping between neighboring sites, etc. 

Next we compute the dressed density matrices. We consider two natural initial states in which the particle is either localized in position space or in momentum space, corresponding to the bare density matrices:
\begin{eqnarray}
\rho_m &=& |m\rangle\langle m|,\ \ \ \rho_k = |k\rangle\langle k|.
\end{eqnarray}
We shall distinguish between the two density operators by the subindex $m$ or $k$.
In the rotating frame, the two operators transform to
\begin{eqnarray}
\rho_m^\text{rot}(t) &=& |m\rangle\langle m|,\nonumber\\
\rho_k^\text{rot}(t) &=& \frac{1}{N_s}\sum_{mn}\mathrm e^{ik(m-n)}\mathrm \mathrm e^{-i \zeta \sin\frac{\Phi(m-n)}{2}\sin\left(\Omega t - \Phi{m+n-1\over 2}\right) }|m\rangle\langle n|.
\end{eqnarray}
Here $N_s$ is the number of lattice sites. In the infinite-frequency limit, averaging over $t$ leads to
\begin{eqnarray}
\overline{\rho}_{\text{eff},m}^{(0)} &=& |m\rangle\langle m|,\nonumber\\
\overline{\rho}_{\text{eff},k}^{(0)} &=& \frac{1}{N_s}\sum_{mn}\mathrm e^{ik(m-n)}\mathcal{J}_0\left(\zeta\sin\frac{\Phi(m-n)}{2}\right)|m\rangle\langle n|.\nonumber\\
\end{eqnarray}
As expected, the Fock-state density matrix, which commutes with the driving protocol, is not modified in the infinite-frequency limit and hence it still represents a pure state. On the contrary, the momentum-state density matrix gets dressed. In momentum space, this density matrix remains diagonal:
\begin{equation}
\overline{\rho}_{\text{eff},k}^{(0)} = \sum_q \sum_{l=-\infty}^\infty \mathrm e^{i(k-q)l}\mathcal{J}_0\left(\zeta \sin\frac{l\Phi}{2}\right)|q\rangle\langle q|.
\end{equation} 
and it represents a mixed state. We mention in passing that finite-frequency corrections to the density matrices result in a mixed state even for a pure Fock-state.

\begin{figure}[h]
	\makebox[\textwidth]{%
		\resizebox*{0.5\columnwidth}{!}{\includegraphics{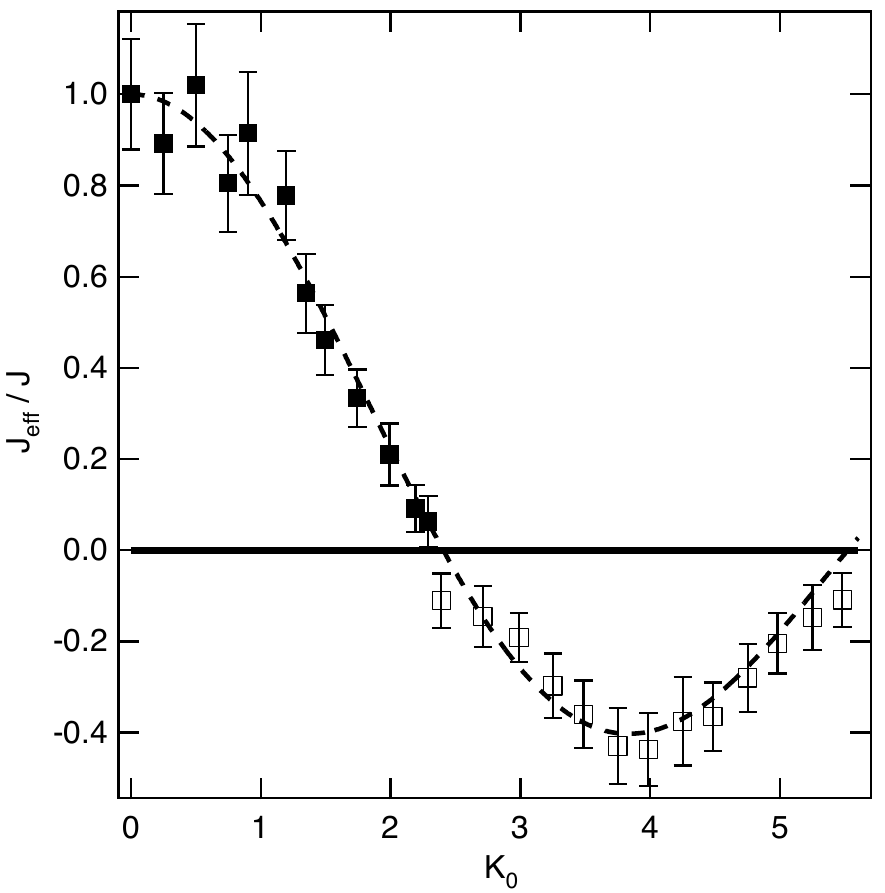}}
		\hfill    
		\resizebox*{0.5\columnwidth}{!}{\includegraphics{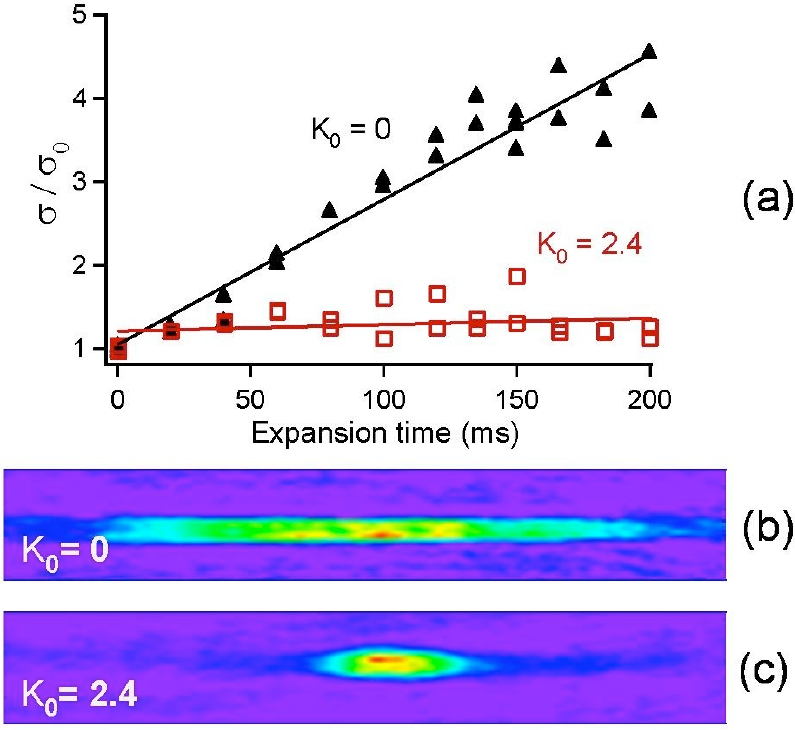}}}
	\caption{\label{fig:zenesini}(Color online). Left panel (taken from Ref.~\cite{ciampini_11}). Reproduced by per- mission of World Scientific. Observation of dynamical localisation with cold atoms. The magnitude of the effective tunnelling coupling $J_\text{eff}$ is extracted from an in situ image, while the relative sign is determined through the interference pattern using a TOF image. Right panel (taken from Ref.~\cite{zenesini_10}). Reproduced by permission of Springer. In situ images reveal the change in the condensate width $\sigma$ during free expansion of a BEC in an optical lattice. (a) Normalised expansion width $\sigma/\sigma_0$ of the atomic cloud versus expansion time for $K_0 = 0$ (black triangles) and $K_0 = 2.4$ (red squares). (b) CCD in situ images of the condensate cloud for $K_0 = 0$ and $K_0 =  2.4$. The parameters on the figre are related to those in the main text by $J= J_0$, $K_0 = \zeta$ and $J_\text{eff} = J_0\mathcal{J}_0(\zeta)$. For more details, see Refs.~\cite{eckardt_05,lignier_07, zenesini_09,zenesini_10,creffield_10,ciampini_11,arimondo_12}. }
\end{figure}

\emph{Experimental Observation of Dynamical Localisation in Cold Atom Systems.} Let us briefly describe some recent experimental setups where the renormalisation of the hopping amplitude, $J_0\to J_0\mathcal{J}_0(\zeta)$, has been observed. For example, in Refs.~\cite{lignier_07,sias_08} the dynamical localization of a strongly driven chain of $^{87}$Rb atoms was observed. First, the atoms are cooled down to form a Bose-Einstein condensate (BEC). With the help of acousto-optical modulators, the 1D optical lattice is moved back and forth, c.f.~Fig.~\ref{fig:1D_lattice}, creating a periodic net force of the form $V_0 \sum_m m\cos\Omega t\;  n_m$ on the atoms in the wells (recall that $V_0$ has to scale with the frequency, i.e.~$V_0=\zeta\Omega$). According to the predictions of Floquet's theory, the system is expected to exhibit dynamical localisation~\cite{eckardt_08,zenesini_09} when the  effective hopping approaches zero. This can be achieved by tuning to one of the zeros of the Bessel function [the first zero occurring at $\zeta\approx 2.4$, c.f.~Eqs.~\eqref{zeta} and \eqref{eq:H_F^1_bosons_1D}]. The tunneling can be measured experimentally by turning down the confining potential along the lattice direction and allowing the atom cloud to expand in the lattice~\cite{creffield_10} as it is shown in the left panel of Fig.~\ref{fig:zenesini} taken from Ref.~\cite{ciampini_11}. The right panel in this figure shows an image of the cloud taken with a CCD camera from a similar experiment~\cite{zenesini_10}. From this image one can extract the in situ width of the atom cloud after the expansion. It is evident that the expansion is very slow near the zero of the Bessel function (plot c) indicating dynamical localization. The resulting data showed an excellent agreement with the theoretical predictions.

To investigate the coherence of the BEC in the driven system, both the confining potential and the lattice beams are switched off. The atom cloud undergoes a free fall, and the degree of phase coherence is determined from the visibility of the interference pattern after time-of-flight imaging. It was shown that the system starts losing its coherent behavior when the effective hopping approaches zero. Phase coherence is restored soon after passing through the zero of the Bessel function when the effective hopping changes sign. 
	
In the same experiments, the authors also investigated closely the regions of parameter space of the shaken Bose-Hubbard model which correspond to dynamical localisation~\cite{eckardt_05,zenesini_09,zenesini_10}. There they found loss of coherence and attributed this to the Mott-insulator-to-superfluid transition. By performing time-of-flight measurements, the momentum distribution of the atom cloud was mapped out for different values of the driving amplitude. Far away from the zeros of the Bessel function, where the hopping is expected to be large compared to the atom-atom interactions, the experiments found a momentum distribution with well-defined peaks at quasimomentum $q=0$, indicating that the system is in the phase-coherent superfluid state. However, when the value of the driving amplitude is tuned to the zero of the Bessel function the visibility in the corresponding interference pattern is reduced drastically. The atoms lose phase coherence and the system is believed to enter the Mott insulating phase. Past the zero of the Bessel function, the hopping amplitude changes sign, since the Bessel function becomes negative, and the lowest Bloch band gets inverted. In agreement with theory, the position of the momentum peaks in the experiment reappears at quasimomentum $q = \pi$ at the edge of the Brillouin zone and the phase coherence in the system is being restored.

\subsection{\label{subsec:harper} Cold Atoms Realization of the Harper-Hofstadter Hamiltonian.}

We now extend the model from the previous section adding a second spatial dimension and a magnetic field gradient along this new direction. This setup was first proposed in Refs.~\cite{jaksch_03,mueller_04} for the simulation of the Harper-Hofstadter Hamiltonian with cold atoms and was recently realized experimentally~\cite{aidelsburger_13,miyake_13,aidelsburger_14,kennedy_15}. After giving an overview of the infinite-frequency limit, we discuss the leading $\Omega^{-1}$-corrections using both the Magnus and the High-Frequency expansion. These corrections, as well as the dressing of the operators, may be important for the existing experimental setups. The discussion of their effect on the dynamics goes beyond the scope of the review and they are discussed in a different work~\cite{bukov_14}. 

\begin{figure}
	\centering
	\subfigure{
		\resizebox*{10cm}{!}{\includegraphics[width = 0.6\columnwidth]{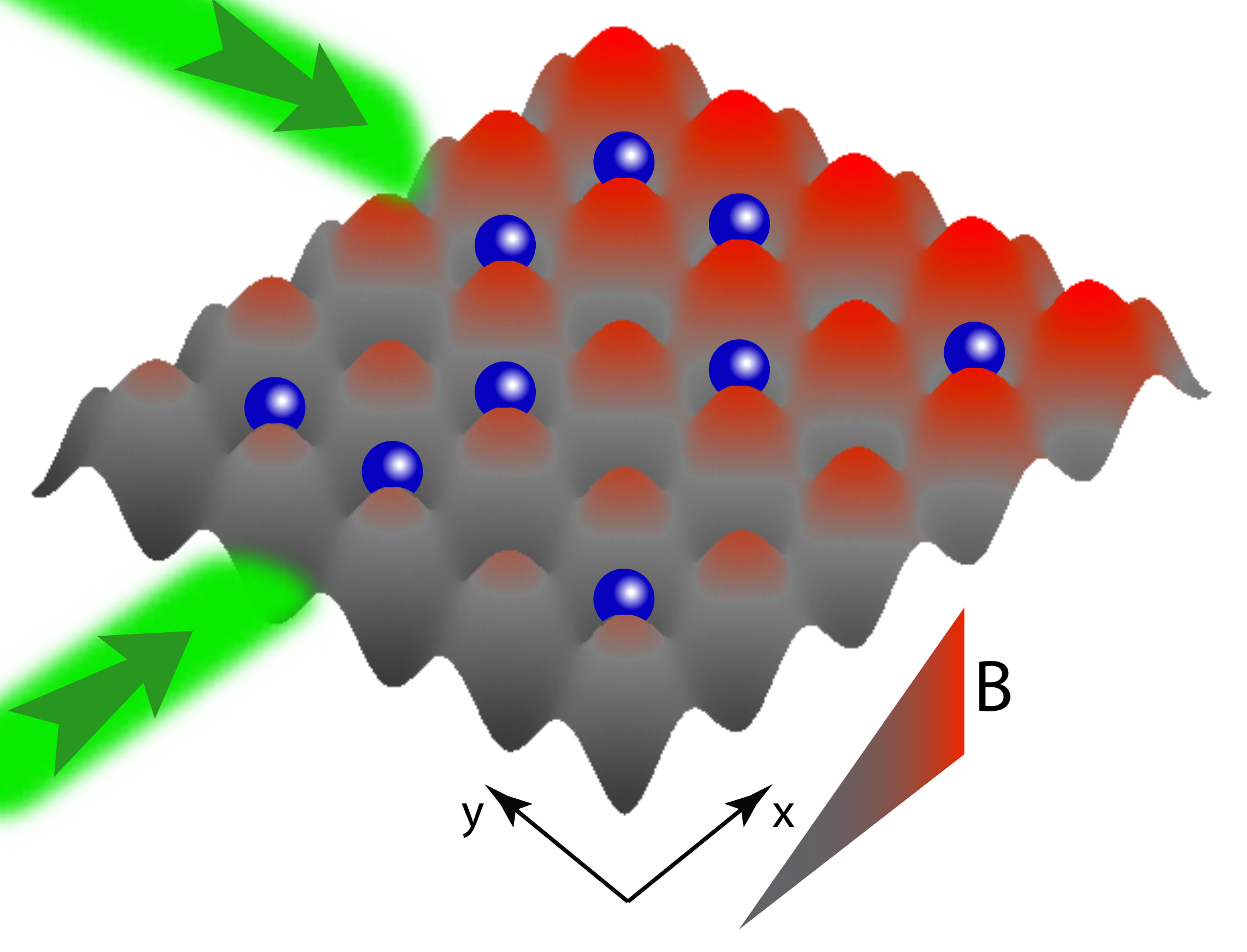}}}
	\hspace{6pt}
	\subfigure{
		\resizebox*{10cm}{!}{\includegraphics[width = 0.6\columnwidth]{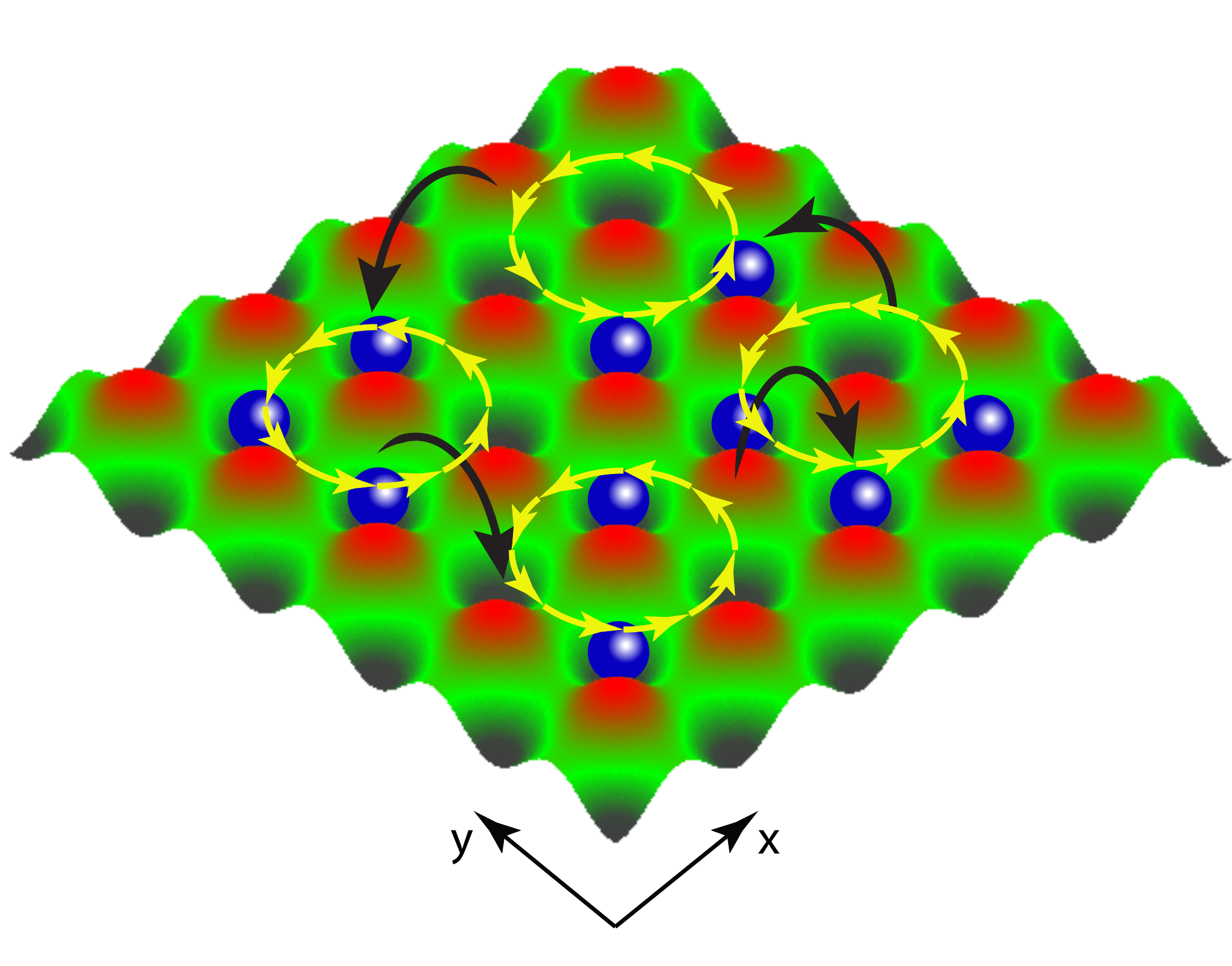}}}
	\caption{\label{fig:harper}(Color online). The Floquet realization of the Harper-Hofstadter model. Electrically neutral bosons are loaded in an optical lattice and subject to a Zeeman magnetic field gradient of value $\Omega$, which plays the role of the external potential along the $x$-direction. In addition, two Raman lasers of resonant frequency $\Omega$, with site-dependent phase lags, create a second running lattice. In the high-frequency limit, when the amplitude of the Raman lasers scales with the frequency, one generates an effective orbital magnetic field, realizing the bosonic Harper-Hofstadter Hamiltonian~\cite{harper_55,hofstadter_76}. Figure taken from Ref.~\cite{bukov_14}. }
\end{figure}

Consider a bosonic system on a square lattice subject to a linear potential along the $x$-direction and a periodic driving. In two recent experiments, the linear potential was achieved using either an static Zeeman magnetic field gradient~\cite{aidelsburger_13} or gravity~\cite{miyake_13}. In both cases this creates a constant force on the system. The periodic driving was realized by using a running (dynamical) optical lattice (c.f.~Fig.~\ref{fig:harper}). The system is described by the following Hamiltonian:
\be
H(t) =H_0+ H_1(t),
\ee
where
\begin{eqnarray}
H_0 &=& -\sum_{m,n}\bigg[ J_x\left(a^\dagger_{m+1,n}a_{mn} + \text{h.c.}\right)+ J_y\left(a^\dagger_{m,n+1}a_{mn} + \text{h.c.}\right)\bigg]\nonumber\\
&& +\frac{U}{2}\sum_{m,n} n_{mn}(n_{mn}-1),\nonumber\\
H_1(t) &=&\Omega\sum_{m,n}\bigg[\frac{\lambda}{2}\sin\left(\Omega t - \phi_{mn} + \frac{\Phi_\square}{2}\right) 
+ \Omega m\bigg] n_{mn}.
\label{eq:harper_h0}
\end{eqnarray}
Here $J_x$ and $J_y$ denote the hopping amplitude, and $V_0=\Omega\lambda$ is the strength of the 
dynamical (running) lattice which, as in the previous example, should scale linearly with the driving frequency.
The field gradient along the $x$-direction is resonant with $\Omega$ (see the term $\Omega\, m\, n_{mn}$ in $H_1$). The phase $\phi_{mn}$ is spatially inhomogeneous $\phi_{mn}= \Phi_\square(n+m)$ and makes it impossible to find a Floquet gauge (i.e.~a choice of the initial time of the stroboscopic period) for which the driving is symmetric. Breaking time-reversal symmetry ultimately allows for a synthetic static magnetic field to appear in the infinite-frequency Hamiltonian.

\begin{figure}
	\centering
	\includegraphics[width = 0.7\columnwidth]{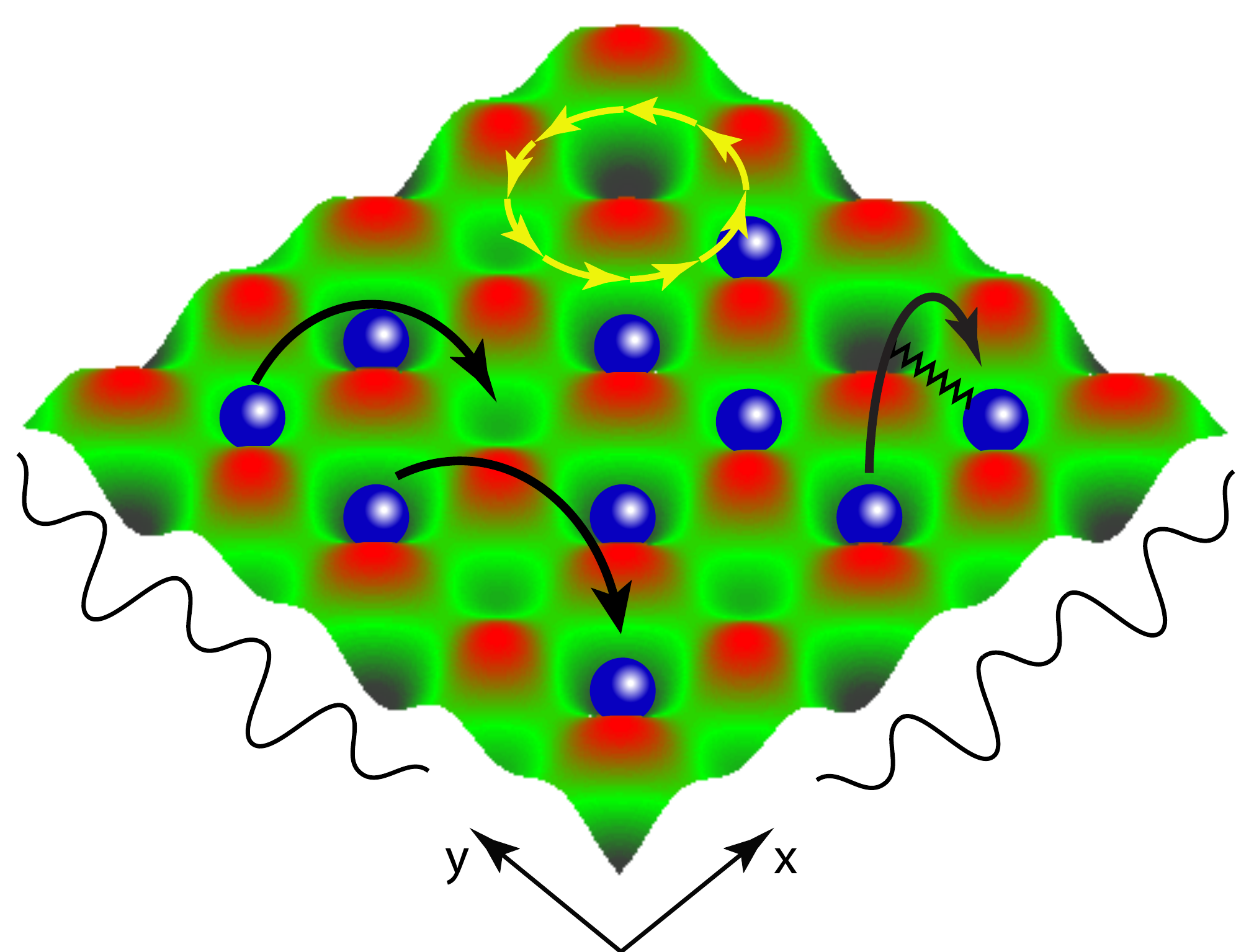}
	\caption{\label{fig:harper_corr}(Color online) The leading corrections in $\Omega^{-1}$ to the Harper-Hofstadter Hamiltonian: second-neighbor hopping including along the diagonal (solid black lines), interaction-dependent hopping (solid black lines connected to zig-zag lines) and a site-dependent chemical potential (indicated by the thin black lines on the side and the green color gradient). The interaction dependent hopping does not influence the Floquet spectrum to order $\Omega^{-1}$ but is important for the correct description of the stroboscopic evolution. In the effective Floquet picture these terms show up the in kick operator instead (see main text). }
\end{figure}

Upon applying a transformation to the rotating frame~\cite{miyake_13}, the 
Hamiltonian takes the form 
\be
H^\text{rot}(t) = W(t) + W^\dagger(t) + H_\text{int},
\label{in_rot}
\ee
where
\begin{eqnarray}
W(t) =-\sum_{m,n} \bigg[J_x \mathrm e^{-i\zeta\sin(\Omega t - \phi_{nm}) +i\Omega 
	t }a^\dagger_{m+1,n}a_{mn}
+ J_y \mathrm e^{-i\zeta\sin(\Omega t - \phi_{nm})}a^\dagger_{m,n+1}a_{mn}\bigg],\nonumber\\
\end{eqnarray}
and $\zeta = \lambda\sin(\Phi_\square/2)$. 
To zeroth order, the Floquet Hamiltonian coincides with the effective Hamiltonian and it is obtained by averaging Eq.~\eqref{in_rot} over a period:
\begin{eqnarray}
H_F^{(0)} = H_\text{eff}^{(0)} &=& -K\sum_{m,n} \left(\mathrm e^{i\phi_{mn}}a^\dagger_{m+1,n}a_{mn} + 
\text{h.c.}\right)
-J\sum_{m,n}\left(a^\dagger_{m,n+1}a_{mn} + \text{h.c.}\right)\nonumber\\
&& +\frac{U}{2}\sum_{m,n} n_{mn}(n_{mn}-1).
\label{eq:harper_Heff}
\end{eqnarray}
The effective hoppings are given by $K = J_x\mathcal{J}_1(\zeta)$, $J = J_y\mathcal{J}_0(\zeta)$, and $\mathcal{J}_\nu$ is the $\nu$-th Bessel function. 
The next order in the Magnus expansion delivers the leading finite-frequency corrections to the stroboscopic Floquet Hamiltonian:
	\begin{eqnarray}
	&&H_F^{(1)}[0] =\ -\sum_{m,n} 
	\left(\frac{J_x^2}{\Omega}{^{\rightarrow}C}^n_{m,m+2}(\zeta)a^\dagger_{m+2,n}a_{mn} 
	+ \frac{J_y^2}{\Omega}{^{\uparrow}C}^{n,n+2}_{m}(\zeta)a^\dagger_{m,n+2}a_{mn} + 
	\text{h.c.}\right)\nonumber\\
	&&\ -\sum_{m,n} 
	\left(\frac{J_xJ_y}{\Omega}{^{\nearrow}D}^{n,n+1}_{m,m+1}(\zeta)a^\dagger_{m+1,n+1}
	a_{mn} + 
	\frac{J_xJ_y}{\Omega}{^{\nwarrow}D}^{n,n+1}_{m,m-1}(\zeta)a^\dagger_{m-1,n+1}a_{mn} 
	+ \text{h.c.}\right)\nonumber\\
	&&\ +\sum_{m,n}\left( \frac{J_x^2}{\Omega}{^{\rightarrow}E}^n_{m,m+1}(\zeta)(n_{m,n} 
	- n_{m+1,n}) + \frac{J_y^2}{\Omega}{^{\uparrow}E}^{n,n+1}_{m}(\zeta)(n_{mn} - 
	n_{m,n+1})\right)\nonumber\\
	&&\ -\sum_{m,n} 
	\Bigg(\frac{J_xU}{\Omega}{^{\rightarrow}B}^{n}_{m,m+1}(\zeta)a^\dagger_{m+1,n}a_{mn}(n_{mn} - 
	n_{m+1,n}+1) \nonumber\\
	&&\ \ \ \ \ \ \ \ \ \ + 
	\frac{J_yU}{\Omega}{^{\uparrow}B}^{n,n+1}_{m}(\zeta)a^\dagger_{m,n+1}a_{mn}(n_{mn} - n_{m,n+1}+1) 
	+ \text{h.c.} \Bigg).
	\label{eq:HH_corr}
	\end{eqnarray}
The arrows on the corresponding hopping coefficient indicate the direction of the hopping. The complex-valued functions $B(\zeta)$, $C(\zeta)$, $D(\zeta)$ and $E(\zeta)$  are defined on the bonds of the lattice. They are obtained from the time-ordered integrals in the Magnus expansion, and 
are given in Appendix~\ref{app:DK_corr_coeffs_2D}. We see that, when we include the $\Omega^{-1}$-corrections, the Floquet Hamiltonian becomes quite complex. These corrections introduce effective static potentials (periodic for rational $\Phi_\square/\pi)$ along both directions of the lattice, second-nearest-neighbor hopping both across the diagonals and along the lattice directions, and interaction-dependent hopping (see Fig.~\ref{fig:harper_corr}). The consequences of these corrections for the single-particle dynamics, as well as the dressing of the density matrix and observables, was discussed in a different work~\cite{bukov_14}.

Similarly the $\Omega^{-1}$-corrections to the effective Hamiltonian are obtained from the first order in the High-Frequency Expansion:
\begin{eqnarray}
	&&H_\text{eff}^{(1)} =\ -\sum_{m,n} 
	\left(\frac{J_x^2}{\Omega}{^{\rightarrow}\tilde C}^n_{m,m+2}(\zeta)a^\dagger_{m+2,n}a_{mn} 
	+ \frac{J_y^2}{\Omega}{^{\uparrow}\tilde C}^{n,n+2}_{m}(\zeta)a^\dagger_{m,n+2}a_{mn} + 
	\text{h.c.}\right)\nonumber\\
	&&\ -\sum_{m,n} 
	\left(\frac{J_xJ_y}{\Omega}{^{\nearrow}\tilde D}^{n,n+1}_{m,m+1}(\zeta)a^\dagger_{m+1,n+1}
	a_{mn} + 
	\frac{J_xJ_y}{\Omega}{^{\nwarrow}\tilde D}^{n,n+1}_{m,m-1}(\zeta)a^\dagger_{m-1,n+1}a_{mn} 
	+ \text{h.c.}\right)\nonumber\\
	&&\ +\sum_{m,n}\left( \frac{J_x^2}{\Omega}{^{\rightarrow}\tilde E}^n_{m,m+1}(\zeta)(n_{m,n} 
	- n_{m+1,n}) + \frac{J_y^2}{\Omega}{^{\uparrow}\tilde E}^{n,n+1}_{m}(\zeta)(n_{mn} - 
	n_{m,n+1})\right).
	\label{eq:HH_corr_Heff}
\end{eqnarray}
The effective Hamiltonian is similar to the stroboscopic Hamiltonian. However, the coefficients defining the renormalised parameters in the effective Hamiltonian are, in general, different from those for the stroboscopic Hamiltonian, and are denoted by a tilde. They are defined in Appendix~\ref{app:DK_corr_coeffs_2D_Heff} and are Floquet-gauge invariant, i.e.~do not depend on the phase of the drive. The main qualitative difference between the stroboscopic and effective expansions is the absence of interaction-dependent hopping terms in $H_\text{eff}^{(1)}$ which are instead present in $H_F^{(1)}[0]$. This means that those terms modify the Floquet spectrum (and all other invariants under a change of basis) at the order $\Omega^{-2}$, i.e.~beyond te validity if the current approximation. In the effective picture these terms appear in the kick operator affecting the initial density matrix and observables to the order $1/\Omega$. In particular

\begin{eqnarray}
K_{\rm eff}^{\rm rot, (1)}(t) &=& \sum_{m,n} \bigg[J_x \kappa_x(t) a^\dagger_{m+1,n}a_{mn}
+ J_y \kappa_y(t)a^\dagger_{m,n+1}a_{mn} + \text{h.c.}\bigg], \nonumber \\
K_{F}^{\rm rot, (1)}[0](t) &=& \sum_{m,n} \bigg[J_x \left[\kappa_x(t)- \kappa_x(0)\right] a^\dagger_{m+1,n}a_{mn}
+ J_y \left[\kappa_y(t) - \kappa_y(0)\right] a^\dagger_{m,n+1}a_{mn} + \text{h.c.}\bigg],\nonumber\\
\end{eqnarray}
where
\begin{eqnarray}
\kappa_x(t) &=& \frac{1}{2}\int_t^{T+t}\mathrm{d}t'\left(1+2\frac{t-t'}{T}\right)\mathrm e^{-i\zeta\sin(\Omega t' - \phi_{nm}) +i\Omega t' },\nonumber\\
\kappa_y(t) &=& \frac{1}{2}\int_t^{T+t}\mathrm{d}t'\left(1+2\frac{t-t'}{T}\right)\mathrm e^{-i\zeta\sin(\Omega t' - \phi_{nm})}.
\nonumber
\end{eqnarray}
Applying Eq.~\eqref{eq:F_vs_eff_1st_order} in the rotating frame, we have 
	\begin{equation}
	H_F^{(1)}[0] = H_\text{eff}^{(1)} -i \left[ K^{\rm rot, (1)}_\text{eff}(0),H_\text{eff}^{(0)} \right].\nonumber
	\end{equation}
	Therefore, whenever one chooses to work in the effective picture, the interaction-dependent hopping terms are implicitly contained in the kick operator $K_\text{eff}(0)$.  
	
\begin{figure}
	\centering
	\includegraphics[width = 0.6\columnwidth]{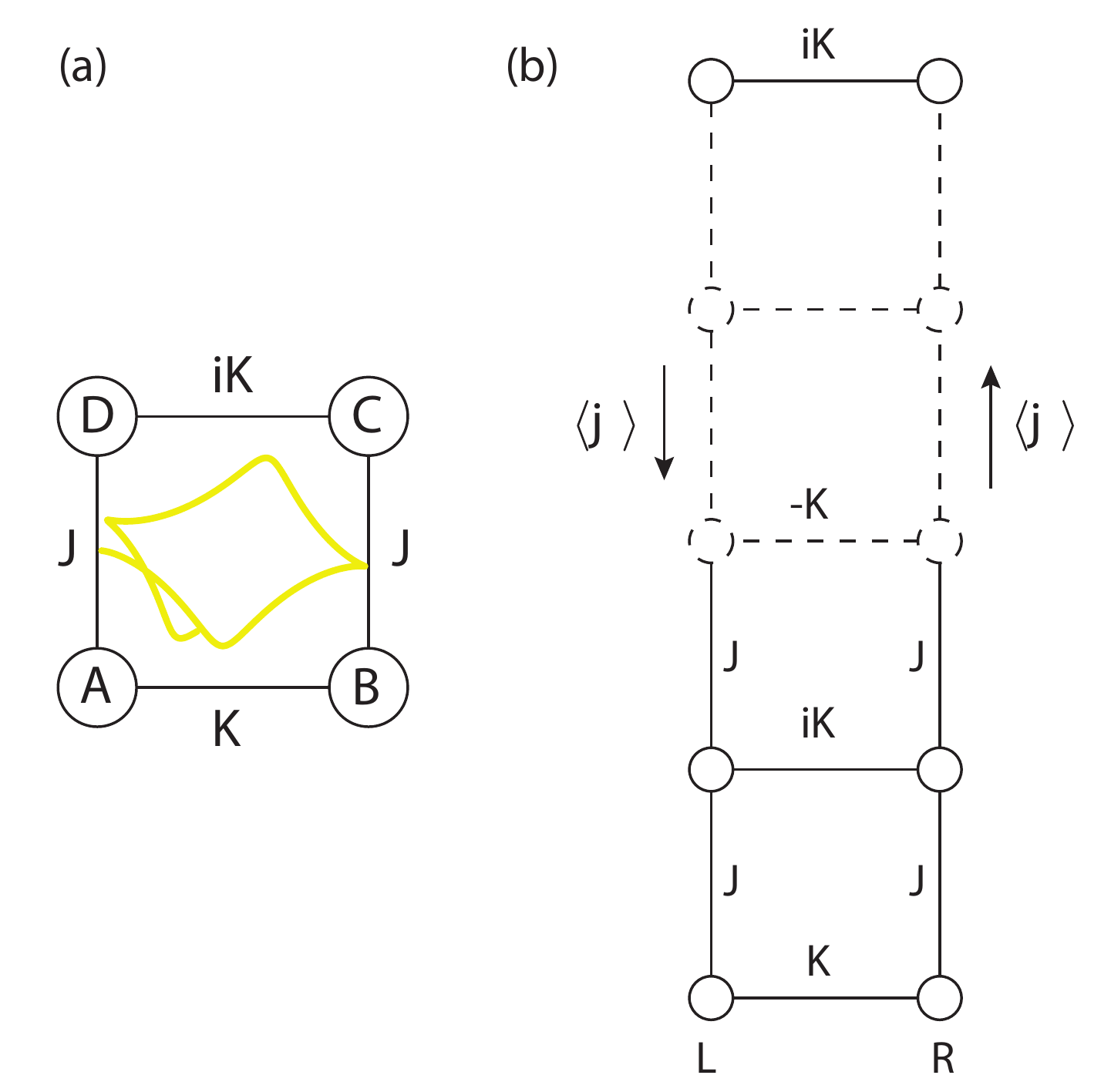}
	\caption{\label{fig:ladder_plaquette1}(Color online) (a) The plaquette geometry used to study the quantum cyclotron orbits (yellow) in the Harper-Hofstadter Hamiltonian. (b) The ladder geometry with the chiral currents used to study the transition between the vortex and the Meissner phases. Figure taken from Ref.~\cite{bukov_14}. }
\end{figure}

Before we close the discussion on the theoretical Floquet realisation of the Harper-Hofstadter model, we mention that a different method of engineering artificial gauge fields using a high-frequency periodic perturbation was proposed in Ref.~\cite{kolovski_11} (but see also Ref.~\cite{creffield_13}), based on an oscillating field gradient, where $H_1(t) = \sum_{mn} m (\Omega + V_0\cos\Omega t\;  ) n_{mn}$. Moreover, in Ref.~\cite{verdeny_13} the flow-equation method, which is an alternative to the Magnus expansion, has been used to compute the finite-frequency corrections to the Floquet Hamiltonian. As expected, this method reproduces the same results as the Magnus expansion. Recently, the stability~\cite{choudhury_14} of a related Bose-Hubbard model under a periodic driving, and scattering properties of periodically-driven lattice systems~\cite{bilitewski_14} have been studied too.

\emph{Experimental realisation of the Harper-Hofstadter model}. The Harper-Hofstadter Model has been realised experimentally using cold atoms in optical lattices~\cite{aidelsburger_13,miyake_13,atala_14,aidelsburger_14,kennedy_15}. First, Rb atoms are cooled down to form a BEC and loaded in a 2D optical lattice. Then a field gradient is applied along the $x$-direction, such that tunnelling along the $x$-direction is suppressed. The latter is then restored by a running lattice, which consists of two additional laser beams which interfere at an angle with respect to one another, c.f.~Fig.~\ref{fig:harper}. The resulting running lattice leads to a periodic on-site modulation with a site-dependent phase. The frequency of the running lattice is chosen to match the magnetic field gradient, realizing the Hamiltonian~\eqref{eq:harper_h0} with $\Phi_\square=\pi/2$. This flux can be controlled by the angle between the running lattice beams. In the infinite-frequency limit, the flux is equivalent to a very strong static magnetic field (see Eq.~\eqref{eq:harper_Heff}). 
	
\begin{figure}
	\centering
	\includegraphics[width = 1\columnwidth]{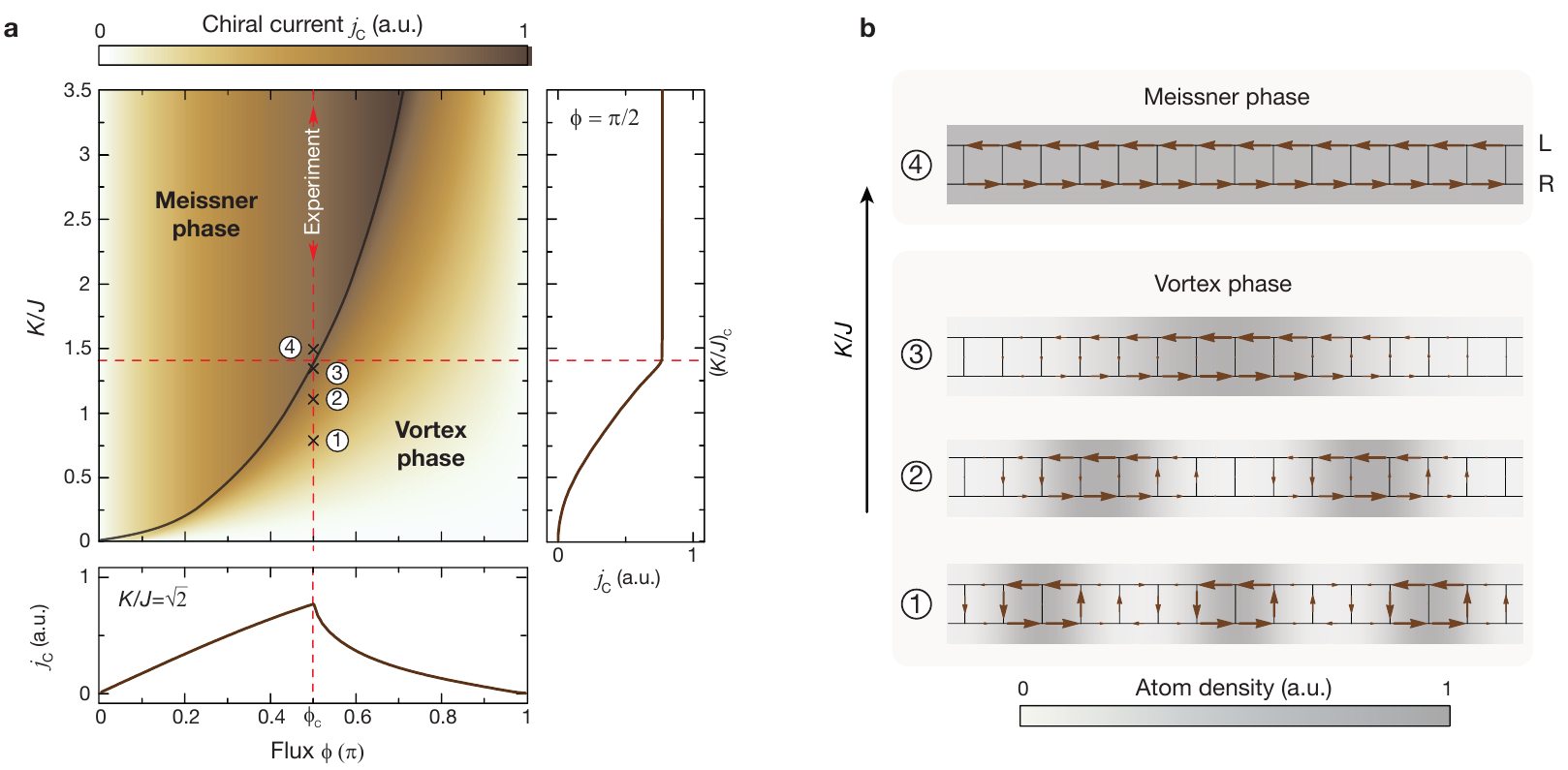}
	\caption{\label{fig:atala}(Color online) (a) Phase diagram of the Harper-Hofstadter model on a ladder in $(K/J,\phi)$- space. The insets show the chiral current $j_c$ as a function of the ratio of the effective hoppings $K/J$ and the flux per plaquette $\phi$. (b) Pictorial representation of the Meissner and vortex phases for several values of the ratio $K/J$. The parameters are related to those in the main text by $\phi = \Phi_\square$. Reprinted by permission from Macmillan Publishers Ltd: [Nature Physics] (\cite{atala_14}), copyright (2014).}
\end{figure}	

In the experiment of Ref.~\cite{aidelsburger_13} the authors additionally introduced a static superlattice potential, which effectively divided the 2D lattice in $2\times 2$ plaquettes (see Fig.~\ref{fig:ladder_plaquette1} (a)), and completely suppressed the tunnelling among different plaquettes. Then they studied the analogue of the classical cyclotron orbit and found a good agreement with the predictions of the effective Hamiltonian~\eqref{eq:harper_Heff}. In another experiment~\cite{atala_14}, the superlattice was switched only along the $x$-direction, such that the 2D lattice was divided into many ladders along the $y$-axis, c.f.~Fig.~\ref{fig:ladder_plaquette1} (b), and the tunneling between different ladders was completely suppressed. 
The atoms in the ground state of the system move along the edges of the ladders in the form of chiral currents. Depending on the ratio between the effective hopping parameters in the $x$- and $y$-directions, a phase transition between a vortex phase and a Meissner phase was found at which the chiral current undergoes a cusp singularity. For the flux $\Phi_\square=\pi/2$ this transition occurs at the critical ratio $(K/J)_c=\sqrt{2}$ (recall that $K$ and $J$ are the effective hopping along the $x$ and $y$ directions, c.f. Eq.~\eqref{eq:harper_Heff}). For $K/J<(K/J)_c$  (vortex phase) the current increases and the vortex density decreases with $K$ until it hits zero (in the thermodynamic limit) at the critical ratio. For $K/J>(K/J)_c$ (Meissner phase) the current at a fixed value of $J$ is independent of $K$, see Fig.~\ref{fig:atala}. Effectively the authors performed an FNS measurement of the current by projecting the system into an array of decoupled double wells along the horizontal direction. Then they fit the Josephson density oscillations in a double well to a simple formula with the chiral current entering through the initial conditions (see Ref.~\cite{atala_14} for details).

In a follow-up experiment, the realisation of the Harper-Hofstadter model has been optimised. The field gradient has been replaced by a superlattice, and the running lattice has been modified accordingly~\cite{aidelsburger_14}. This experiment measured the Chern number of the lowest band by detecting the differential drift of the atom cloud in momentum space, which arises due to the Berry curvature of the band~\cite{price_12}. The Chern number was measured to be close to unity with excellent precision. This most recent experiment also takes into account the relevant first order corrections to the time-averaged Hamiltonian.

\subsection{\label{subsec:Floquet_top_ins} The Periodically Driven Fermi-Hubbard Model. Floquet Topological Insulators.}

\begin{figure}
	\centering
	\subfigure{
		\resizebox*{10cm}{!}{\includegraphics[width = 0.6\columnwidth]{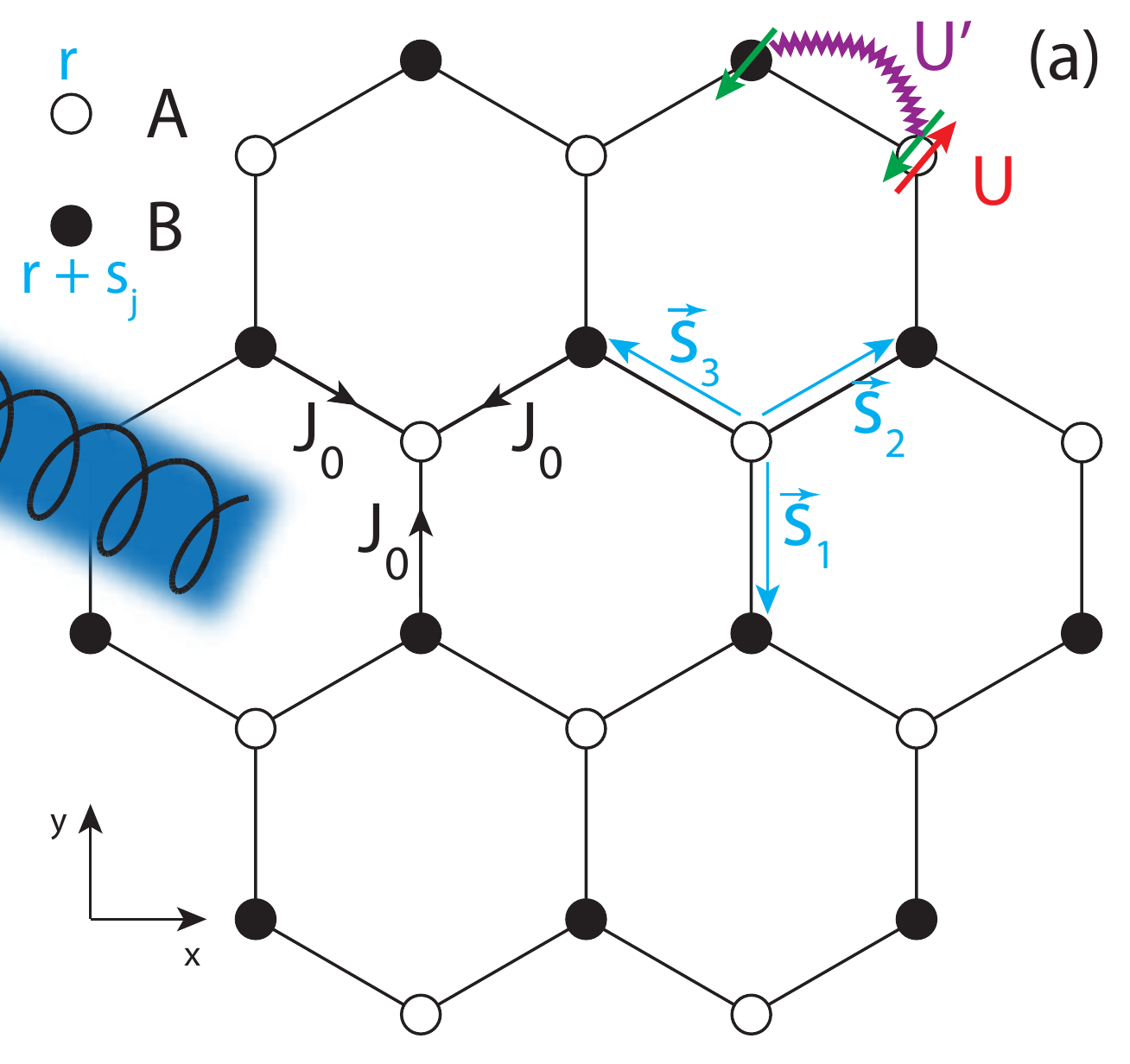}}}
	\hspace{6pt}
	\subfigure{
		\resizebox*{10cm}{!}{\includegraphics[width = 0.6\columnwidth]{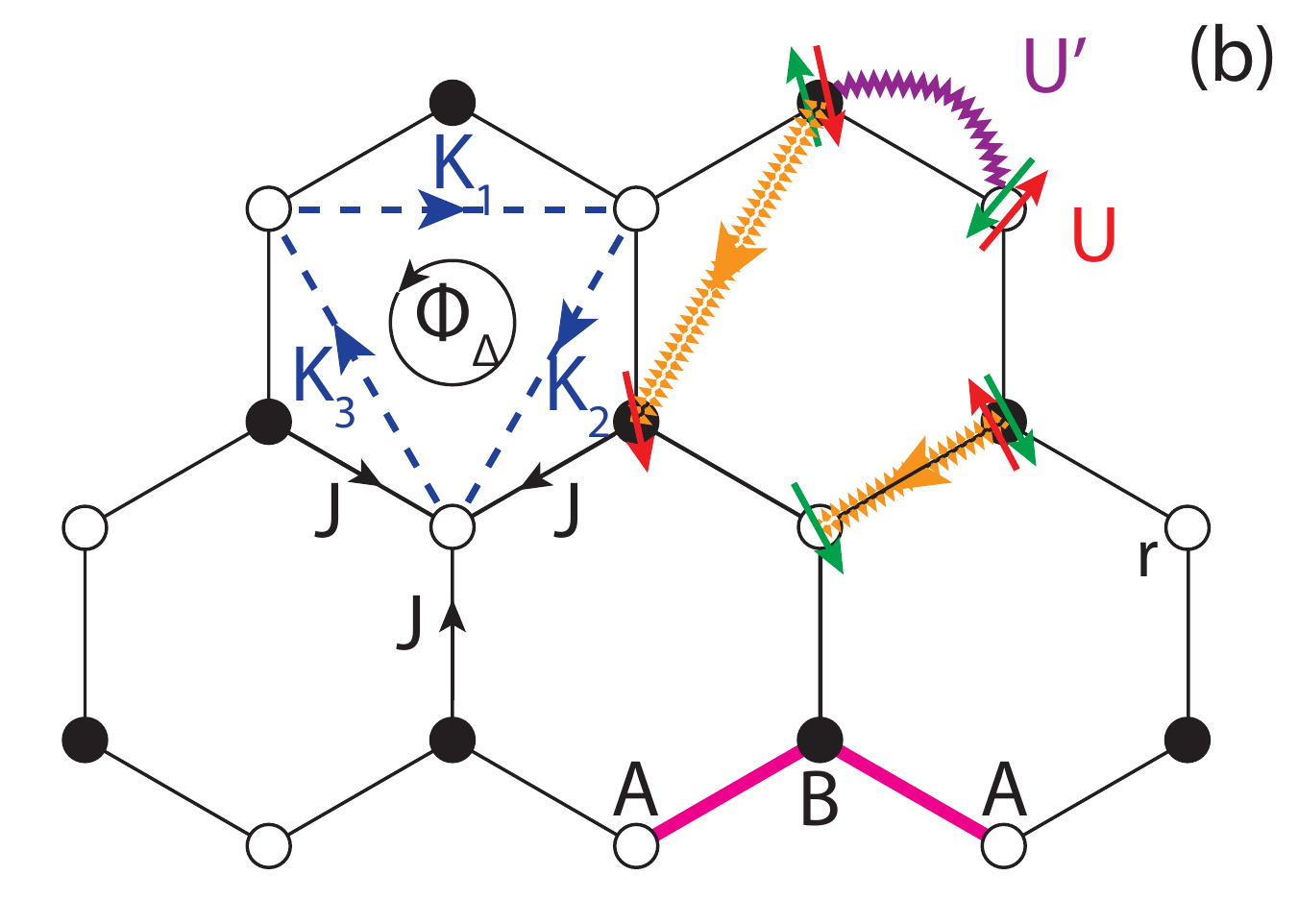}}}
	\caption{\label{fig:haldane_model}(Color online). Floquet realization of the Haldane-Fermi-Hubbard Model (see also Refs.~\cite{kitagawa_11, jotzu_14}). (a) A circularly polarized electric field is shone on a sheet of graphene. The non-driven model includes spinful fermions with hopping matrix elements $J_0$, on-site interactions $U$ between spin-up (dark green arrows) and spin-down (dark red arrows) fermions, as well as nn interactions $U'$ (purple full zigzag line) between either spin species. (b) To zeroth order the Floquet Hamiltonian has the same form as the non-driven Hamiltonian but with renormalized parameters. The leading $\Omega^{-1}$-corrections include complex next-nearest-neighbor hopping elements $K_j$ (dashed blue lines), such that the flux through a sublattice is $\Phi_\triangle = -\pi/2$. If the system is interacting, to the same order in perturbation theory, an interaction-dependent hopping is induced (orange lines) in the stroboscopic Floquet Hamiltonian $H_F^{(1)}[0]$.These interaction-dependent hopping terms enter the Floquet spectrum starting from order $\Omega^{-2}$ and are, therefore, absent in the effective Floquet Hamiltonian, $H_\text{eff}^{(1)}$ (see text). However, in the effective picture the interaction-dependent hopping shows up in the kick operator $K_{\rm eff}^{{\rm rot}, (1)}$ (see Eq.~\eqref{eq:fermions_Heff}), and should be included in dressing the observables and the initial density matrix for a correct description of both stroboscopic and non-stroboscopic dynamics.
 }
\end{figure}

In this section, we analyze a spinful fermionic system on a bipartite lattice driven by a periodic external field which couples to the atomic density. First, we shall describe the setup in general, and later we restrict our attention to the case of graphene subject to a circularly polarized electric field. Let $z$ be the lattice coordination number, and let $A$ denote the set of all points in the sublattice $A$. We label the points on the sublattice $A$ by a vector $\bm{r}$. Let us define the vectors $\bm{s}_j$ ($j = 1,\dots,z$) to connect a fixed point on the sublattice $A$ to all its adjacent neighbors on the sublattice $B$. The vectors $\bm{s}_j$ point from $A$ to $B$ (c.f.~Fig.~\ref{fig:haldane_model}, Panel (a)). To simplify the notation we skip the bold notation for vectors in the subscripts of operators.

The system is described by the Hamiltonian
\begin{eqnarray}
H(t)&=&\ H_0 + H_1(t) \nonumber \\
	H_1(t) &=&\ \Omega\frac{\zeta}{z}\sum_{r\in A}\sum_{j = 1}^{z}\sum_{\sigma} \left(f_{r}(t)n^a_{r,\sigma}+ f_{r+s_j}(t)n^b_{r+s_j,\sigma}\right) \nonumber\\
H_0 &=& H_\text{kin} + H_\text{int},\nonumber\\
H_\text{kin} &=&\ -J_0\sum_{r\in A}\sum_{j=1}^z\sum_{\sigma}\left( a^\dagger_{r,\sigma}\,b_{r + s_j,\sigma} + \text{h.c.} \right) \nonumber\\
H_\text{int}&=&\ \frac{U}{2z}\sum_{r\in A}\sum_{j=1}^z\sum_{\sigma}\left(n^a_{r,\sigma}\,n^a_{r,\bar{\sigma}} + n^b_{r+s_j,\sigma}\,n^b_{r+s_j,\bar{\sigma}}\right)\nonumber\\
&&+ U'\sum_{r\in A}\sum_{j=1}^z\sum_{\sigma,\sigma'} n^a_{r,\sigma}n^b_{r+s_j,\sigma'}.
\end{eqnarray}  
where $\sigma = \uparrow,\downarrow$ indicates the spin with the convention $\bar\uparrow =\downarrow$ and $\bar\downarrow =\uparrow$. The factors $1/z$ are introduced to avoid over-counting. The operators $a^\dagger_{r,\sigma}$ and $b^\dagger_{r+s_j,\sigma}$ create a fermion of spin $\sigma$ on sublattices $A$ and $B$, respectively. In the Hamiltonian, $n^a_{r,\sigma} = a^\dagger_{r,\sigma}a_{r,\sigma}$ and $n^b_{r+s_j,\sigma} = b^\dagger_{r+s_j,\sigma}b_{r+s_j,\sigma}$ denote the number operators on sublattices $A$ and $B$. The bare tight-binding hopping is $J_0$, the on-site interaction strength which couples fermions of opposite spin is $U$, while the next-nearest neighbor interaction is $U'$. The driving protocol $f_j(t) = f_j(t+T)$ is periodic and site-dependent. The driving potential has the amplitude $V_0=\Omega\zeta$. 

In the rotating frame the Hamiltonian reads:
\begin{eqnarray}
H^\text{rot}(t) &=& W(t) + W^\dagger(t) + H_{\text{int}}, \label{eq:Haldane_rotating}\\
W(t) &=& -J_0\sum_{r\in A,\sigma}\sum_{j=1}^z \lambda_j(t) a^\dagger_{r,\sigma}b_{r + s_j,\sigma}\nonumber\\
\lambda_j(t) &=&\ \exp\left(i\zeta\Omega\int_{t_0}^t\mathrm{d}t' \left( f_{r+s_j}(t') - f_r(t')\right) \right) \nonumber.
\end{eqnarray} 
To zeroth order in the inverse-frequency expansion, the  Floquet Hamiltonian is given by the time-average of the Hamiltonian above (similarly to the bosonic case described in Sec.~\ref{subsec:harper}):
\begin{eqnarray}
H_F^{(0)} = H_\text{eff}^{(0)} &=&\ \overline{W(t)} + \overline{W^\dagger(t)} + H_{\text{int}}.
\label{eq:graphene_Heff}
\end{eqnarray}
It has the same form of Eq.~\eqref{eq:Haldane_rotating} but with renormalized hopping parameters. The leading $\Omega^{-1}$-corrections to the Floquet Hamiltonian in the Floquet gauge $t_0=0$ are given by
	\begin{eqnarray}
	H_F^{(1)}[0] =&&\ \frac{J_0^2}{\Omega}\sum_{r\in A}\sum_{j=1}^z\sum_{\sigma}\bigg[F_{jj}(\zeta)\left(n^a_{r\sigma}-n^b_{r+s_j,\sigma}\right)\bigg]\nonumber\\
	&&+ \frac{J_0^2}{\Omega}\sum_{r\in A}\sum_{\sigma}\sum_{j>k=1}^z\bigg[F_{jk}(\zeta)\left( a^\dagger_{r\sigma}a_{r+s_j-s_k,\sigma} - b^\dagger_{r+s_k}b_{r+s_j}\right) + \text{h.c.}\bigg]\nonumber\\
	&&+\ \frac{J_0U}{2\Omega} \sum_{r\in A}
	\sum_{j=1}^z\sum_{\sigma}\bigg[G_j(\zeta)\left( n^a_{r\sigma} - n^b_{r+s_j,\sigma}\right)a^\dagger_{r\bar\sigma}b_{r+s_j,\bar\sigma} + \text{h.c.}\bigg] \nonumber\\
	&&-\ \frac{J_0U'}{\Omega}\sum_{r\in A}\sum_{j,k=1}^z\sum_{\sigma,\sigma'}\bigg[G_j(\zeta)\left(n^a_{r+s_j-s_k,\sigma}a^\dagger_{r\sigma'}b_{r+s_j,\sigma'} - a^\dagger_{r\sigma'}b_{r+s_j,\sigma'}n^b_{r+s_k,\sigma} \right) + \text{h.c.}\bigg],\nonumber\\
	\label{eq:fermions_eff}
	\end{eqnarray}	
where $\zeta = V_0/\Omega$ is the ratio of the driving amplitude and the driving frequency.	The stroboscopic kick operator is given by
\begin{eqnarray}
K_F^{\rm rot, (1)}[0](t) &=& J_0\sum_{r\in A,\sigma}\sum_{j=1}^z [\kappa_j(t) - \kappa_j(0)] a^\dagger_{r,\sigma}b_{r + s_j,\sigma}+ \text{h.c.},\nonumber\\ 
\kappa_j(t) &=& \frac{1}{2}\int_t^{T+t}\mathrm{d}t'\left(1+2\frac{t-t'}{T}\right)\lambda_j(t').
\end{eqnarray}
For comparison, we also show the leading $\Omega^{-1}$-corrections to the effective Hamiltonian:
	\begin{eqnarray}
	H_\text{eff}^{(1)} &=& \frac{J_0^2}{\Omega}\sum_{r\in A}\sum_{j=1}^z\sum_{\sigma}\bigg[\tilde F_{jj}(\zeta)\left(n^a_{r\sigma}-n^b_{r+s_j,\sigma}\right)\bigg]\nonumber\\
	&&+ \frac{J_0^2}{\Omega}\sum_{r\in A}\sum_{\sigma}\sum_{j>k=1}^z\bigg[\tilde F_{jk}(\zeta)\left( a^\dagger_{r\sigma}a_{r+s_j-s_k,\sigma} - b^\dagger_{r+s_k}b_{r+s_j}\right) + \text{h.c.}\bigg],\nonumber\\
	K_\text{eff}^{\rm rot, (1)}(t) &=& J_0\sum_{r\in A,\sigma}\sum_{j=1}^z \kappa_j(t) a^\dagger_{r,\sigma}b_{r + s_j,\sigma} + \text{h.c.}
	\label{eq:fermions_Heff}
	\end{eqnarray}
One readily sees that the first-order correction to both the stroboscopic and effective Hamiltonian contains a static potential and a next-nearest-neighbor (nnn) hopping. The nnn hopping terms in $H_F^{(1)}[0]$, in general, have a Floquet-gauge dependent magnitude and direction while the hopping elements of $H_\text{eff}^{(1)}$ are Floquet-gauge invariant. Furthermore to order $\Omega^{-1}$, the interaction-dependent hopping terms enter the stroboscopic Floquet Hamiltonian, but not the effective Hamiltonian. Similarly to the Harper-Hofstadter model discussed in the previous section, the interaction-dependent hopping in the effective picture is encoded in the operator $K_{\rm eff}^\text{rot}$, via the relation $H_F^{(1)}[0] = H_\text{eff}^{ (1)} -i \left[ K^{\rm rot, (1)}_\text{eff}(0),H_\text{eff}^{(0)} \right]$. We note in passing that interaction-dependent hopping terms also appear in the Floquet spectrum of the Fermi-Hubbard model, when one drives the interaction term~\cite{di_liberto_14}. 

The effective parameters of the two expansions can be obtained from the following integrals
\begin{eqnarray}
F_{jk}[0](\zeta) =&&\ \frac{1}{4\pi i}\int_0^{2\pi}\mathrm{d}\tau_1\int_0^{\tau_1}\mathrm{d}\tau_2\,\left[\lambda_j(\tau_1)\lambda_k^*(\tau_2) - (1\leftrightarrow 2)\right],\nonumber\\
G_j[0](\zeta) =&&\ \frac{1}{4\pi i}\int_0^{2\pi}\mathrm{d}\tau_1\int_0^{\tau_1}\mathrm{d}\tau_2\,\left[ \lambda_j(\tau_1) - \lambda_j(\tau_2)\right],\nonumber\\
\tilde F_{jk}(\zeta) =&&\ \frac{1}{4\pi i}\int_0^{2\pi}\mathrm{d}\tau_1\int_0^{\tau_1}\mathrm{d}\tau_2\,\left(1-\frac{\tau_1-\tau_2}{\pi}\right) \left[ \lambda_j(\tau_1)\lambda_k^*(\tau_2) - (1\leftrightarrow 2) \right],
\label{eq:F_jk1}
\end{eqnarray}
where $\tau_i = \Omega t_i$. 

We would like to make a few remarks about a possible overlap of this model, as part of the DK class, with the Dirac class defined in Sec.~\ref{sec:diraclimit}. 
The overlap is possible because the lattice models considered here can have relativistic low energy dispersion, e.g.~if we consider a graphene-type honeycomb lattice (see below). However, we work in the limit where the amplitude of the driving protocol scales with the driving frequency $\Omega$ which is considered to be higher than the single-particle bandwidth. In this limit, the low-energy relativistic description of the spectrum inadequate. In order to realize the Dirac class in e.g.~graphene, one has to make sure that all involved energy scales, including the lattice potential, are small compared to the band width, so that only the linear part of the dispersion relation is important. The relation between lattice and continuum models is discussed in detail in App.~\ref{app:lattice_vs_cont}. In Sec.~\ref{sec:diraclimit}, we also used a symmetry argument to argue that there are no $\Omega^{-1}$-corrections to the infinite-frequency Floquet Hamiltonian in the Dirac class. That argument relied on the linear polarization of the driving protocol and does not apply to e.g.~a circularly polarized protocol, where  the phase of the driving depends on the direction. Such a protocol was suggested to realize a Floquet Chern Insulator~\cite{kitagawa_11} and we will briefly discuss it next.

\emph{Circularly Polarized Drive and the Floquet Realization of Haldane's Model.}

We now focus on graphene, where two triangular lattices build up the hexagonal structure, and consider the situation in which the driving frequency is higher than the band width. This scheme has been suggested theoretically to induce topological properties in graphene~\cite{kitagawa_11,rechtsman_13,yan_14,perez-piskunow_14,usaj_14,morell_12,calvo_11,sentef_14,dalessio_14_Chern,perez-piskunow_15}, and turn it into a Chern insulator. The topological properties of the quasi-energy spectrum of Floquet systems, in general, depend on the lattice geometry~\cite{verdeny_15}. Moreover, in Refs.~\cite{grushin_14,anisimovas_15} it was shown that Floquet Chern insulators with sufficiently strong nearest-neighbour interactions exhibit the phenomenon fractionalisation at fractional fillings. Cold atom experiments managed to realize a fermionic system with topological bands in the laboratory~\cite{jotzu_14}. 
As in Ref.~\cite{kitagawa_11} we consider a circularly polarized electric field. The driving protocol in this case reads as
\begin{equation}
f_r(t) = \bm{E}(t)\cdot \bm{r},\ \ \bm{E}(t) = V_0(\cos\Omega t,\sin\Omega t ).
\label{eq:fermions_f_r(t)}
\end{equation}
	where, in agreement with the general discussion, the amplitude of the electric fields need to scale with the driving frequency $\Omega$, that is $V_0=\zeta\Omega$.
For a honeycomb lattice, the unit vectors ${\bm s}_j$ point from the sublattice A to B (see Fig.~\ref{fig:haldane_model}, Panel (a)):
\[
{\bm s}_1 = (0,-1), \; {\bm s}_{2,3} = {1\over 2} (\pm\sqrt{3},1).
\]
In the rotating frame, this leads to
\[
\lambda_j(t) = \exp\left(i{\bm s}_j\cdot{\bm A}(t)\right), \; {\bm A}(t) = \frac{V_0}{\Omega}(\sin\Omega t\; ,-\cos\Omega t\; ).
\]
where ${\bm A}(t)$ is the vector potential describing the electric field. 
One can show that all three renormalized nn hopping amplitudes in $H_F^{(0)}$ and $H_\text{eff}^{(0)}$ are real and equal in magnitude. As in the bosonic case, they are given by $J_j = J_0\mathcal{J}_0(\zeta)$, where $\zeta =  V_0/\Omega$ is kept constant in the high-frequency limit and $\mathcal{J}_0$ is the Bessel function of first kind. To order $\Omega^{-1}$ we find that the site-dependent chemical potential vanishes identically for the circularly polarized drive owing to $\sum_j F_{jj} =0$, while the next-nearest-neighbor terms are finite and complex. As proposed in Ref.~\cite{kitagawa_11} they lead to a  topological band structure in the Floquet spectrum. For the case of a circularly polarized drive, we further obtain that the next-nearest-neighbor hopping elements in the effective picture are imaginary and equal in magnitude (while in the stroboscopic picture they are complex numbers whose magnitude and direction depend on the Floquet gauge), such that they lead to opposite fluxes of $\Phi_\triangle =\mp\pi/2$ penetrating the two sublattices $A$ and $B$ (see Fig.~\ref{fig:haldane_model}, Panel (b)). At half-filling, the model realizes Haldane's Chern insulator~\cite{haldane_88}.

\begin{figure}[h]
	\makebox[\textwidth]{%
		\resizebox*{0.45\columnwidth}{!}{\includegraphics{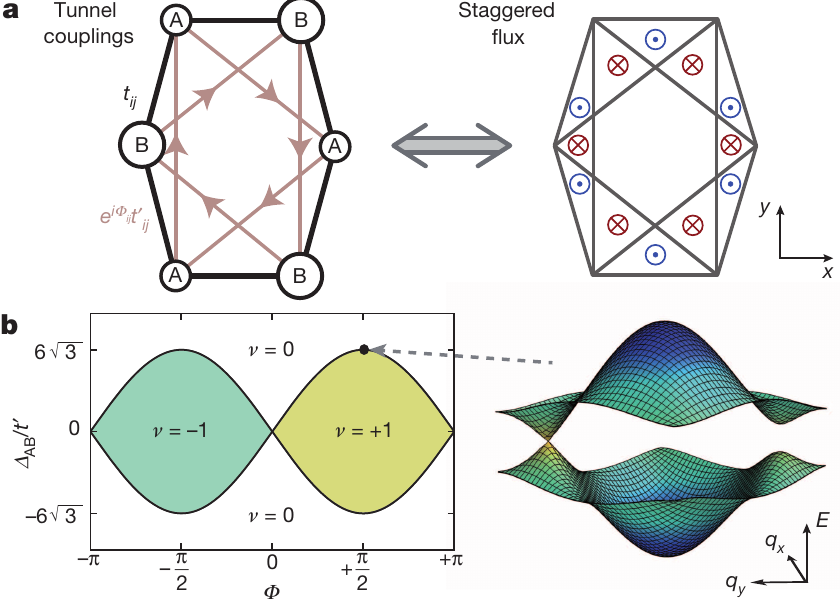}}
		\hfill    
		\resizebox*{0.55\columnwidth}{!}{\includegraphics{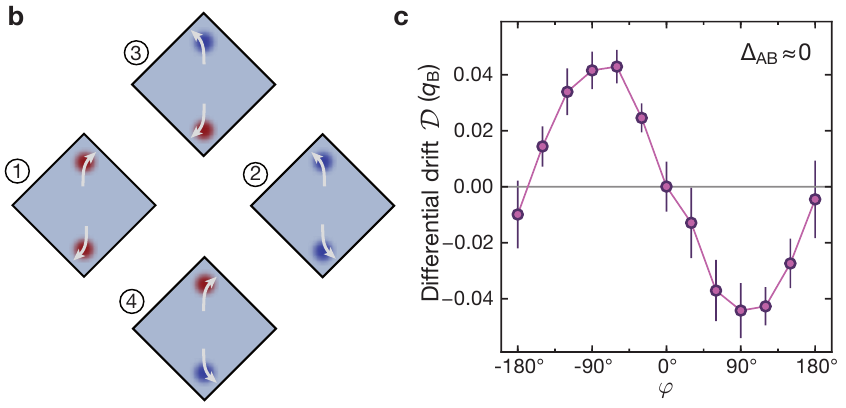}}}
	\caption{\label{fig:jotzu}(Color online). Left panel: Cold atom realisation of Haldane's model. (a) The brick-wall lattice unit cell and the dynamically generated staggered flux pattern. (b) Topological phase diagram of Haldane's model and dispersion relation for the brick-wall lattice. Right panel: Differential drift of the fermions in the first Brillouin zone. (b) Pictorial representation of the differential drift near the Dirac cones for fixed parameters. (c) Differential drift vs.~sublattice flux $\phi$ for a near zero staggered potential $\Delta_\text{AB}\approx 0$ Hz. The parameters are related to those in the main text by $\varphi = \Phi_\triangle$. Reprinted by permission from Macmillan Publishers Ltd: [Nature] (\cite{jotzu_14}), copyright (2014).}
\end{figure}

\emph{Experimental realisation of Haldane's model.} Haldane's model has been realised using ultracold fermionic $^{40}$K atoms in a brick-wall (almost hexagonal) optical lattice~\cite{jotzu_14}. A superlattice induced an energy off-set between the two sublattices which resulted in a staggered potential $\Delta_\text{AB}$. By mechanically shaking the lattice position along the $x$- and $y$-direction using piezo-electric actuators, the lattice sites were moved on elliptical trajectories which mimic the application of elliptically polarised electric field in the plane of the lattice, and break time-reversal symmetry. As discussed in the previous paragraphs, this leads to complex-valued nnn hopping terms between sites of the same sublattice. As a result, the Dirac cones open up a topological band gap, which is reflected in the non-zero and opposite Chern numbers of the two lowest bands, see Fig.~\ref{fig:jotzu}, left panel. 
	
In a topologically non-trivial band, atoms moving in the Brillouin zone acquire a Berry phase. This, in turn, results in a force, perpendicular to the direction of movement, pretty much like the Lorentz force acts on a charged particle moving in a real-space region of nonzero \emph{orbital} magnetic field\footnote{Note that an orbital magnetic field leads to cyclotron orbits, while a static magnetic field gradient (a Zeeman field) acts as an external potential and is responsible for the hyperfine splitting of atoms.}. By turning on a Zeeman magnetic field gradient which acts as an external potential on the atoms~\cite{price_12}, a constant force is applied on the atoms, leading to Bloch oscillations. Hence, the atoms are brought to explore the region of the Brillouin zone near the two Dirac cones, where the Berry curvature and, therefore, the Lorentz-like force the atoms experience, is the strongest. The experiment measured the motion of the centre of mass in the presence of the topological gap. Reversing the sign of the magnetic field gradient flips the sign of the force the atoms feel, and the drift is experienced in the opposite direction. Subtracting the two drifts from one another defines the differential drift which is proportional to the strength of the Berry curvature near the topological gaps, see Fig.~\ref{fig:jotzu}, right panel. 
A fraction of the atoms passing near the band gaps undergoes a Landau-Zener transition and transfers to the upper band. The precise band population can be extracted from the integrated column density in the absorption image following a band mapping technique. This provides a test for the existence of the Floquet-engineered dispersion relation of Haldane's model.

In the same experiments, using a Feshbach resonance the authors turned on the interaction between atoms in different hyper-fine states and briefly studied the interacting model. In general, the topological phases of the interacting Fermi-Hubbard model are expected to be strongly susceptible to heating effects. In the experiment conducted in Ref.~\cite{jotzu_14}, the authors observed a $25\%$ increase in entropy for the driven interacting system, when compared to the non-driven interacting case. Heating effects in ultracold fermionic systems in the high-frequency limit are a subject of current research~\cite{jotzu_15}.   

\subsection{\label{spins} Periodically Driven Spin Systems.}

As a final model in the DK class, we consider a spin-$1/2$ system on a lattice of arbitrary dimension, driven by a time-periodic, linearly-polarized external magnetic field. As we discussed in Sec.~\ref{sec:magnus_rotframe} the Magnus (High-Frequency) expansion works both for quantum and classical systems. So with minor modifications the results of this section apply equally to driven classical spins models. The effect of resonant driving on benchmark properties, such as the Rabi oscillations, was investigated~\cite{khomitsky_12}. A topological Floquet spin system was realised in Ref.~\cite{iadecola_15_spins}. Here, we assume that the magnetic field on each lattice site $m$ points along a fixed-in-time, but site-dependent direction. The magnitude of the magnetic field is allowed to vary from one lattice site to another. In agreement with the discussion in the introduction to the DK class, we assume that the amplitude of the magnetic field scales linearly with the frequency of the drive $\Omega$. The Hamiltonian in the lab frame reads as:
\be
H(t)= H_0+ \Omega\sum_m f_m(t) \, \bm{n}_m \cdot \bm{\sigma}_m,
\ee
where $H_0$ is time-independent and can include arbitrary spin-spin interactions, $\bm{\sigma}_m=(\sigma_m^x,\sigma_m^y,\sigma_m^z)$ is the vector of the three Pauli matrices on the $m$-th site, $\bm{n}_m$ is a time-independent unit vector, and $f_m(t)$ is a periodic function with period $T=2\pi/\Omega$. 

In the high-frequency limit, the Floquet Hamiltonian is equal to the time-average of the Hamiltonian in the rotating frame:
\be
H^\text{rot}(t)=V^\dagger(t) H_0 V(t), \label{spin-rot}
\ee
where
\[
\begin{split}
V(t)&=\exp\left[-i \sum_m \Delta_m(t)\,\bm{n}_m \cdot \bm{\sigma}_m\right], \\
\Delta_m(t)&=\Omega \int_{t_0}^t dt' f_m(t').
\end{split}
\]
The lower limit in the integral above can be used to change the Floquet gauge when going to the rotating frame.  Since spins at different sites commute, the operator $V(t)$ factorizes, and can be written as  
\[
V(t)=\prod_m V_m(t), \quad V_m(t)=\exp\left[-i \Delta_m(t)\, \bm{n}_m \cdot \bm{\sigma}_m\right],
\]
where $V_m(t)$ is the operator rotating the spin at the site $m$ by an angle $\theta_m(t)=2 \Delta_m(t)$ around the direction $\bm{n}_m$.  
Using Eq.\eqref{spin-rot} it is easy to see that the Hamiltonian in the rotating frame is given by the Hamiltonian $H_0$ with the substitution $\bm{\sigma}_m\longrightarrow \bm{\sigma}^\text{rot}_m(t) =  V_m^\dagger(t) \bm{\sigma}_m V_m(t)$ or explicitly:

\begin{equation}
	\bm{\sigma}_m \longrightarrow  \cos\theta_m\,\bm{\sigma}_m+\sin\theta_m\,\bm{n}_m \times \bm{\sigma}_m + (1-\cos\theta_m)\, \left(\bm{n}_m \otimes \bm{n}_m\right) \bm{\sigma}_m,
	\label{rotation}
\end{equation}
where $\times$ and $\otimes$ indicate the vector and tensor product. The entries of the matrix $M_m\equiv\bm{n}_m \otimes \bm{n}_m$ are defined by $(M_m)_{\alpha\beta}=n_m^{\alpha} n_m^{\beta}$.

\begin{figure}
	\centering
	\includegraphics[width = 0.9\columnwidth]{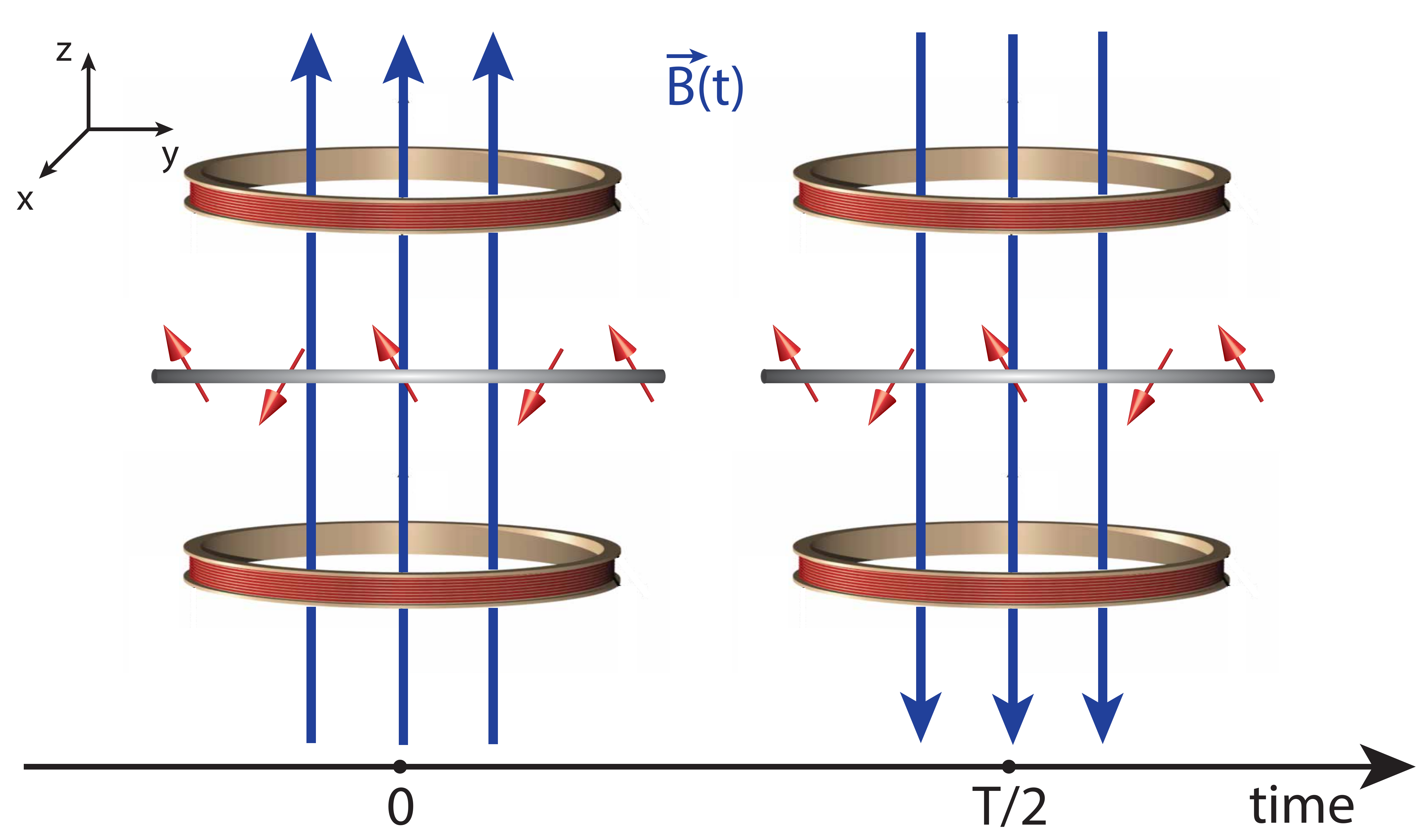}
	\caption{\label{fig:spins}(Color online) The flip-flop model: a periodically modulated, spatially uniform magnetic field $\bm B(t)$ is applied to a spin chain, c.f.~Eq.~\eqref{eq:flip_flop_H}.}
\end{figure}

We now consider two specific examples. First let us assume that:
\be
\begin{split}
	H(t)&=H_0+ \alpha\,\Omega\,\cos\Omega t\;  \sum_m \sigma^z_m \\
	H_0&=J_0\sum_{\langle m,n\rangle} \left(\sigma^x_m\sigma^x_n-\sigma^y_m\sigma^y_n\right)
\end{split}
\label{eq:flip_flop_H}
\ee
Here the driving corresponds to a uniform magnetic field along the $z$-direction, $\bm{n}_m=(0,0,1)$, with  oscillating intensity, $f_m(t) = \alpha\cos\Omega t$, c.f.~Fig.~\ref{fig:spins}.
Using Eqs.~\eqref{spin-rot} and \eqref{rotation} we find that the Hamiltonian in the rotating frame is obtained from $H_0$ via the substitutions:
\be
\begin{split}
	\sigma^x_m &\longrightarrow \cos\theta(t)\,\sigma^x_m-\sin\theta(t)\,\sigma^y_m, \\
	\sigma^y_m &\longrightarrow \cos\theta(t)\,\sigma^y_m+\sin\theta(t)\,\sigma^x_m, \\
	\sigma^z_m &\longrightarrow \sigma^z_m,
\end{split}
\label{explicit}
\ee
where $\theta(t)=2 \Delta(t)= 2\alpha\sin\Omega t\; $. After some algebra, we have:

\begin{equation}
H^\text{rot}(t)=J_0\sum_{\langle m,n\rangle} \cos (2\theta(t)) \left(\sigma^x_m\sigma^x_n-\sigma^y_m\sigma^y_n\right) 
-\sin(2\theta(t))  \left(\sigma^x_m\sigma^y_n+\sigma^y_m\sigma^x_n\right).
\end{equation}
In the infinite-frequency limit, the Floquet Hamiltonian can be calculated as:
\be
H^{(0)}_{F}=H^{(0)}_\text{eff}=\frac{1}{T}\int_0^T \mathrm{d} t H^\text{rot}(t)=\mathcal{J}_0(4\alpha) H_0,
\label{hrot-spin}
\ee 
where $\mathcal{J}_0$ is the Bessel function and we have used the mathematical identities:
\[
\frac{1}{T}\int_0^T \mathrm{d} t\cos(4\alpha\sin\Omega t\; )=\mathcal{J}_0(4\alpha),\quad \frac{1}{T}\int_0^T \mathrm{d} t\sin(4\alpha\sin\Omega t\; )=0.
\]
By choosing $\alpha$ to coincide with the zero of the Bessel function, periodically driven spin systems can exhibit dynamical freezing~\cite{das_10,hedge_13}.

We derive the leading $\Omega^{-1}$-corrections by computing the next term in the Magnus (High-Frequency) expansion. We choose to focus on a $1D$ chain for simplicity:
\begin{eqnarray}
H^{(1)}_{F}[0] &=& G\frac{J_0^2}{\Omega} \sum_m \left( \sigma_{m-1}^x\sigma_{m}^z\sigma_{m+1}^y+\sigma_{m-1}^y\sigma_{m}^z\sigma_{m+1}^x\right),\nonumber\\
H^{(1)}_{\text{eff}} &=& {\bf 0},\nonumber\\
K^{\text{rot},(1)}_F[0](t) &=& J_0\sum_{\langle m,n\rangle} [\kappa_c(t) - \kappa_c(0)] \left(\sigma^x_m\sigma^x_n-\sigma^y_m\sigma^y_n\right) 
-[\kappa_s(t) - \kappa_s(0)]  \left(\sigma^x_m\sigma^y_n+\sigma^y_m\sigma^x_n\right),\nonumber\\
K^{\text{rot},(1)}_\text{eff}(t) &=& J_0\sum_{\langle m,n\rangle} \kappa_c(t) \left(\sigma^x_m\sigma^x_n-\sigma^y_m\sigma^y_n\right) 
-\kappa_s(t)  \left(\sigma^x_m\sigma^y_n+\sigma^y_m\sigma^x_n\right),
\end{eqnarray}
where $G$, $\tilde G$, $\kappa_c(t)$ and $\kappa_s(t)$ are of order one and given by the expression:
\begin{eqnarray}
G &=& \frac{1}{\pi}\int_0^{2\pi}d\tau_1 \int_0^{\tau_1} d\tau_2  \left[\sin\left(4\alpha\sin(\tau_1)\right)\cos\left(4\alpha\sin(\tau_2)\right) - \left(\tau_1\longleftrightarrow \tau_2\right)\right],\nonumber\\
\kappa_c(t) &=& -\frac{1}{2}\int_t^{T+t}\mathrm{d}t'\left(1+2\frac{t-t'}{T}\right)\cos (2\theta(t')),\nonumber\\
\kappa_s(t) &=& -\frac{1}{2}\int_t^{T+t}\mathrm{d}t'\left(1+2\frac{t-t'}{T}\right)\sin (2\theta(t')). \nonumber
\end{eqnarray}
We thus see that, in this example, the infinite-frequency limit results in a renormalization of the spin-spin interactions of the bare Hamiltonian, while the first subleading correction in $\Omega^{-1}$ introduces $3$-spin interaction terms in the stroboscopic Floquet Hamiltonian. In the basis of $\sigma_z$, these terms play a role similar to the interaction-dependent hopping in Eqs.~\eqref{eq:HH_corr} and \eqref{eq:fermions_eff}. They induce next-nearest-neighbor spin flip processes, whose amplitude depends on the direction of the spin at the middle-site. The effective Floquet Hamiltonian does not contain these terms, since they are encoded in the kick operator $K_\text{eff}(t)$ via Eq.~\eqref{eq:F_vs_eff_1st_order}.

Let now us analyze another, slightly more complicated example on a two-dimensional lattice. The system is driven by a linearly-polarized magnetic field along the $z$-direction
\begin{equation}
H(t)=H_0+ \alpha\,\Omega\,\cos\Omega t\;  \sum_{m,n}\,m\,\sigma^z_{m,n},
\label{eq:spins_HH_drive}
\end{equation}
where $H_0$ is a standard XY-Hamiltonian:

\begin{equation}
	H_0=\sum_{m,n} J_y  \left( \sigma^x_{m,n}\sigma^x_{m,n+1}+\sigma^y_{m,n}\sigma^y_{m,n+1} \right)  
	+ J_x  \left(\sigma^x_{m,n}\sigma^x_{m+1,n}+\sigma^y_{m,n}\sigma^y_{m+1,n}\right)
	\label{standard_XY}
\end{equation}
and $J_x$ and $J_y$ are the bare coupling along the $x$ and $y$ directions.
In analogy with the previous example, we find $\theta_{m,n}(t)=2\Delta_{m,n}(t)=2 m \alpha \sin\Omega t\; $. Using the transformation in Eq.~\eqref{explicit} we arrive at:

\begin{equation}
H^\text{rot}(t)=\sum_{m,n} J_y  \left( \sigma^x_{m,n}\sigma^x_{m,n+1}+\sigma^y_{m,n}\sigma^y_{m,n+1} \right) 
+ h(t) J_x  \left(\sigma^x_{m,n}\sigma^x_{m+1,n}+\sigma^y_{m,n}\sigma^y_{m+1,n}\right),
\end{equation}
where we defined
\[
h(t)\equiv\cos(\theta_{m,n}(t)-\theta_{m+1,n}(t))=\cos\left(2\alpha\sin\Omega t\; \right).
\]
Observe that if the magnetic field were uniform, i.e.~if there were no magnetic gradients, then $h(t)\equiv 1$ and $H^\text{rot}(t)=H_0$. This is not surprising since, in this case, the driving would commute with $H_0$. In the infinite-frequency limit, the Floquet Hamiltonian reads as:

\begin{eqnarray*}
H^{(0)}_{F}=H^{(0)}_\text{eff}&=&\sum_{m,n} J_y \left( \sigma^x_{m,n}\sigma^x_{m,n+1}+\sigma^y_{m,n}\sigma^y_{m,n+1} \right) \\
&+& \mathcal{J}_0(2\alpha) J_x \left(\sigma^x_{m,n}\sigma^x_{m+1,n}+\sigma^y_{m,n}\sigma^y_{m+1,n}\right).
\end{eqnarray*}
This expression shows that, for $\Omega\to\infty$, the coupling strength along the $x$ direction is renormalized, while the one along the $y$ direction is not. By changing the value of $\alpha$ the Bessel function $\mathcal{J}_0(2\alpha)$ can be tuned to zero or even take negative values, in the same spirit as the original work by Dunlap and Kenkre~\cite{dunlap_86,dunlap_88}. This opens up possibilities for studying dimensional crossovers, effectively tuning the spin system between the $1d$ and the $2d$ regime,
and dynamically switching between ferromagnetic and anti-ferromagnetic couplings.  

Finally, notice that a close analogue to the Harper-Hofstadter Hamiltonian can be realized for spins by choosing the static Hamiltonian on a two-dimensional lattice as in Eq.~\eqref{standard_XY} and the periodic driving:
\begin{eqnarray}
	H_1(t)=\Omega \sum_{m,m} f_{m,n}(t) \sigma^z_{m,n},\ \ f_{m,n}(t)= m + \alpha \cos(\Omega t + \phi_{n,m}),
\end{eqnarray}
where $\phi_{m,n} = \Phi_\square(n+m)$ (see Sec.~\ref{subsec:harper} for details). 
The calculation of the dominant and subleading correction to the  Floquet Hamiltonian follows closely the steps shown above and in Sec.~\ref{subsec:harper}. In the infinite-frequency limit, this leads to complex interaction amplitudes with a flux $\Phi_\square$ per plaquette. Hence, one can expect to observe nontrivial spin-wave dynamics.

\begin{figure}[h]
	\centering
	\includegraphics[width=.9\columnwidth]{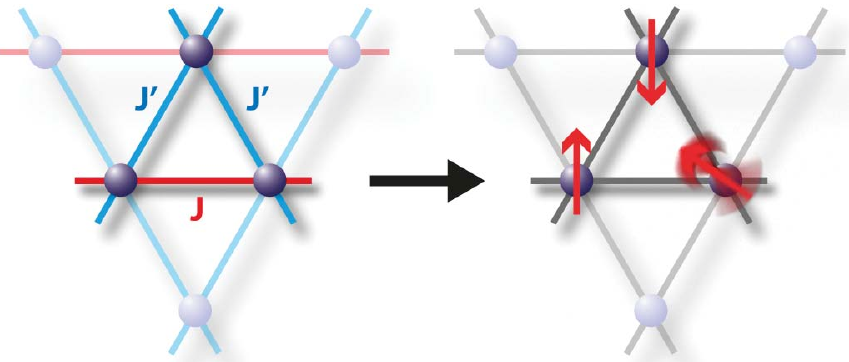}
	\caption{\label{fig:struck_0}(Color online). Realisation of the classical XY-model on a frustrated triangular lattice using ultracold bosons. By applying a periodic driving, it is possible to establish independent control over the two spin interactions $J,J'$. From Ref.~\cite{struck_11}. Reprinted with permission from AAAS.}
	\label{fig:struck0}
\end{figure}

\emph{Cold Atom Experiments with Spins Systems.} We now briefly mention some recent experimental realisations of classical spin systems using periodically driven cold atoms~\cite{eckardt_10,struck_11,struck_12,struck_13,yan_14}.  
In Ref.~\cite{struck_11} the authors employed a quantum system to simulate classical magnetism. A weakly-interacting $^{87}$Rb Bose gas was loaded in a two-dimensional triangular lattice. In the superfluid regime where phase fluctuations are suppressed and for high filling numbers, the system is effectively described by the classical XY-model

\begin{equation}
	H_0 = -J\sum_{\langle ij\rangle}\cos(\varphi_i - \varphi_j) + {U\over 2}\sum_j (S^z_j)^2 = - J\sum_{\langle ij\rangle}  {\bm S}_i\cdot{\bm S}_j + {U\over 2}\sum_j (S^z_j)^2,\nonumber
\end{equation}      
where the effective spin interaction $J$ is proportional to the boson hopping matrix element. $U$ is the effective local interaction related to the Hubbard coupling in the Bose-Hubbard model, and ${\bm S}_i$ is a unit vector confined to the $xy$-plane such that $S_j^x=\cos\varphi_j,\; S_j^y=\sin\varphi_j$, which represents the classical spin or rotor variable. As we saw in Secs.~\ref{subsec:Floquet_measurement_bosons},~\ref{subsec:harper},~\ref{subsec:foquet_circ_pol_drive}, and \ref{spins}, it is possible to modify the hopping matrix elements along the bonds by applying a periodic modulation. Mechanically moving the lattice along an elliptical orbit is equivalent to applying the force ${\bm F}(t) = F_c\cos\Omega t\ {\bm e}_c + F_s\sin\Omega t\ {\bm e}_s$ where ${\bm e}_{c/s}$ are two orthonormal vectors in the lattice plane and $F_{c/s}$ are experimentally controlled amplitudes~\cite{eckardt_10}. This driving protocol can be taken into account by the following effective spin Hamiltonian
\begin{equation}
H(t) = -J\sum_{\langle ij\rangle}  {\bm S}_i\cdot{\bm S}_j + \sum_j {\bm F}(t)\cdot{\bm r_j}S^z_j+{U\over 2}\sum_j (S^z_j)^2.
\end{equation}
The setup is very similar to the realisation of Haldane's model with circularly polarised electric field, c.f.~Eq.~\eqref{eq:fermions_f_r(t)}. Using Eq.~\eqref{explicit} together with the discussion after Eq.~\eqref{eq:spins_HH_drive} and the identification ${\bm S}_j\leftrightarrow{\bm\sigma}_j$, we can transform to the rotating frame. This results in a modification of the hopping matrix elements $J\to J\mathcal{J}_0(\zeta_{ij})$ with $\zeta_{ij} = \Omega^{-1}\sqrt{\left(F_c {\bm e}_c\cdot{\bm r}_{ij}\right)^2 + \left(F_s {\bm e}_s\cdot{\bm r}_{ij}\right)^2}$, where the vectors ${\bm r}_{ij} = {\bm r}_{i}- {\bm r}_{j}$ connect nearest-neighbouring sites. Consequently, as a result of the periodic shaking, it is possible to establish control over the spin-interactions on the three bonds of the triangular plaquette. The infinite-frequency  Floquet Hamiltonian is
\begin{equation}
		H_F^{(0)} = H_\text{eff}^{(0)} = -\sum_{\langle ij\rangle}J_{ij}{\bm S}_i\cdot{\bm S}_j+{U\over 2}\sum_j (S^z_j)^2,
\end{equation}
where $J_{23} = J_{31} = J'$ and $J_{12} = J$, c.f.~Fig.~\ref{fig:struck_0}. In the original paper~\cite{struck_11} the last term did not appear in the Hamiltonian because the interactions were tuned to a small value and also because they do not affect the thermal phase diagram in the classical limit (large filling). By tuning the driving amplitudes $F_c$ and $F_s$, it is possible to reach regimes in which the spin-interactions $J,J'$ flip sign independently. This opens up the way towards studying a rich phase diagram where spin order competes with frustration due to the lattice geometry.	

\begin{figure}[h]
		\makebox[\textwidth]{%
			\resizebox*{0.5\columnwidth}{!}{\includegraphics{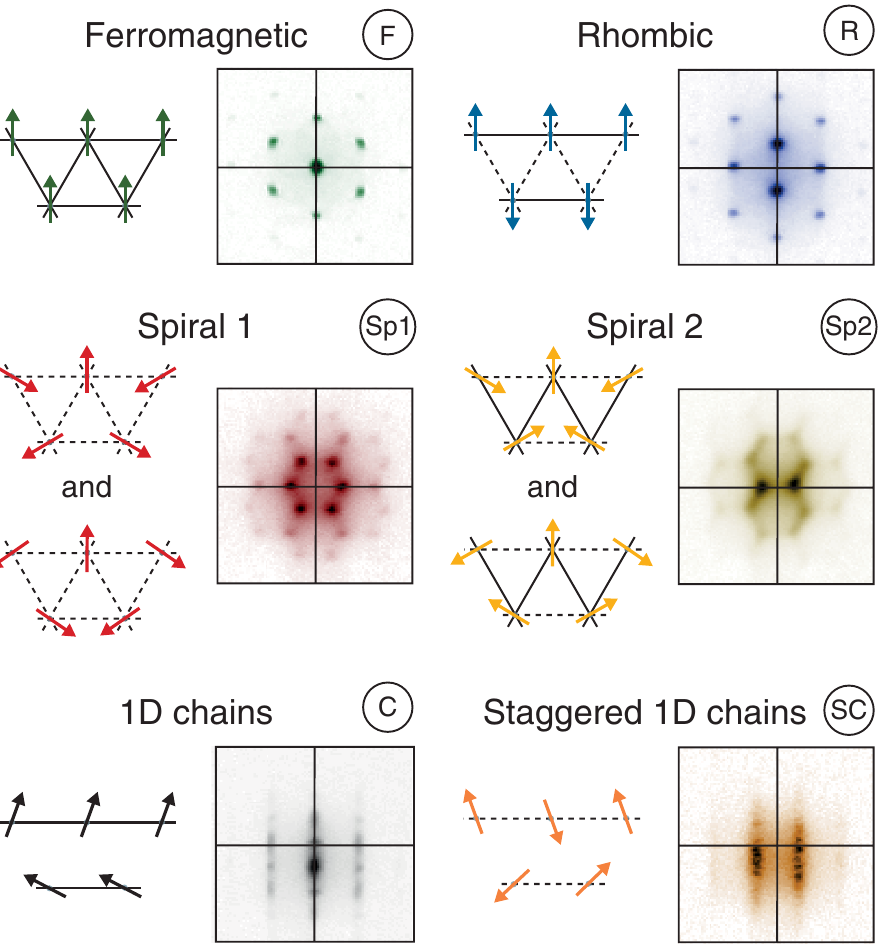}}
			\hfill    
			\resizebox*{0.5\columnwidth}{!}{\includegraphics{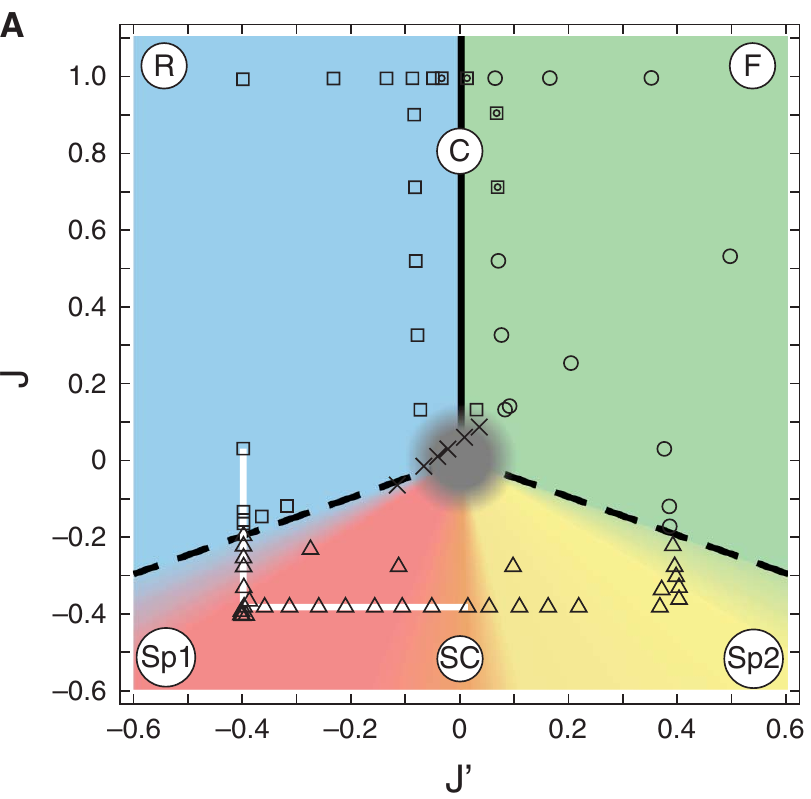}}}
		\caption{\label{fig:struck}(Color online). Left panel: Ground states of the frustrated XY model on a triangular lattice for different values of the hopping parameters $J$ and $J'$ (see right panel). Right panel: Phase diagram of the frustrated XY model on a triangular lattice in the $(J,J')$-plane. The symbols refer to the states in the left panel. The solid line is a first-order, while the dashed lines represent a second-order phase transition. In the grey region where tunnelling is suppressed the bosonic system is strongly interacting and the interference pattern is destroyed. From Ref.~\cite{struck_11}. Reprinted with permission from AAAS.}
\end{figure} 
	
As a main tool to study the phase diagram, c.f.~Fig.~\ref{fig:struck}, the authors performed time-of-flight measurements which gives access to the momentum distribution of the superfluid Bose gas. By assigning a unique momentum distribution to the ground states of the candidate ordered phases, the phase diagram was mapped out with a very high precision. It was even possible to observe spontaneous symmetry breaking directly in the case where the two degenerate ground states lead to different time-of-flight images. For $J,J'>0$ the system was found in a rhombic state (R), while for $J>0$, $J'<0$ it undergoes a first-order phase transition to a ferromagnet (F). On the transition boundary ($J'=0$, $J>0$) ferromagnetic chains build up in the ground state. Frustration effects become relevant when $J<-|J'|/2$, where the system undergoes a second-order phase transition to two different spiral states, (SP2) and (SP1), depending on whether it is approached from the ferromagnetic ($J'<0$) or the rhombic ($J'>0$) side, respectively. These spiral states are connected by a continuous crossover at $J'=0$ and $J<0$, where the ground state displays the order of staggered spin-chains (SC). For more details, see Ref.~\cite{struck_11}.       
	
In a subsequent experiment~\cite{struck_13}, the interplay between the continuous U(1) symmetry of the XY-model in the presence of a $\mathbb{Z}_2$ Ising-like artificial gauge field was studied. Once again, $^{87}$Rb was loaded in a 3D triangular lattice (weakly confined along the vertical direction) which realised the classical $XY$-model. In addition, an artificial magnetic field in the form of complex Peierls phases was imprinted in the hopping amplitudes $J_{ij}$, created by shaking the positions of the lattice wells according to an by elliptically polarized polychromatic modulation which breaks time-reversal symmetry at the level of the time-average Hamiltonian. The model realises a nonzero net flux which penetrates the triangular plaquettes in an alternating fashion. This flux leads to mass currents along the plaquettes whose direction, clockwise or counter-clockwise, constitutes the classical Ising variable, which was indirectly measured through the occupation of the momentum modes. In addition from such measurements the authors were able to identify a thermal phase transition between an anti-ferromagnetic and a paramagnetic phase.


\section{Summary and Outlook.}
\label{sec:conclusions} 

Periodically driven systems in the high-frequency limit can be used to engineer interesting effective Hamiltonians, which are very difficult or impossible to realize in equilibrium systems. They provide an important step towards the simulation of quantum condensed matter systems, and can be used to test predictions of physical theories in new regimes.

In this review, we  have presented a systematic analysis of the high-frequency regimes in periodically driven (Floquet) systems. We have identified both the infinite-frequency and first leading correction ($\Omega^{-1}$) to the stroboscopic and effective Floquet Hamiltonians using the Magnus and the High-Frequency expansions. We have precisely defined the Floquet stroboscopic (FS) and Floquet non-stroboscopic (FNS) dynamics and computed the dressed operators and the dressed density matrices required to correctly describe both these measurement schemes. We also discussed the Floquet gauge structure associated with the choice of the stroboscopic time, and how one can translate between the stroboscopic and the effective picture. The Floquet non-stroboscopic dynamics (FNS), which suits very well the current experimental techniques, often opens up the possibility of measuring Floquet gauge-invariant physical observables like the proper Floquet current.

As the main tools to study the high-frequency limit, we employed the Magnus and the High-Frequency expansions. We showed that they can be used to reliably calculate the leading corrections, to the infinite-frequency  Floquet Hamiltonian. When applied to time-independent Hamiltonians in the rotating frame, one can use them to eliminate a high-energy scale from the problem and derive an effective dressed low-energy Hamiltonian with renormalized parameters similarly to the Schrieffer-Wolff transformation. Moreover, we discussed how one can naturally extend this transformation to driven setups and identified new terms in the dressed Hamiltonian, which appear due to the driving, and which lead to heating and other non-equilibrium effects. We briefly mentioned the convergence of the Magnus expansion, which is not guaranteed for interacting many-body systems in the thermodynamic limit, and the relation between this mathematical question and (the lack of) heating in periodically driven systems. These important issues are not yet settled. We hope that they will be resolved in future experiments and theoretical work.

\begin{figure}
	\centering
	\includegraphics[width = 1\columnwidth]{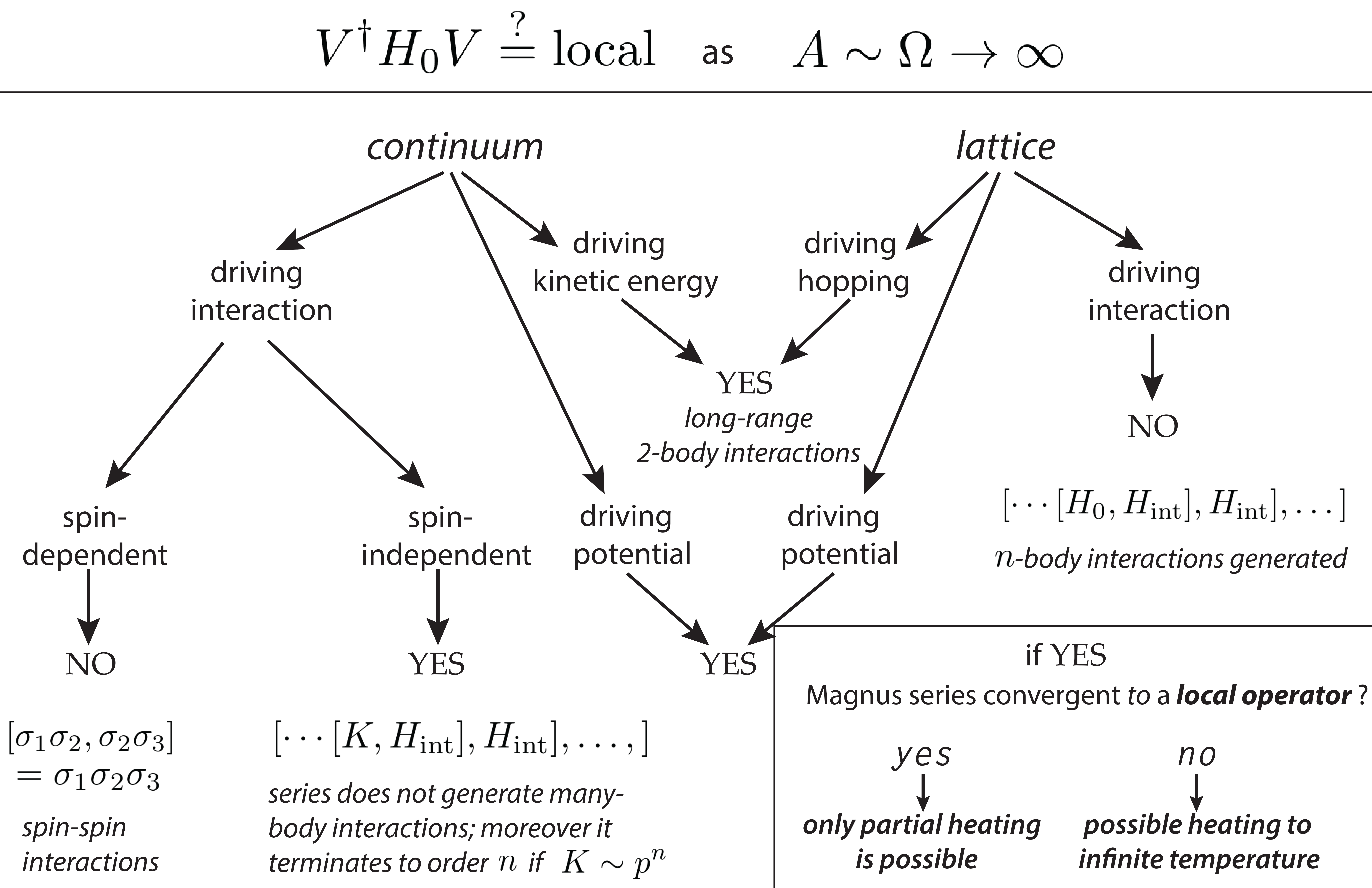}
		\caption{\label{fig:summary_fig}(Color online) The figure summarises the scenarios discussed in this review: globally periodically driven continuum and lattice models. Depending on whether one drives the interaction or the external potential the Magnus Hamiltonian can be local (YES) or a non-local one (NO). A local Magnus Hamiltonian is a sum of spatially local terms and can include only few-body interactions. Different scenarios might appear in locally driven systems. For example, driving any local in space term like the local hopping strength or the local interaction coupling does not produce any long-range terms in the infinite-frequency Floquet Hamiltonian.  }
\end{figure}    

A prerequisite for finding non-trivial high-frequency limits is a strong coupling of the driving protocol to the system, in the form of a driving amplitude which scales with a power of the driving frequency. Often times, a systematic way of studying the inverse-frequency expansion of the  Floquet Hamiltonian is to first go to the rotating frame w.r.t.~the driving Hamiltonian. We proved that this amounts to the resummation of an infinite lab-frame subseries and demonstrated this on several examples. Moreover, we identified three classes of universal high-frequency driving protocols leading to well-defined local Floquet Hamiltonians (c.f.~Fig.~\ref{fig:summary_fig}), but there may be more. For each class, we have calculated the form of the effective Floquet Hamiltonian which differs significantly from the time-averaged one.  

The Kapitza class is characterized by a kinetic energy term which is quadratic in
momentum, and a driving amplitude which scales linearly in $\Omega$. We gave examples of both a single- and many-body systems which realize this limit. The Dirac class is benchmarked by a linear kinetic energy term which requires adding a spin structure via the Pauli matrices. One can periodically drive either an external magnetic field, in which case the amplitude should scale as $\Omega$ or, alternatively, the drive can couple to an external potential but then the driving amplitude is required to scale as $\Omega^2$. The Dunlap-Kenkre (DK) class applies to lattice systems with an arbitrary dispersion relation, where, one drives an external single-particle scalar potential, whose amplitude scales linearly with $\Omega$. We illustrated all three classes with various examples and discussed recent experimental progress made with ultracold atoms. 

While classical few body Floquet systems, such as the Kapitza pendulum and its variations, found a multitude of interesting and useful applications, the experimental realization and systematic theoretical analysis of  many-particle periodically driven systems is very recent. We discussed several realizations of Floquet systems both in cold atoms and in solid state materials, where new, hard to achieve otherwise, regimes have been accessed using a periodic modulation. This lead to the emergence of a new research direction, dubbed ``Floquet engineering'', which has the potential to develop systems with unique properties in the near future. Floquet systems constitute a playground for studying many different phenomena such as  information and entanglement propagation in the absence of conservation laws, finding non-equilibrium optimum quantum annealing protocols, designing materials with tunable properties, and many more. There are also many open conceptual problems in Floquet systems, which we mentioned only briefly in this review but which are obviously important for our overall understanding of driven systems. In particular, the nature of steady states in open Floquet systems, i.e.~Floquet systems coupled to a thermal bath, non-adiabatic response in Floquet systems with slowly changing external parameters, robustness and universality of topological Floquet phases, nature and classification of phase transitions in driven open and isolated systems and others. We hope that these and other questions will be understood in the near future.

\section*{Acknowledgements}
The authors would like to thank M.~Aidelsburger, A.~George, N.~Goldman, A.~Grushin, P.~Hauke, M.~Heyl, D.~Huse, T.~Iadecola, G.~Jotzu, S.~Kehrein, M.~Kolodrubetz, A.~Rosch, J.~Struck, P.~Weinberg and P.~Zoller for insightful and 
interesting discussions. Special thanks go to A.~Eckardt for drawing our attention to the subtle difference between the Magnus expansion and the High-Frequency expansion, and to G.~Jotzu for proof-reading the manuscript. We are very grateful to the experimental teams in the groups of E.~Arimondo, I.~Bloch, T.~Esslinger and K.~Sengstock for kindly allowing us to include some of the figures from their previous papers in this review. Last but not least, we would also like to thank J.~G.~Wright and B.~S.~Shastry 
for developing the software DiracQ~\cite{diracQ} used to verify the commutator 
algebra calculations in this work. This work was supported by AFOSR FA9550-13-1-0039, ARO W911NF1410540, NSF DMR-1206410, and BSF 2010318.


\bibliographystyle{unsrt}
\bibliography{Magnus_paper}

	


\appendices

\section{\label{app:ME_HEF_derivation} Outline of the Derivation of the Inverse-Frequency Expansions. }
In this appendix, we briefly summarise the most important steps in the derivation of the Magnus Expansion (ME) for the stroboscopic Floquet Hamiltonian and the High-Frequency Expansion (HFE) for effective Hamiltonian.

\subsection{\label{app:ME_derivation} The Magnus Expansion.}

The evolution operator for a single period, starting at time $t_0$, is given by 
\begin{equation}
U(T+t_0,t_0) = \mathcal{T}_t\exp\left(-i\int_{t_0}^{t_0+T}\mathrm{d}t H(t)\right) = \exp\left( -i H_F[t_0]T\right),
\end{equation}
where we used Floquet's theorem for the second equality. Inverting this equation,
\begin{equation}
H_F[t_0] = \frac{i}{T}\log\left[ \mathcal{T}_t\exp\left(-i\int_{t_0}^{t_0+T}\mathrm{d}t H(t)\right) \right],
\end{equation}
and we find the unique stroboscopic Floquet Hamiltonian\footnote{The uniqueness of the evolution operator is provided by the uniqueness of the solution to Schr\"odinger's equation. The uniqueness of $H_F[t_0]$ is up to the quasi-energy spectrum folding ambiguity.}. Now, we can expand the RHS in powers of the inverse frequency $\Omega = 2\pi/T$ using the Baker-Campbell-Hausdorff lemma. It can be verified that this immediately results in Eqs.~\eqref{eq:magnus_series} from the main text.

To obtain the stroboscopic kick operator $K_F[t_0](t) = i\log P(t,t_0)$, we invert Floquet's theorem $U(t+t_0,t_0) = P(t,t_0)\exp\left(-iH_F[t_0](t-t_0)\right)$:
\begin{eqnarray}
K_F[t_0](t) = i\log\left[ \mathcal{T}_t\exp\left(-i\int_{t_0}^{t_0+t}\mathrm{d}t' H(t')\right)\exp\left(iH_F[t_0](t-t_0)\right)  \right]. 
\end{eqnarray}
Expanding the RHS in powers of $\Omega^{-1}$ yields Eqs.~\eqref{eq:kick_operator_ME}. Since $K_F[t_0](t)$ is constructed order by order in the inverse frequency, it is also a unique operator.

\subsection{\label{app:HFE_derivation} The High-Frequency Expansion.}

To derive the High-Frequency Expansion, we make use of the results from Appendix~\ref{app:ME_derivation} above. Starting from the relation between the fast-motion operator $P(t,t_0) = \exp\left(-i K_F[t_0](t)\right)$ and the effective kick operator $K_\text{eff}(t)$ in Eq.~\eqref{rel:P_K}, we have
\begin{equation}
K_F[t_0](t) = i\log\left[\exp\left(-i K_\text{eff}(t)\right) \exp\left(i K_\text{eff}(t_0)\right) \right].
\end{equation}
Notice that the $\Omega^{-1}$-expansion of the LHS is already known from Appendix~\ref{app:ME_derivation}. Hence, postulating $K_\text{eff} = \sum_{n=0}^\infty K_\text{eff}^{(n)}$ with $K_\text{eff}^{(n)}\sim\Omega^{-n}$, we can again apply the Baker-Campbell-Hausdorff lemma. By comparing equal powers of the inverse frequency, we arrive at Eq.~\eqref{eq:kick_operator_HFE}. Notice again that $K_\text{eff}$ is an operator, whose uniqueness is inherited by that of $K_F[t_0](t)$. 

Once we have the effective kick operator, we can apply Floquet's theorem again to determine uniquely the effective Hamiltonian $H_\text{eff}$ order by order in $\Omega^{-n}$. Equivalently, one can use the transformation law $H_\text{eff} = \exp\left(iK_\text{eff}(t)\right)H_F[t_0]\exp\left(iK_\text{eff}(t)\right)$ to find the perturbative expansion of the \emph{unique} effective Hamiltonian from the Magnus expansion. For alternative derivations of the high-frequency expansion, see Refs.~\cite{rahav_03_pra,goldman_14,eckardt_15}. We mention in passing that expanding the LHS and the RHS in powers of the inverse frequency and matching the coefficients is essentially the same idea which also lies behind the derivation of the generator of the Schrieffer-Wolff transformation, c.f.~the discussion in Sec.~\ref{subsec:Schrieffer_Wolff}.

\section{\label{app:lattice_vs_cont} Lattice vs.~Continuum Models.}

In this appendix we discuss some subtle differences between the lattice and continuum models discussed in the Secs.~\ref{sec:kapitza},~\ref{sec:diraclimit} and~\ref{sec:driven_external_fields}. In particular, we show how to combine the results of Sec.~\ref{sec:diraclimit} and Sec.~\ref{subsec:Floquet_top_ins}. We demonstrate the relation between the models for one-dimensional non-interacting systems, but the generalisation to higher dimensions including interactions is straightforward.

\emph{Systems with linear dispersion.} Consider first the following static, non-interacting Hamiltonian with linear dispersion

\begin{eqnarray}
H_\text{cont} = \int\mathrm{d}x \frac{J_0}{2}\left(-i\psi^\dagger(x)\partial_x\psi(x) + \text{h.c.}\right).
\label{eq:lat_vs_cont_H0_fermions_cont}
\end{eqnarray}
To discretise the model, we put it on a lattice with lattice constant $a$. The corresponding lattice Hamiltonian is given by
\begin{equation}
H_\text{latt} = \sum_x\frac{J_0}{2a}(-i\psi^\dagger(x)\psi(x+a) + \text{h.c.}).
\label{eq:lat_vs_cont_H0_fermions_latt}
\end{equation} 
If one goes to momentum space, the dispersion relation is $\varepsilon(k) = \frac{J_0}{a} \sin(ak)$, and in the long-wavelength limit, $ak\ll 1$, we conveniently recover the continuum linear dispersion $\varepsilon_k\approx J_0 k$. In particular, it follows that in lattice theories with linear dispersion, the hopping matrix element should scale as $1/a$. If one, on the other hand, starts with a lattice theory, one can recover the Hamiltonian~\eqref{eq:lat_vs_cont_H0_fermions_cont} from Eq.~\eqref{eq:lat_vs_cont_H0_fermions_latt} in the limit $a\to 0$ by using the identity $\psi(x+a) = \psi(x) + a\partial_x\psi(x) + \mathcal{O}(a^2)$, and collecting powers of $a$.

Let us now add to this Hamiltonian a time-dependent electric field with the amplitude $V_0$ and frequency $\Omega$: 
\begin{eqnarray}
H(t) = \int\mathrm{d}x\frac{J_0}{2}\left(-i\psi^\dagger(x)\partial_x\psi(x) + \text{h.c.}\right) + V_0\,\cos(\Omega t)\,x\,\psi^\dagger(x)\psi(x).
\end{eqnarray} 
In Sec.~\ref{subsec:dirac:driven_H1(t)}, we showed that the zeroth-order Floquet Hamiltonian for this relativistic continuum theory is not affected by the drive if we keep $\zeta=V_0/\Omega$ independent of $\Omega$, i.e.~scale the electric field amplitude $V_0$ linearly with the frequency:
\begin{equation}
H_{F,\text{cont}}^{(0)} = \int\mathrm{d}x\frac{J_0}{2}\left(-i\psi^\dagger(x)\partial_x\psi(x) + \text{h.c.}\right).
\end{equation}
On the contrary, in Sec.~\ref{subsec:Floquet_top_ins} we considered the same Hamiltonian on the lattice, and found the following non-trivial zeroth-order Floquet Hamiltonian:
\begin{equation}
H_{F,\text{latt}}^{(0)} = \sum_x\frac{J_0}{2a}\mathcal{J}_0(\zeta a)\left(-i \psi^\dagger(x)\psi(x+a) + \text{h.c.}\right).
\label{eq:lat_vs_cont_H(t)_fermions_latt}
\end{equation} 
At first sight, the two results seem contradictory. To find the proper continuum theory, we expand Eq.~\eqref{eq:lat_vs_cont_H(t)_fermions_latt} in powers of the inverse lattice constant. Using the identity $\mathcal{J}_0(\zeta a) = 1 + \mathcal{O}(a^2)$, we see that the low-energy effective Hamiltonian is independent of $\zeta$. Consequently, all the non-trivial effects introduced by the driving vanish in the long-wavelength limit and, therefore, the lattice and continuum models are consistent and yield the same result. A similar derivation applies to higher order corrections in the Magnus expansion. From Eq.~(\ref{eq:lat_vs_cont_H(t)_fermions_latt}) we also see the condition under which the continuum approximation holds:
\be
\zeta a\ll 1\; \Leftrightarrow \; V_0 a\ll \Omega.
\ee
where, the equivalence of these two conditions follows from the definition $\zeta=V_0/\Omega$. 
The product $V_0 a$ is the maximum energy difference generated by the driving potential between two lattice sites. So the continuum approximation holds only in the limit when this difference is small compared to the driving frequency. Once this condition is violated, the full lattice dispersion has to be taken into account and the continuum approximation breaks down.

\emph{Systems with quadratic dispersion.} We now show the correspondence between the continuum and lattice theories for systems with quadratic dispersion. The non-driven continuum and lattice Hamiltonians read as
\begin{eqnarray}
H_\text{cont} &=& \int\mathrm{d}x\phi^\dagger(x)(-\partial_x^2)\phi(x),\nonumber\\
H_\text{latt} &=& -\frac{J_0}{a^2}\sum_x\left( \phi(x)^\dagger\phi(x+a) - 2\phi(x)^\dagger\phi(x) + \phi(x+a)^\dagger\phi(x) \right). 
\end{eqnarray}
Notice that in the case of a quadratic dispersion, the hopping matrix element scales as $1/a^2$. Now consider the driven model
\begin{equation}
H(t) = \int\mathrm{d}x\phi^\dagger(x)(-\partial_x^2)\phi(x) + V_0\cos\Omega t\;  f(x)\phi^\dagger(x)\phi(x).
\end{equation}
Recall that the continuum model fits into the Kapitza class, c.f.~Sec.~\ref{sec:kapitza}, while the lattice model is part of the DK class, Sec.~\ref{sec:driven_external_fields}. A careful reader might be worried that in the former case, in the limit $\Omega\to\infty$, we found an emergent effective potential leading to dynamical stabilisation whereas, in the latter case, we obtained the following modification to the hopping matrix element:
\begin{equation}
H_{F,\text{latt}}^{(0)} = - \frac{J_0}{a^2}\sum_x \mathcal{J}_0\left(\zeta f(x+a) - \zeta f(x)\right)\left(\phi^\dagger(x+a)\phi(x) + \text{h.c.}\right)+{2J_0\over a^2}\sum_x \phi(x)^\dagger\phi(x).
\end{equation}
where, as usual, $\zeta=V_0/\Omega$. To reconcile the two approaches, again we take the limit $a\to 0$. In doing so, we write $f(x+a) - f(x) = af'(x) + \mathcal{O}(a^2)$, and use the expansion $\mathcal{J}_0(z) = 1 - z^2/4 + \mathcal{O}(z^4)$. The results is
\begin{equation}
H_{F,\text{cont}}^{(0)} = \int\mathrm{d}x \phi^\dagger(x)\left(-\partial_x^2 + \frac{\zeta^2}{4}[f'(x)]^2\right)\phi(x).
\end{equation}
We therefore see that indeed the continuum theory features an emergent potential given by $\zeta^2/4[f'(x)]^2$ which establishes the relation between the Kapitza and the DK classes, c.f. Eq.~\eqref{h_f_kapitza}.

\section{\label{app:corr_HF_gen} Corrections to the Stroboscopic Floquet Hamiltonian $H_F[0]$.}

\subsection{\label{app:DK_corr_coeffs_1D}First-order Coefficients for the 1D Driven Boson Model.}

Here, we briefly list the expressions for the nnn hopping, and the staggered potential, found to first order in the Magnus expansion to the model discussed in Sec.~\ref{subsec:Floquet_measurement_bosons}. All the integrals are given in the Floquet gauge $t_0=0$. We recall that 
\[
g^{m,m+1}(\tau;\zeta) = \exp\left[-i\zeta\sin(\tau - \phi_{nm}) \right],
\]
where $\tau = \Omega t$. Then the coefficients to the Hamiltonian $H_F^{(1)}$ given in Eq.~\eqref{eq:H_F^1_bosons_1D} are given by the following time-ordered integrals
	\begin{eqnarray}
	\mathcal{C}^{m,m+2}(\zeta) =&&\ \frac{1}{4\pi i}\int_0^{2\pi}\mathrm{d}\tau_1\int_0^{\tau_1}\mathrm{d}\tau_2\left[g^{m,m+1}(\tau_1)g^{m+1,m+2}(\tau_2) - (1\leftrightarrow 2) \right],\nonumber\\
	\mathcal{G}^{m,m+1}(\zeta) =&&\ \frac{1}{2\pi}\int_0^{2\pi}\mathrm{d}\tau_1\int_0^{\tau_1}\mathrm{d}\tau_2\text{Im}\left\{ \left(g^{m,m+1}(\tau_1)\right)^* g^{m,m+1}(\tau_2) \right\}.
	\end{eqnarray}
We mention that these expressions are the same as the corresponding one for nnn hopping along the $y$-direction ${^{\uparrow}C}_0^{m,m+2}(\zeta)$, and a staggered potential along the $y$-direction and ${^{\uparrow}E}_{0}^{m,m+1}(\zeta)$ found in the 2D extension of the model from Sec.~\ref{subsec:harper} (see below).

\subsection{\label{app:DK_corr_coeffs_2D}First-order Coefficients for the Harper-Hofstadter Model.}

In this appendix we discuss the parameters of the leading correction, Eq.~\eqref{eq:HH_corr}. Let us define two auxiliary functions $f$ and $g$ by
\begin{eqnarray}
f_{m,m+1}^n(\tau;\zeta) &=& \exp\left[-i\zeta\sin(\tau - \phi_{nm}) +i\tau \right]\nonumber\\
g_m^{n,n+1}(\tau;\zeta) &=& \exp\left[-i\zeta\sin(\tau - \phi_{nm}) \right]. 
\end{eqnarray} 
The coefficients $B$, $C$, $D$, and $E$ in Eq.~\eqref{eq:HH_corr} are given by the following time-ordered integrals:

	\begin{eqnarray}
	{^{\rightarrow}B}^n_{m,m+1}(\zeta) =&&\ \frac{1}{4\pi i}\int_0^{2\pi}\mathrm{d}\tau_1\int_0^{\tau_1}\mathrm{d}\tau_2\left[f^n_{m,m+1}(\tau_1) - f^n_{m,m+1}(\tau_2) \right],\nonumber\\
	{^{\uparrow}B}_m^{n,n+1}(\zeta) =&&\ \frac{1}{4\pi i}\int_0^{2\pi}\mathrm{d}\tau_1\int_0^{\tau_1}\mathrm{d}\tau_2\left[g_m^{n,n+1}(\tau_1) - g_m^{n,n+1}(\tau_2) \right].
	\end{eqnarray} 
	\begin{eqnarray}
	{^{\rightarrow}C}^n_{m,m+2}(\zeta) =&&\ \frac{1}{4\pi i}\int_0^{2\pi}\mathrm{d}\tau_1\int_0^{\tau_1}\mathrm{d}\tau_2\left[f^n_{m,m+1}(\tau_1)f^n_{m+1,m+2}(\tau_2) - (1\leftrightarrow 2) \right],\nonumber\\
	{^{\uparrow}C}_m^{n,n+2}(\zeta) =&&\ \frac{1}{4\pi i}\int_0^{2\pi}\mathrm{d}\tau_1\int_0^{\tau_1}\mathrm{d}\tau_2\left[g_m^{n,n+1}(\tau_1)g_m^{n+1,n+2}(\tau_2) - (1\leftrightarrow 2) \right],\nonumber
	\end{eqnarray} 
	\begin{eqnarray}
	{^{\nearrow}D}_{m,m+1}^{n,n+1}(\zeta) =&&\ \frac{1}{4\pi i}\int_0^{2\pi}\mathrm{d}\tau_1\int_0^{\tau_1}\mathrm{d}\tau_2\bigg[f^n_{m,m+1}(\tau_1)g_{m+1}^{n,n+1}(\tau_2) \nonumber\\
	&& + f^{n+1}_{m,m+1}(\tau_2)g_{m}^{n,n+1}(\tau_1) - (1\leftrightarrow 2) \bigg],\nonumber\\
	{^{\nwarrow}D}_{m,m-1}^{n,n+1}(\zeta) =&&\ \frac{1}{4\pi i}\int_0^{2\pi}\mathrm{d}\tau_1\int_0^{\tau_1}\mathrm{d}\tau_2\bigg[\left(f^n_{m-1,m}(\tau_1)\right)^*g_{m-1}^{n,n+1}(\tau_2)\nonumber\\
	&& + \left(f^{n+1}_{m-1,m}(\tau_2)\right)^*g_{m}^{n,n+1}(\tau_1) - (1\leftrightarrow 2) \bigg],\nonumber
	\end{eqnarray} 
	\begin{eqnarray}
	{^{\rightarrow}E}^n_{m,m+1}(\zeta) =&&\  \frac{1}{2\pi}\int_0^{2\pi}\mathrm{d}\tau_1\int_0^{\tau_1}\mathrm{d}\tau_2\text{Im}\left\{ \left(f^n_{m,m+1}(\tau_1)\right)^* f^n_{m,m+1}(\tau_2) \right\},\nonumber\\
	{^{\uparrow}E}_{m}^{n,n+1}(\zeta) =&&\  \frac{1}{2\pi}\int_0^{2\pi}\mathrm{d}\tau_1\int_0^{\tau_1}\mathrm{d}\tau_2\text{Im}\left\{ \left(g_m^{n,n+1}(\tau_1)\right)^* g_m^{n,n+1}(\tau_2) \right\},\nonumber
	\end{eqnarray} 

All the coefficients are defined on the bonds between sites, labelled by $(m,n)$. Apart from $E$, the coefficients $B$, $C$, and $D$ are complex numbers, and hence modify the properties of the artificial magnetic field. Furthermore, the diagonal hoppings ${^{\nearrow}D}$ and ${^{\nwarrow}D}$ are different, due to broken rotational symmetry.

\section{\label{app:corr_Heff_gen} Corrections to the Effective Hamiltonian $H_\text{eff}$.}
	
	\subsection{\label{app:DK_corr_coeffs_1D_Heff}First-order Coefficients for the 1D Driven Boson Model.}
	
	In this appendix, we list the expressions for the nnn hopping, and the staggered potential, found to first order in the High-Frequency expansion to the model discussed in Sec.~\ref{subsec:Floquet_measurement_bosons}. In order to distinguish them from those in the Magnus expansion, we use an extra tilde in the notation. Formally, the difference is the factor $\left(1 - \frac{\tau_1-\tau_2}{\pi} \right)$ in the integrands, which ensures that the expressions are Floquet-gauge independent. We recall that 
	\[
	g^{m,m+1}(\tau;\zeta) = \exp\left[-i\zeta\sin(\tau - \phi_{nm}) \right],
	\]
	where $\tau = \Omega t$. Then the coefficients to the Hamiltonian $H_\text{eff}^{(1)}$ given in Eq.~\eqref{eq:H_F^1_bosons_1D} are given by the following time-ordered integrals
	\begin{eqnarray}
	\mathcal{\tilde C}^{m,m+2}(\zeta) =&&\ \frac{1}{4\pi i}\int_0^{2\pi}\mathrm{d}\tau_1\int_0^{\tau_1}\mathrm{d}\tau_2 \left(1 - \frac{\tau_1-\tau_2}{\pi} \right) \left[g^{m,m+1}(\tau_1)g^{m+1,m+2}(\tau_2) - (1\leftrightarrow 2) \right],\nonumber\\
	\mathcal{\tilde G}^{m,m+1}(\zeta) =&&\ \frac{1}{2\pi}\int_0^{2\pi}\mathrm{d}\tau_1\int_0^{\tau_1}\mathrm{d}\tau_2 \left(1 - \frac{\tau_1-\tau_2}{\pi} \right) \text{Im}\left\{ \left(g^{m,m+1}(\tau_1)\right)^* g^{m,m+1}(\tau_2) \right\}.
	\end{eqnarray}
	We mention that these expressions are the same as the corresponding one for nnn hopping along the $y$-direction ${^{\uparrow}\tilde C}_0^{m,m+2}(\zeta)$, and a staggered potential along the $y$-direction and ${^{\uparrow}\tilde E}_{0}^{m,m+1}(\zeta)$ found in the 2D extension of the model from Sec.~\ref{subsec:harper} (see below).

	\subsection{\label{app:DK_corr_coeffs_2D_Heff}First-order Coefficients for the Harper-Hofstadter Model.}

	In this appendix we discuss the parameters of the leading correction, Eq.~\eqref{eq:HH_corr}. Let us define two auxiliary functions $f$ and $g$ by
	\begin{eqnarray}
	f_{m,m+1}^n(\tau;\zeta) &=& \exp\left[-i\zeta\sin(\tau - \phi_{nm}) +i\tau \right]\nonumber\\
	g_m^{n,n+1}(\tau;\zeta) &=& \exp\left[-i\zeta\sin(\tau - \phi_{nm}) \right]. 
	\end{eqnarray} 
	The coefficients $\tilde C$, $\tilde D$, and $\tilde E$ in Eq.~\eqref{eq:HH_corr} are given by the following time-ordered integrals:

	\begin{eqnarray}
	{^{\rightarrow}\tilde C}^n_{m,m+2}(\zeta) =&&\ \frac{1}{4\pi i}\int_0^{2\pi}\mathrm{d}\tau_1\int_0^{\tau_1}\mathrm{d}\tau_2 \left(1 - \frac{\tau_1-\tau_2}{\pi} \right) \left[f^n_{m,m+1}(\tau_1)f^n_{m+1,m+2}(\tau_2) - (1\leftrightarrow 2) \right],\nonumber\\
	{^{\uparrow}\tilde C}_m^{n,n+2}(\zeta) =&&\ \frac{1}{4\pi i}\int_0^{2\pi}\mathrm{d}\tau_1\int_0^{\tau_1}\mathrm{d}\tau_2 \left(1 - \frac{\tau_1-\tau_2}{\pi} \right) \left[g_m^{n,n+1}(\tau_1)g_m^{n+1,n+2}(\tau_2) - (1\leftrightarrow 2) \right],\nonumber
	\end{eqnarray} 
	\begin{eqnarray}
	{^{\nearrow}\tilde D}_{m,m+1}^{n,n+1}(\zeta) =&&\ \frac{1}{4\pi i}\int_0^{2\pi}\mathrm{d}\tau_1\int_0^{\tau_1}\mathrm{d}\tau_2 \left(1 - \frac{\tau_1-\tau_2}{\pi} \right) \bigg[f^n_{m,m+1}(\tau_1)g_{m+1}^{n,n+1}(\tau_2) \nonumber\\
	&& + f^{n+1}_{m,m+1}(\tau_2)g_{m}^{n,n+1}(\tau_1) - (1\leftrightarrow 2) \bigg],\nonumber\\
	{^{\nwarrow}\tilde D}_{m,m-1}^{n,n+1}(\zeta) =&&\ \frac{1}{4\pi i}\int_0^{2\pi}\mathrm{d}\tau_1\int_0^{\tau_1}\mathrm{d}\tau_2 \left(1 - \frac{\tau_1-\tau_2}{\pi} \right) \bigg[\left(f^n_{m-1,m}(\tau_1)\right)^*g_{m-1}^{n,n+1}(\tau_2)\nonumber\\
	&& + \left(f^{n+1}_{m-1,m}(\tau_2)\right)^*g_{m}^{n,n+1}(\tau_1) - (1\leftrightarrow 2) \bigg],\nonumber
	\end{eqnarray} 
	\begin{eqnarray}
	{^{\rightarrow}\tilde E}^n_{m,m+1}(\zeta) =&&\  \frac{1}{2\pi}\int_0^{2\pi}\mathrm{d}\tau_1\int_0^{\tau_1}\mathrm{d}\tau_2 \left(1 - \frac{\tau_1-\tau_2}{\pi} \right) \text{Im}\left\{ \left(f^n_{m,m+1}(\tau_1)\right)^* f^n_{m,m+1}(\tau_2) \right\},\nonumber\\
	{^{\uparrow}\tilde E}_{m}^{n,n+1}(\zeta) =&&\  \frac{1}{2\pi}\int_0^{2\pi}\mathrm{d}\tau_1\int_0^{\tau_1}\mathrm{d}\tau_2 \left(1 - \frac{\tau_1-\tau_2}{\pi} \right) \text{Im}\left\{ \left(g_m^{n,n+1}(\tau_1)\right)^* g_m^{n,n+1}(\tau_2) \right\}.\nonumber
	\end{eqnarray} 
	
	All the coefficients are defined on the bonds between sites, labelled by $(m,n)$. Apart from $\tilde E$, the coefficients $\tilde C$, and $\tilde D$ are complex numbers, and hence modify the properties of the artificial magnetic field.

\end{document}